\documentclass[10pt,preprint]{emulateapj}

\newcommand \msun{\hbox{$\hbox{M}_{\odot}$}}

\begin{document}

\title{\textit{Spitzer} and HHT observations of starless cores: masses
  and environments}
  
\author{Amelia M. Stutz\altaffilmark{1,2}, 
  George H. Rieke\altaffilmark{2},
  John H. Bieging\altaffilmark{2}, 
  Zoltan Balog\altaffilmark{1,2}, 
  Fabian Heitsch\altaffilmark{3},
  Miju Kang\altaffilmark{4,5,2},
  William L. Peters\altaffilmark{2},
  Yancy L. Shirley\altaffilmark{2},
  Michael W. Werner\altaffilmark{6}}

\altaffiltext{1}{Max Planck Institute for Astronomy, K\"onigstuhl 17,
  D-69117 Heidelberg, Germany; stutz@mpia.de}

\altaffiltext{2}{Department of Astronomy and Steward Observatory,
  University of Arizona, 933 North Cherry Avenue, Tucson, AZ 85721,
  USA}

\altaffiltext{3}{Department of Physics and Astronomy, University of
  North Carolina Chapel Hill, CB 3255, Phillips Hall, Chapel Hill, NC
  27599, USA}

\altaffiltext{4}{Korea Astronomy and Space Science Institute, Hwaam
  61--1, Yuseong, Daejeon 305--348, South Korea}

\altaffiltext{5}{Department of Astronomy and Space Science, Chungnam
  National University, Daejeon 305--348, South Korea}

\altaffiltext{6}{Jet Propulsion Lab, California Institute of
  Technology, 4800 Oak Grove Drive, Pasadena, CA 91109, USA}

\begin{abstract}

We present {\it Spitzer} observations of a sample of 12 starless cores
selected to have prominent 24~\micron\ shadows.  The {\it Spitzer}
images show 8 and 24~\micron\ shadows and in some cases
70~\micron\ shadows; these spatially resolved absorption features
trace the densest regions of the cores.  We have carried out a
$^{12}$CO (2-1) and $^{13}$CO (2-1) mapping survey of these cores with
the Heinrich Hertz Telescope (HHT).  We use the shadow features to
derive optical depth maps.  We derive molecular masses for the cores
and the surrounding environment; we find that the 24~\micron\ shadow
masses are always greater than or equal to the molecular masses
derived in the same region, a discrepancy likely caused by CO
freeze--out onto dust grains.  We combine this sample with two
additional cores that we studied previously to bring the total sample
to 14 cores.  Using a simple Jeans mass criterion we find that
$\sim\!2/3$ of the cores selected to have prominent
24~\micron\ shadows are collapsing or near collapse, a result that is
supported by millimeter line observations.  Of this subset at least
half have indications of 70~\micron\ shadows.  All cores observed to
produce absorption features at 70~\micron\ are close to collapse.  We
conclude that 24~\micron\ shadows, and even more so the
70~\micron\ ones, are useful markers of cloud cores that are
approaching collapse.

\end{abstract}


\section{Introduction}

Stars are born in cold cloud cores \citep[e.g.,][]{bergin07}, where
gas and dust are compressed to high enough densities to cause the core
to start collapsing. One of the most critical questions in astronomy
is what the physical conditions for core formation and collapse are.
Therefore, observations of cores in the rare stage during, or just
prior to, collapse are critical to constrain theories of the very
earliest stages of star-formation
\citep[e.g.,][]{shu87,ballesteros03,myers05}.

The initial process of collapse, or the transition from a stable core
to a core with an embedded protostar, happens rapidly
\citep{hayashi66}; additionally, the cores are necessarily dense and
cold. These conditions pose significant observational challenges to
identify cores close to collapse. Because of the temperature ranges
involved, $\sim\!10$~K \citep[e.g.,][]{lemme96,caselli99,hotzel02},
and high densities, $\sim$10$^5$ cm$^{-3}$ \citep[e.g.,][]{bacmann00},
cold cloud cores can best be observed at far-infrared, submillimeter,
and millimeter wavelengths \citep[e.g.,][]{stutz09}.  Such
observations of dense cores show that most of them are close to
equilibrium and not collapsing \citep[e.g.,][]{lada08}. Furthermore,
if the cores are not supported by thermal and turbulent pressure
alone, then a modest magnetic field can halt collapse
\citep[e.g.,][]{kandori05,stutz07,alves08}.  

Millimeter line observations are the traditional way to search for
collapsing cores \citep[e.g.,][]{walker86}; however, there are
ambiguities in the interpretation of such measurements
\citep[e.g.,][]{walker88,menten87,mundy90,narayanan98}. Here we use
mid-infrared shadows as an alternative observational approach to the
study of starless cores, one that does not depend on the
interpretation of millimeter line profiles and how these profiles
relate to the underlying velocity field of the core
material. Additionally, our method is sensitive to higher column
densities than traditional studies of starless cores by background
star extinction, usually limited to A$_V$ $\le$ 30 magnitudes
\citep[e.g.,][]{alves01,kandori05}.

Cores that are isolated from regions of massive star formation are
very useful to understand the initial stages of collapse into
stars. These cores are free from complicating factors, e.g., massive
core fragmentation, feed-back from high-mass stars, and protostellar
outflows \citep[e.g.,][]{lada03,dewit05,fallscheer09}, to name a few,
that can have large effects on a study aimed at understanding how
individual low mass stars form. We present a sample of 12 relatively
isolated cores that were selected from {\it Spitzer} MIPS observations
of Bok globules and other star-forming regions to have prominent 24
$\mu$m shadows, spatially resolved absorption features caused by the
dense core material viewed in absorption against the interstellar
radiation field. We also include 2 previously studied cores, CB190
\citep{stutz07} and L429 \citep{stutz09}, for a final sample of 14
cores.  Such objects with 24 $\mu$m shadows always show a counterpart
8 $\mu$m shadow, analogous to the IR absorption features produced by
more distant and massive structures termed infrared dark clouds
\citep[IRDCs; e.g.,][]{perault96,butler09,peretto09,lee09} and the ISO
7 $\mu$m absorption features studied by \citet{bacmann00}. We also
present Heinrich Hertz Telescope (HHT) $^{12}$CO (2-1) and $^{13}$CO
(2-1) on-the-fly (OTF) maps of regions $\sim$ 10$'$ on a side
surrounding each core.

Our search method is efficient at identifying cores that are near
collapse.  In \S~2 we describe the observations and data processing;
in \S~3 we derive optical depth maps from the 8 and 24 $\mu$m images;
in \S~4 we present our mass measurements and we describe the stability
analysis that we apply to the sample of cores; and finally, in \S 5 we
summarize our main conclusions.

\section{Observations and Processing}

\subsection{{\it Spitzer} IRAC and MIPS data}

The 12 cores presented here were selected from two observing programs:
{\it Spitzer} GTO program ID 53 and GO program ID 30384.  Core names,
MIPS and IRAC AOR numbers, and corresponding program IDs are listed in
Table~1.  The IRAC data consist of observations in all four channels
from programs 139 (P.I.\ N.\ Evans), 94 (P.I.\ C.\ Lawrence), and
20386 (P.I.\ P.\ Myers).  The frames were processed using the {\it
  Spitzer} Science Center (SSC) IRAC Pipeline v14.0, and mosaics were
created from the basic calibrated data (BCD) frames using a custom IDL
program; see \citet{guter08} for details.  The MIPS observations at
24, 70, and 160~\micron\ are from programs 53 (P.I. G.\ Rieke) and
30384 (P.I. T.\ Bourke).  All observations were carried out in scan
map mode and were reduced using the Data Analysis Tool
\citep[DAT;][]{gordon05}.  These data were processed according to
steps outlined in \citet{stutz07}; please refer to that publication
for further details.  The data for each core are shown in
Figures~\ref{fig:img1} through~\ref{fig:img12}.

\subsection{$^{12}$CO (2-1) and $^{13}$CO (2-1) Data}

These cores were mapped in the J = 2--1 transitions of $^{12}$CO
(230.538 GHz) and $^{13}$CO (220.399 GHz) with the 10--m diameter
Heinrich Hertz Telescope (HHT) on Mt. Graham, Arizona.  Observations
taken in 2007 (CB27 and CB44) were obtained using a receiver that was
a dual polarization SIS mixer system operating in double-sideband mode
with a 4 -- 6 GHz IF band.  The observations taken after 2007 were
obtained using a prototype 1.3mm ALMA sideband separating receiver
with a 4 - 6 GHz IF band.  The $^{12}$CO (2--1) line was placed in the
upper sideband and the $^{13}$CO (2--1) one in the lower sideband.
The spectrometers were filter banks with 128 channels of 250 KHz width
and separation.  At the observed frequencies the spectral resolution
was $\sim\!0.3$~km~s$^{-1}$, the band-width was
$\sim\!40.0$~km~s$^{-1}$, and the angular resolution of the telescope
was 32$\arcsec$ (FWHM).

\begin{figure}
  \begin{center}
    \scalebox{0.52}{\includegraphics{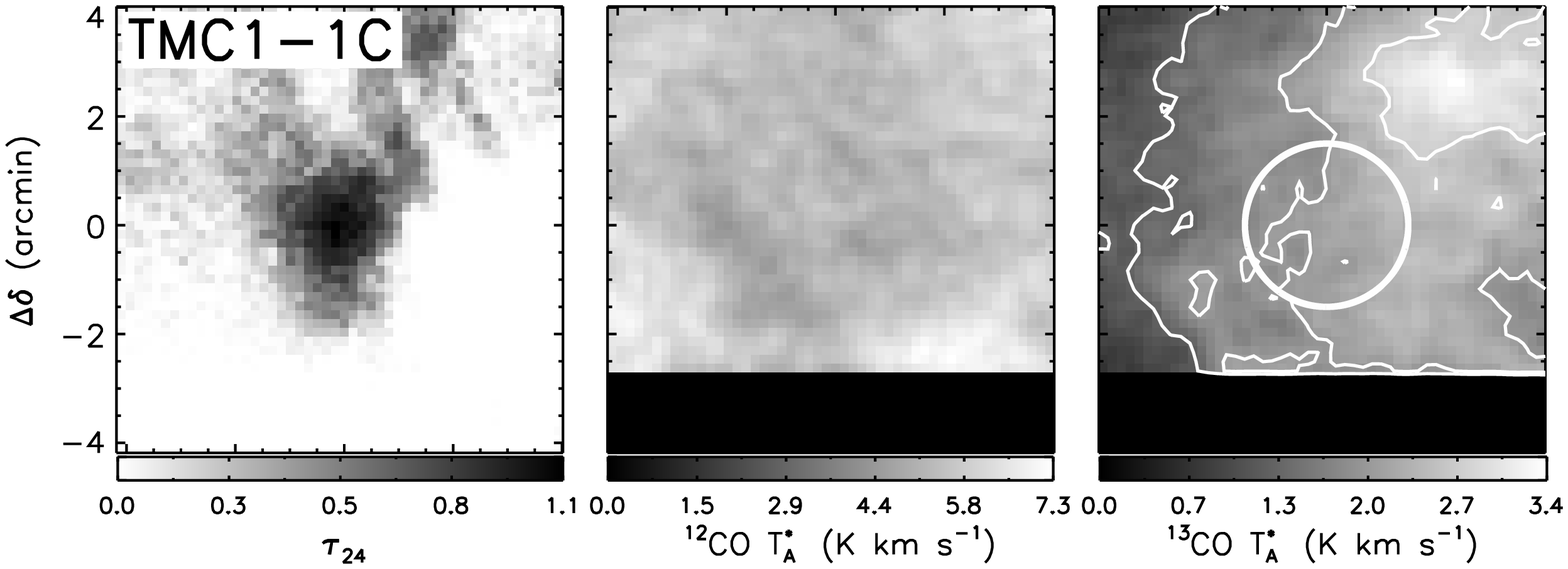}}
    \scalebox{0.52}{\includegraphics{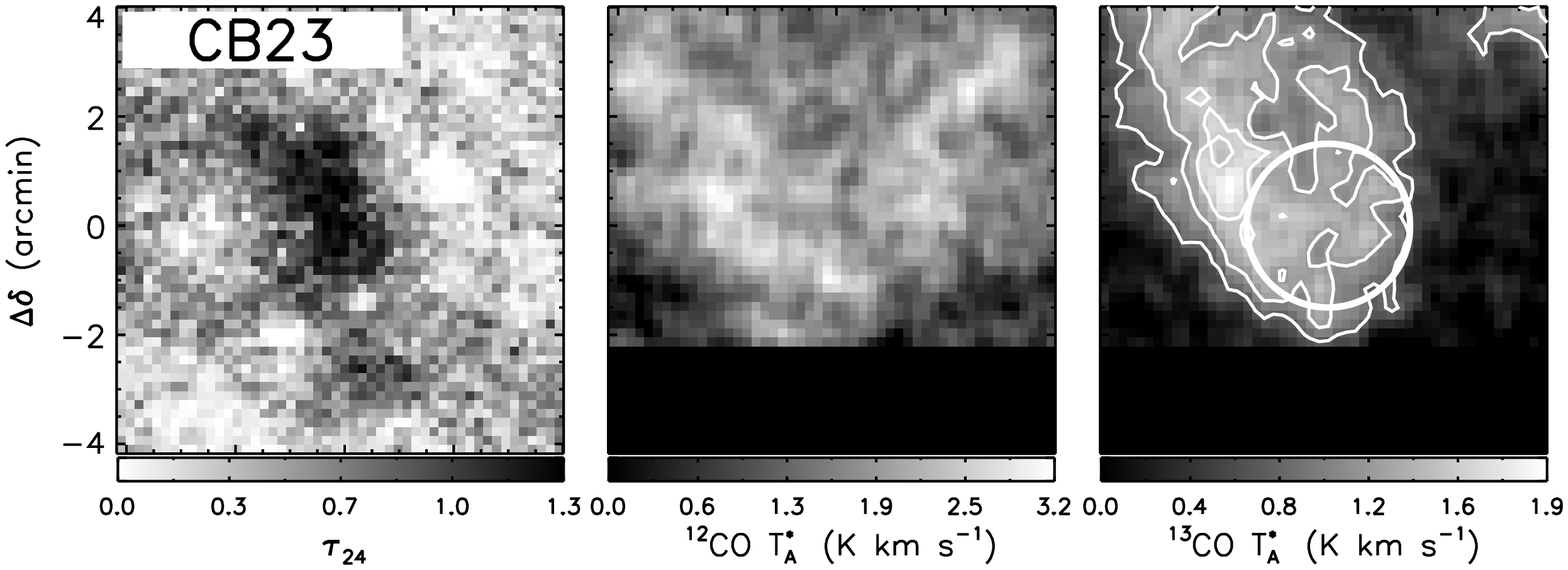}}
    \scalebox{0.52}{\includegraphics{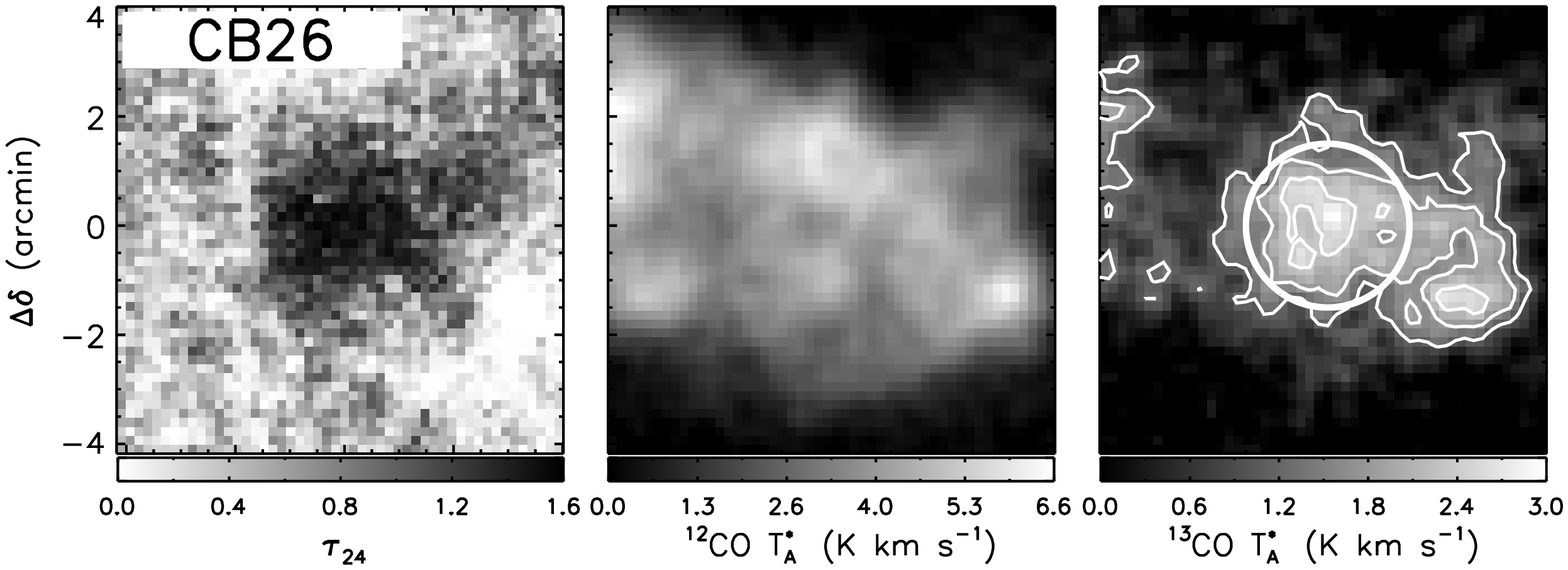}}
    \scalebox{0.52}{\includegraphics{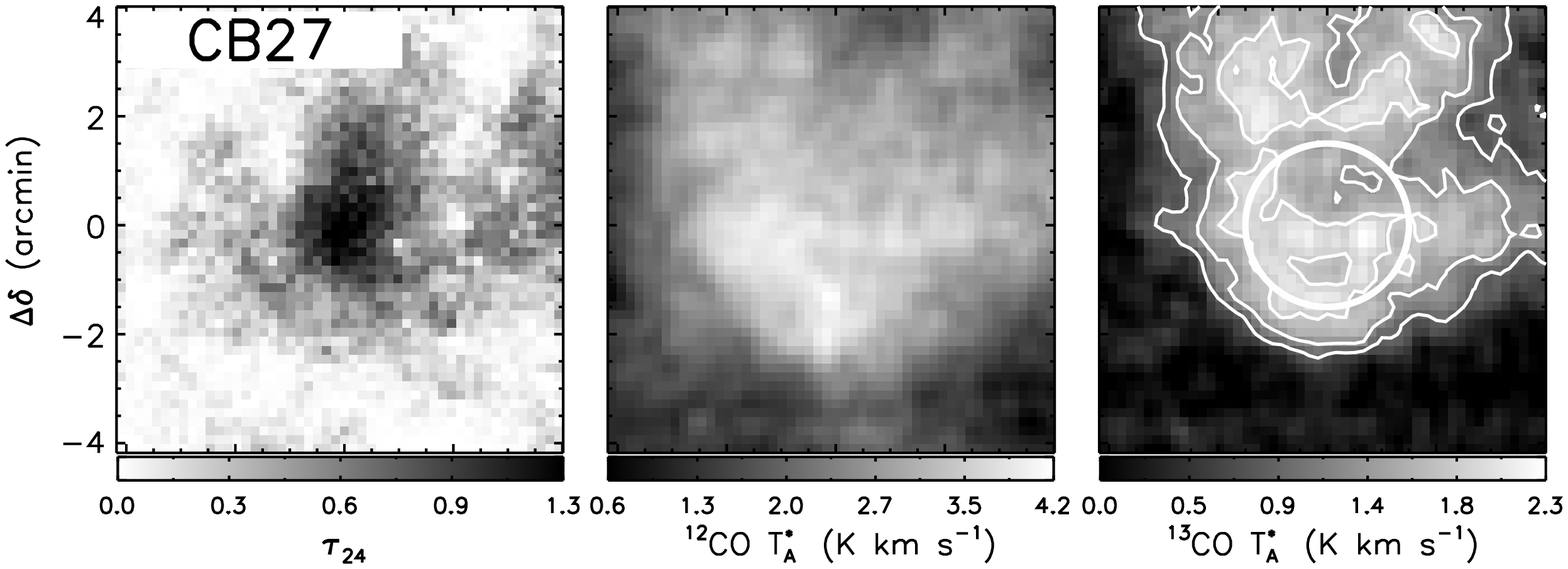}}
    \caption{$\tau_{24}$--maps and $^{12}$CO and $^{13}$CO integrated
      intensity maps, over the velocity ranges listed in Table~2.  The
      black circle ($1\farcm5$ radius) indicates the area used to
      calculate the 24~\micron\ shadow masses; the white circle
      indicates that used to calculate the molecular masses.  The
      $^{13}$CO contours levels are ${0.4, 0.6, 0.8}$ times the
      maximum value indicated int the $^{13}$CO color--bar.  {\it For
        full resolution figures contact stutz@mpia.de.}}
    \label{fig:tauco}
  \end{center}
\end{figure}

\begin{figure}
  \begin{center}
    \scalebox{0.52}{\includegraphics{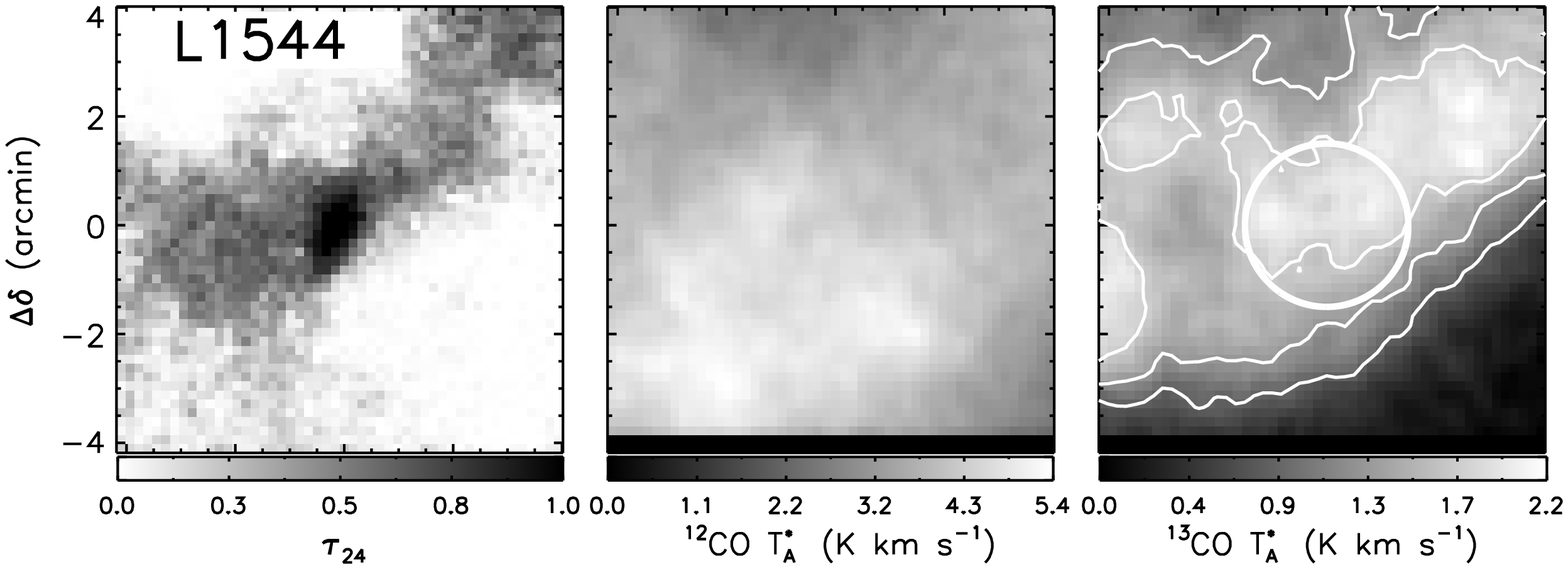}}
    \scalebox{0.52}{\includegraphics{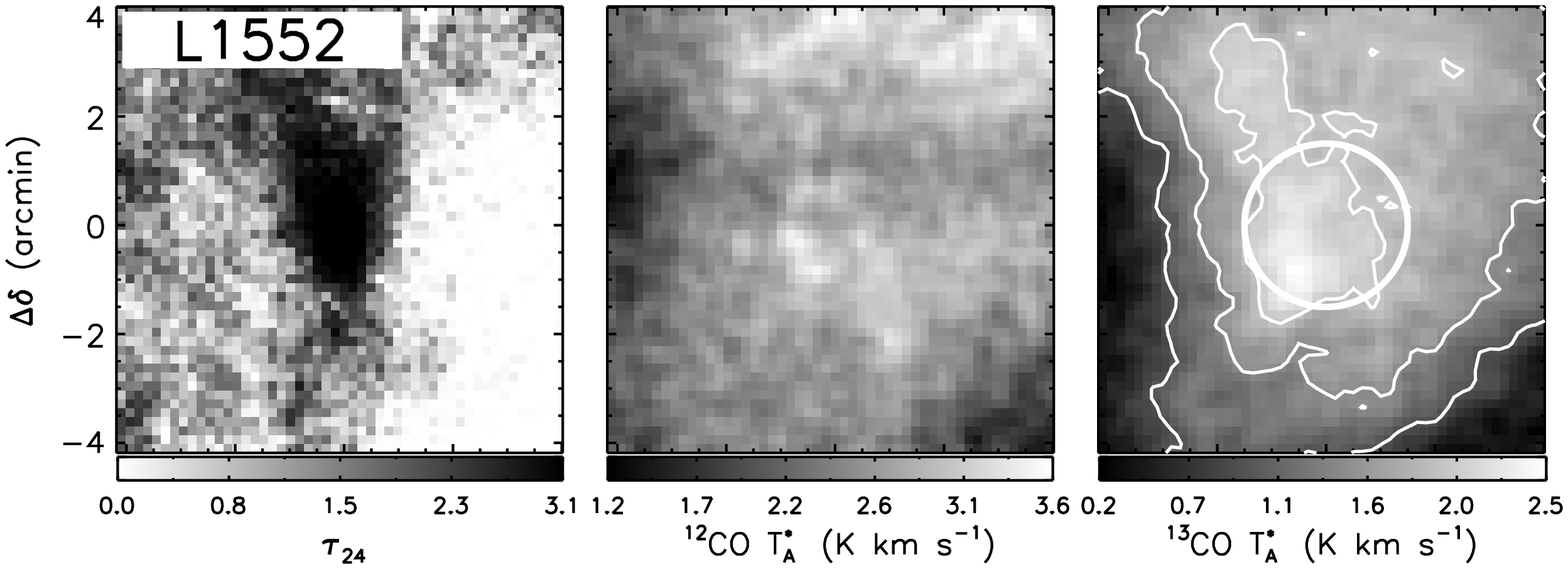}}
    \scalebox{0.52}{\includegraphics{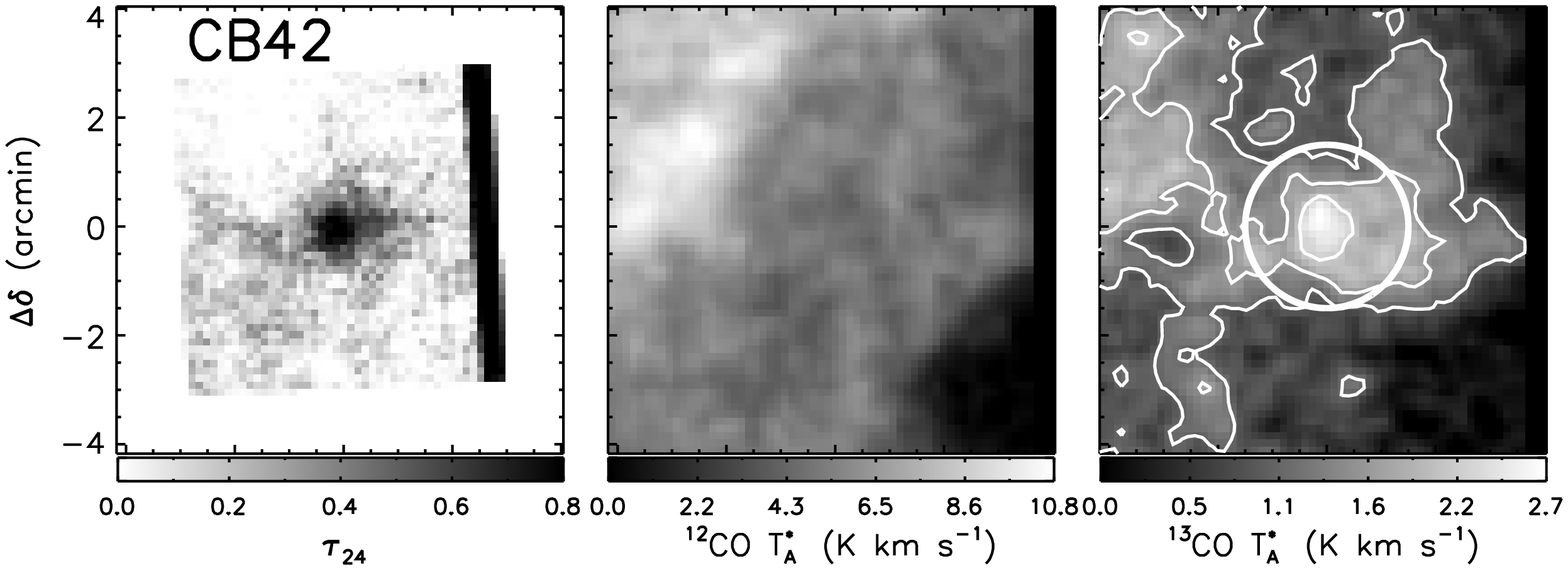}}
    \scalebox{0.52}{\includegraphics{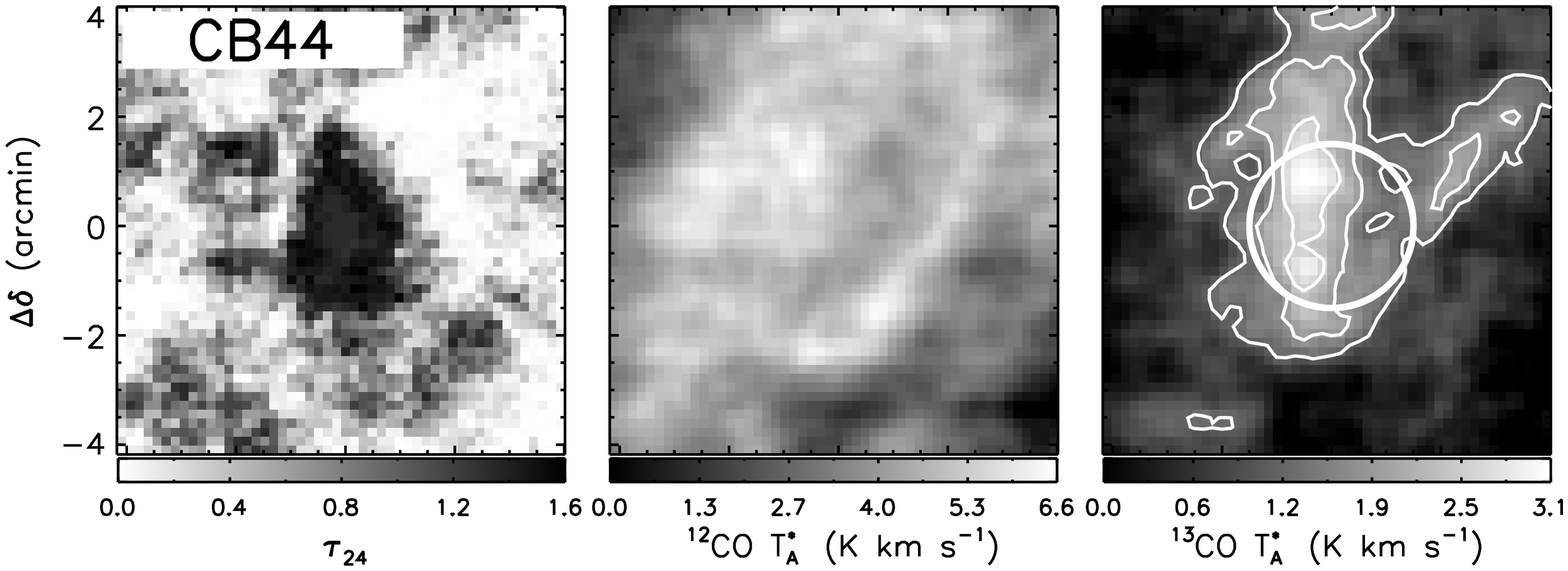}}
    \caption{Same as Figure~\ref{fig:tauco}}
    \label{fig:tauco2}
  \end{center}
\end{figure}

\begin{figure}
  \begin{center}
    \scalebox{0.52}{\includegraphics{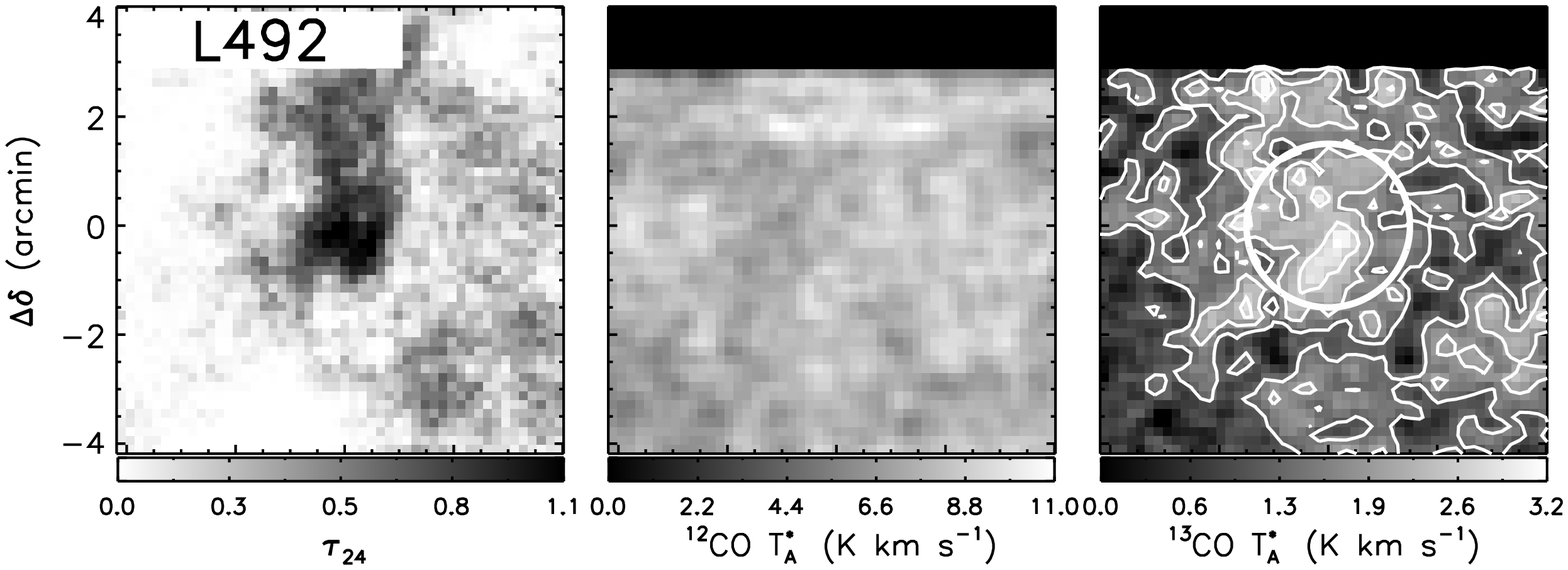}}
    \scalebox{0.52}{\includegraphics{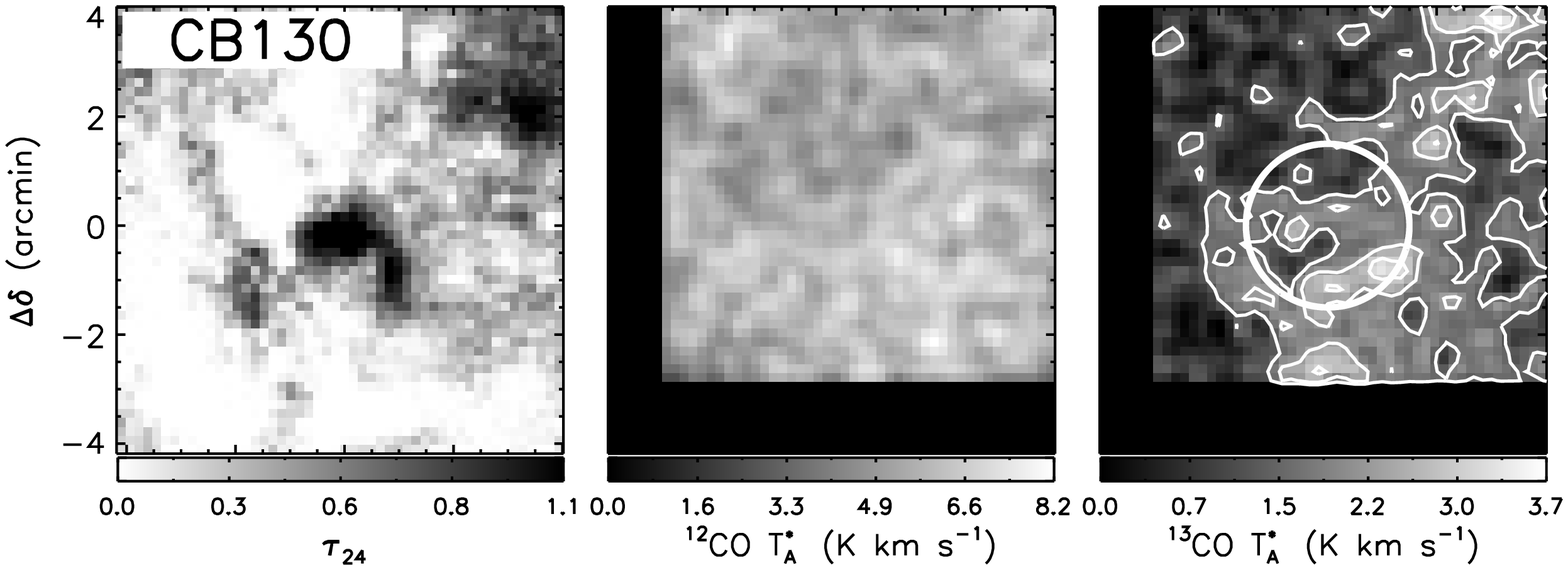}}
    \scalebox{0.52}{\includegraphics{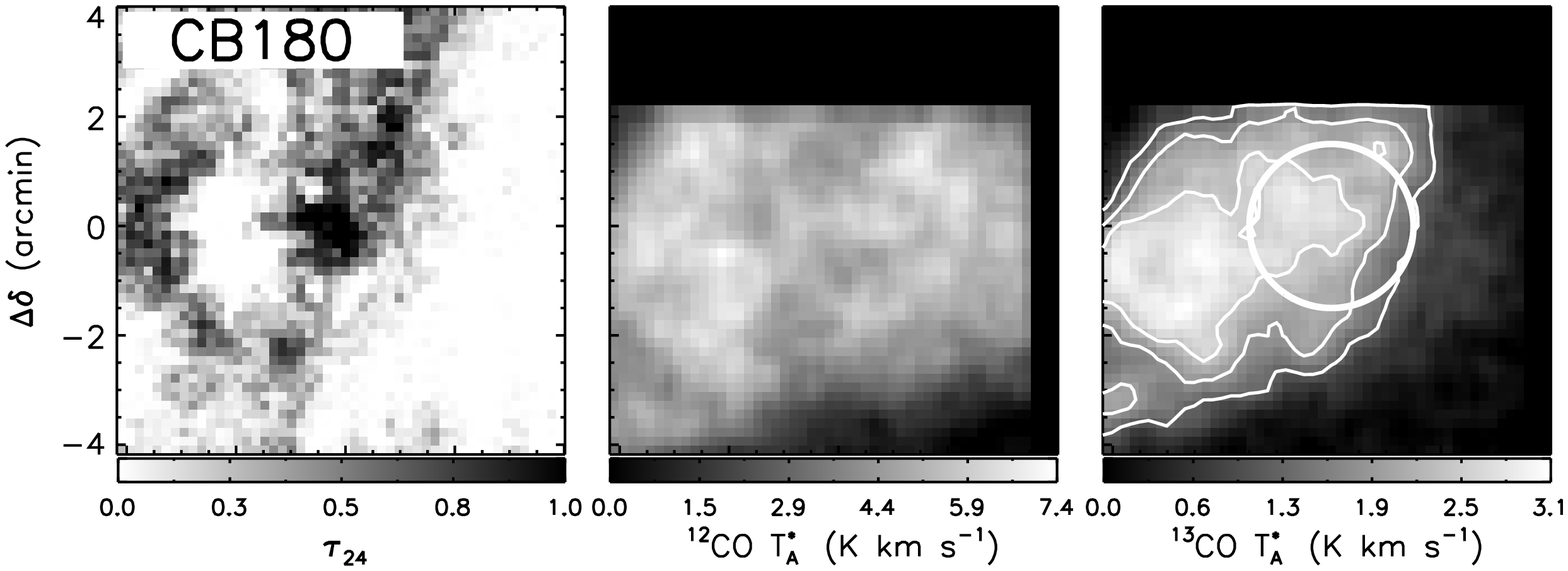}}
    \scalebox{0.52}{\includegraphics{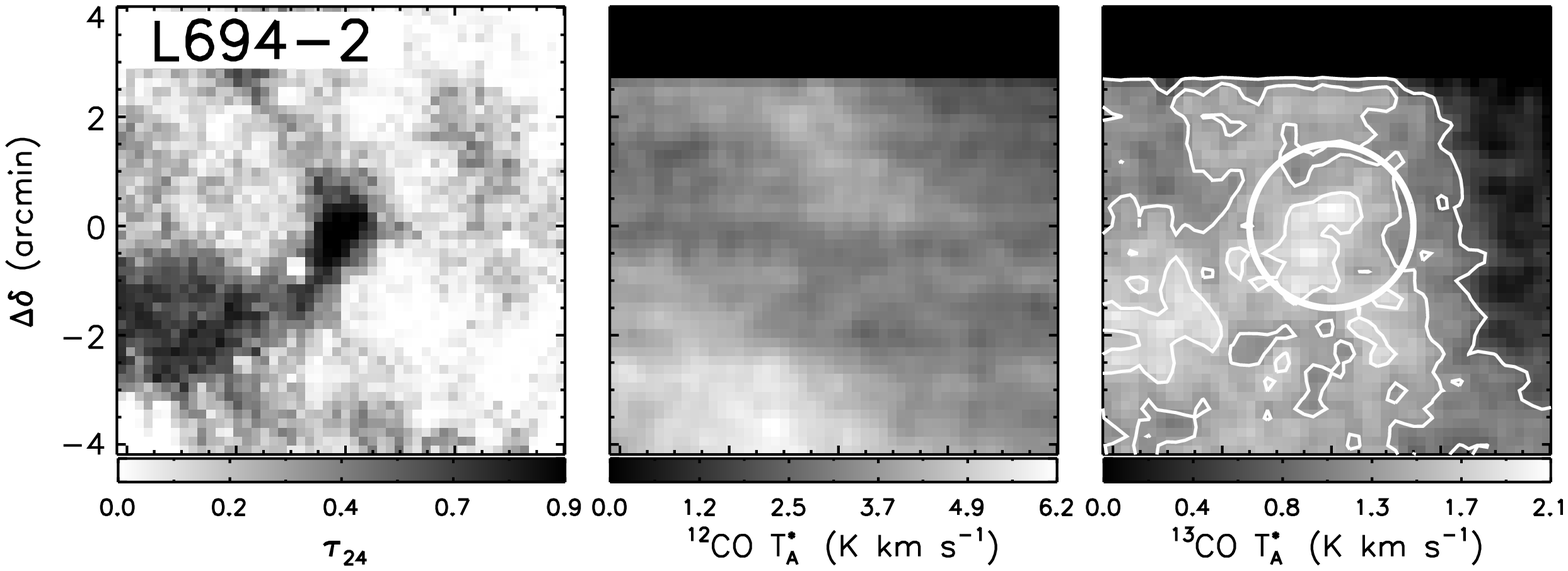}}
    \caption{Same as Figure~\ref{fig:tauco}}
    \label{fig:tauco3}
  \end{center}
\end{figure}

The cores were mapped with on-the-fly (OTF) scanning in RA at
$10\arcsec$~s$^{-1}$, with row spacing of $10\arcsec$ in declination.
System temperatures were calibrated by the standard ambient
temperature load method \citep{kutner81} after every other row of the
map grid.  Atmospheric conditions were generally clear and stable, and
the system temperatures were nearly constant throughout the maps.
Data for each CO isotopologues were processed with the {\it CLASS}
reduction package (from the Observatoire de Grenoble Astrophysics
Group), by removing a linear baseline and convolving the data to a
square grid with $10\arcsec$ grid spacing (equal to one-third the
telescope beamwidth).  The intensity scales for the two polarizations
were determined from observations of standard sources made just before
the OTF maps.  The gridded spectral data cubes were processed with the
{\it Miriad} software package \citep{sault95} for further analysis.
For each sideband the two polarizations were averaged together
yielding the map RMS noise levels listed in Table~2; also listed are
the map center coordinates, dates of observation, sizes, system
temperatures, and velocity ranges which include significant emission.  
The $^{12}$CO and $^{13}$CO integrated intensity images, integrated
over the velocity ranges listed in Table~2, are shown in
Figures~\ref{fig:tauco} through \ref{fig:tauco3}; these images are
centered on the 24~\micron\ shadow coordinates.

\section{Analysis}

\subsection{Optical Depth Maps}

In this section we calculate the 24~\micron\ and 8~\micron\ optical
depths for the shadow features seen in
Figures~\ref{fig:img1}--\ref{fig:img12}.  We assume that the cores are
optically thin and that there is no foreground emission originating
from the surface layers.  We apply an analysis analogous to that
presented in \citet{stutz07} at 24~\micron\ and \citet{stutz08b} at
8~\micron; however, we update the analysis presented in that work with
a more robust calculation of the unobscured background flux level
($f_0$ in what follows).  

\subsubsection{$24$~\micron\ Optical Depth Calculation}

Under the above assumptions, the optical depth at 24~\micron\ is
\begin{equation}
\tau_{24} = -ln \left( \frac{f' - f_{\rm DC}}{f_0' - f_{\rm DC}}
\right). 
\end{equation}
Here $f_{\rm DC}$ is the contribution to the image flux level of all
foreground sources, such as zodiacal emission and instrumental
background; $f'$ and $f_0'$ are the shadow flux and the intrinsic,
unobscured flux, respectively.  In what follows $f$ and $f_0$ indicate
the DC--subtracted values of the shadow fluxes: $f = f' - f_{\rm DC}$
and $f_0 = f_0' - f_{\rm DC}$.  The contribution of $f_{\rm DC}$ to
the overall image flux level is assumed to be constant over image
areas large compared to the size of the shadow while $f_0$ is
determined in the region immediately adjacent to the shadow.  Further
details on the determination of these two quantities is provided in
the following discussion.  However, before proceeding, we note that
the shadow depth, or contrast, in the absence of foreground emission
from the surface of the shadow itself, is the result of the depth of
the absorption feature relative to the quantity $(f_0' - f_{\rm DC})$;
a stronger background allows for robust sensitivity to larger $\tau$.
Furthermore, an over-estimate of $f_{\rm DC}$ will artificially boost
the shadow contrast and cause an over--estimate of $\tau$ and all
dependent quantities.  Rather than following the procedure applied in
\citet{stutz09}, where the darkest region of the L429 shadow was used
to measure f$_{\rm DC}$, the approach implemented here takes advantage
of dark regions in the {\it Spitzer} images which are not associated
with any absorption features.  These regions provide an independent
estimate of the foreground emission; furthermore, when shadow pixels
are darker than, or have a lower value than, the f$_{\rm DC}$ level,
we exclude them from our analysis because they are most likely not
sensitive to the total column of material in that pixel, i.e., they
have become saturated.

{\it Image DC level ($f_{\rm DC}$) determination}.  To measure $f_{\rm
DC}$, we select regions in the 24~\micron\ image with no diffuse
emission.  These regions also have little to no detected emission at
70 and 160~\micron.  As a further check on the cleanliness of the
background estimate, when possible we select the background region to
overlap partially with, or be adjacent to, $^{12}$CO and $^{13}$CO map
areas that are free from molecular line emission.  Additionally, we
avoid the areas near the edges of the images due to poor
signal--to--noise and sparse coverage.  The background region box size
is $5\arcmin$ on a side, corresponding to $\sim 59,000$ mosaic pixels,
which are subsampled by $2\times2$ relative to the physical pixels.
We set $f_{\rm DC}$ equal to the 0.5 percentile pixel value; as
discussed above, this low value is conservative because a higher value
may over--estimate the shadow contrast relative to $f_0'$.  The error
in $f_{\rm DC}$, $\sigma_{\rm DC}$, is determined by simulating $\sim
59,000$ pixels drawn from a normal distribution with the same mean and
standard deviation as the background region; the 0.5 percentile pixel
value is stored and the process is repeated $10^4$ times.  The
standard deviation of the resulting distribution values is our
estimate of $\sigma_{\rm DC}$ \citep{stutz07}.  The pixel error map
for the background subtracted images is then calculated using the DAT
\citep{gordon05} pixel error map, multiplied by a factor of 2 to take
into account pixel--to--pixel correlations due to the sub--sampling,
with $\sigma_{\rm DC}$ added in quadrature.

\begin{figure}
  \begin{center}
    \scalebox{0.5}{\includegraphics{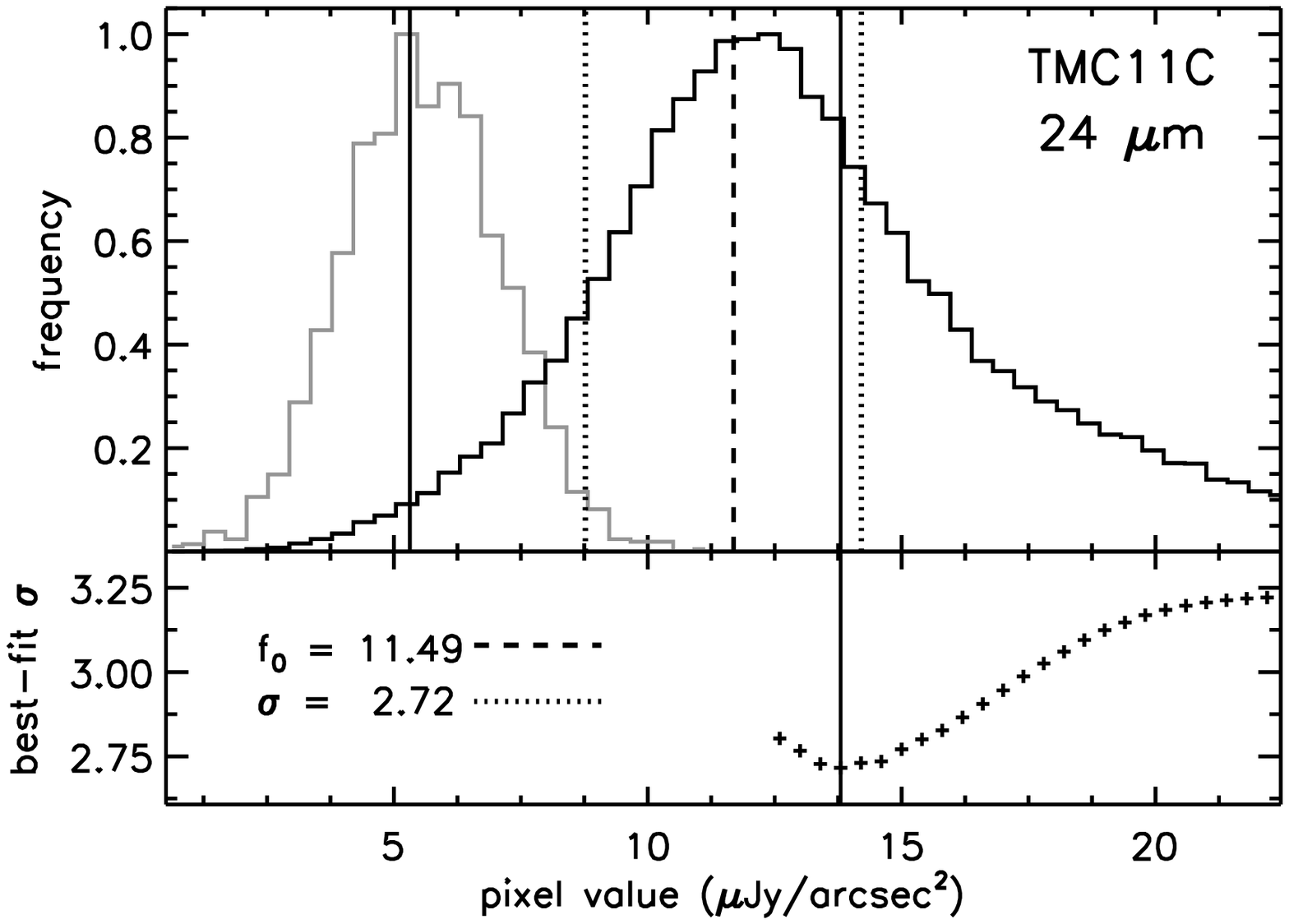}\includegraphics{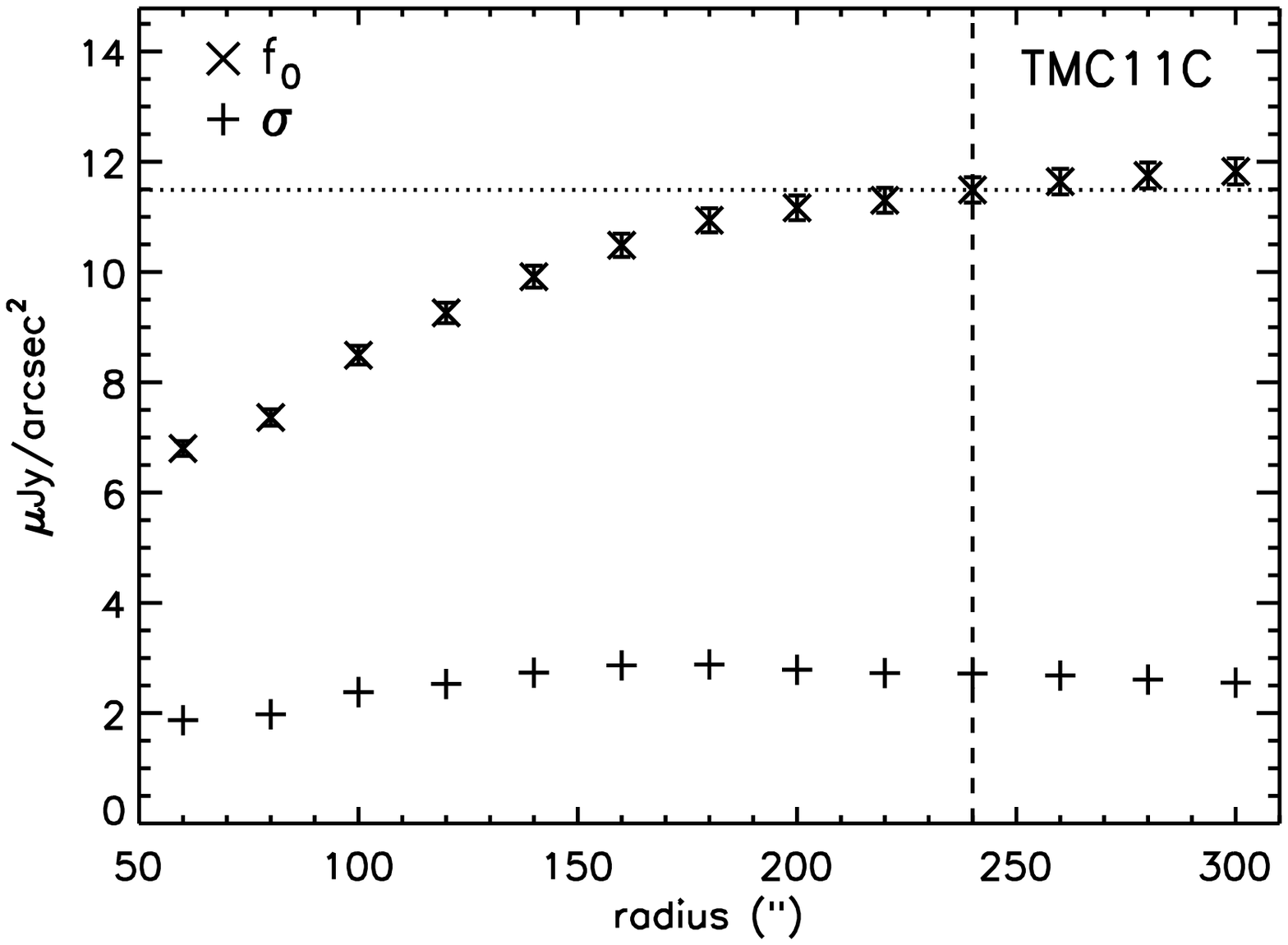}}
    \scalebox{0.5}{\includegraphics{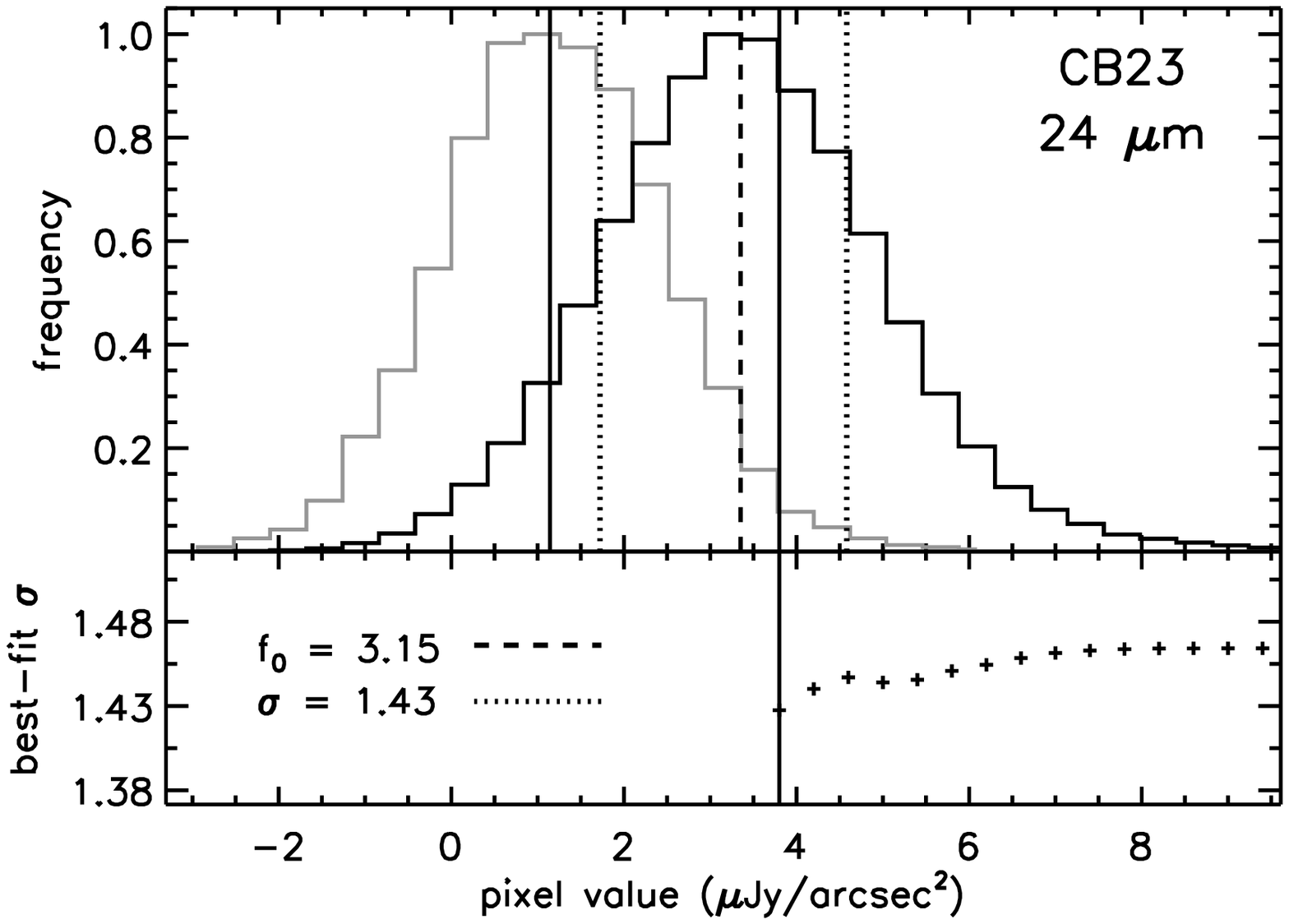}\includegraphics{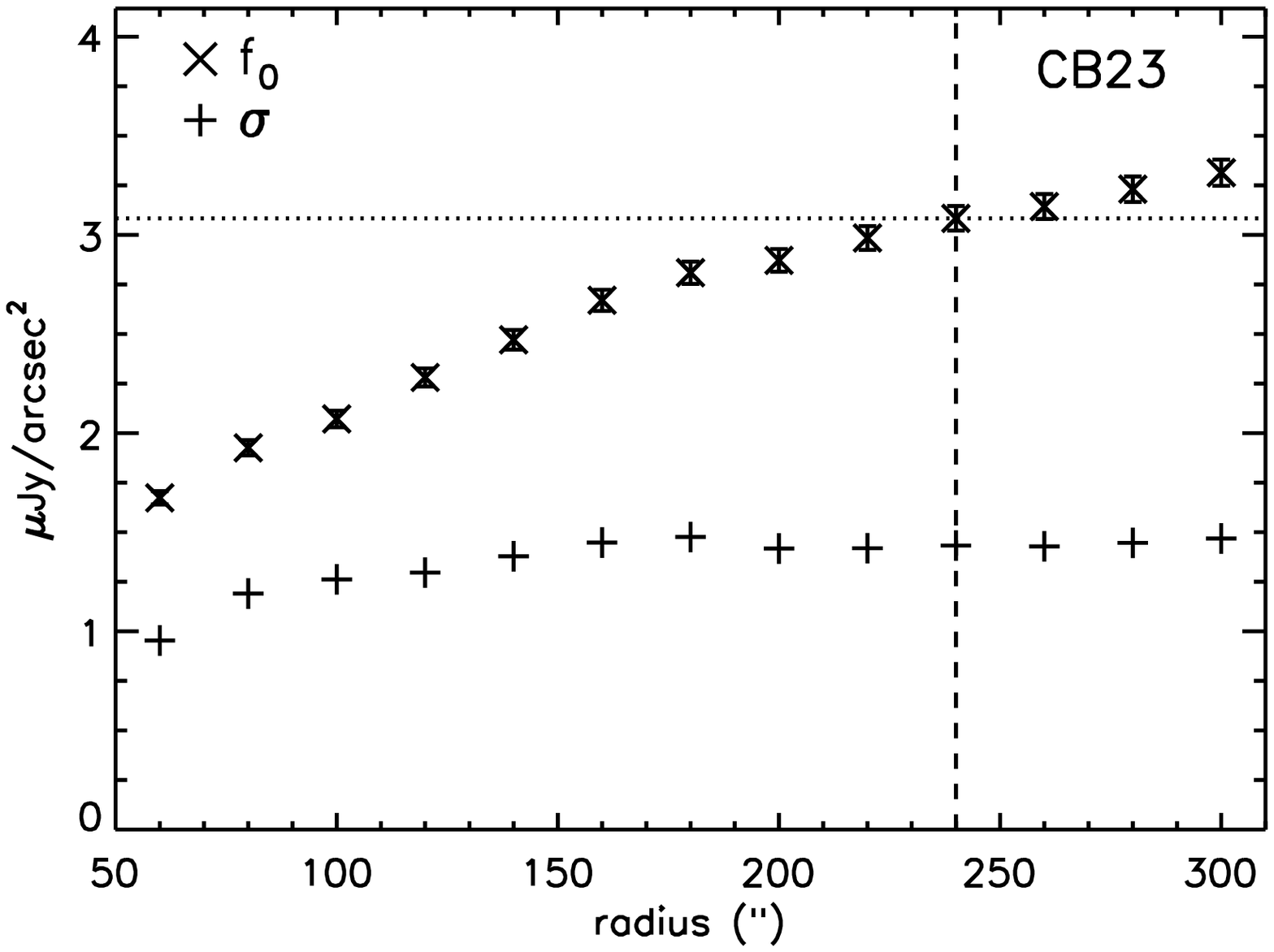}}
    \scalebox{0.5}{\includegraphics{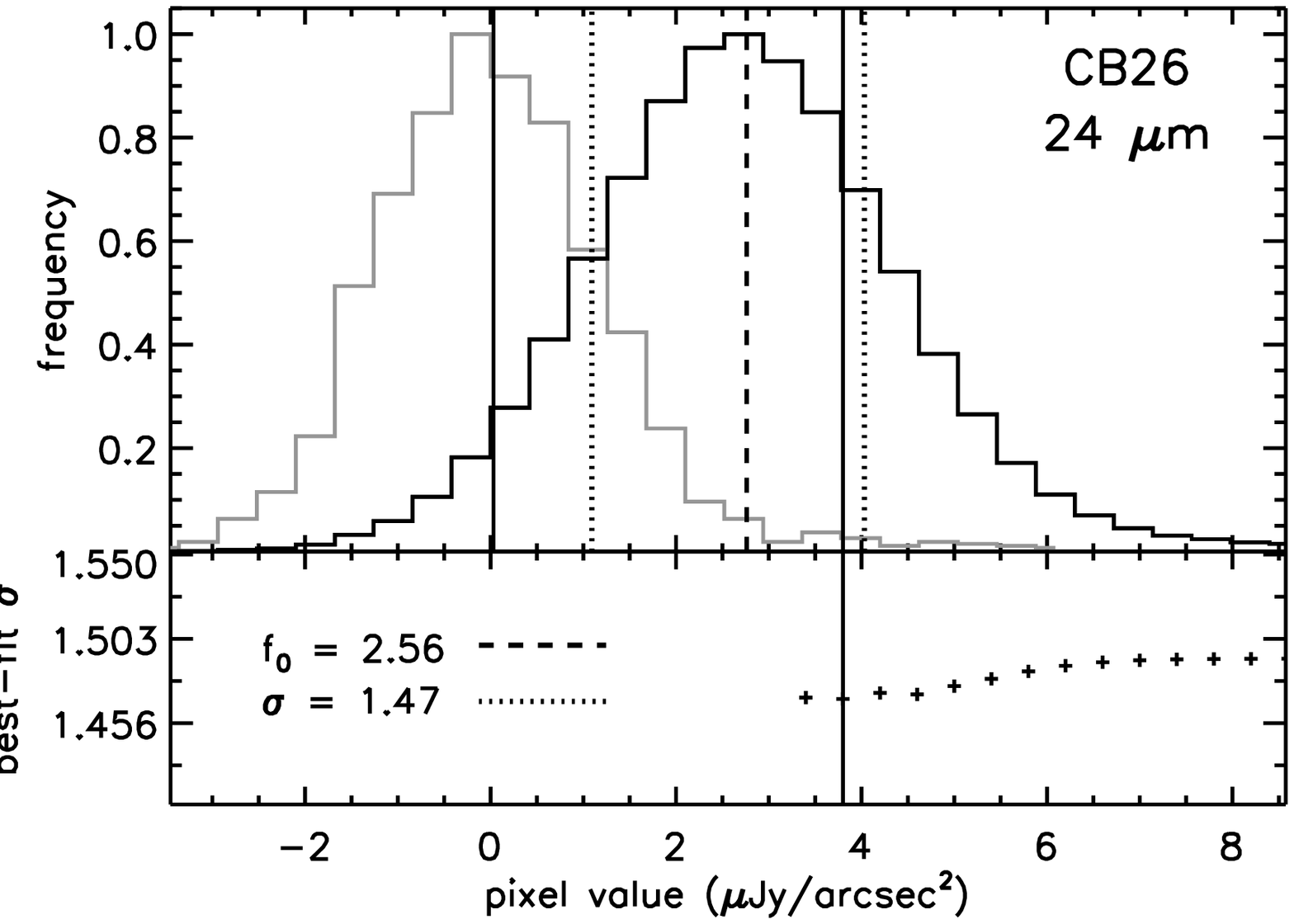}\includegraphics{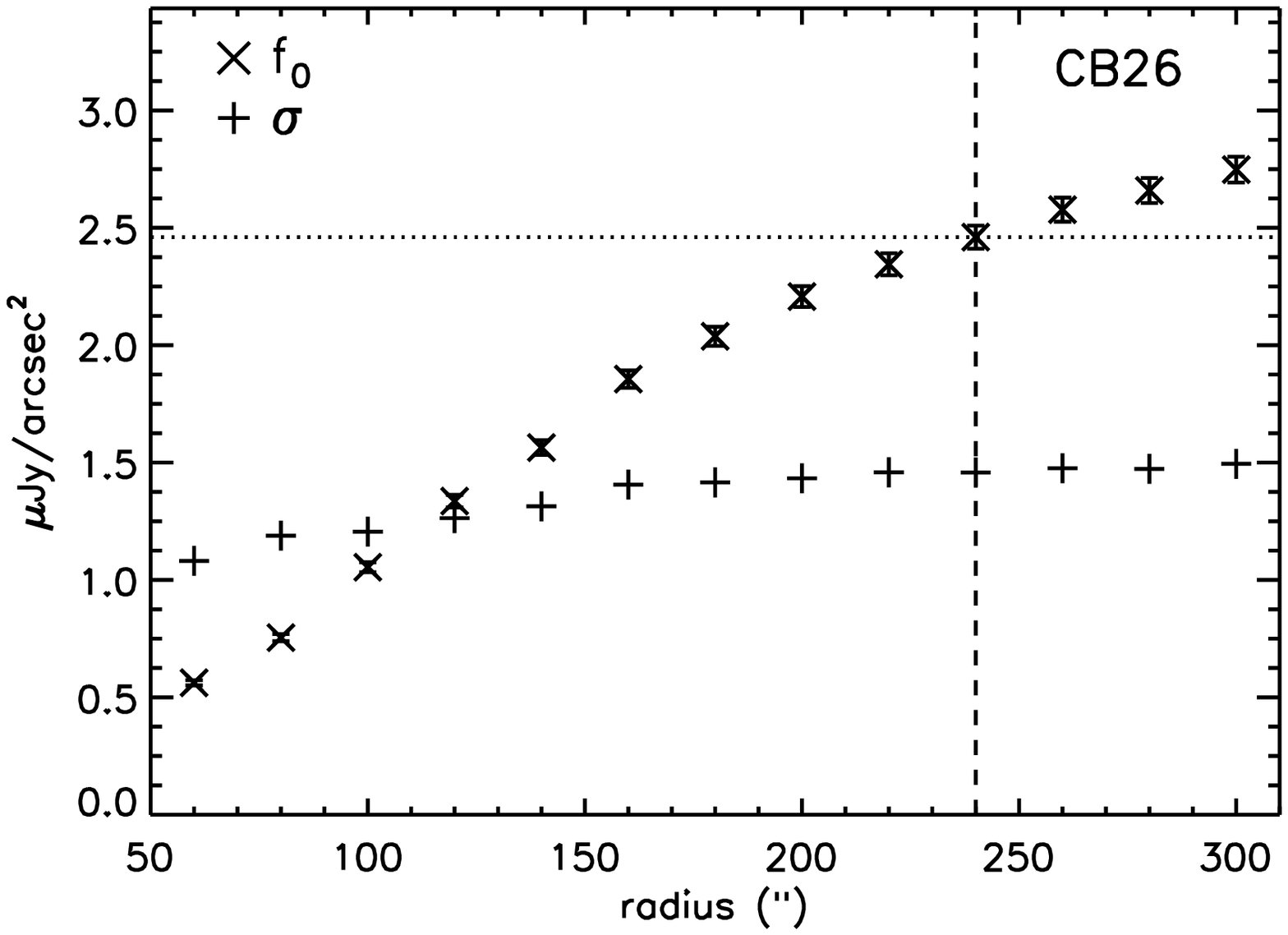}}
    \caption{{\it Left Column, top panel --- } Black histogram:
      24~\micron\ pixel value distribution for background subtracted
      image region of $r_0 = 240\arcsec$ radius (except for CB42),
      $r_0$ is indicated in the right column as the dashed line, and
      is centered on the shadow.  Grey histogram: 24~\micron\ pixel
      value distribution for $30\arcsec$ region centered on the
      shadow.  Solid lines: minimum and maximum black histogram bins
      used to fit for the minimum $\sigma$.  The dashed line indicates
      the best--fit mean, or $f_0$, value at the minimum best--fit
      value for $\sigma$, indicated by the dotted line.  {\it Left
        Column, bottom panel --- } Best--fit values of $\sigma$
      plotted against the maximum bin value included in the fit.  The
      solid line indicates the minimum value of $\sigma$.  Best--fit
      Gaussian parameters $\sigma$ and the mean ($f_0$) are
      indicated. Right Column: Best--fit $f_0$ and minimum $\sigma$ as
      a function of aperture size ($r_0$).}
    \label{fig:f01}
  \end{center}
\end{figure}

\begin{figure}
  \begin{center}
    \scalebox{0.5}{\includegraphics{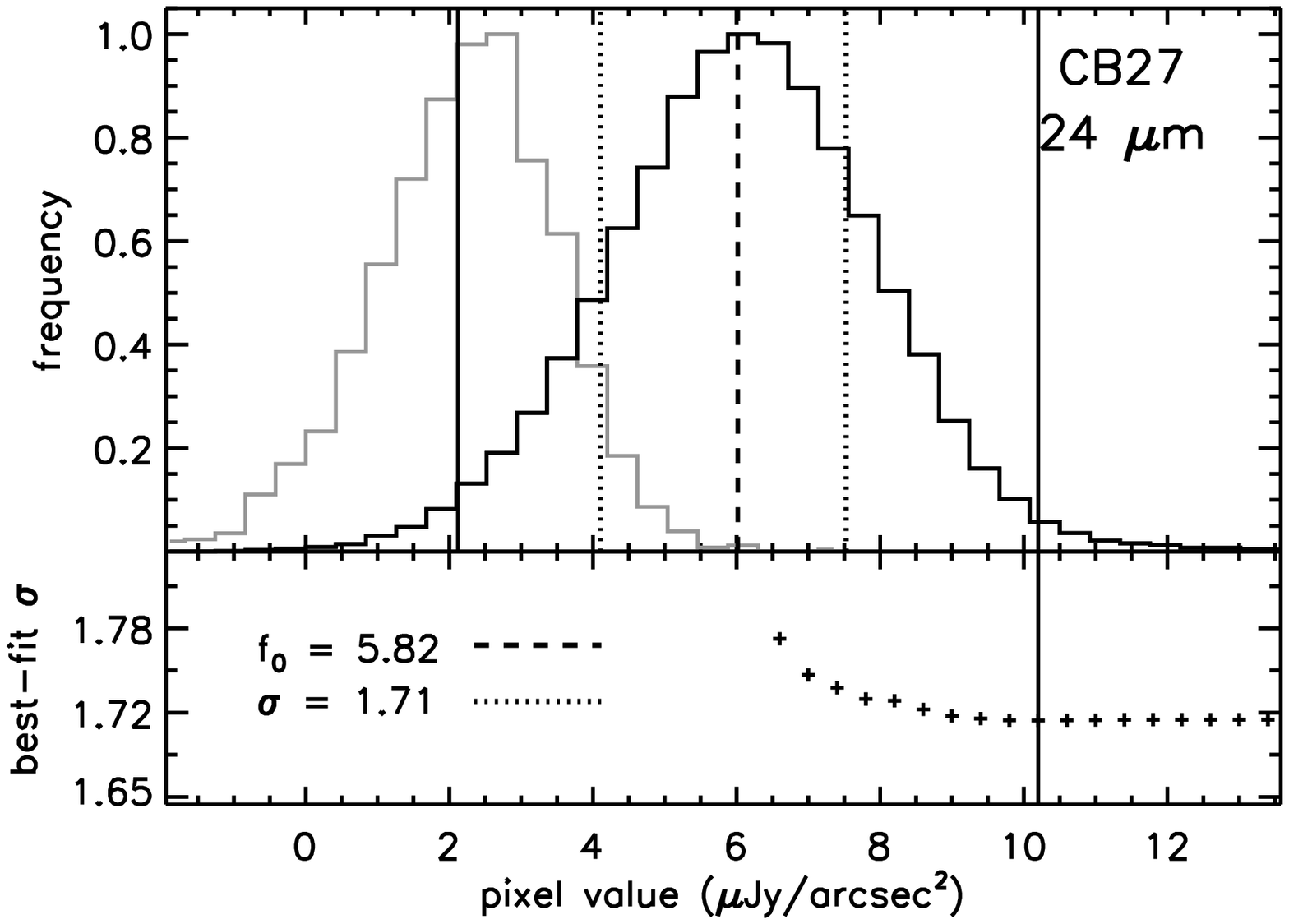}\includegraphics{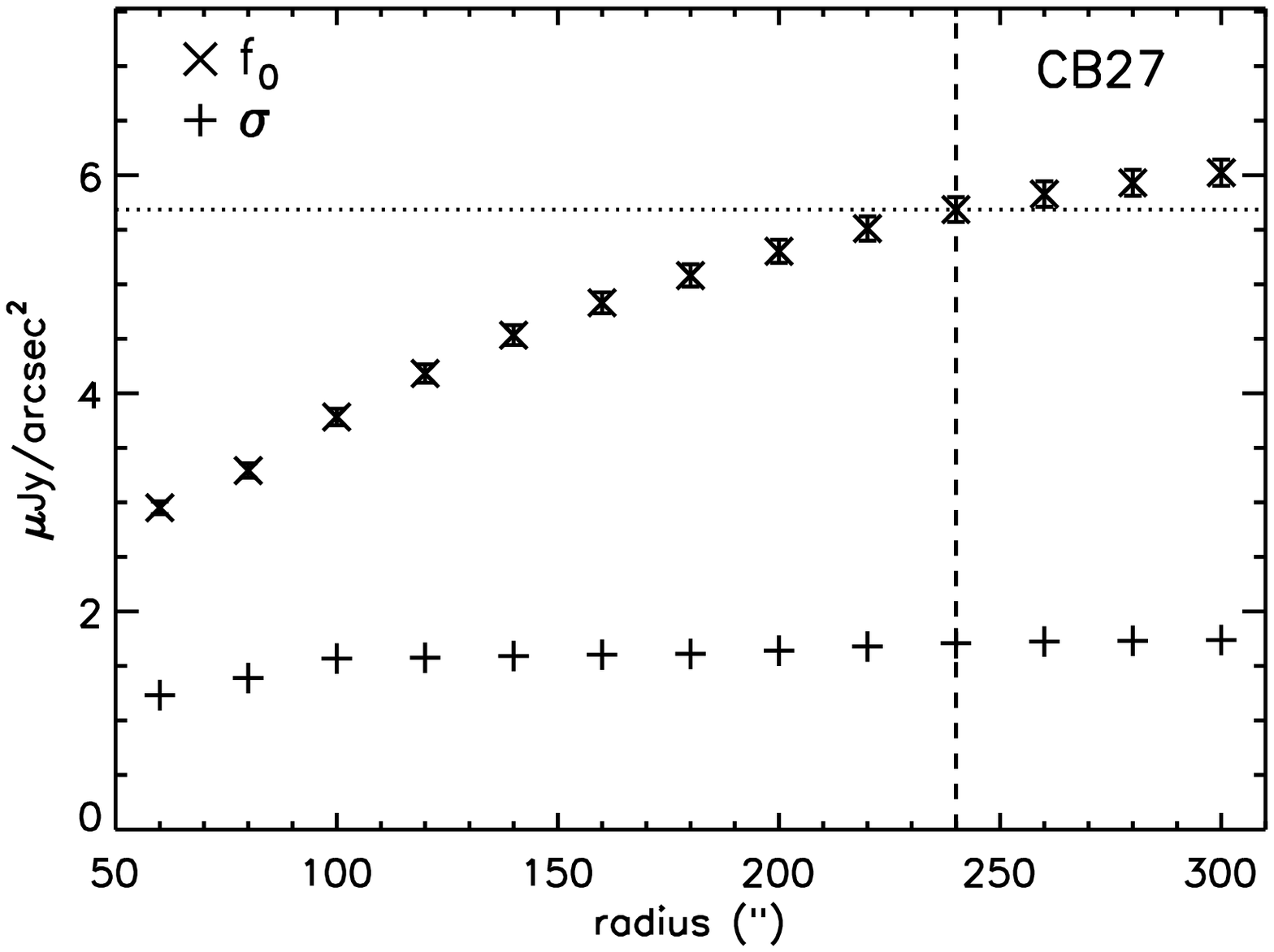}}
    \scalebox{0.5}{\includegraphics{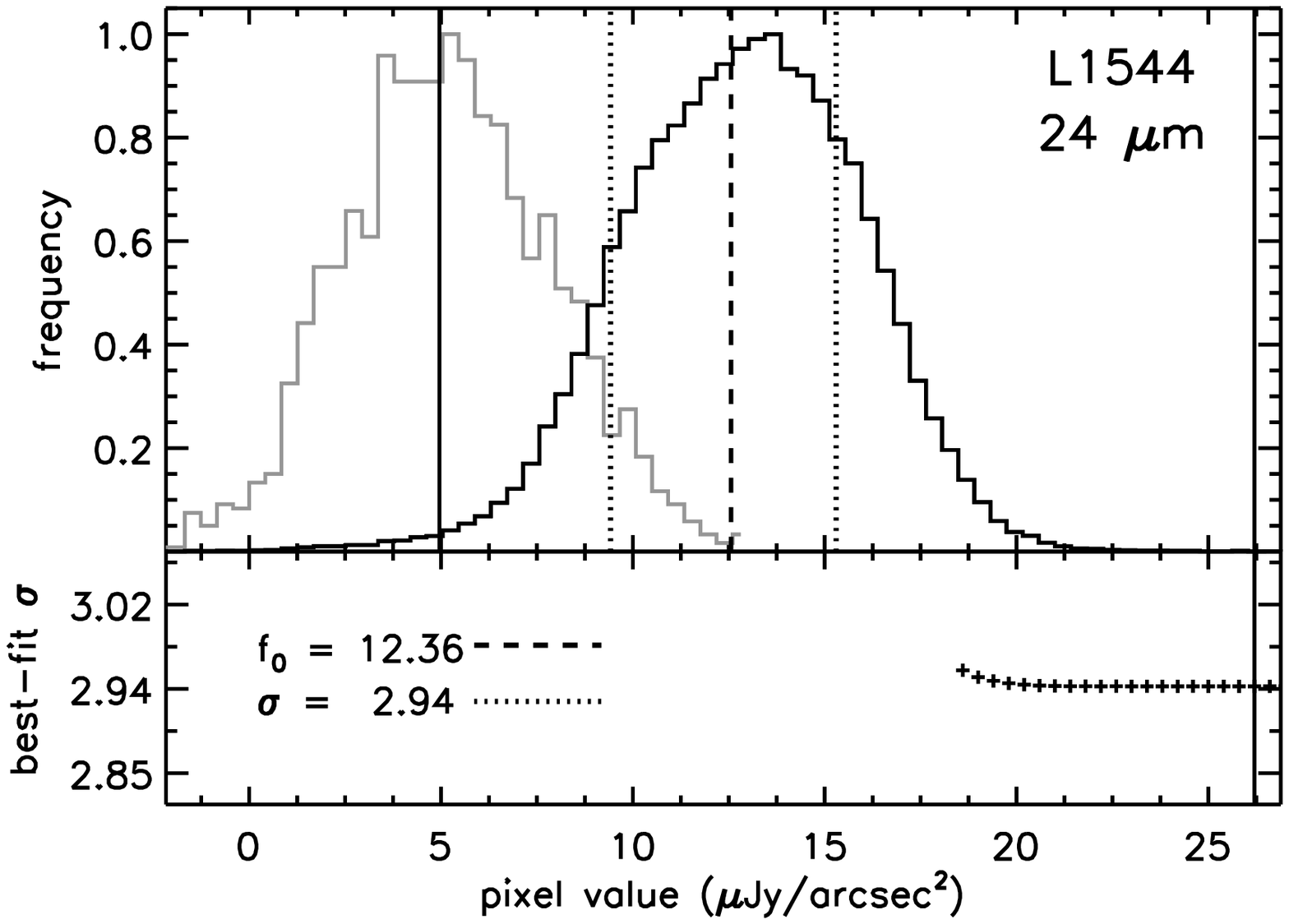}\includegraphics{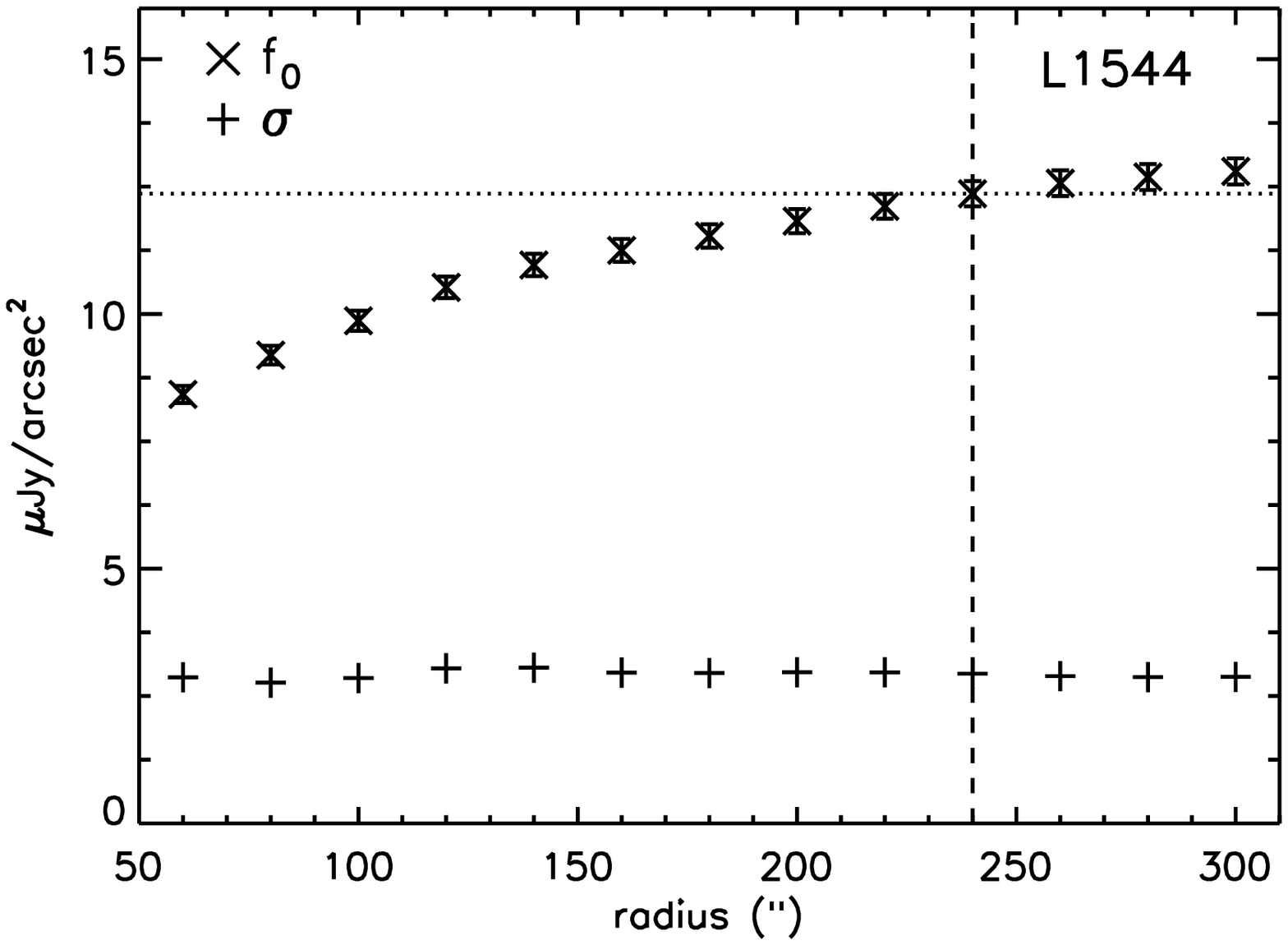}}
    \scalebox{0.5}{\includegraphics{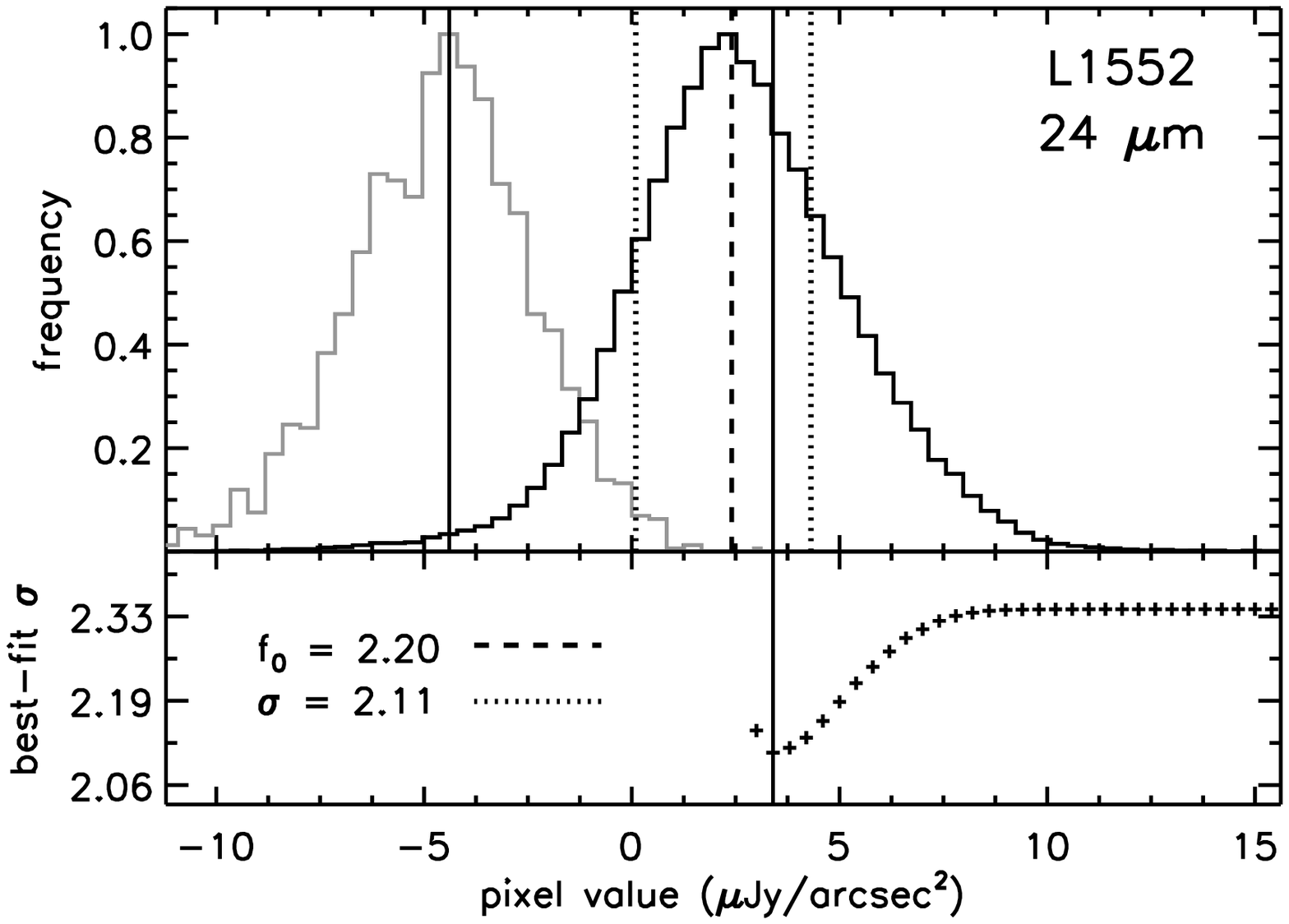}\includegraphics{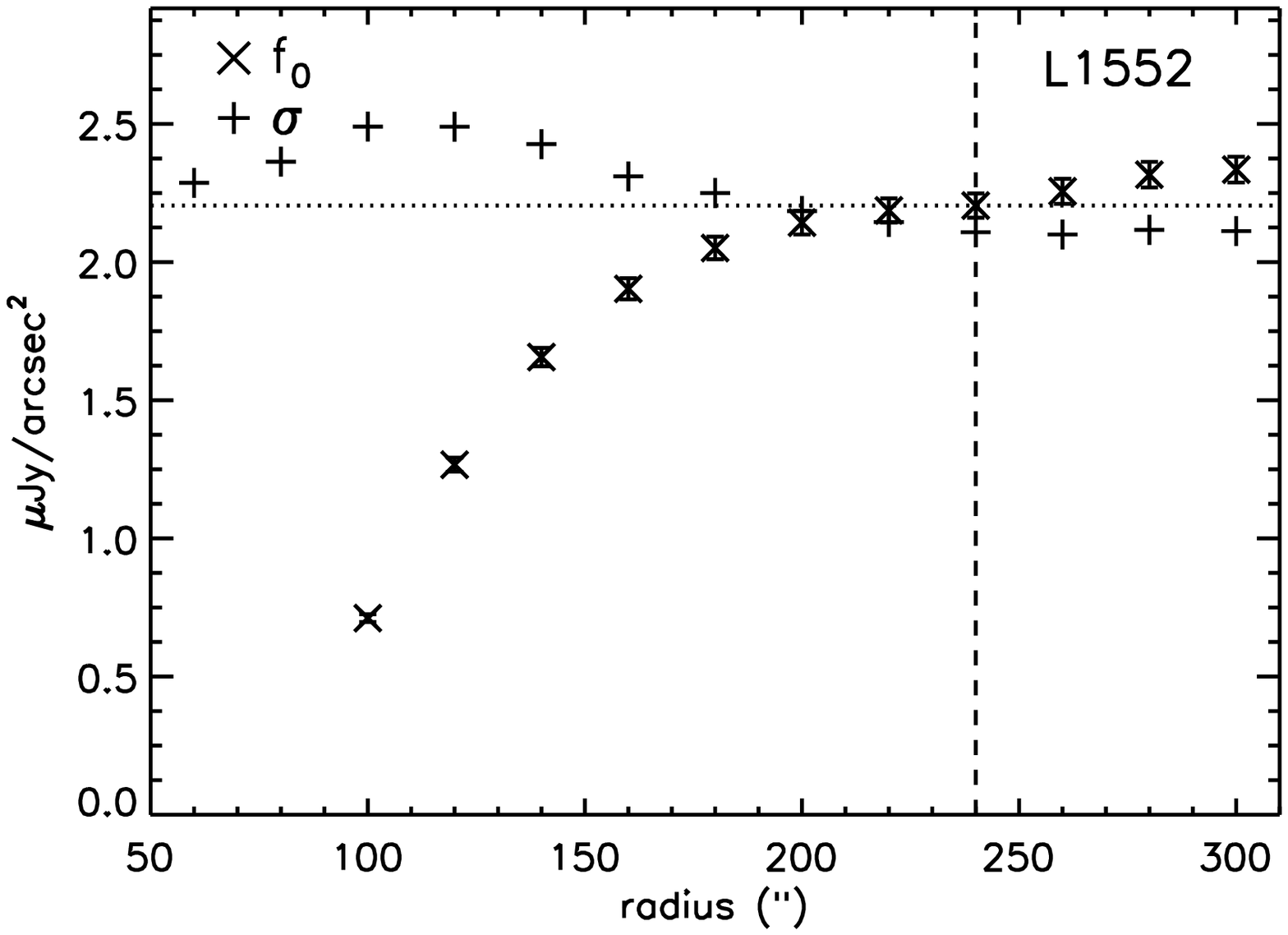}}
    \caption{Same as Figure~\ref{fig:f01}}
    \label{fig:f02}
  \end{center}
\end{figure}

\begin{figure}
  \begin{center}
    \scalebox{0.5}{\includegraphics{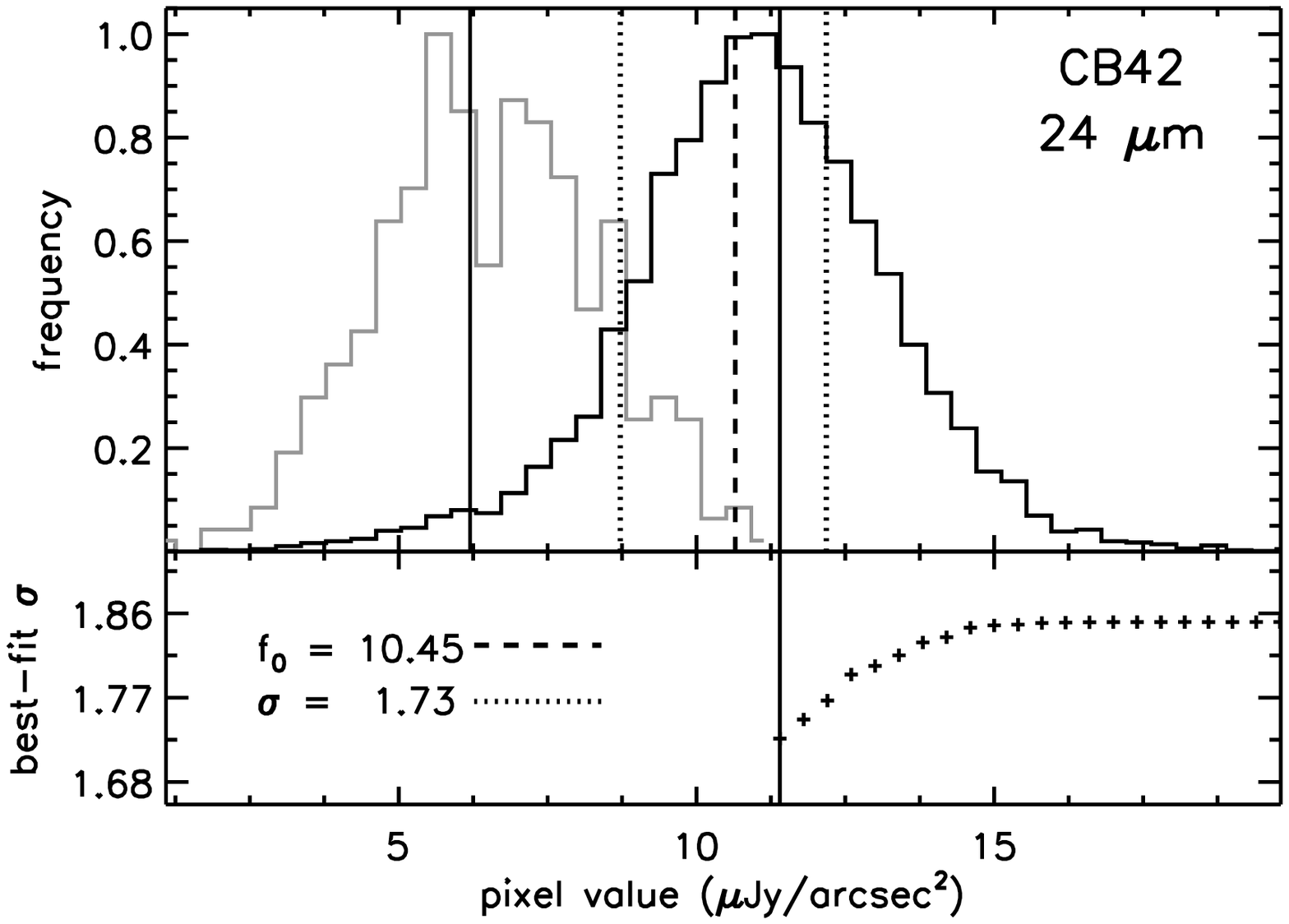}\includegraphics{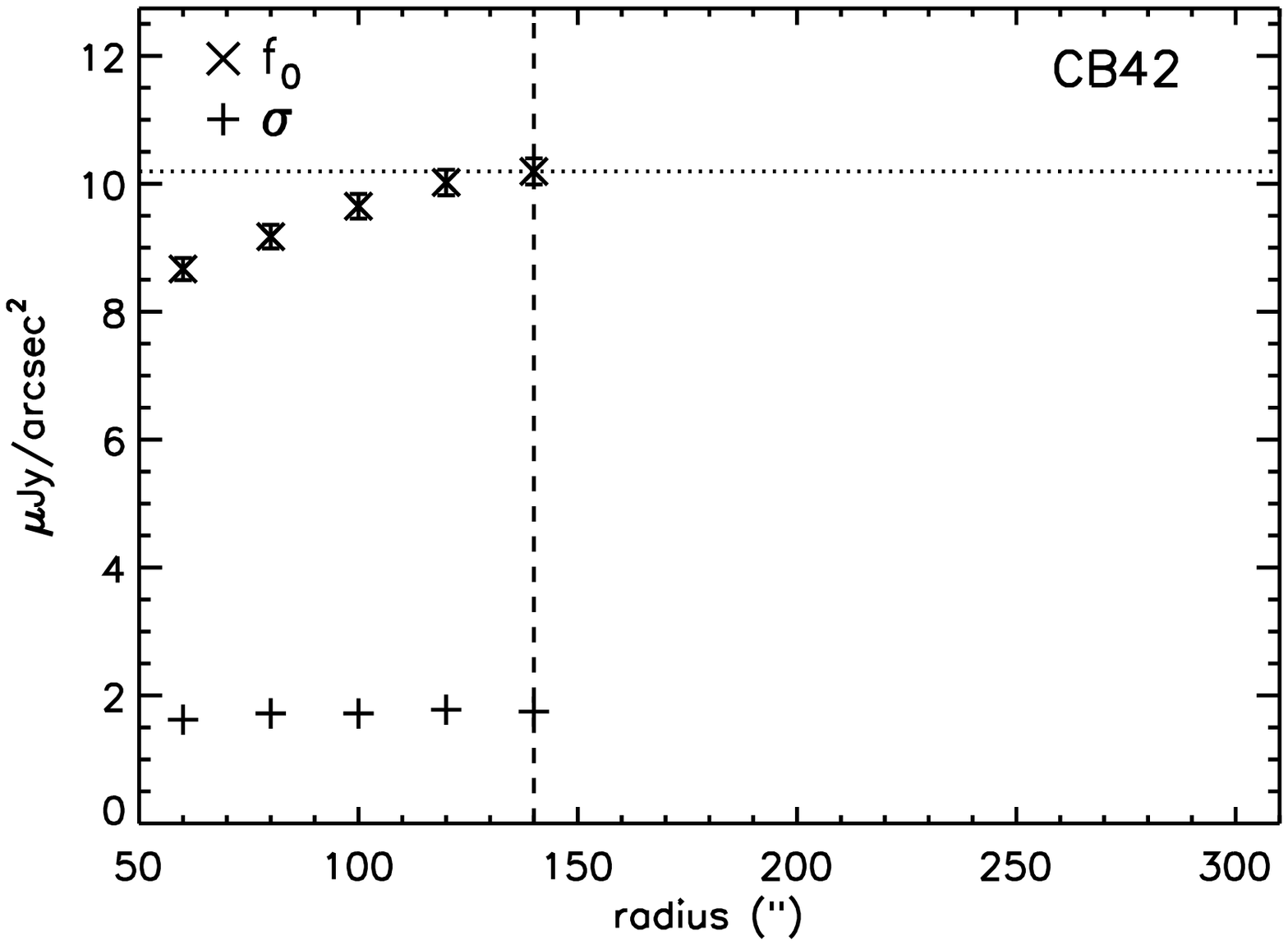}}
    \scalebox{0.5}{\includegraphics{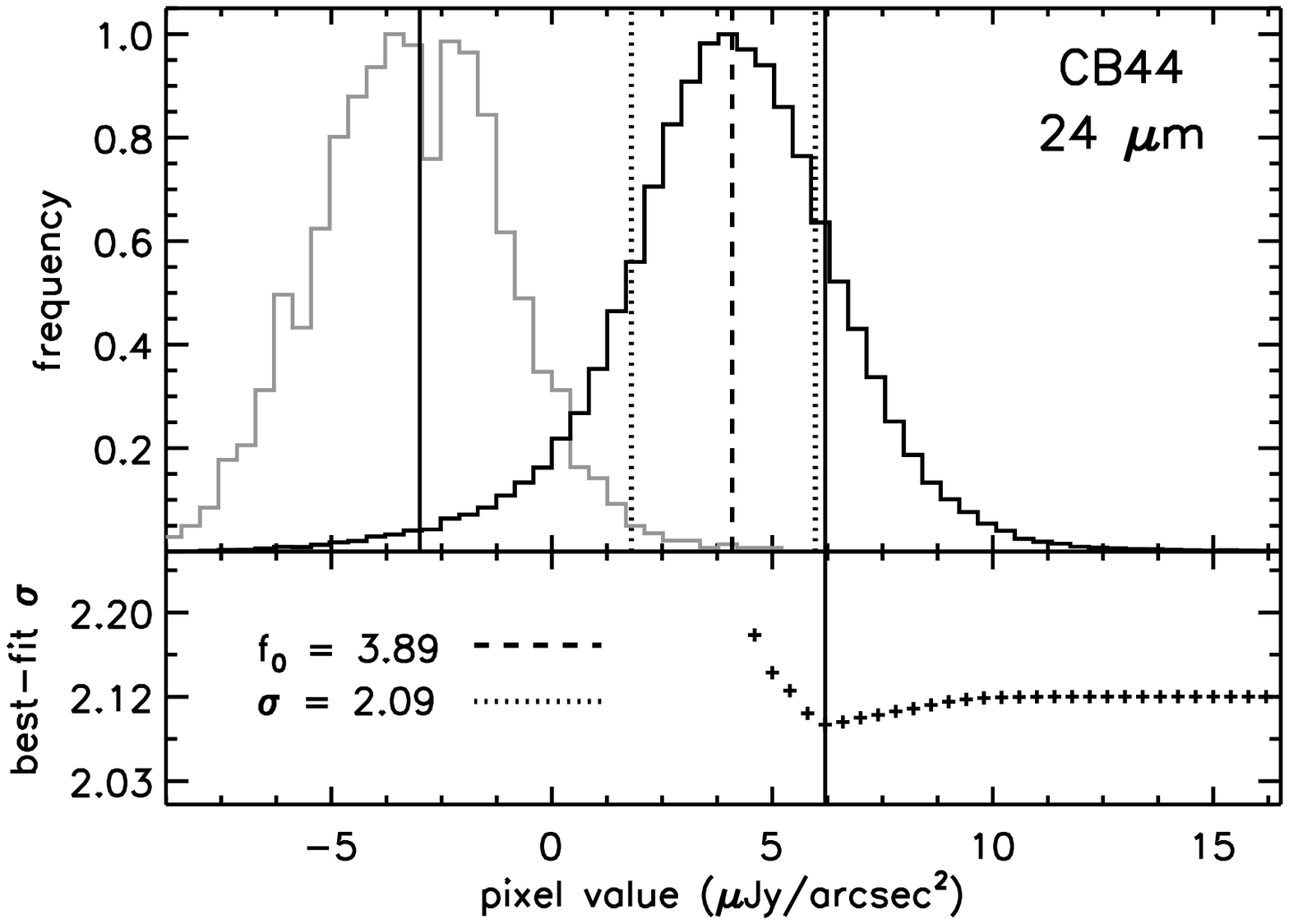}\includegraphics{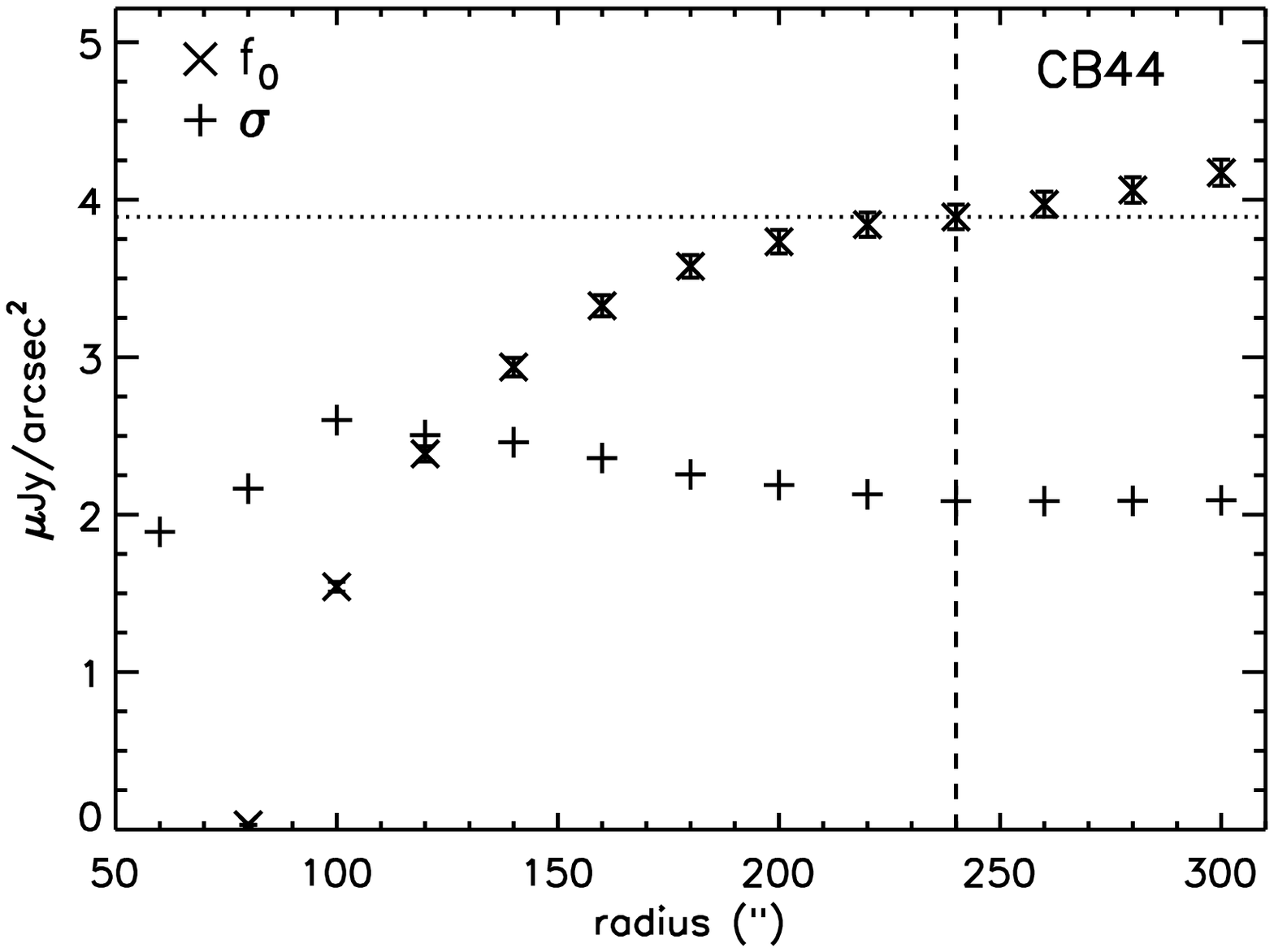}}
    \scalebox{0.5}{\includegraphics{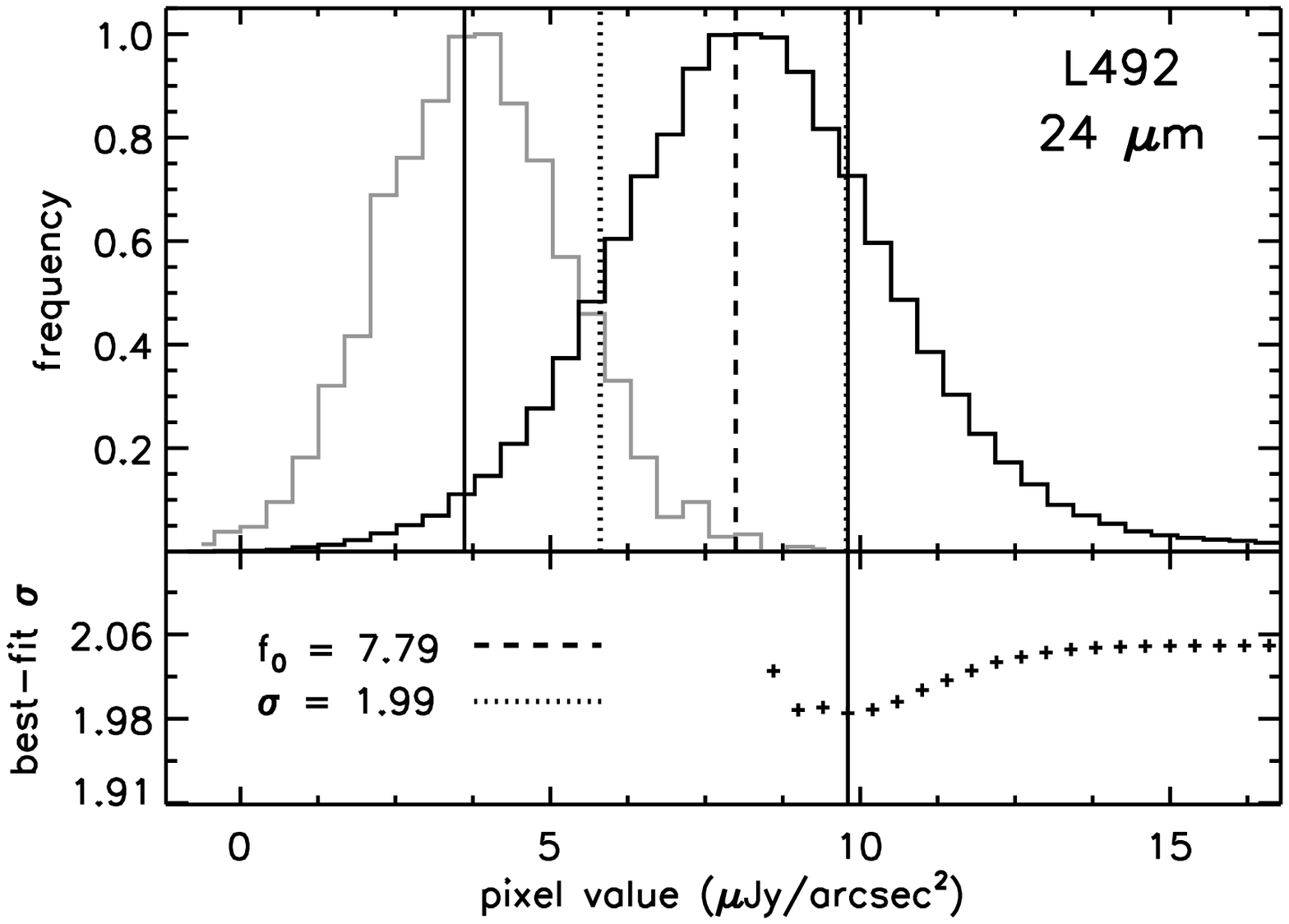}\includegraphics{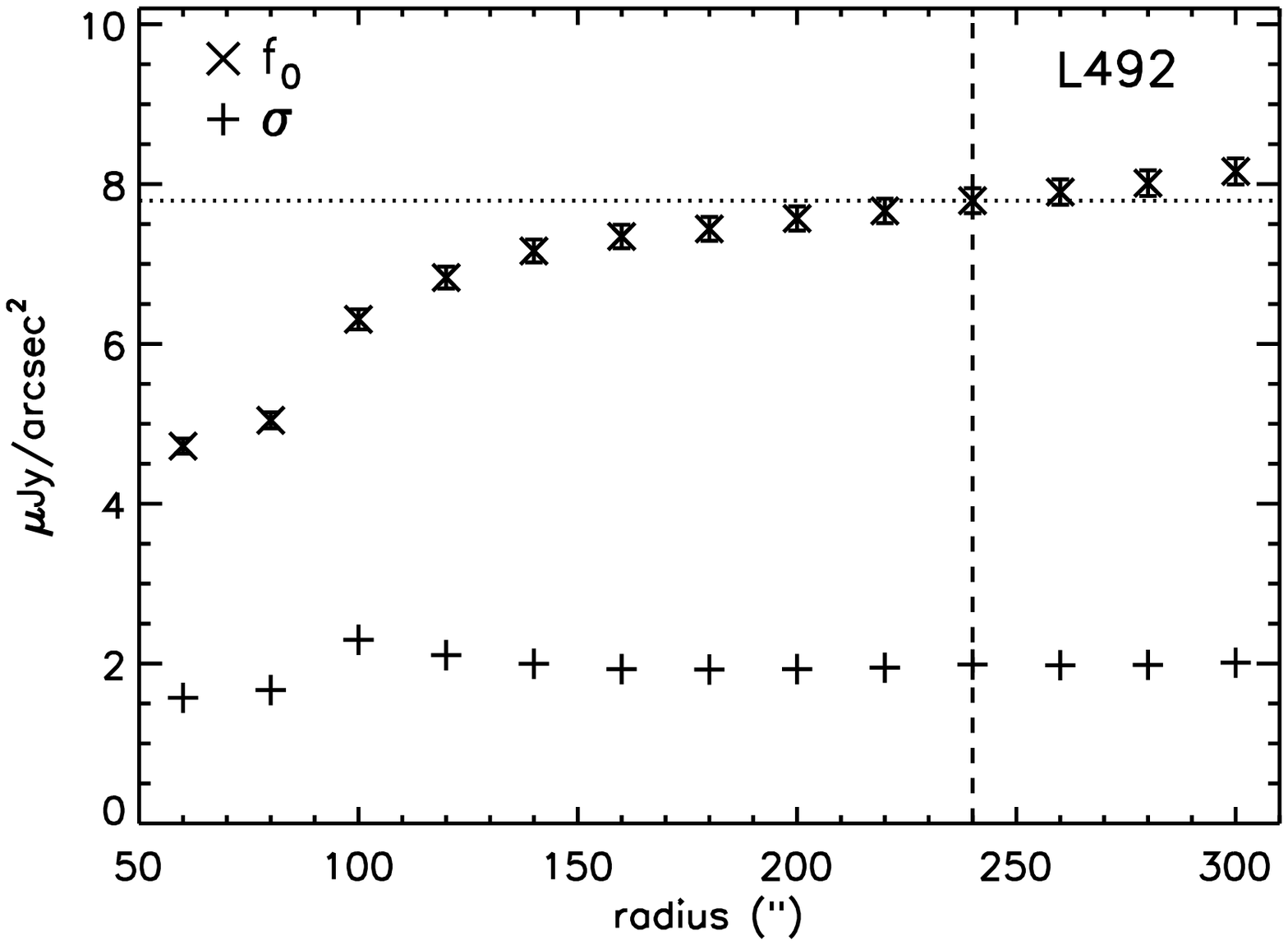}}
    \caption{Same as Figure~\ref{fig:f01}}
    \label{fig:f03}
  \end{center}
\end{figure}

\begin{figure}
  \begin{center}
    \scalebox{0.5}{\includegraphics{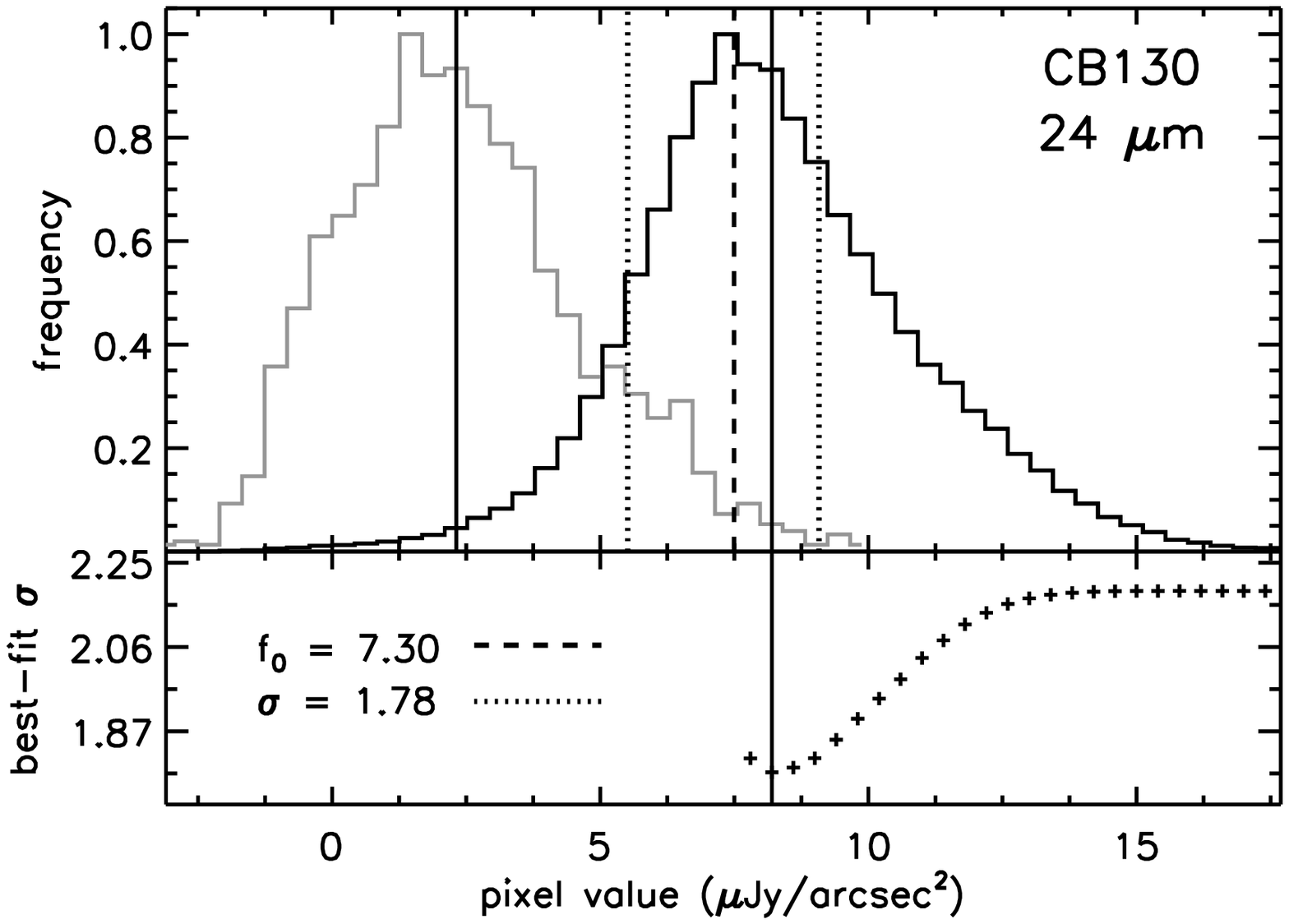}\includegraphics{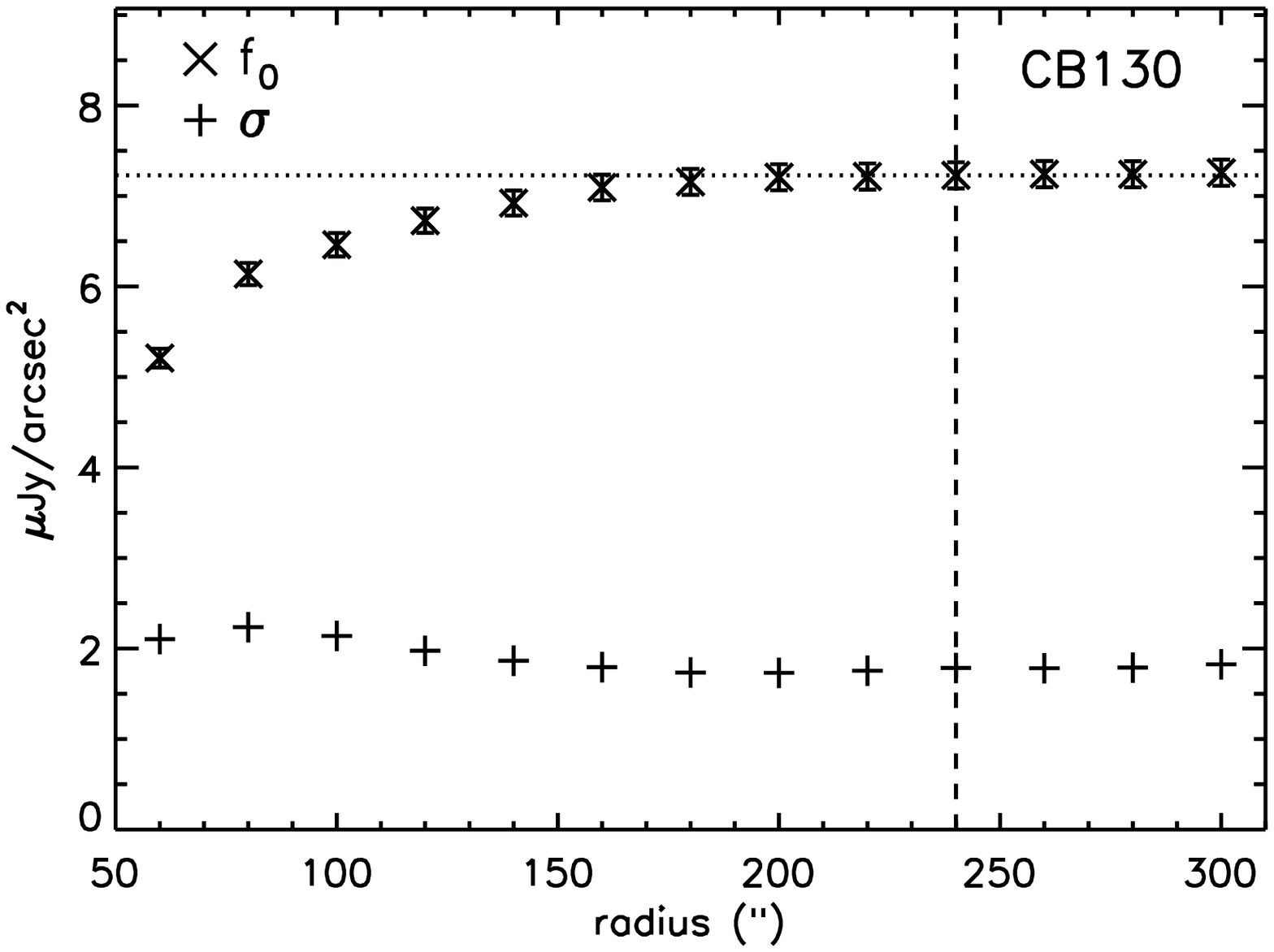}}
    \scalebox{0.5}{\includegraphics{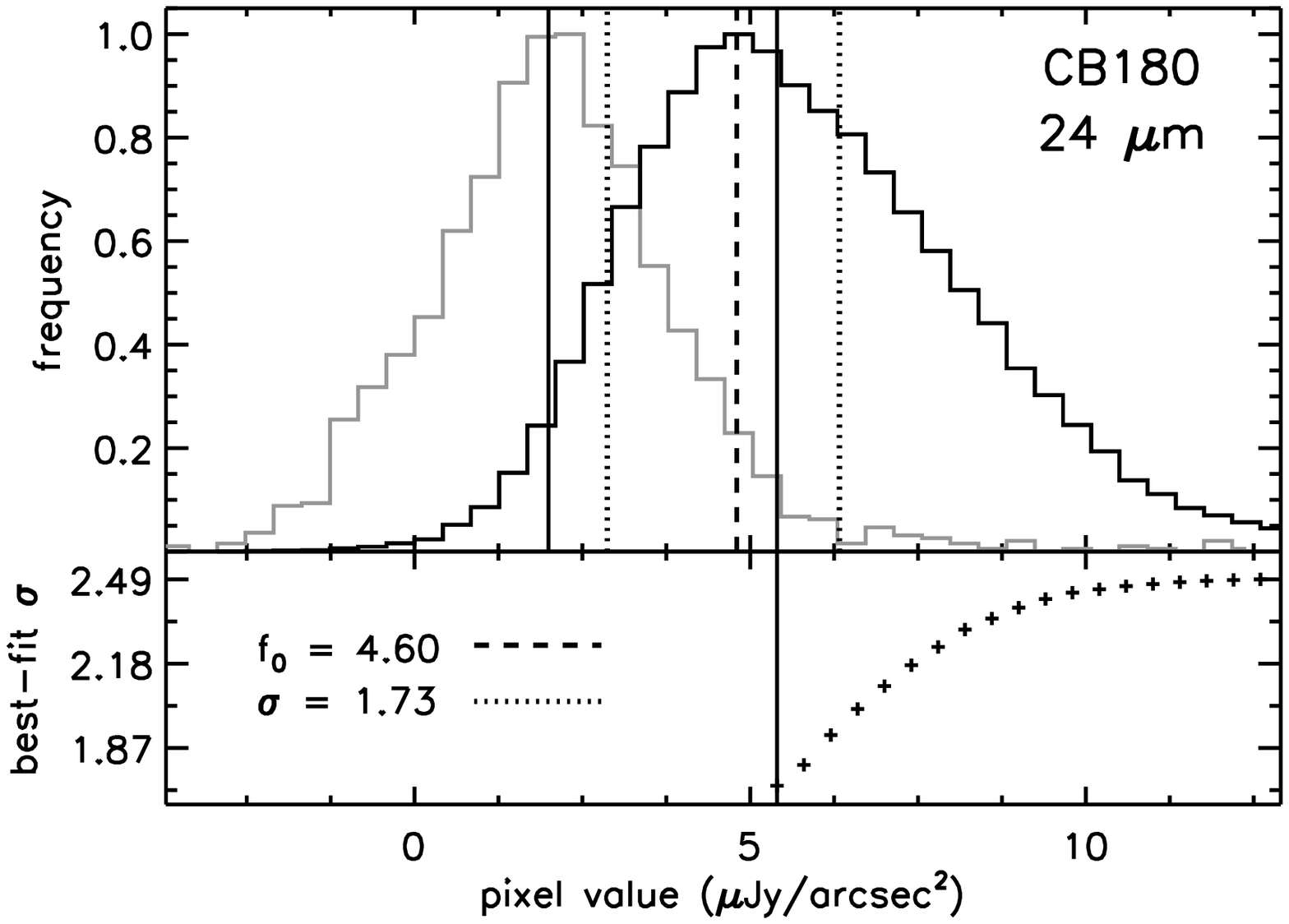}\includegraphics{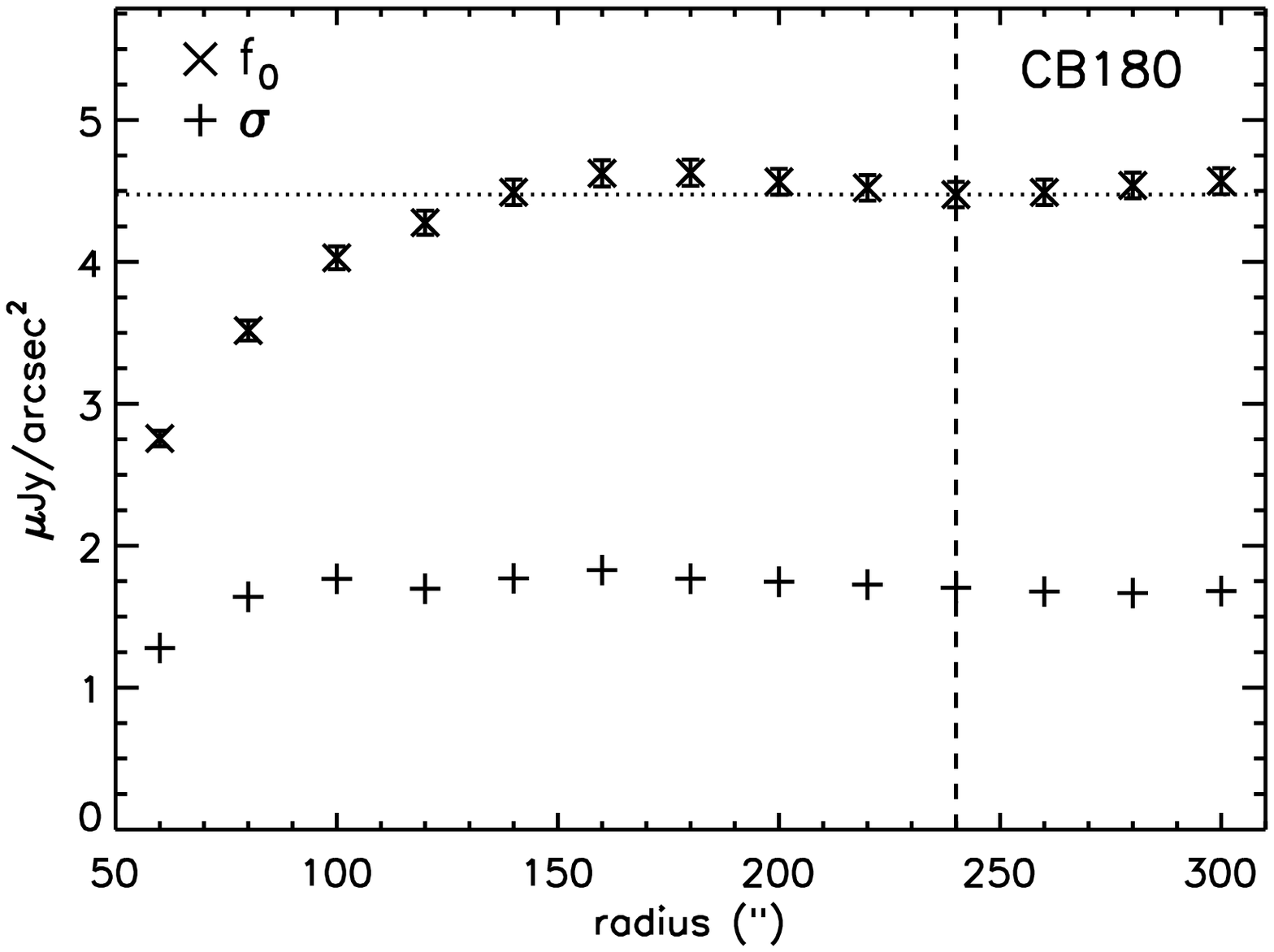}}
    \scalebox{0.5}{\includegraphics{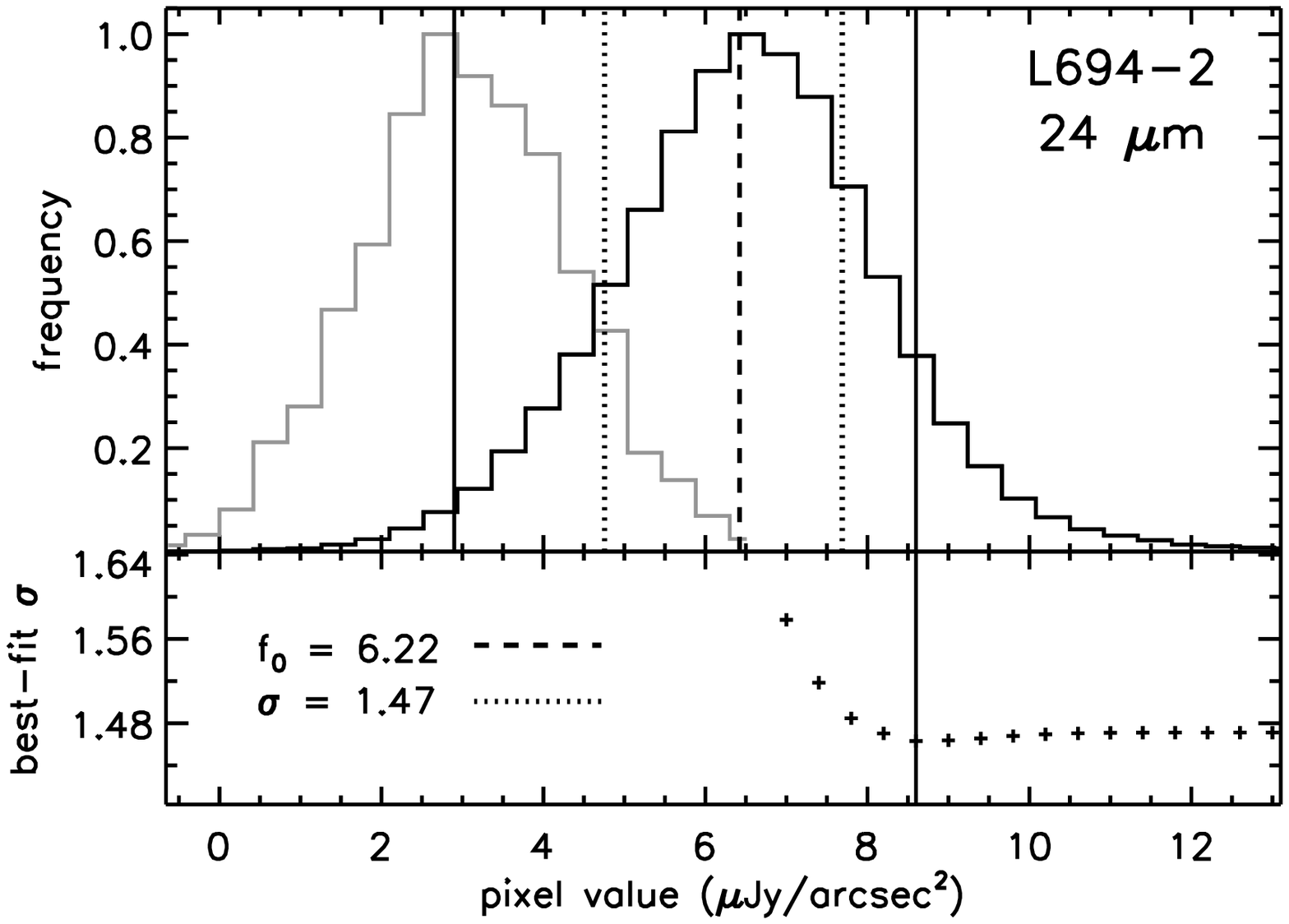}\includegraphics{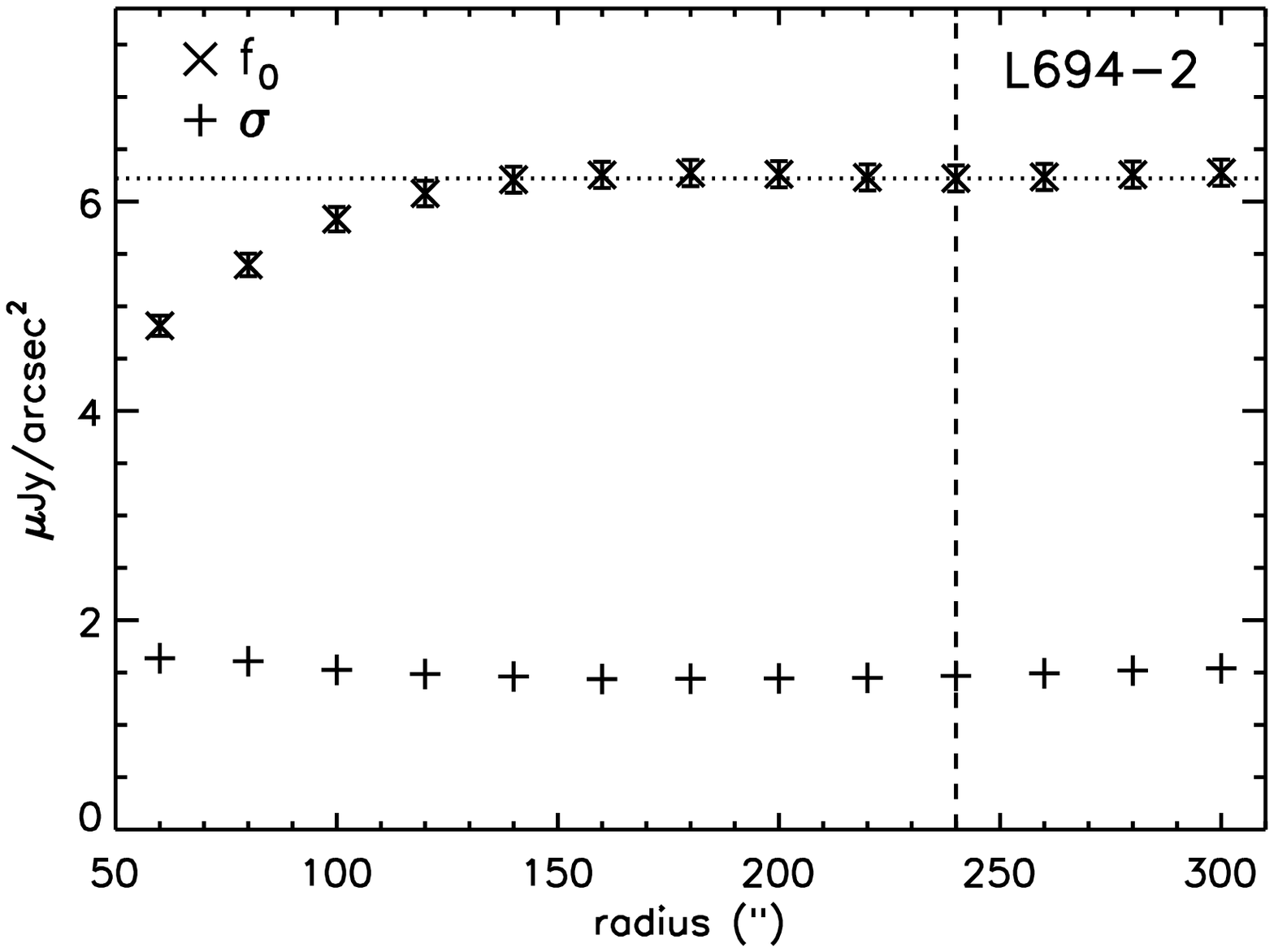}}
    \caption{Same as Figure~\ref{fig:f01}}
     \label{fig:f04}
  \end{center}
\end{figure}

{\it Unobscured flux ($f_0$) determination}.  We apply iterative
Gaussian fitting to the image pixel value distributions to estimate
$f_0$, the local, unobscured flux near the shadow, similar to the
method developed by \citet{stutz08a} to reconstruct and analyze the
noise properties in deep {\it Hubble} Space Telescope ACS images.  
In summary, we adjust the upper cutoff of the pixel value distribution
so that the Gaussian fit is least affected by the non-Gaussian tail of
the high pixel value distribution.  This is done by finding the upper
pixel value cutoff that has the narrowest Gaussian width. 
This fitting method is constructed to minimize the effects of sources
or shadows on the estimate of $f_0$.  We proceed by constructing a
pixel value histogram for all pixels within $r_0$ of the shadow using
a bin size of 0.4~$\mu$Jy/arcsec$^2$.  We then clip the distribution
by excluding all bins with values below the mean pixel value in a
$30\arcsec$ radius aperture centered on the shadow.  The shadow
pixel--value distribution in this region is indicated as a grey
histogram in Figures~\ref{fig:f01}--\ref{fig:f04}; the mean value of
the shadow flux is indicated with a solid line.  Excluding these bins
avoids a possible bias of the best--fit mean pixel value towards lower
values.  Next we fit the histogram data to an unweighted Gaussian
model, using standard IDL packages, and excluding all bins with values
greater than the maximum bin plus one.  We repeat this fitting
procedure as we include successively higher value bins one at a time.
The resulting best--fit Gaussian width ($\sigma$) values are shown in
Figures~\ref{fig:f01}--\ref{fig:f04} in the lower-left panel for each
object, plotted as a function of the maximum bin value included in the
fit.  The solid line indicates the minimum $\sigma$ from the
distribution of Gaussian widths; the mean value of the fit
corresponding to the minimum value of sigma is the adopted value of
$f_0$.  These two parameters are listed in the figure.

To determine the effect the choice of pixel histogram radius ($r_0$)
has on our derived value of $f_0$ we repeat this procedure over a
range of $r_0$ values from $r_0 = 60\arcsec$ to $300\arcsec$, in
$20\arcsec$ steps; the resulting values of $f_0$ and $\sigma$ are
shown in the right--hand--side panel of
Figures~\ref{fig:f01}--\ref{fig:f04}.  In most cases the $f_0$
profiles flatten out by $r_0 = 240\arcsec$, as indicated by the dashed
line.  This value is used for all objects except CB42, which lies near
the edge of the MIPS mosaic; since the $f_0$ profile for CB42 appears
to flatten out by $140\arcsec$, the largest value of $r_0$ that we
test for this core, we assume this value is valid.  In the cases of
CB26 and CB27 we do not measure as abrupt a flattening.  For these
cores we still use $r_0 = 240\arcsec$ as a conservative estimate;
larger values for $r_0$ may over--estimate $f_0$ and the derived
optical depths.  Furthermore, ideally we calculate $f_0$ as close as
possible to the shadow features; as the distance from the core is
increased, other effects can bias the determination of $f_0$, such as
diffuse emission not associated with the core.  We estimate the
uncertainty in $f_0$ to be $\sigma(f_0) = 0.02 \times f_0$.  The
Gaussian error in the mean gives a SNR for $f_0$ of $\gtrsim 230$ for
all cores; we consider this SNR to be optimistic.

Using our calculated value of $f_0$ and the background subtracted
image, we calculate $\tau_{24}$ in each pixel and the associated error
$\sigma_{\tau,24}$, using standard propagation of errors, including
$\sigma_{\rm DC}$ and $\sigma(f_0)$.  The pixel--to--pixel
correlations are significant and we do not take this into account in
the $\tau$ pixel error maps; consequently, the error maps
underestimate the true errors and we do not use them in the subsequent
analysis.  In the cases where the values of $f$ near the center of the
cores are negative, causing $\tau_{24}$ to diverge, we set the value
of $\tau_{24}$ equal to the mean $\langle \tau_{24} \rangle$ in an
aperture with radius $10\arcsec$ or larger, depending on the size of
divergent region, centered on the core.  For most (8/12) cores, the
number of divergent pixels is less than 90, corresponding to an
effective circular area with radius smaller than $6\farcs5$; for
objects CB130 and CB180 the divergent region has an effective radius
of $\sim\!7\farcs5$, for CB44 the region has a radius of about
$25\arcsec$, and for L1552 the region radius is about $20\arcsec$.
This approach is conservative when analyzing core stability because it
effectively underestimates the 24~\micron\ shadow masses.
Furthermore, an alternative approach is to estimate $f_{DC}$ as a
fraction of the shadow pixel values; while this method would
circumvent the divergent pixel issue, it would yeild larger core
masses, potentially biasing our collapse analysis.  The divergent
pixels found with our method are flagged and accounted for in the
subsequent analysis.  The 24~\micron\ $\tau$ pixel maps are shown in
the left column of Figures~\ref{fig:tauco} through \ref{fig:tauco3}.

\subsubsection{8~\micron\ Optical Depth Calculation}

\begin{figure}
  \begin{center}
    \scalebox{0.5}{\includegraphics{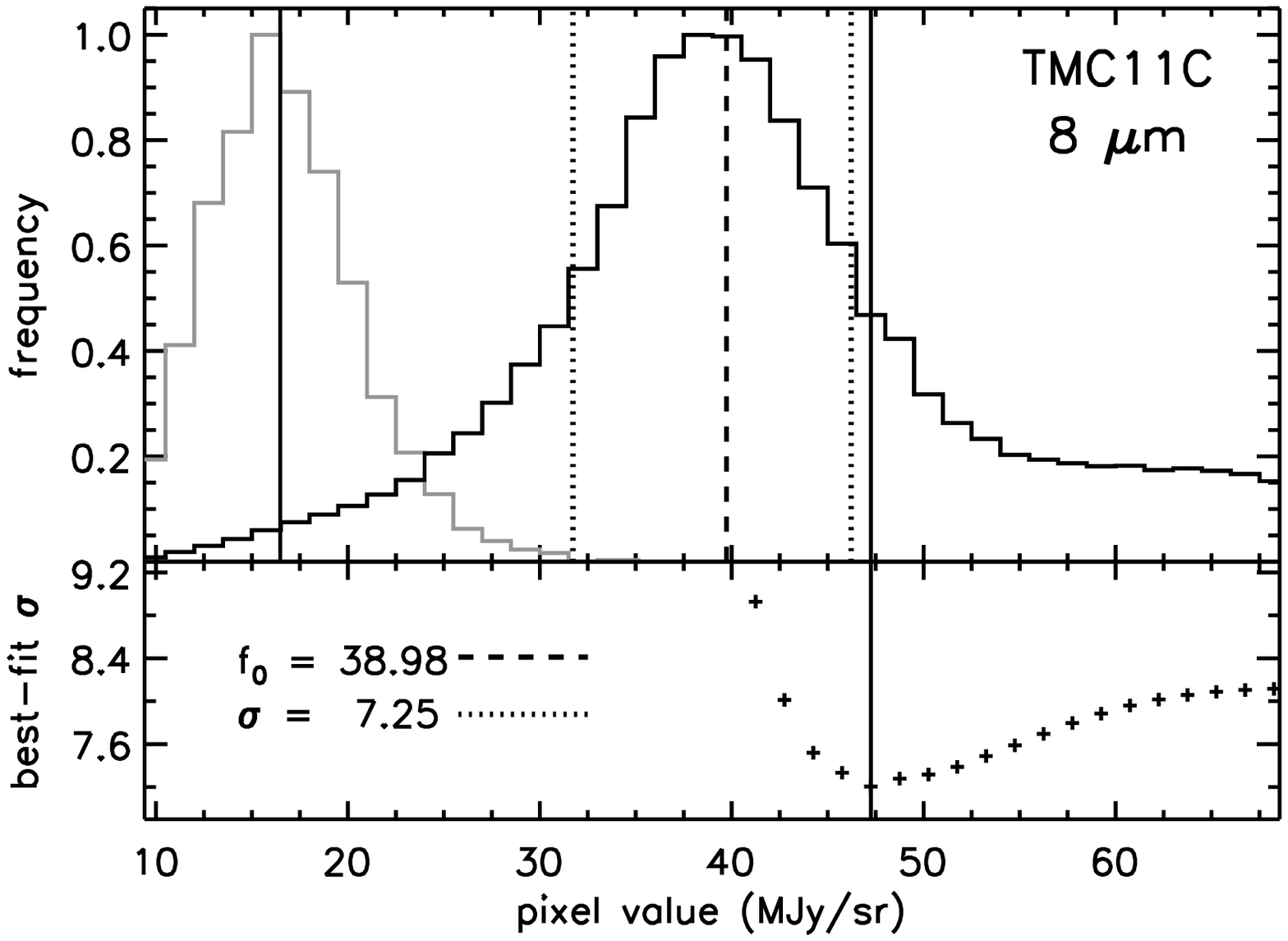}\includegraphics{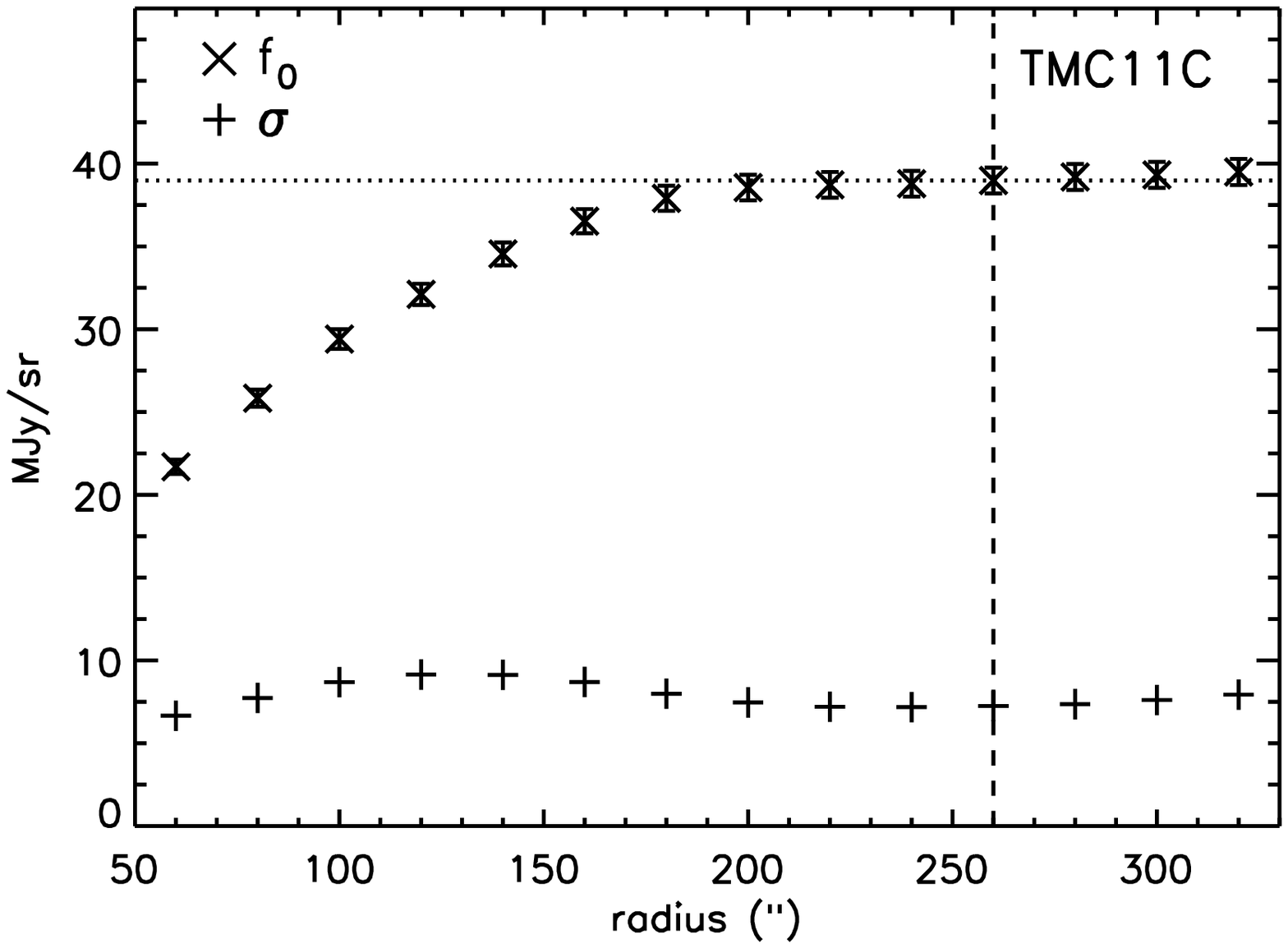}}

    \scalebox{0.5}{\includegraphics{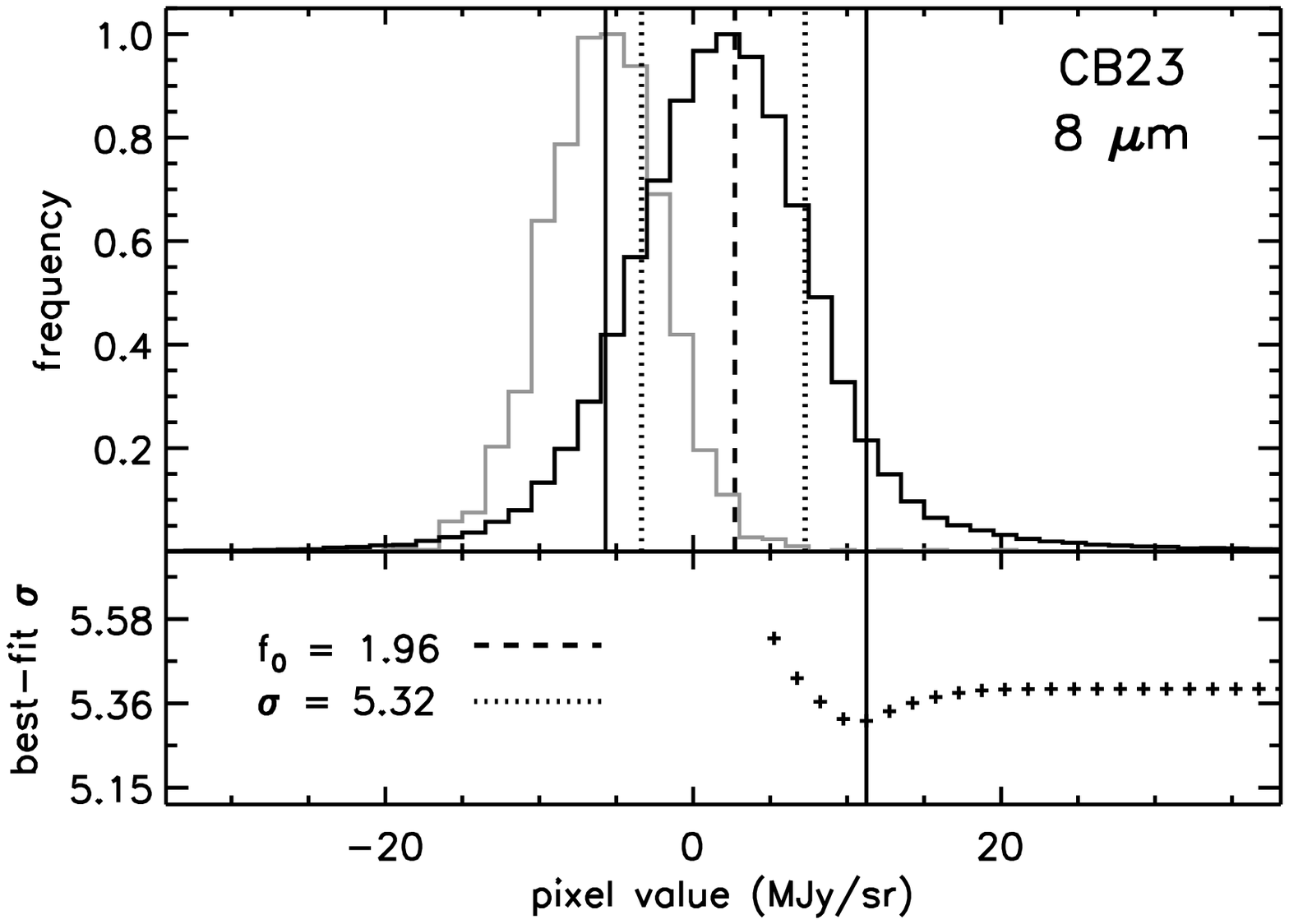}\includegraphics{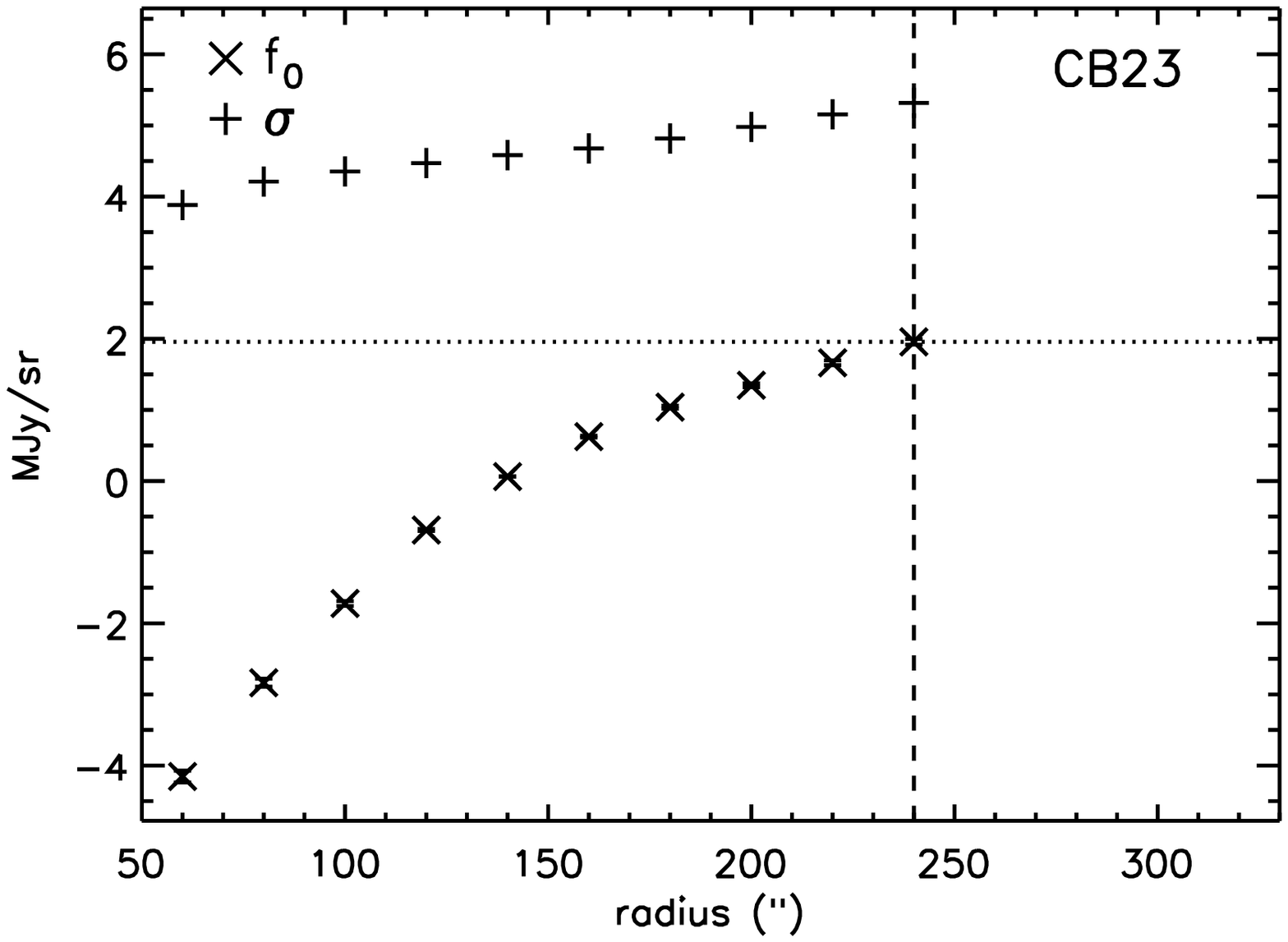}}

    \scalebox{0.5}{\includegraphics{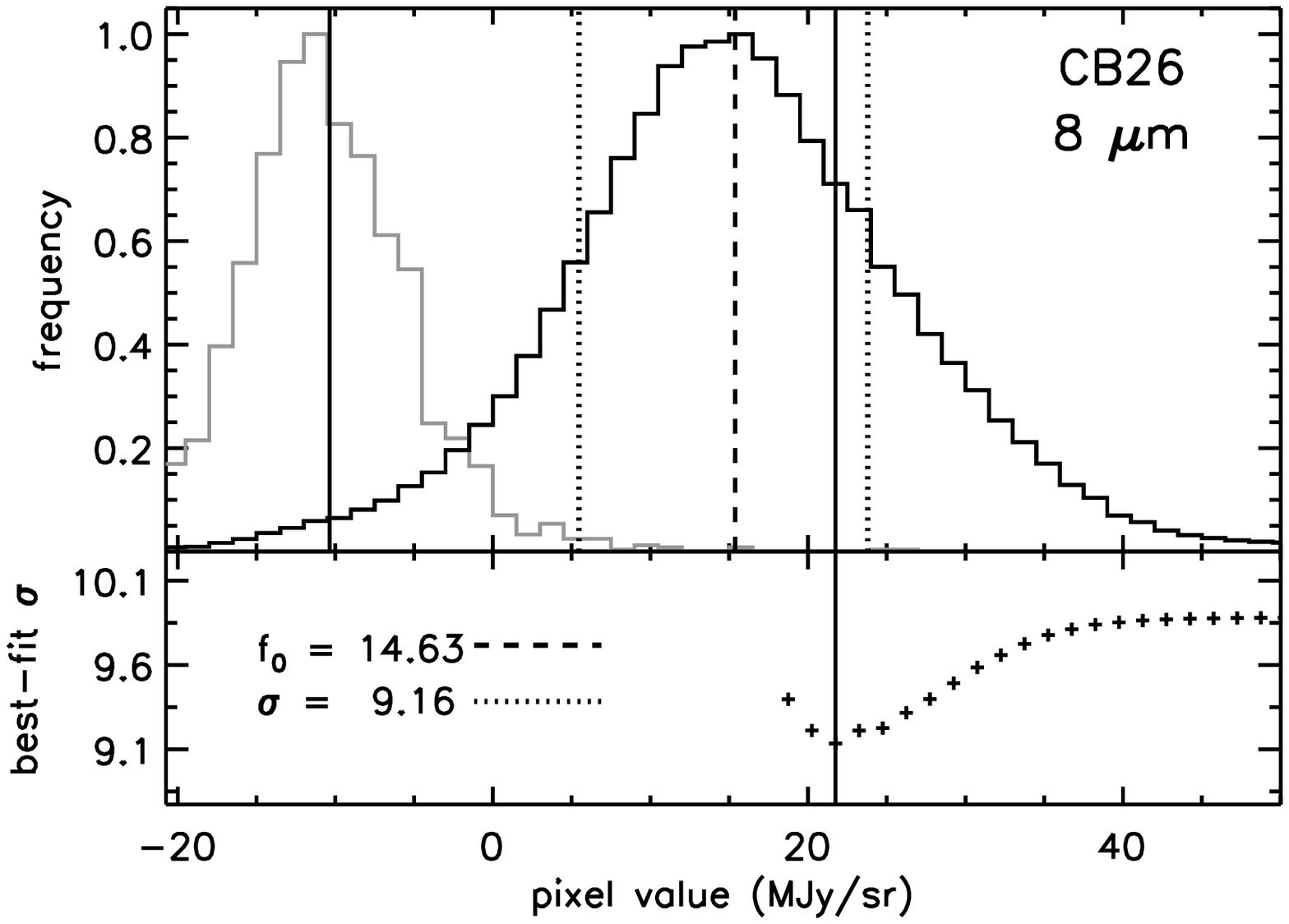}\includegraphics{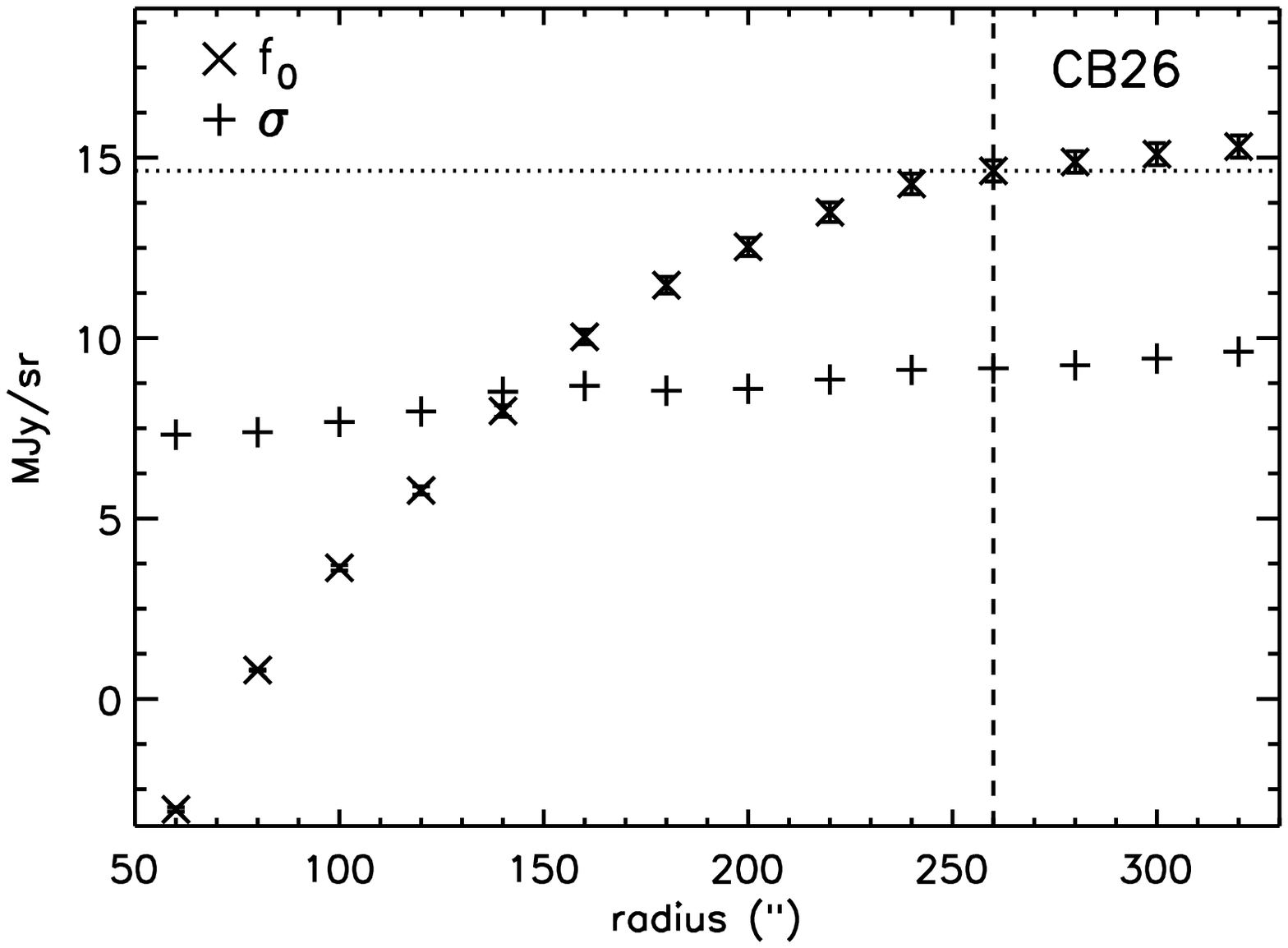}}
    \caption{{\it Left Column, top panel --- } Black histogram:
      8~\micron\ pixel value distribution for background subtracted
      image region of $r_0 = 240\arcsec$ radius (except for CB42),
      $r_0$ is indicated in the right column as the dashed line, and
      is centered on the shadow.  Grey histogram: 8~\micron\ pixel
      value distribution for $30\arcsec$ region centered on the
      shadow.  Solid lines: minimum and maximum black histogram bins
      used to fit for the minimum $\sigma$.  The dashed line indicates
      the best--fit mean, or $f_0$, value at the minimum best--fit
      value for $\sigma$, indicated by the dotted line.  {\it Left
        Column, bottom panel --- } Best--fit values of $\sigma$
      plotted against the maximum bin value included in the fit.  The
      solid line indicates the minimum value of $\sigma$.  Best--fit
      Gaussian parameters $\sigma$ and the mean ($f_0$) are indicated.
      {\it Right Column --- } Best--fit $f_0$ and minimum $\sigma$ as
      a function of aperture size ( $r_0$).}
    \label{fig:f081}
  \end{center}
\end{figure}

\begin{figure}
  \begin{center}
    \scalebox{0.5}{\includegraphics{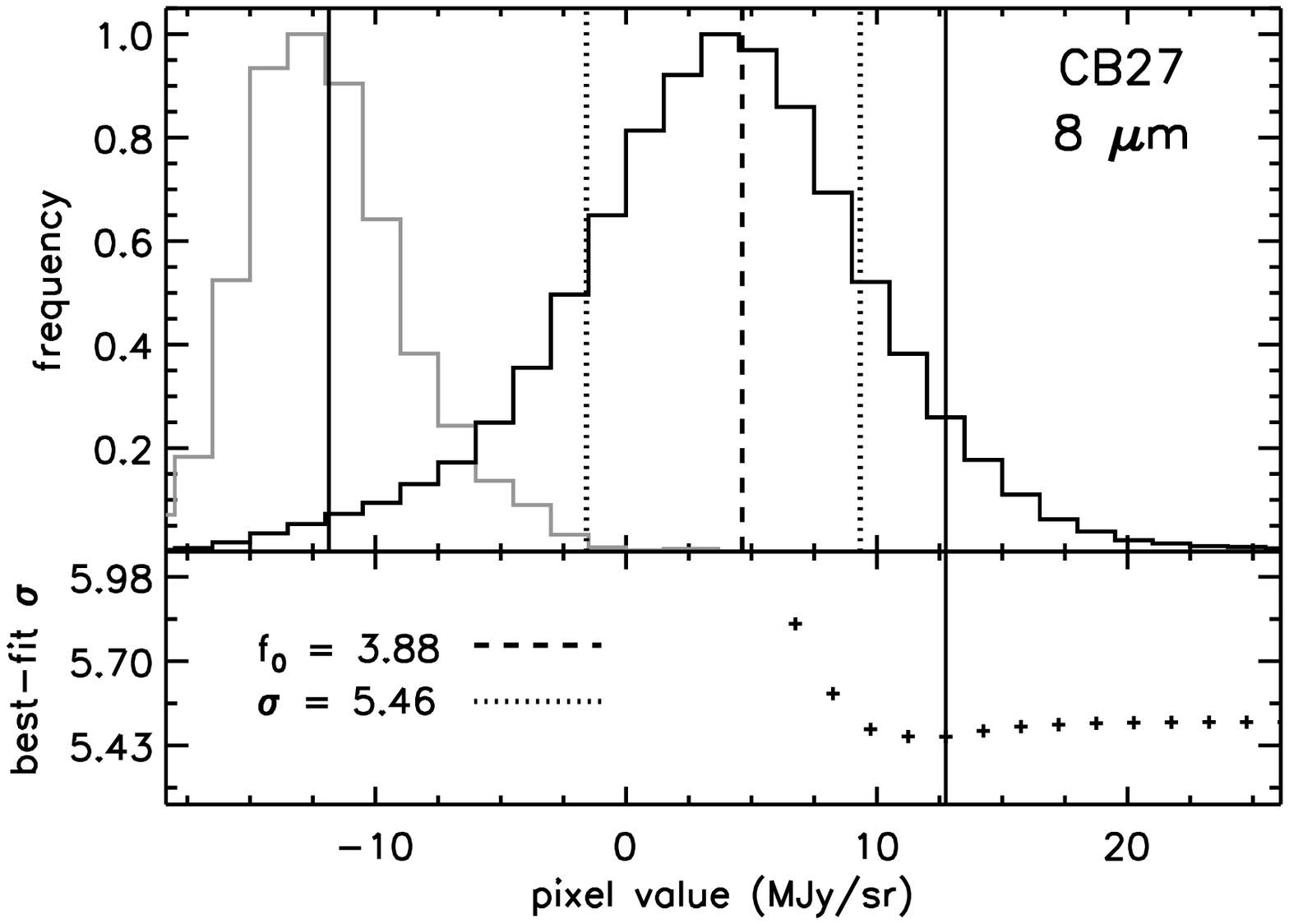}\includegraphics{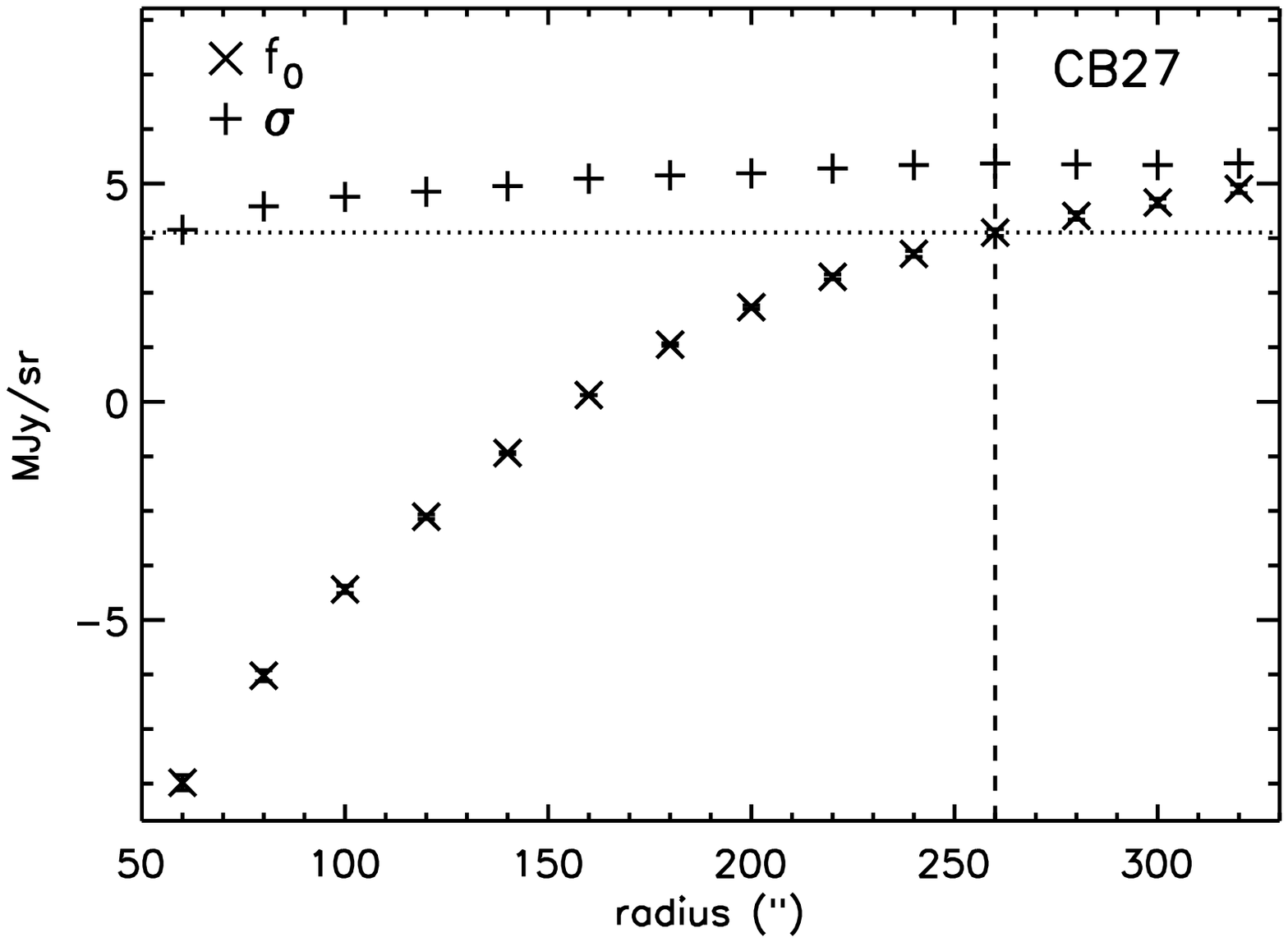}}

    \scalebox{0.5}{\includegraphics{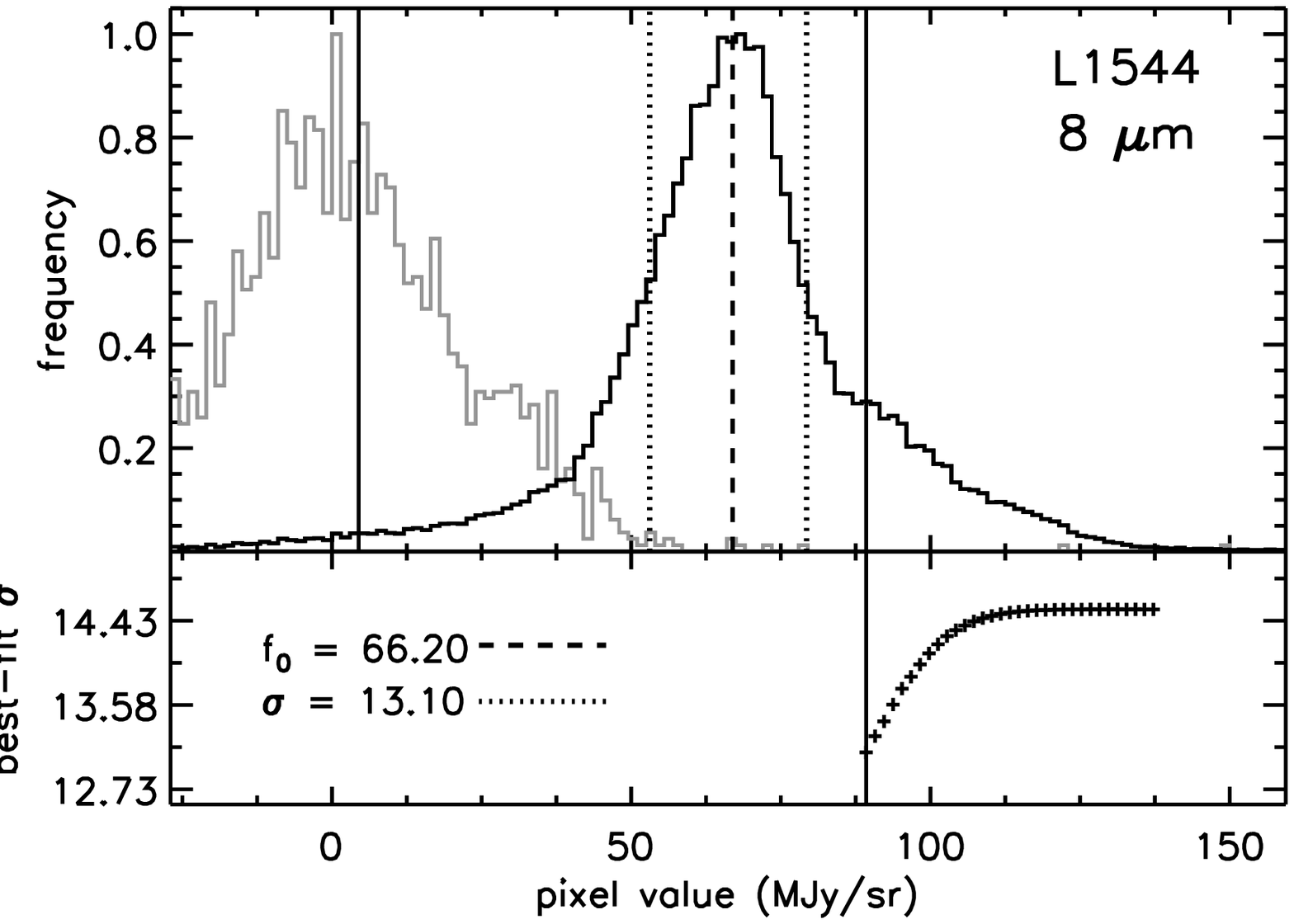}\includegraphics{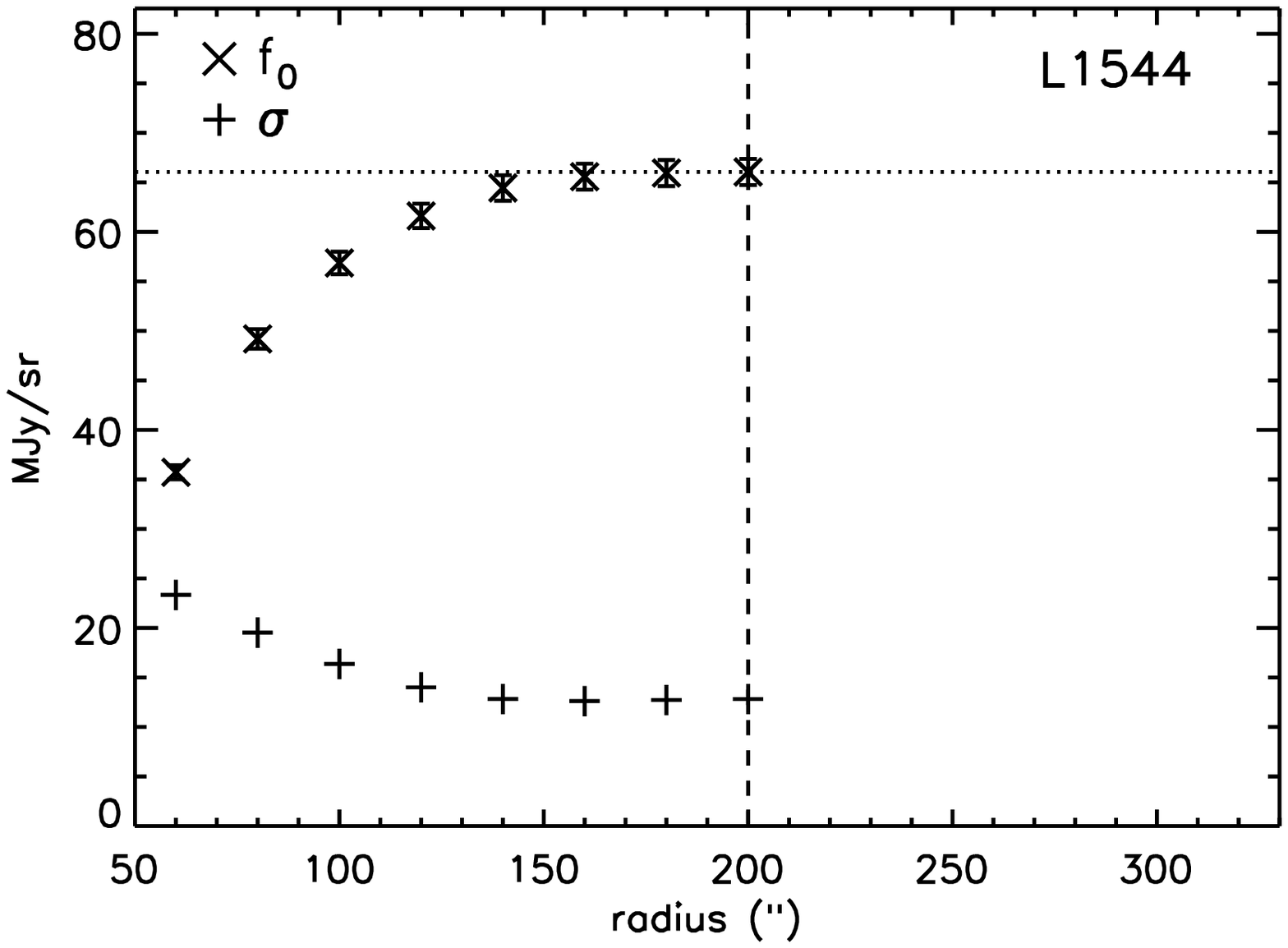}}

    \scalebox{0.5}{\includegraphics{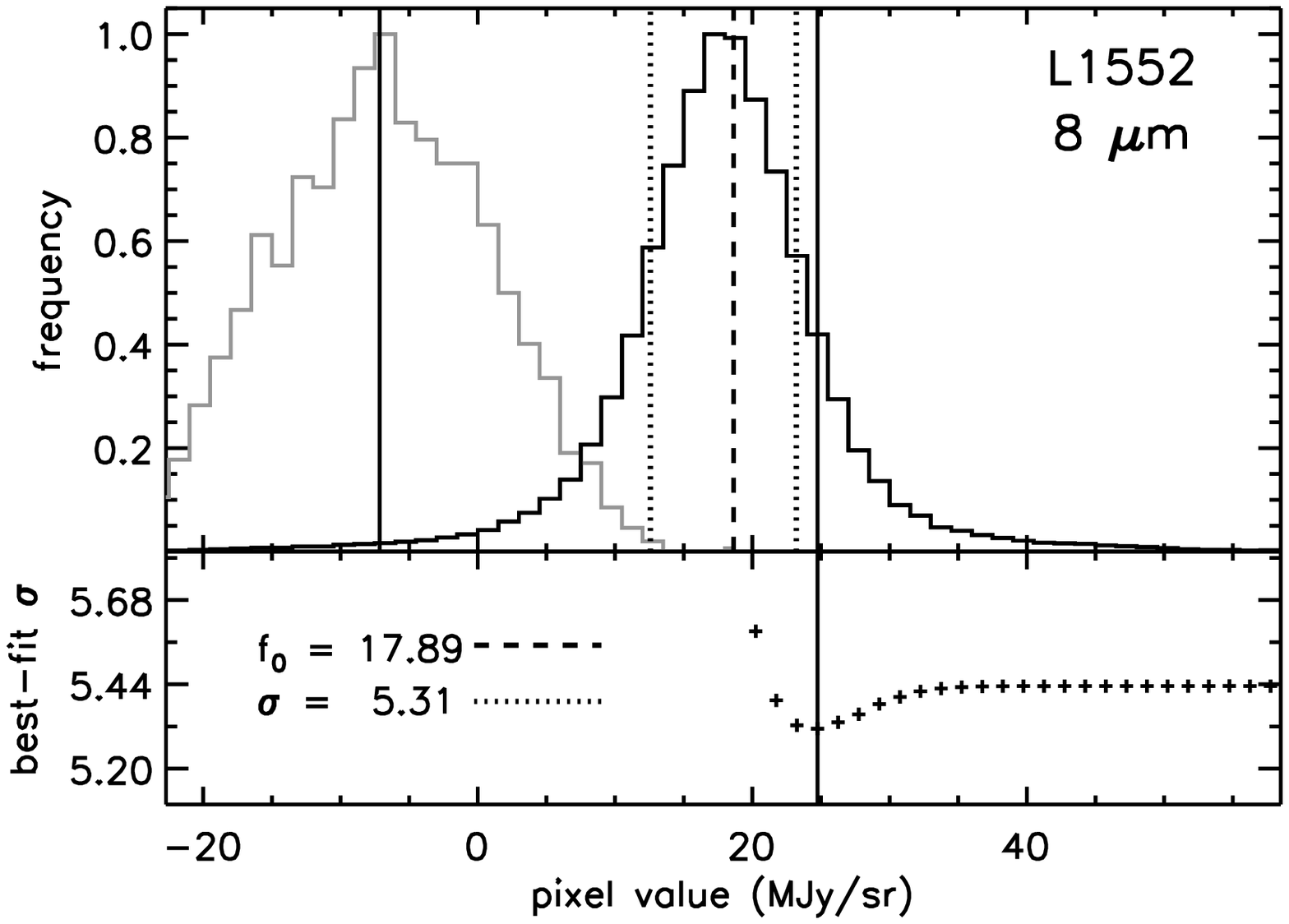}\includegraphics{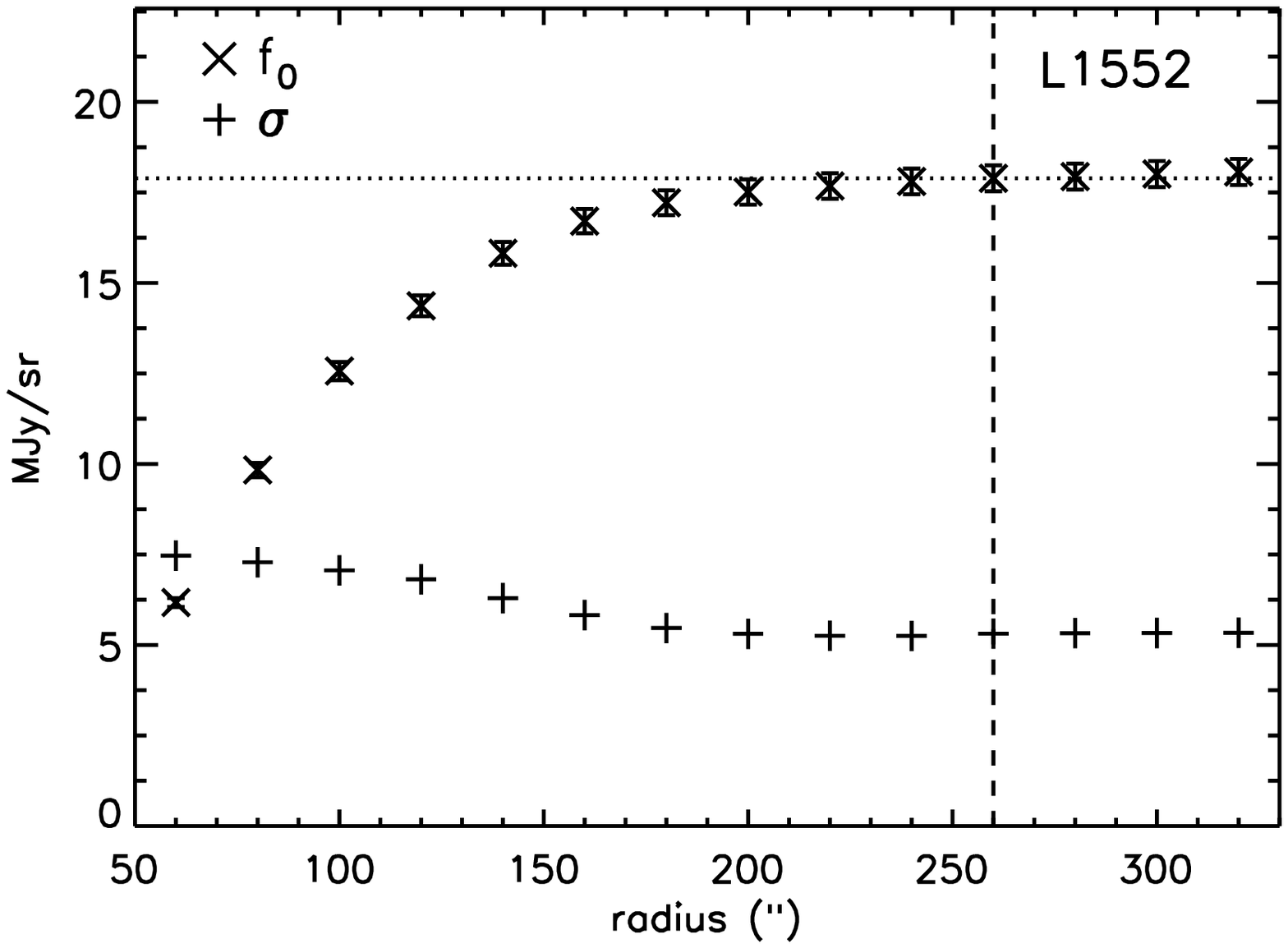}}
    \caption{Same as Figure~\ref{fig:f081}}
    \label{fig:f082}
  \end{center}
\end{figure}

\begin{figure}
  \begin{center}
    \scalebox{0.5}{\includegraphics{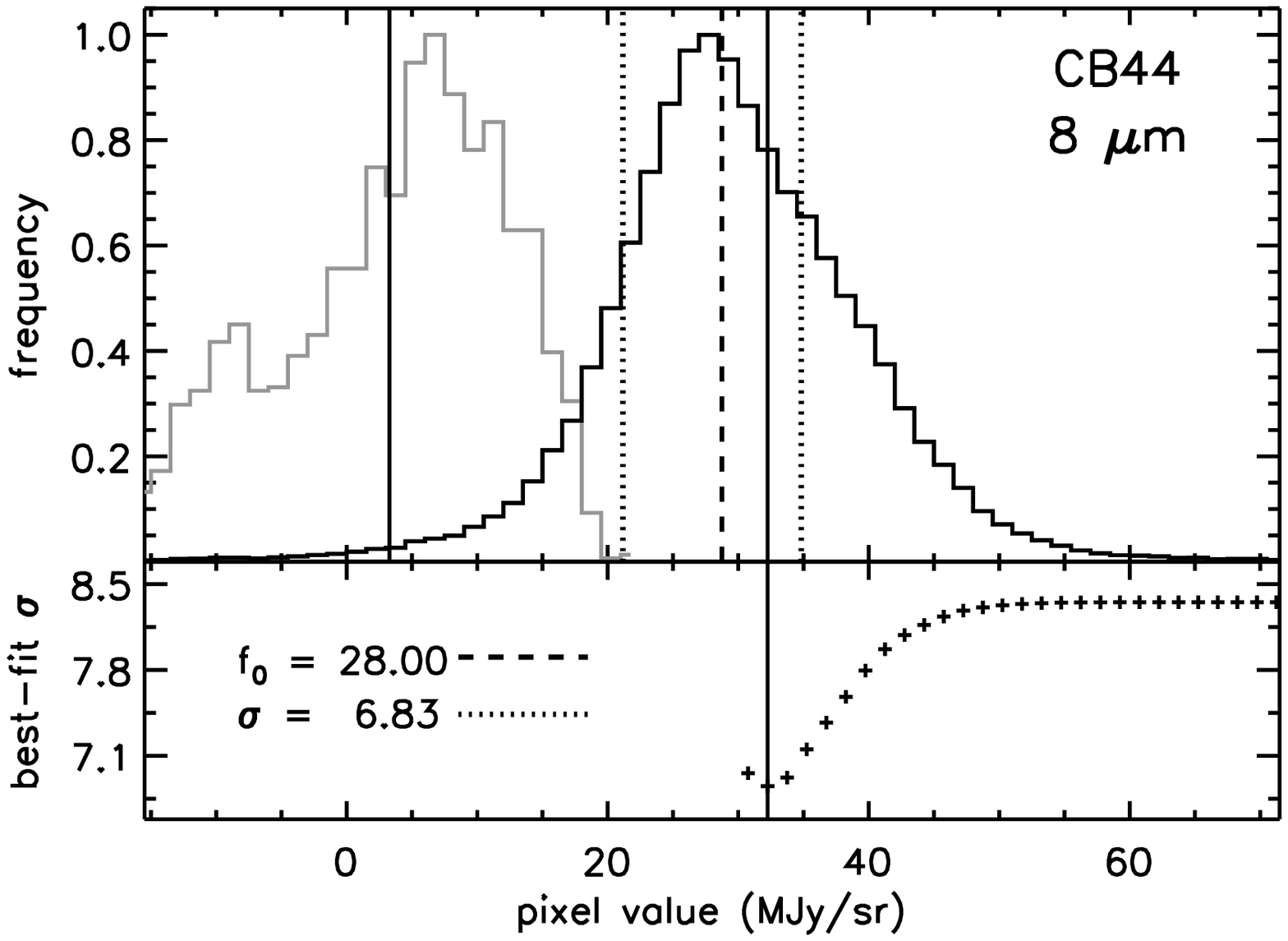}\includegraphics{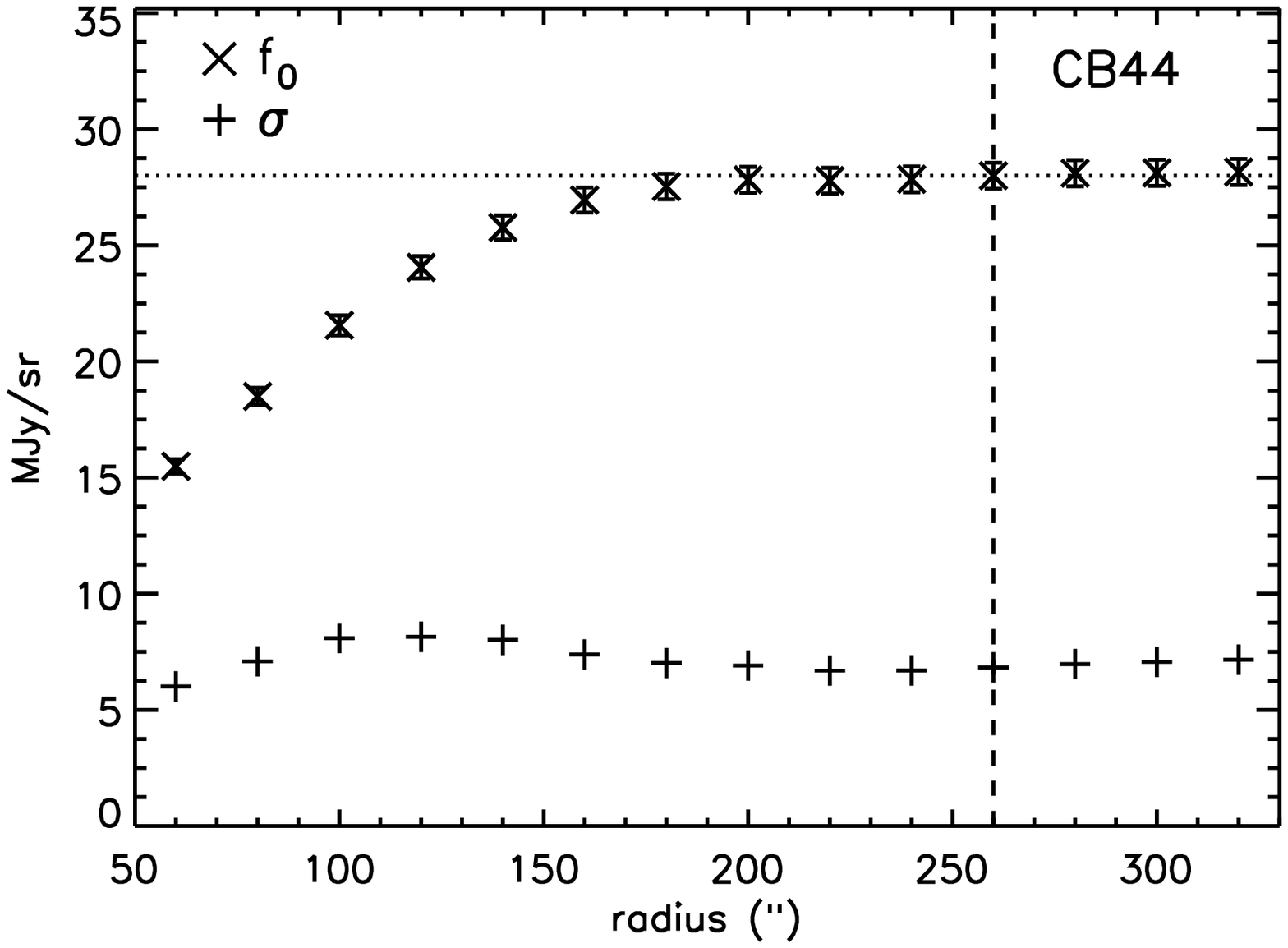}}

    \scalebox{0.5}{\includegraphics{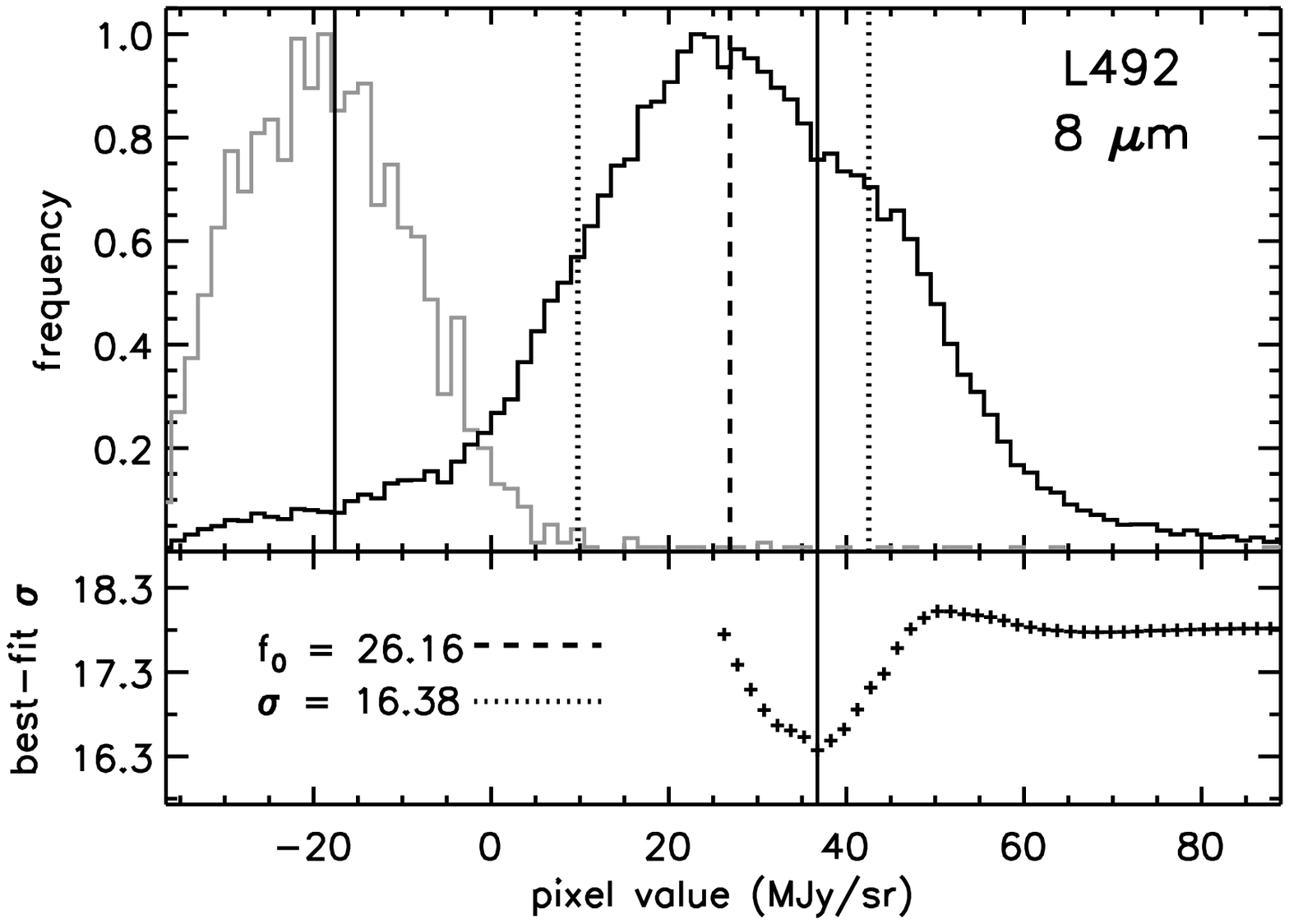}\includegraphics{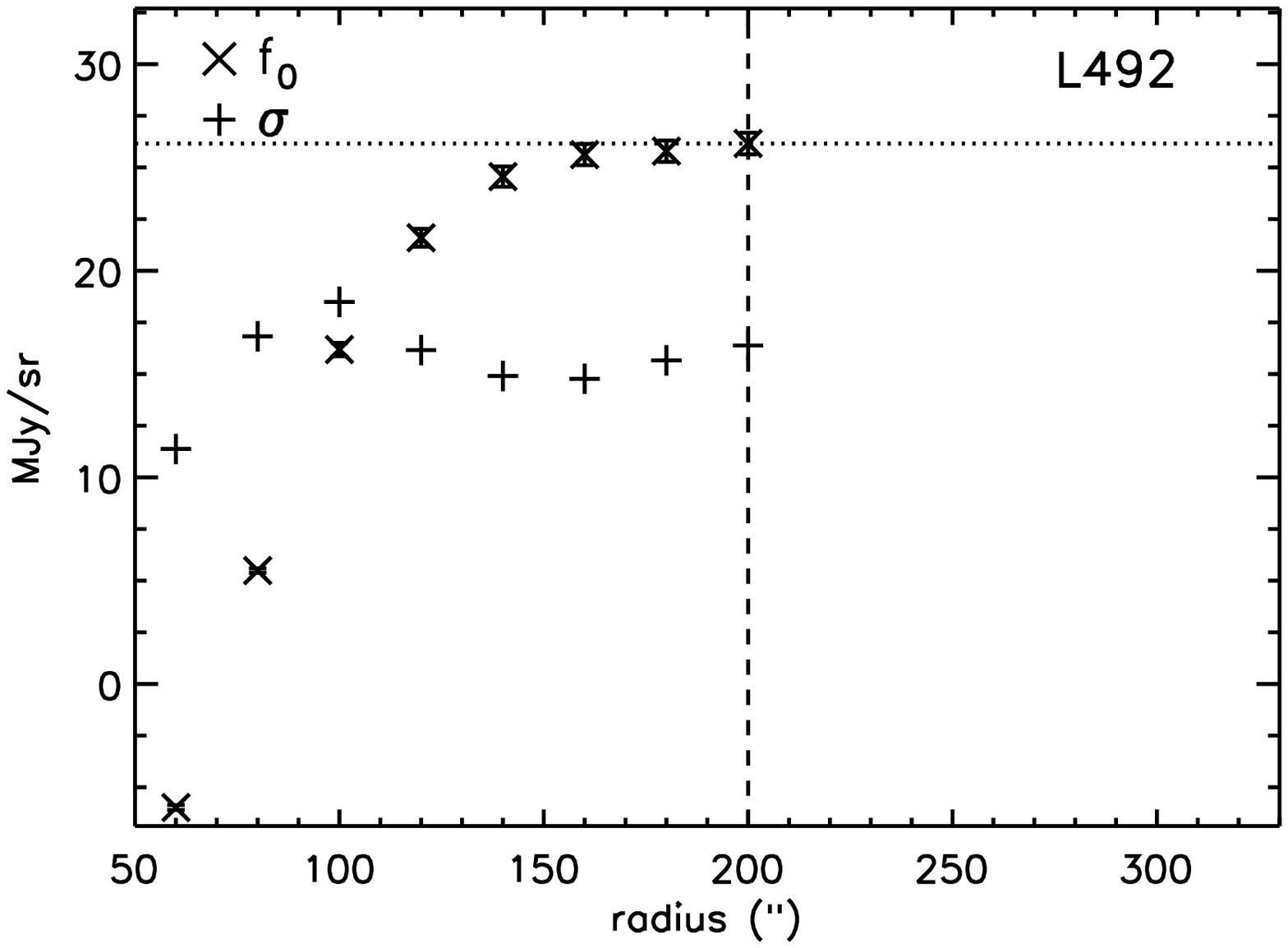}}

    \scalebox{0.5}{\includegraphics{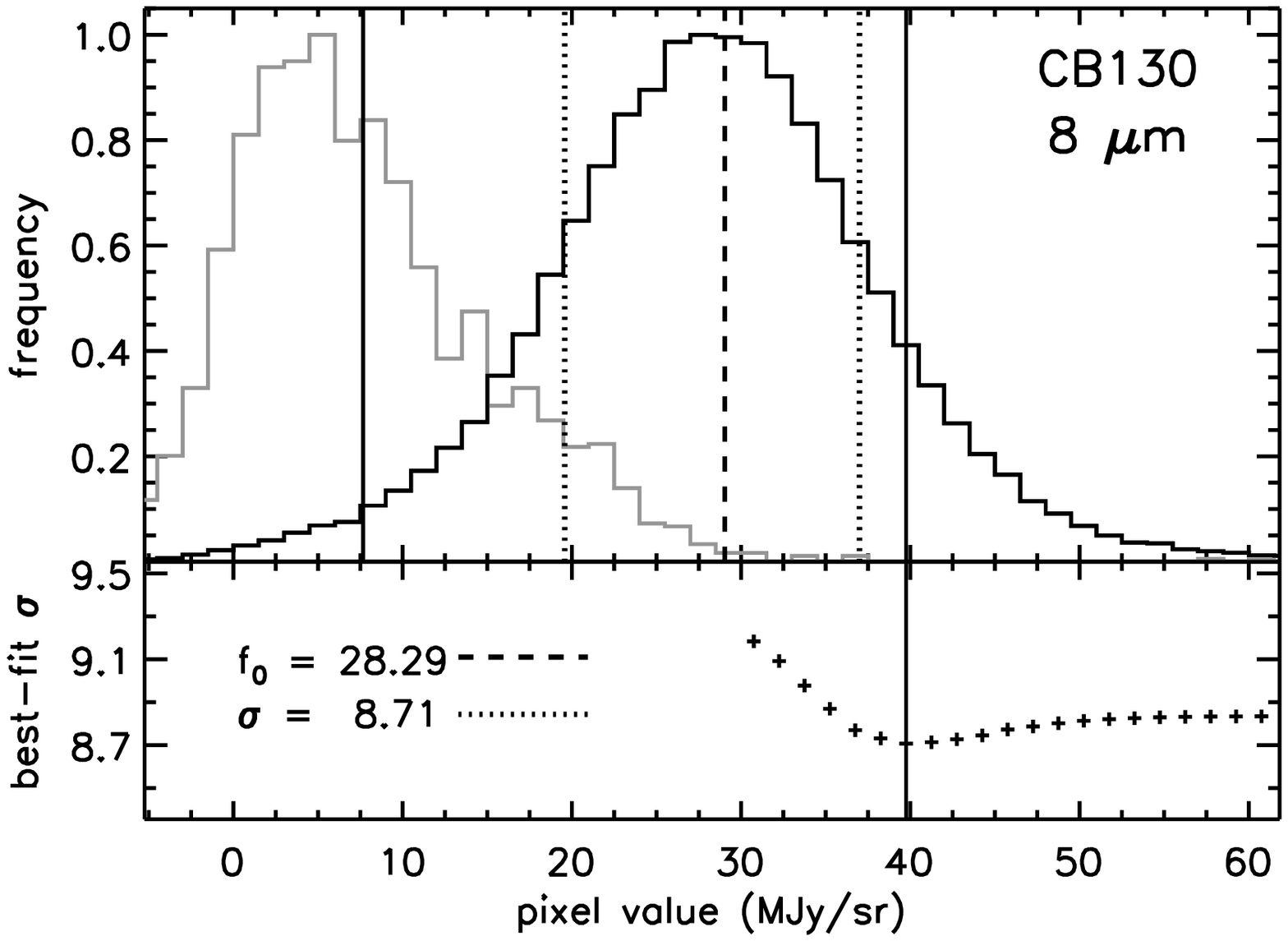}\includegraphics{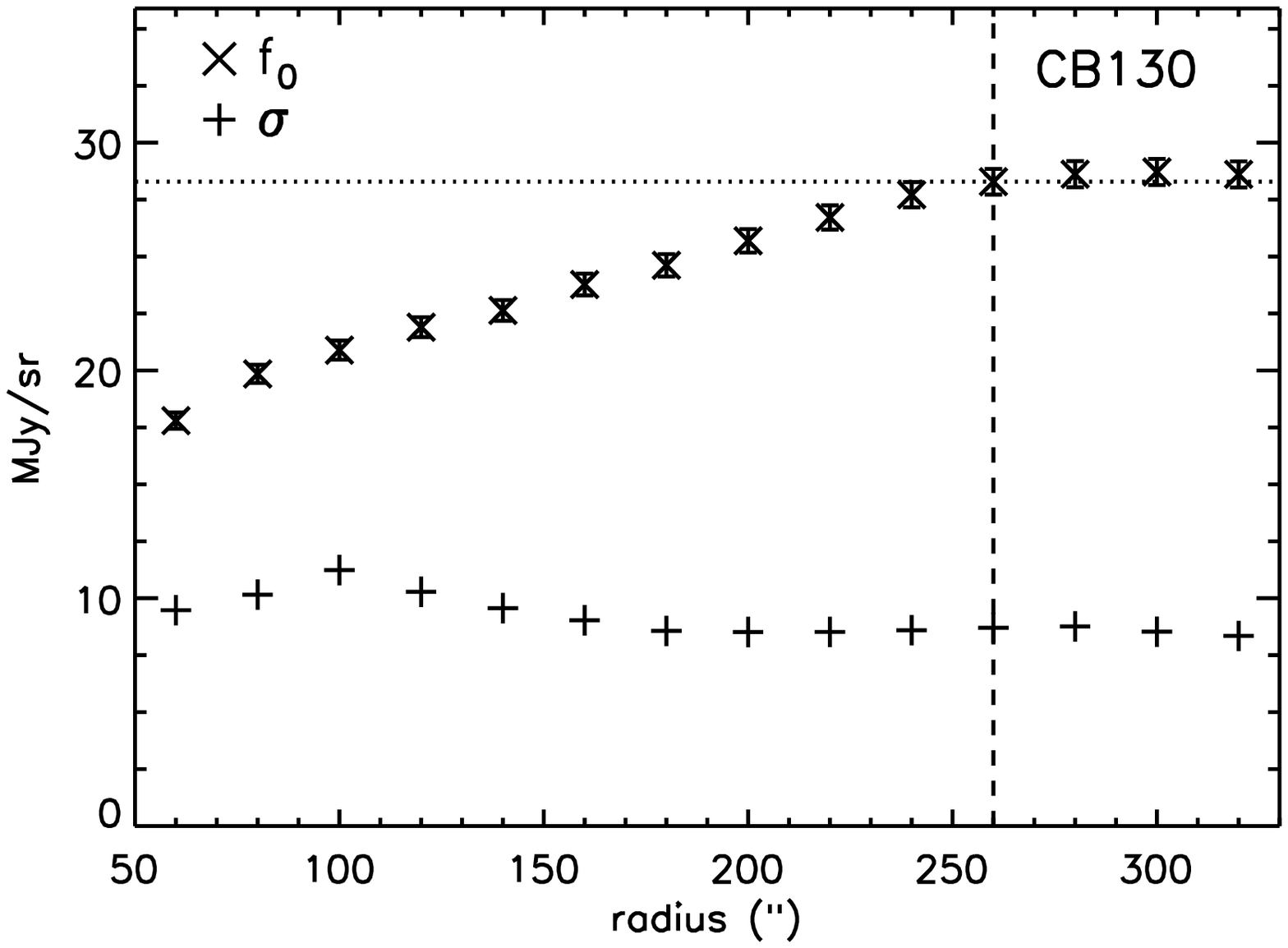}}
    \caption{Same as Figure~\ref{fig:f081}}
    \label{fig:f083}
  \end{center}
\end{figure}

\begin{figure}
  \begin{center}
    \scalebox{0.5}{\includegraphics{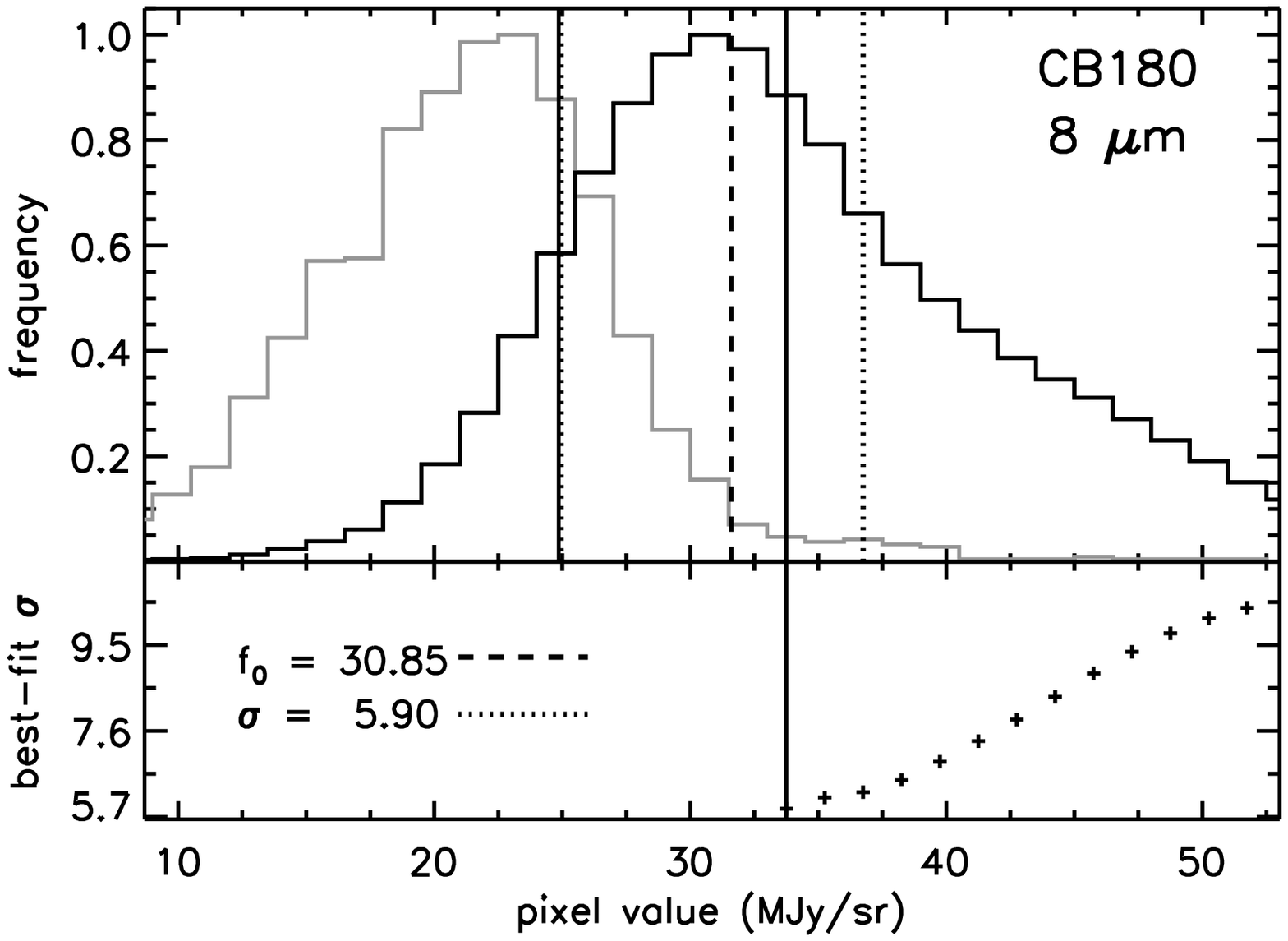}\includegraphics{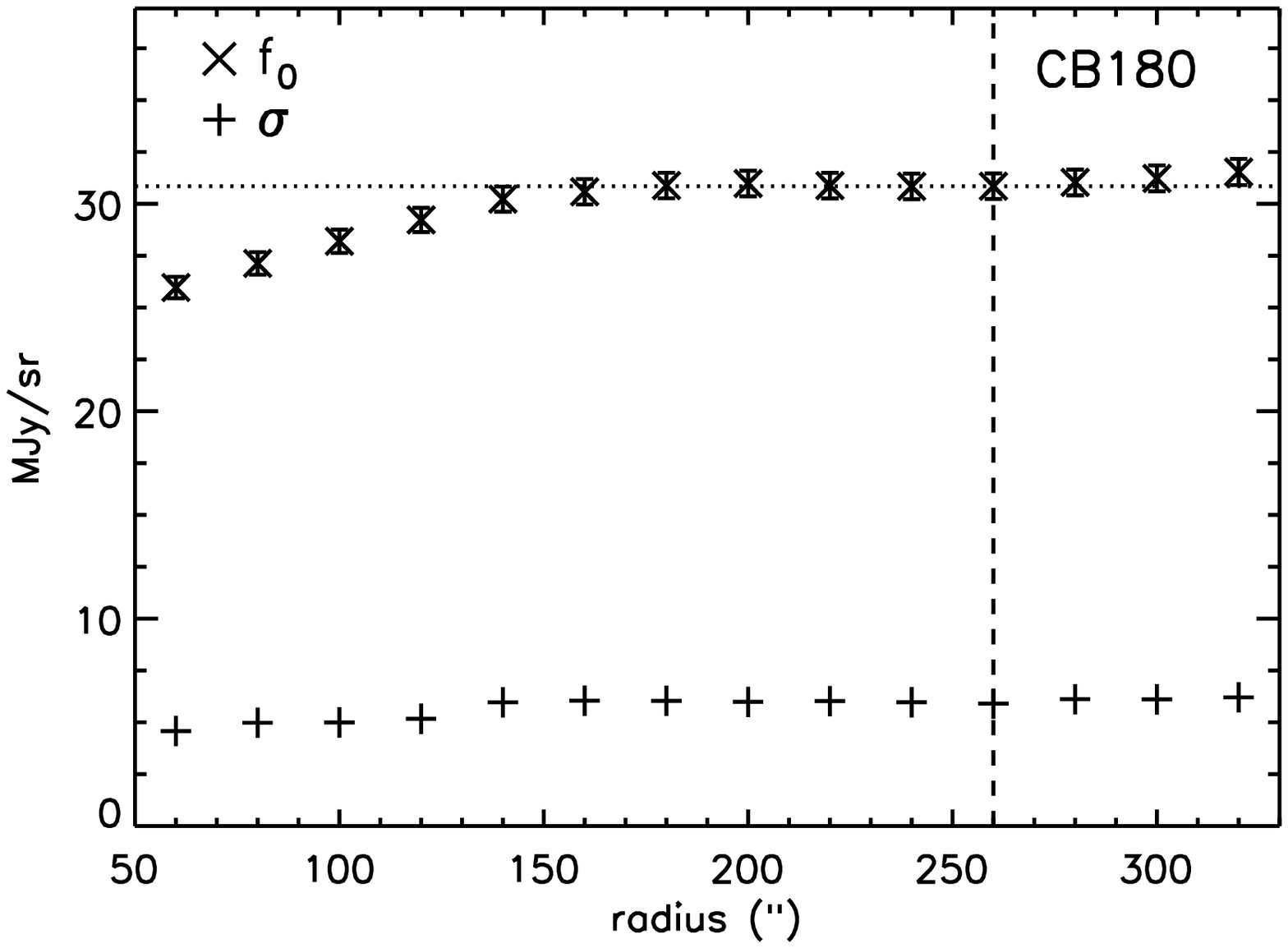}}
 
   \scalebox{0.5}{\includegraphics{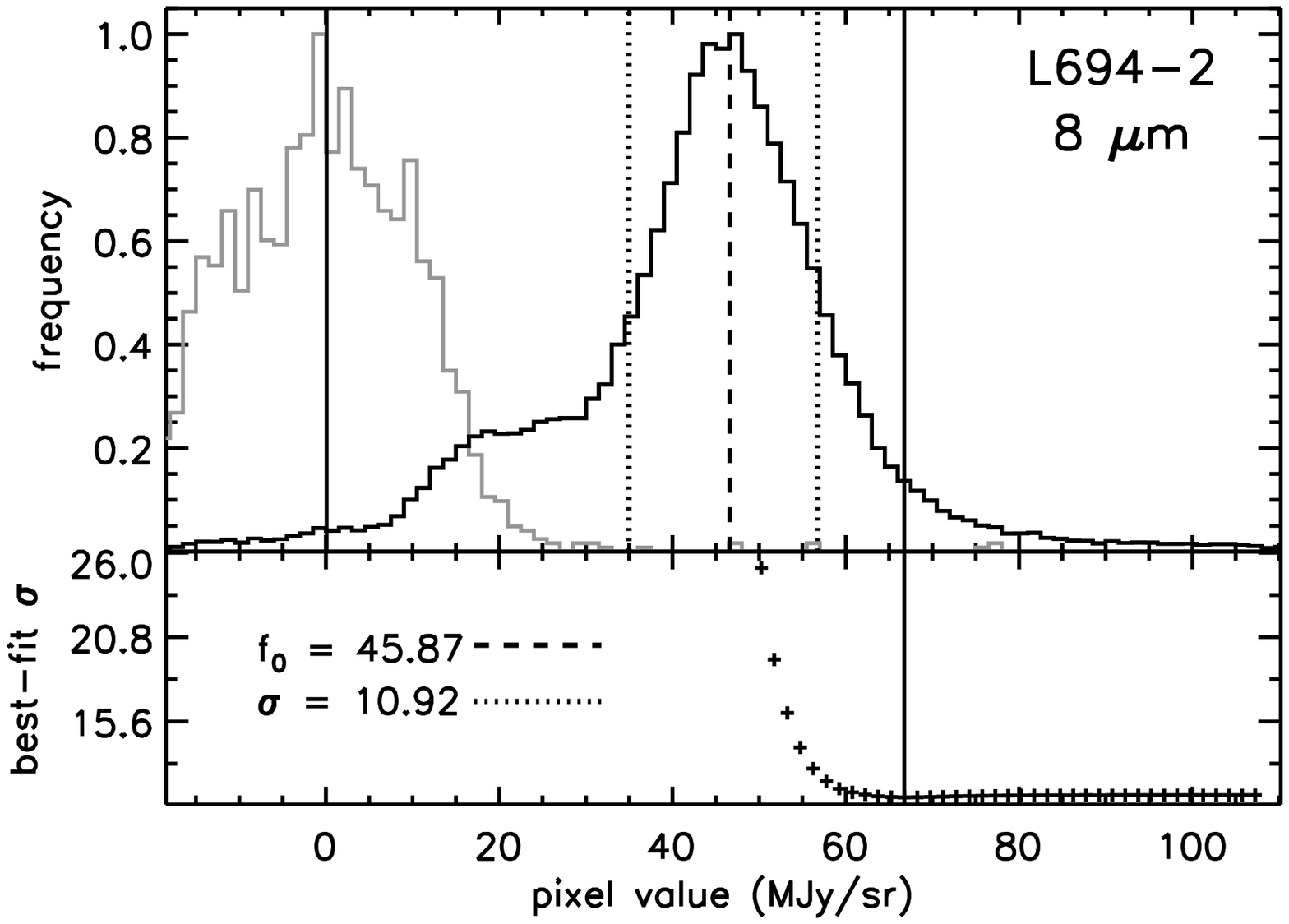}\includegraphics{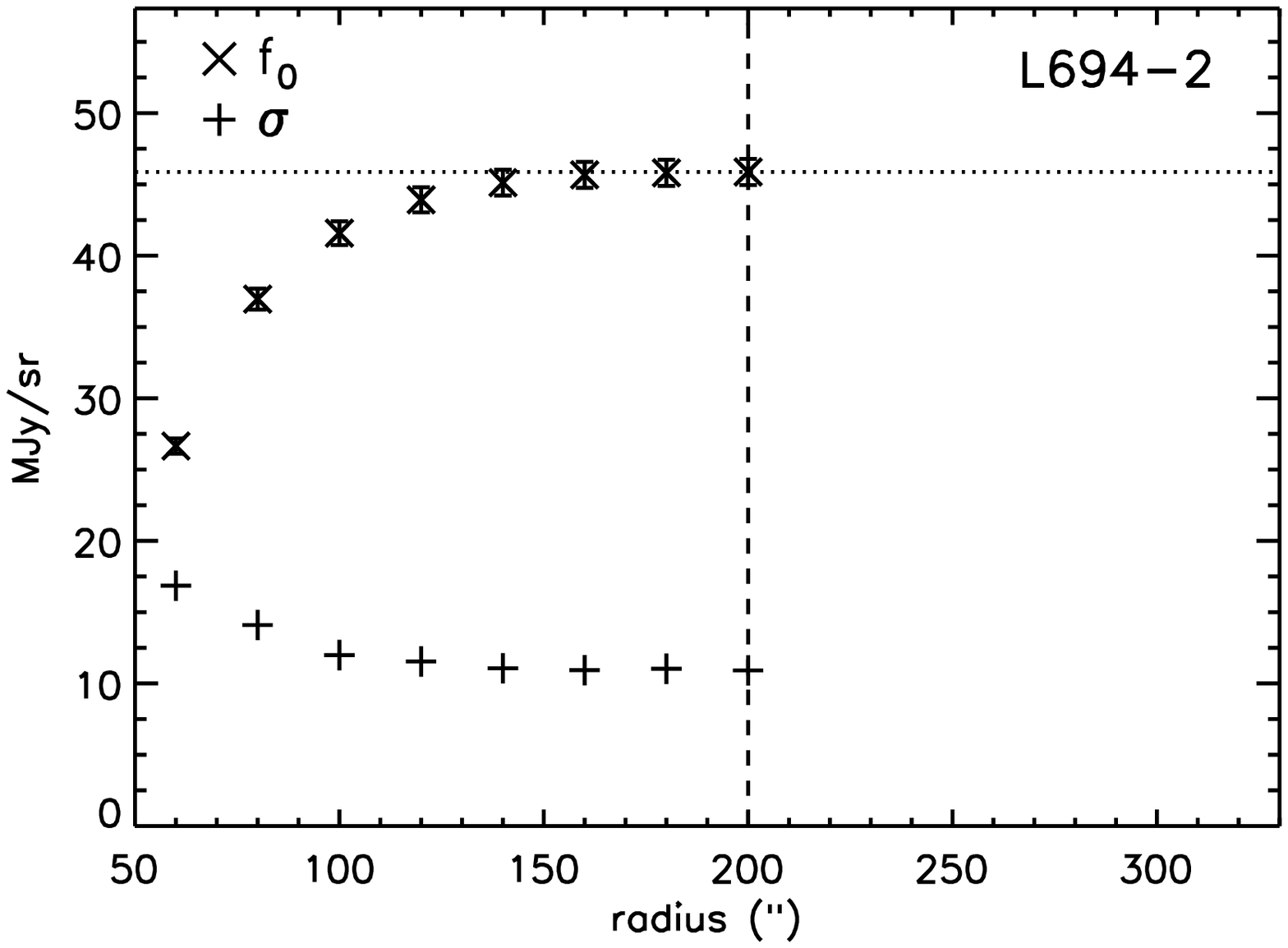}}
    \caption{Same as Figure~\ref{fig:f081}}
     \label{fig:f084}
  \end{center}
\end{figure}

We apply the same techniques described above for the 24~\micron\
images to the 8~\micron\ images, with the exceptions described
here\footnote{There is no IRAC coverage of source CB42}. The results
are illustrated in Figures~\ref{fig:f081}--\ref{fig:f084}. For the
$f_{\rm DC}$ determination, it is not possible to use the same regions
as those used for the 24~\micron\ images because of the limited
coverage of the 8~\micron\ images; therefore we select new regions
that are near the 24~\micron\ regions when possible.  In all cases the
8~\micron\ $f_{\rm DC}$ regions are selected to include the darkest
regions of the image; we set $f_{\rm DC}$ equal to the 0.1 percentile.
The background region box size is $5\arcmin$ on a side, corresponding
to $\sim\!60,000$ pixels at the plate scale of the IRAC images
($1\farcs22$~pix$^{-1}$).  The 8~\micron\ error is assumed to be 2\%
of the pixel flux value; the background error is estimated as
described above.  These two quantities added in quadrature make the
background--subtracted pixel error map.  The method used to determine
$f_0$ is identical to that described above, the only exception being
that for some objects (CB23, L1544, L492, and L694-2) the maximum
radius for which we can measure the pixel value distribution is
limited by the size of the images.  For these four objects, we chose
the largest radius measurement of $f_0$ for which over $50\%$ of the
area of the annulus contains image pixels.  For these cores the
maximum radius is $r_0 = 240\arcsec$ for CB230, and $r_0 = 200\arcsec$
for the rest.  For cores with sufficiently large images, we observe a
flattening of the $f_0$ profiles at a radius of $r_0 = 260\arcsec$;
the adopted values for $r_0$ are indicated as dashed lines in the
right column of Figures~\ref{fig:f081}--\ref{fig:f084}.  In the cases
of cores CB23 and CB27, we do not observe a flattening; therefore the
$\tau_8$ maps for these sources may be less reliable.

As described above for the 24~\micron\ images, we use our measurements
of $f_0'$, $f'$, and $f_{DC}$ measured form the 8~\micron images to
calculate
$$
\tau_{8} = -ln \left( \frac{f' - f_{\rm DC}}{f_{0}' - f_{\rm DC}} \right)
$$
in each pixel; we
also calculate $\sigma_{\tau,8}$ with standard propagation of errors
and including $\sigma_{\rm DC}$ and $\sigma(f_0)$, where $\sigma(f_0)
= 0.02 \times f_0$.  Divergent pixels are assigned a value according
to the procedure outlined above, are flagged, and are accounted for in
subsequent analysis.

We detect 3.6~\micron\ emission above the sky level in a large
fraction of our cores (TMC1-1C, CB23, CB26, CB27, L1544, L1552, and
L694-2, and marginally in L492, CB130, and CB180).  This emission is
spatially coincident with the shadows, and similar in extent to the
160~\micron\ emission.  The lack of any significant
8~\micron\ aromatic feature indicates that this emission is likely to
be caused by the surface layer of dust grains scattering the
interstellar radiation field, analogous to the near-IR emission
studied by \citet{lehtinen96}.  Such near-IR emission has been used to
study interstellar clouds
\citep[e.g.][]{foster06,padoan06,nakajima08}.  \citet{juvela08} show
that when using $J$, $H$, and $K_S$ imaging this technique is
sensitive to densities of $1 \lesssim A_V \lesssim 15$ magnitudes.
Including the 3.6~\micron\ emission might increase the densities
accessible to this technique, as well as providing a high-spatial
resolution shadow-independent extinction estimator.  We note that for
the cores L1544, CB44, and L694-2 the emission shows a
limb-brightening geometry at 3.6~\micron, where the center of the
shadow region appears darker than the surrounding region.

The 8~\micron\ shadow determinations provide a test of the validity of
those at 24~\micron.  In all cases where a good comparison is
possible, the agreement is satisfactory if we take the optical depths
at the two wavelengths to be approximately equal, $\tau_{24}/\tau_8 =
1.0 \pm 0.1$ \citep[see also][]{flaherty07,chapman09,chapman09b}.
Additional details will be provided in a future publication.

\subsection{Distance measures}

Here we make some remarks on our adopted distances, which are
summarized in Table~3.  

\noindent {\bf TMC1-1C, CB23, CB26, CB27, L1544, L1552 ---}
\citet{kenyon94} derive a distance to the northern portion Taurus of
$140\pm 10$~pc based on extinction of background stars.  This distance
is roughly consistent with recent determinations using trigonometric
parallax with the VLBA \citep{torres09}.

\noindent {\bf CB42 ---} We find no distance estimate for this source
in the literature.  In the direction of CB42, the extent of the local
bubble is $\sim\!100$~pc \citep{lallement03}; we adopt a distance of
$100$~pc, a conservative lower limit distance.

\noindent {\bf CB44 ---} \citet{hilton95} list the \citet{bok74}
assumed distance of 400~pc.  They also list the \citet{tomita79}
estimate of 300~pc, and the \citet{arquilla86} distance estimate of
400 to 600~pc.  We adopt a distance of $400$~pc.

\noindent {\bf L492, CB130 ---} \citet{staizys03} investigate the
distance dependence of the interstellar extinction in the direction of
the Aquila Rift.  These authors find that the distance is $225 \pm
55$~pc to the front edge and that the cloud has a depth of about
80~pc.  Because these cores are viewed in absorption they are unlikely
to be located behind much cloud material; therefore we adopt $225$~pc
as the distance to these cores.

\noindent {\bf CB180 ---} \citet{hilton95} list the \citet{bok74}
assumed distance of 400~pc. \citet{bok74} do not elaborate on their
adopted distances.  This object is near the Scutum association.  We
adopt $400$~pc for the distance to this core.

\noindent {\bf L694-2 ---} \citet{kawamura01} derive a distance of
$230 \pm 30$~pc using star counts.

\section{Discussion}

\subsection{Interpretation of Collapsing Cores}

\begin{figure}
  \begin{center}
    \scalebox{0.85}{\includegraphics{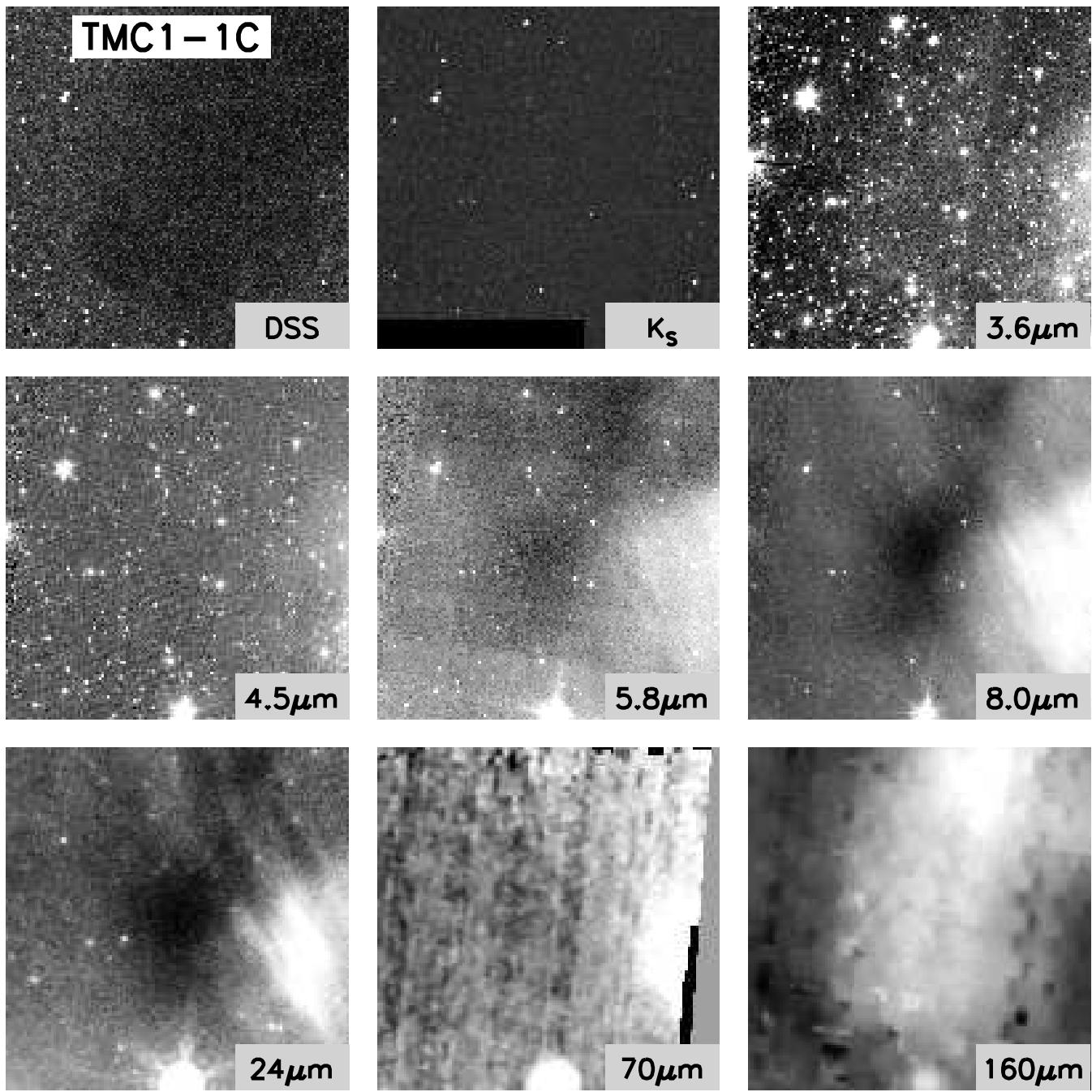}}
    \caption{$8\arcmin \times 8\arcmin$ images at the indicated
      wavelengths, shown on a log scale.  North is up, east is to the
      left. {\it For full resolution figures contact stutz@mpia.de.}}
    \label{fig:img1}
  \end{center}
\end{figure}

\begin{figure}
  \begin{center}
    \scalebox{0.85}{\includegraphics{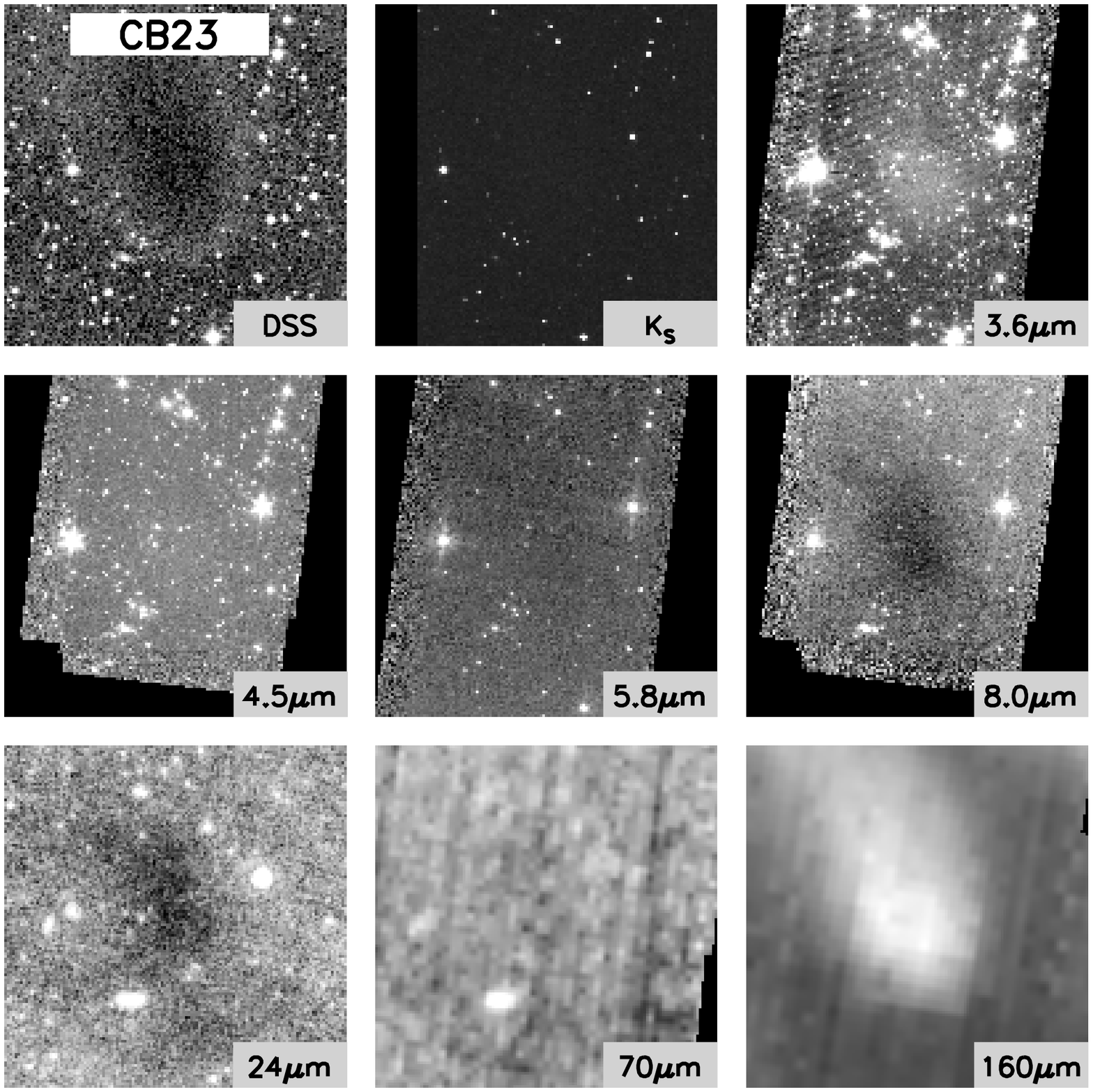}}
    \caption{Same as Figure~\ref{fig:img1}}
    \label{fig:img2}
  \end{center}
\end{figure}

\begin{figure}
  \begin{center}
    \scalebox{0.85}{\includegraphics{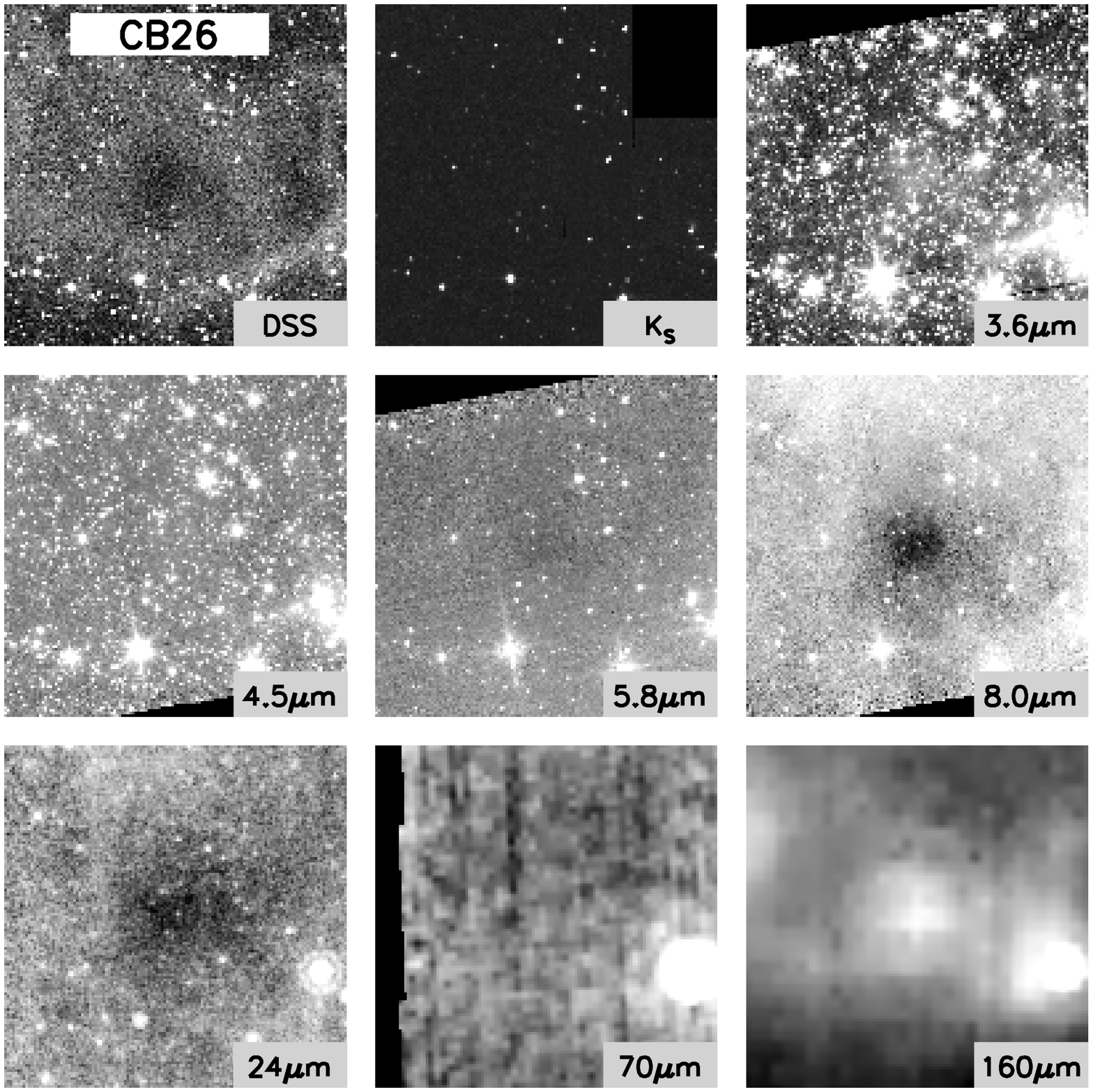}}
    \caption{Same as Figure~\ref{fig:img1}}
    \label{fig:img3}
  \end{center}
\end{figure}

\begin{figure}
  \begin{center}
    \scalebox{0.85}{\includegraphics{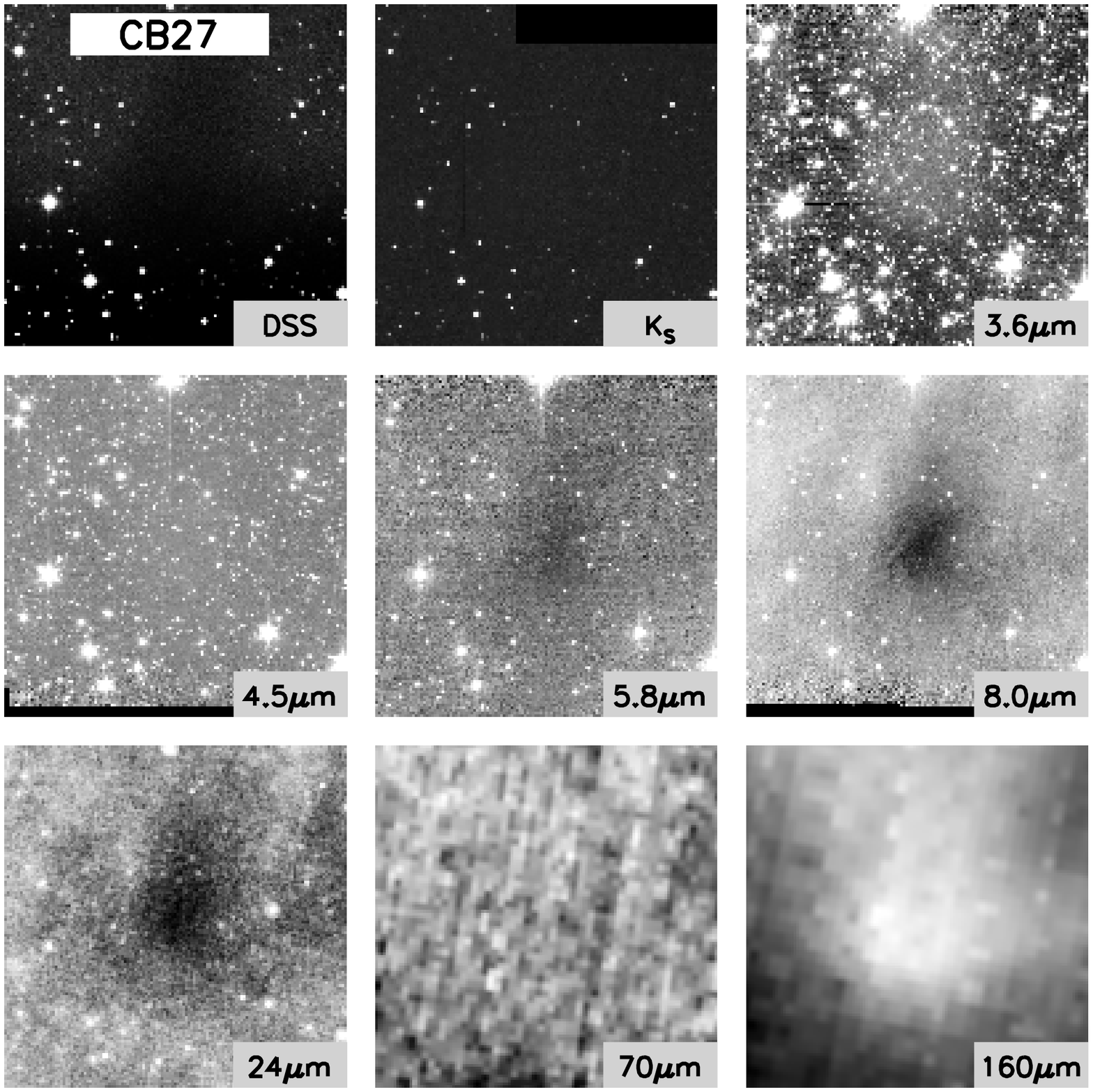}}
    \caption{Same as Figure~\ref{fig:img1}}
    \label{fig:img4}
  \end{center}
\end{figure}

\begin{figure}
  \begin{center}
    \scalebox{0.85}{\includegraphics{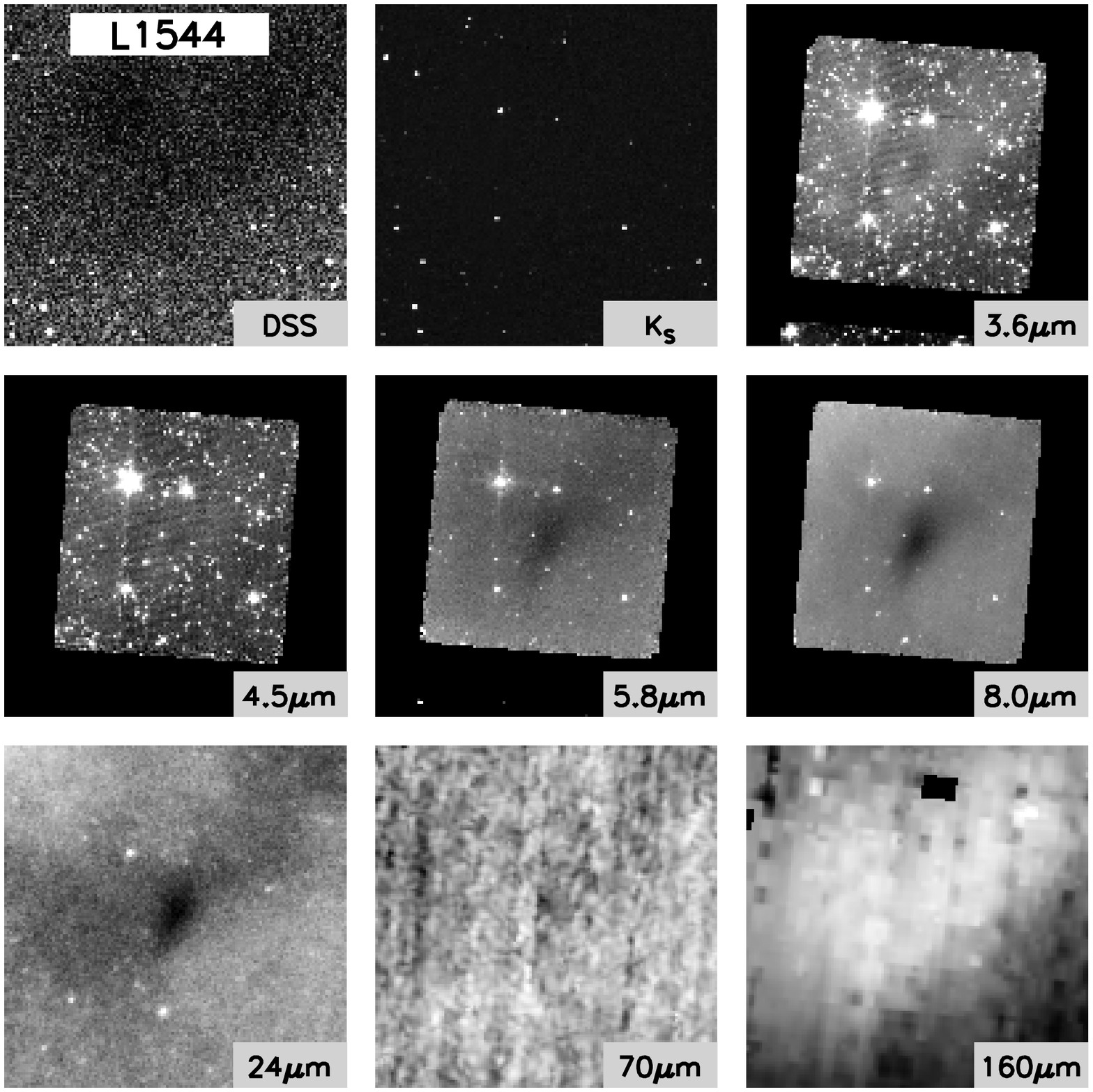}}
    \caption{Same as Figure~\ref{fig:img1}}
    \label{fig:img5}
  \end{center}
\end{figure}

\begin{figure}
  \begin{center}
    \scalebox{0.85}{\includegraphics{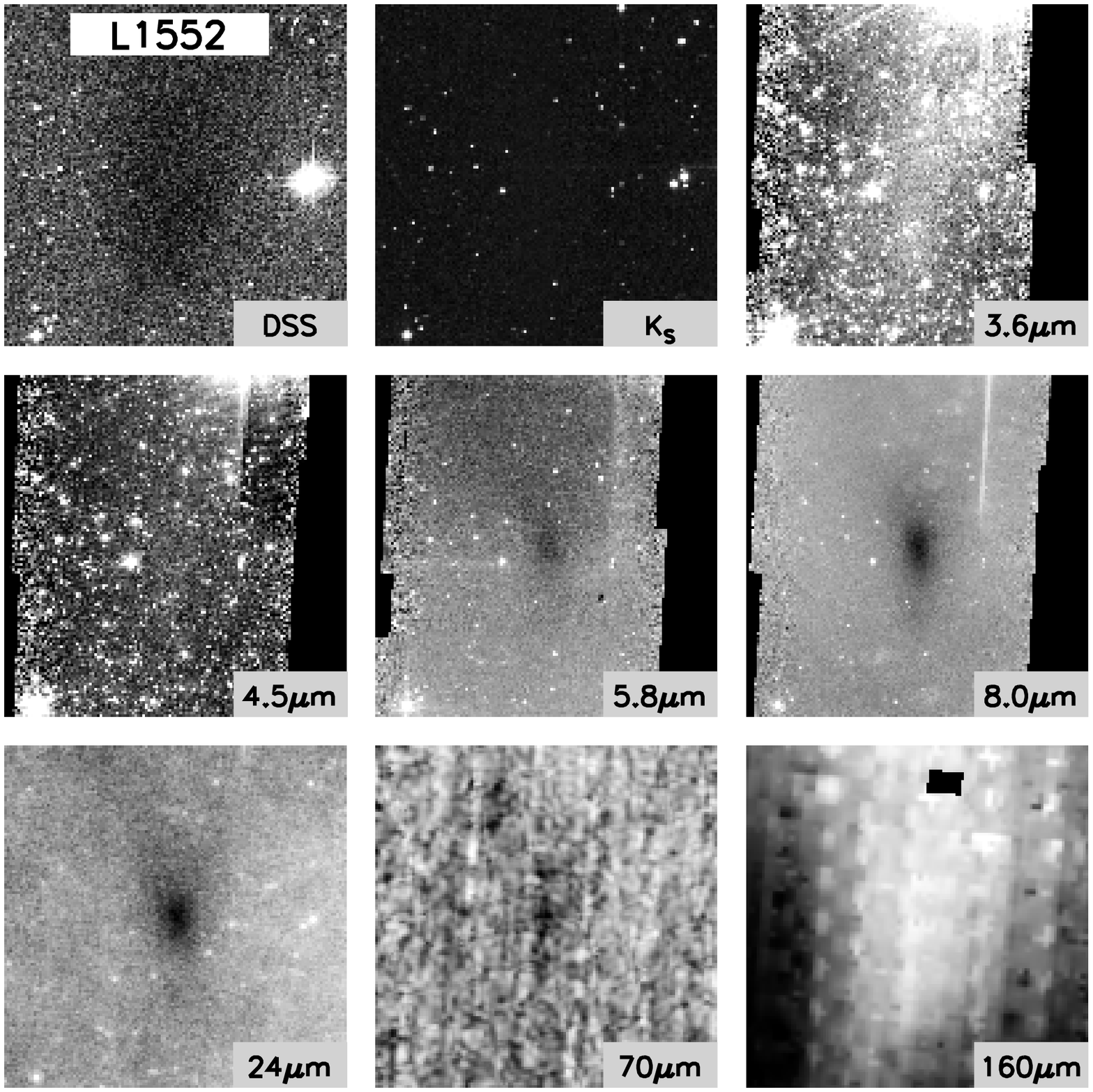}}
    \caption{Same as Figure~\ref{fig:img1}}
    \label{fig:img6}
  \end{center}
\end{figure}

\begin{figure}
  \begin{center}
    \scalebox{0.85}{\includegraphics{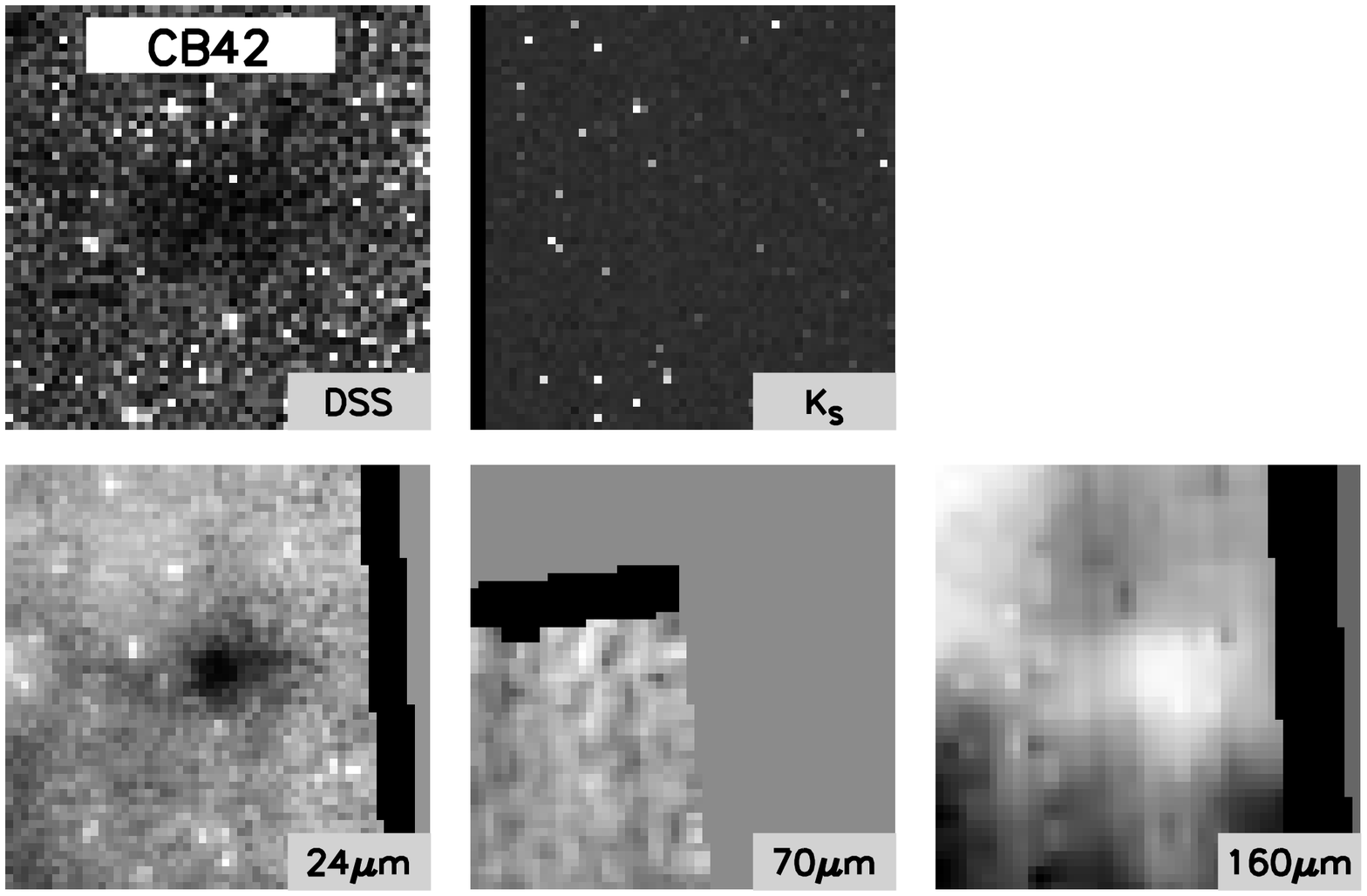}}
    \caption{Same as Figure~\ref{fig:img1}}
    \label{fig:img7}
  \end{center}
\end{figure}

\begin{figure}
  \begin{center}
    \scalebox{0.85}{\includegraphics{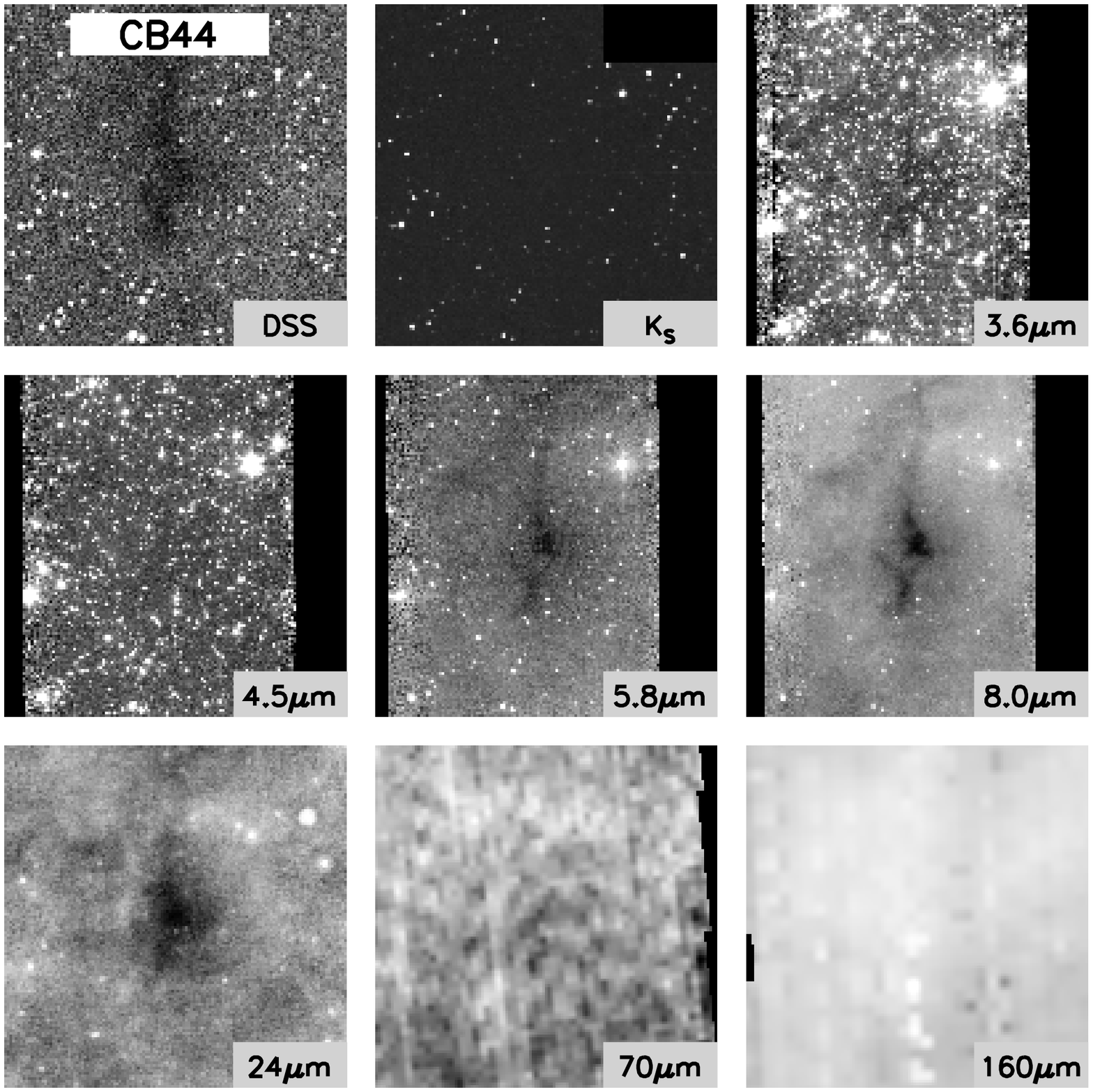}}
    \caption{Same as Figure~\ref{fig:img1}}
    \label{fig:img8}
  \end{center}
\end{figure}

\begin{figure}
  \begin{center}
    \scalebox{0.85}{\includegraphics{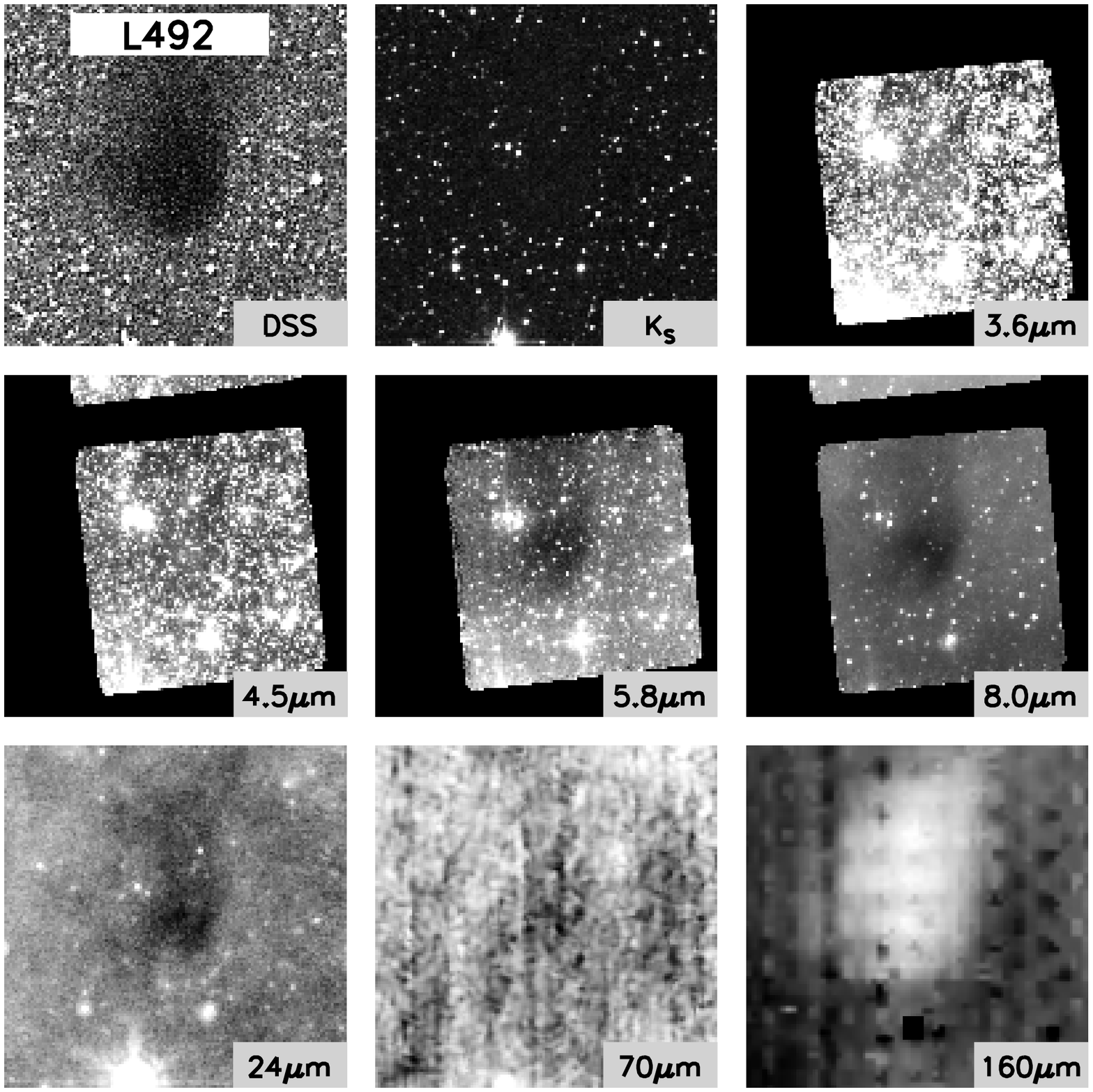}}
    \caption{Same as Figure~\ref{fig:img1}}
    \label{fig:img9}
  \end{center}
\end{figure}

\begin{figure}
  \begin{center}
    \scalebox{0.85}{\includegraphics{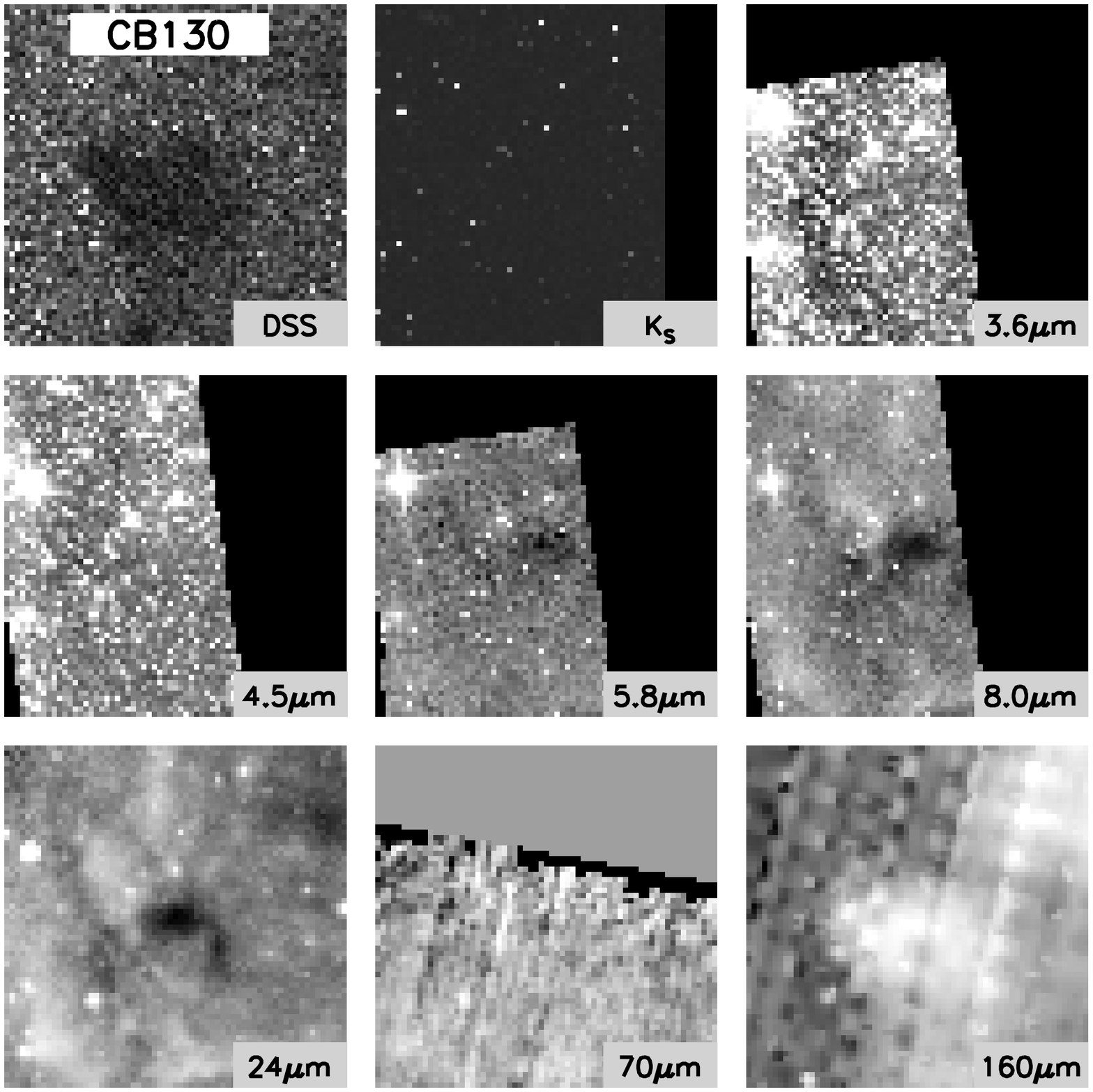}}
    \caption{Same as Figure~\ref{fig:img1}}
    \label{fig:img10}
  \end{center}
\end{figure}

\begin{figure}
  \begin{center}
    \scalebox{0.85}{\includegraphics{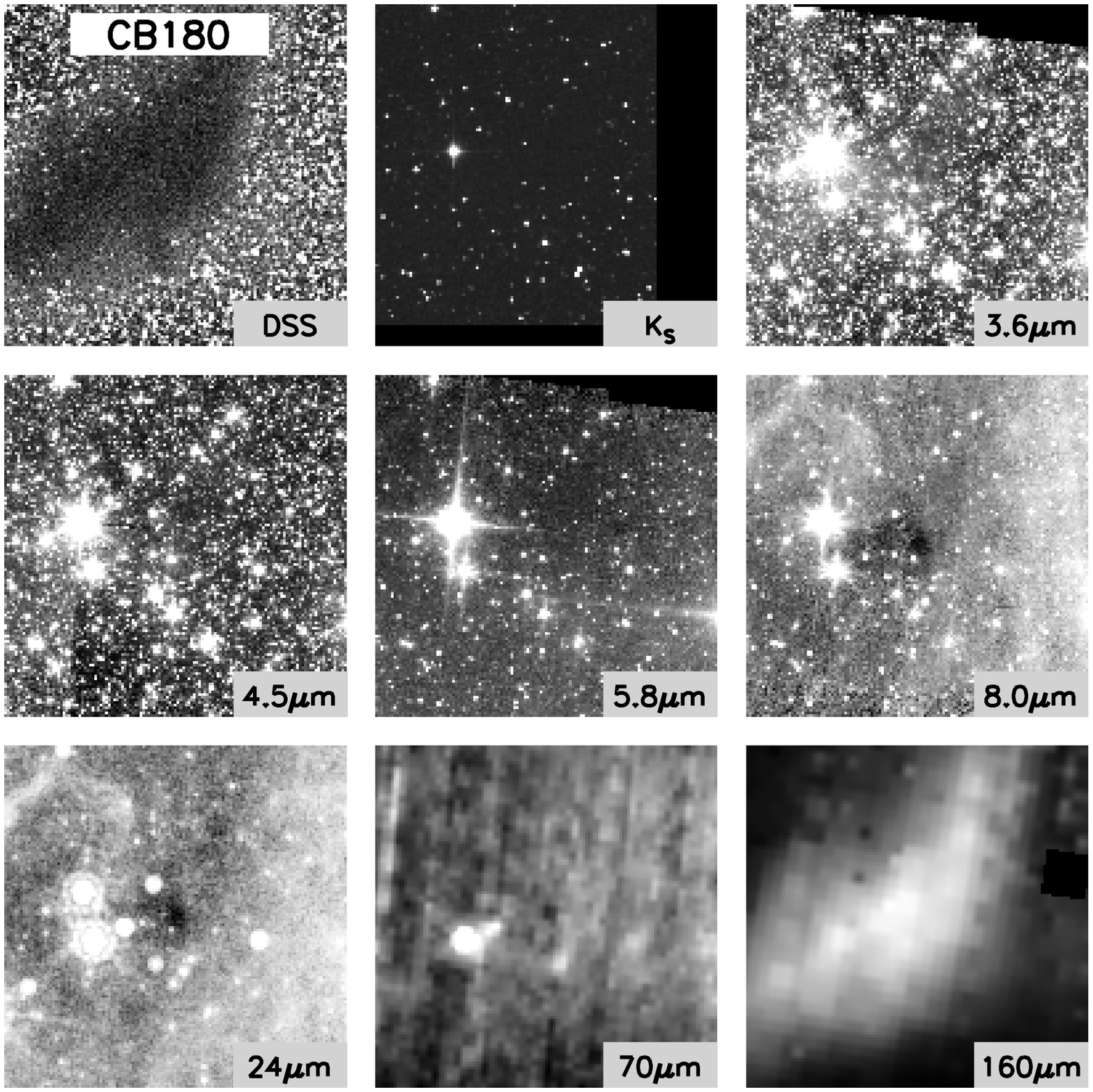}}
    \caption{Same as Figure~\ref{fig:img1}}
    \label{fig:img11}
  \end{center}
\end{figure}

\begin{figure}
  \begin{center}
    \scalebox{0.85}{\includegraphics{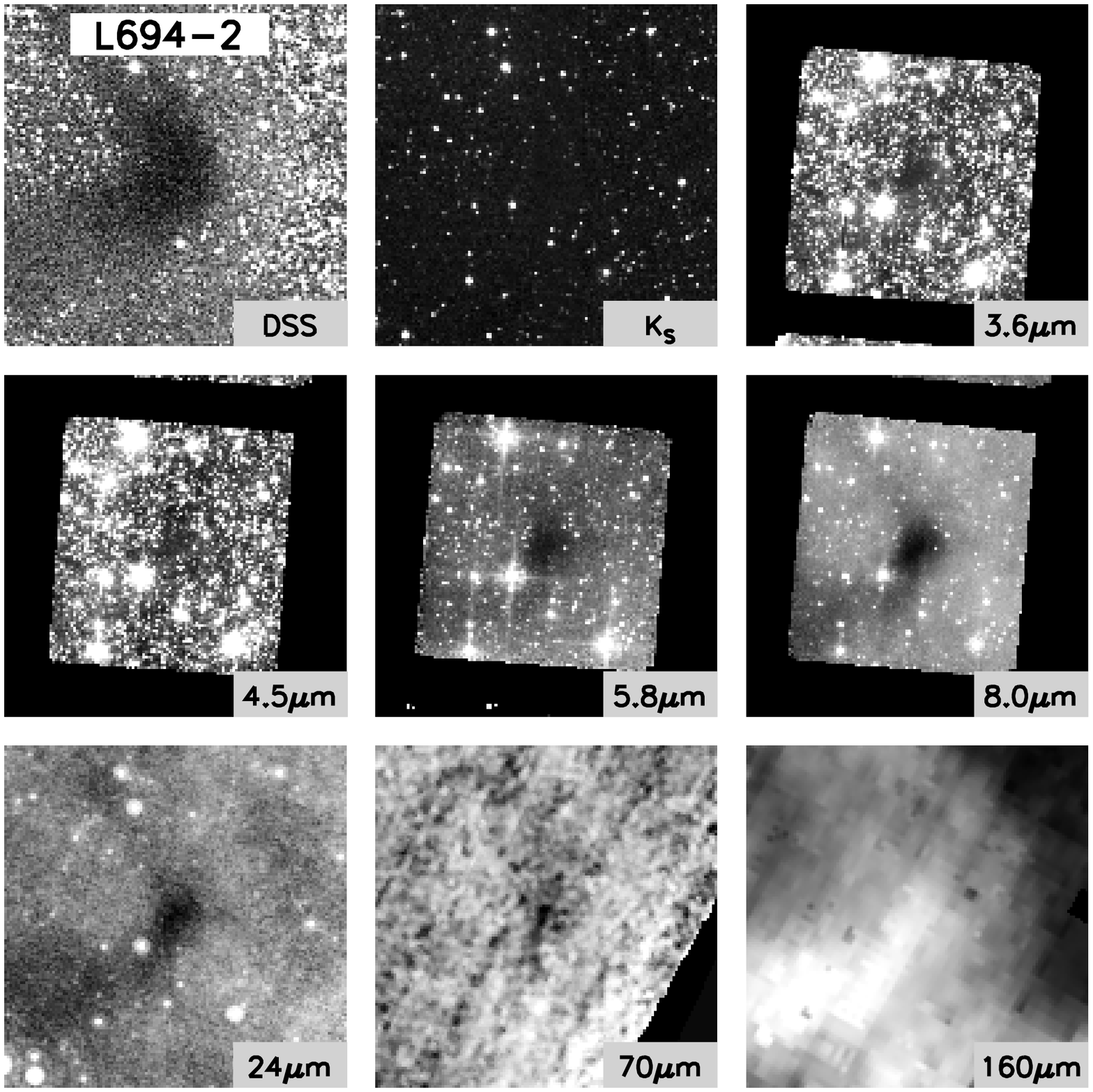}}
    \caption{Same as Figure~\ref{fig:img1}}
    \label{fig:img12}
  \end{center}
\end{figure}

The stability of a cloud core is usually deduced through observations
of millimeter--wave molecular emission lines
\citep[e.g.,][]{gregersen97,mardones97,gregersen00,crapsi05,williams06,sohn07},
but the dynamical state of these lines can be affected by rotation,
turbulence, optical depth, and other effects that can hide or mimic
the indicators for collapse.  We use infrared shadows, together with
millimeter--wave spectroscopy, as an alternative, complementary, and
reliable indicator of impending collapse
\citep[e.g.,][]{stutz07,stutz09}.  These observational techniques can
be used to estimate the masses and dynamical states of the cores.
Conventionally, collapse is identified by fitting such observations to
Bonnor-Ebert model profiles which assume spherical symmetry
\citep[e.g.,][]{evans01,ballesteros03,shirley05}. However, as shown in
Figures~\ref{fig:img1} through \ref{fig:img12}, few of our objects
appear symmetric on the plane of the sky, calling the applicability of
this approach into question. 

Instead, we have probed the collapse state of these cores on the basis
of a simple Jeans mass calculation.  In what follows, we determine
24~\micron\ core masses and sizes.  Given these core properties, and
assuming a temperature of 10~K (reasonable for starless cores) we
derive a Jeans mass for each core:
\begin{equation}
M_J = \left( \frac{5 k T}{G \mu_p m_H} \right )^{3/2} R^{3/2} M^{-1/2}, 
\end{equation}
where $M$ is the 24~\micron\ core mass (M$^{\rm core}_{24}$ in
Table~4), R is the effective radius of the core (R$^{\rm core}_{\rm
  eff}$ in Table~4), measured from the 24~\micron\ image, and we have
assumed that the mean molecular mass per free particle $\mu_p$ is
equal to 2.4.  The sizes of the cores are derived from the area in
which we detected the 24~\micron\ shadows (see \S~4.2.1).  We note
that assuming a temperature of 8~K or 12~K for these cores cause a
$\sim$30\% change in the calculated Jeans mass.
The distance dependence of the Jeans mass is D$^{1/2}$.  In the
following analysis (see Tables 4 and 5) we compare shadow masses, with
a distance dependence of D$^2$, to the Jeans mass.  We caution that
the ratio of the 24~\micron\ masses to the Jeans mass has a steep
distance dependence of D$^{3/2}$.  This Jeans mass analysis avoids
possible biases introduced by imposing more stringent geometrical
assumptions in the core analysis.  However, it only takes thermal
support into account.

Under more realistic conditions, turbulence might also provide support
against collapse.  From \citet{hotzel02}, the energy contributed by
turbulence to the support of the cloud is
\begin{equation}
  E_{turb} = \frac{\mu_p}{AMU}\left( \frac{\Delta V^2}{\Delta V_{th}^2}
  - 1 \right)E_{th},
\end{equation}
where $AMU$ is the atomic mass number of the species from which we
measure the observed line width $\Delta V$, and $\Delta V_{th}$ is the
expected thermal line width for that species, assuming a temperature
of 10~K.  To estimate possible contributions of turbulence to the
support in the cores, we use previously measured $\Delta V$ values for
high density tracers from the literature when available
\citep[e.g.,][]{lee04,crapsi05,sohn07}.  When such measurements are
not available, we use other line widths as estimators, noting that
lower density tracers will overestimate the line width and therefore
the contribution from turbulence to the support.  Often, the turbulent
support is much less than the thermal support and hence can be ignored
in considering the stability against collapse.  In the cases where
turbulent support is potentially significant, we attempt to assess its
contribution relative to the Jeans mass and magnetic field required
for stability.

Magnetic fields may also contribute to the support.  The inferred
magnetic field strengths in cores range from $\lesssim 15$~$\mu$G to
$\sim\!160$~$\mu$G \citep{bergin07}.  From \citet{stahler05}, given a
magnetic field strength $B$ and a cloud radius $R$, the mass that can
be supported is
\begin{equation}
M = 70\msun \left( \frac{B}{10\mu G} \right) \left( \frac{R}{1pc}
\right)^2.
\end{equation}
Using this relation, and a measurement of the core mass and size, we
estimate the minimum magnetic field strength required to support the
core.  If the cores require field strengths above $\sim\!160$~$\mu$G,
we consider it unlikely that the core is supported by a magnetic
field.  Furthermore, if the minimum magnetic field derived in this
fashion is reasonable, but the core mass is significantly greater than
the Jeans mass (by a factor $\gtrsim 2$), we consider it unlikely that
the core is stable against collapse.  Our understanding of the
magnetic field strengths in cores does not allow for strong
constraints.  However, this qualitative assessment gives a useful
estimate of the possible role of the magnetic field.

Finally, rotation can also contribute to the support in cores.  In
column 9 of Table~4 we list the rotational velocity required to fully
support our sample of cores.  Future high--resolution observations
will determine the extent of the role of rotation in these cores.  In
the analysis that follows, we interpret the line widths as caused only
by turbulent and thermal motions; more realistically, turbulent
motions on large scales can be confused with rotation
\citep{burkert00}.  Furthermore, irregular infall, or fragments, can
mimic turbulence; inspection of Figures~\ref{fig:img1}
through~\ref{fig:img12} reveals that a significant fraction of cores
have geometries which are not regular, reminiscent of structures seen
in self-gravitating 2D \citep{burkert04} and 3D \citep{heitsch08}
simulations.  Because of these irregular shapes, we have attempted to
minimize geometrical assumptions in the analysis of our sample of
cores, as discussed above; however, the stability arguments applied
here inherently still assume approximate sphericity.

\subsection{Mass Measurements}

We describe the method used to calculate masses in the cores using the
24~\micron\ shadows, and in the regions surrounding the cores using
the $^{12}$CO and $^{13}$CO data.  The results are summarized in
Table~4.  We also list the Jean mass for each core, assuming all the
cores are at a temperature of 10~K and that the core density is
described by the shadow mass (M$_{24}^{\rm core}$) and the effective
radius (R$_{\rm eff}^{\rm core}$; see below).

\subsubsection{Core Masses: Shadow Masses}

We calculate masses in the cores applying a similar method as that of
\citet{stutz07}.  Core shadow masses, assumed distances, and effective
areas are listed in Table~4.  In these calculations, we use the
24~\micron\ images and 8~\micron\ dust opacities from
\citet{ossenkopf94}: $\kappa_{8} = 1.21 \times 10^3$~cm$^2$~gm$^{-1}$
for thick ice mantles at a density of $10^6$~cm$^{-3}$.  This choice
of dust opacity is motivated by our analysis in \S~3.1.2 where we find
that $r_{24/8} = \tau_{24}/\tau_{8} = 1.0 \pm 0.1$.  This choice of
\citet{ossenkopf94} dust model has a ratio of $\tau_{24}/\tau_{8} =
1.1$, close to our measured value.

The mass in a region with an optical depth $\tau_{24}$ is: 
\begin{equation}
M_{24} = \frac{\Omega D^2}{\kappa_8} \alpha \times \tau_{24},
\end{equation}
where $\Omega$ is the solid angle subtended by the region, $D$ is the
distance, $\alpha (=100)$ is the assumed gas--to--dust ratio, and
$\kappa$ is the assumed dust opacity.  We measure the mean flux in an
aperture of $1\farcm5$ radius, excluding all pixels that have
$\tau_{24} \leq 0.0$ in the maps described above.  Using the
measurements of $f_{\rm DC}$ and $f_0$ and the mean flux per pixel in
the shadow, we then calculate the mean $\tau_{24}$ per pixel.  Using
equation~2 and the number of pixels ($\Omega = N_{pix} \times$ pixel
scale$^2$) included in the mean flux calculation we calculate a shadow
mass.  We use the value of $\Omega$ in subsequent analysis as a
measure of the effective core sizes, where $R_{eff}^{core} = \sqrt
{\Omega/\pi}$.  The uncertainties in the masses measured in this
fashion are of order 15\%, without including distance uncertainties.
We cross--check these masses by calculating core masses directly from
the $\tau$--pixel maps.  We find that the two sets of masses agree
well; however, the mass errors are extremely low for the $\tau$--pixel
map masses.  We attribute this discrepancy to highly correlated
errors, as discussed in \S~3.1, and disregard the small uncertainties
in favor of the $\sim\!15$\% errors.  The agreement between the two
masses indicates that the method we use to estimate $f_{DC}$ (see
\S~3.1.1) does not have a large effect on the calculated core masses,
even accounting for divergent pixel values.

\subsubsection{Envelope Masses: Molecular Masses}

We use the $^{12}$CO (2--1) and $^{13}$CO (2--1) maps to derive the
total column density of hydrogen, N(H) in the area surrounding the
cores.  As listed in Table~2 and shown in Figures~\ref{fig:tauco}
through \ref{fig:tauco3}, the CO maps cover areas that are much larger
than the 24~\micron\ shadow area.  The effect of CO freeze--out onto
dust grains is severe \citep{bacmann02,tafalla02}; therefore, in the
area where we observe 24~\micron\ shadows, the column densities
derived from CO observations are likely to be underestimated.
However, in the regions outside of the shadows, where the densities
are much lower than in the cores themselves, the CO--derived N(H)
column should have a much higher degree of fidelity to the true column
in the observed region.  Therefore, we use the CO maps to derive
masses of the larger environment surrounding the cores, observed as
24~\micron\ shadows\footnote{The $^{13}$CO emission sometimes
appears to be bright along an edge of the shadow region (seen most
clearly in L1552 and CB44) and in one case we clearly observe a
deficit of emission in the region coincident with the shadow (L1544).
These characteristics are likely caused by the freeze--out of CO on to
dust grains, as well as by a non-uniform density and temperature
structure in the cores.}.

To derive masses from the CO data, we follow a similar procedure as
\citet{povich09} \citep[see also][]{kang09} and we apply the escape
probability radiative transfer model for CO from \citet{kulesa05}.
The $^{12}$CO (2--1) and $^{13}$CO (2--1) data cubes are used to
derive the column of CO in each $10\arcsec \times 10\arcsec$ map
pixel, assuming an isotope ratio of $^{12}$CO/$^{13}$CO = 50.  The
model uses the $^{12}$CO and $^{13}$CO observations to solve for the
column density of CO, N(CO), over a model grid of temperatures,
densities, and column densities; given an initial guess for the
density this procedure then fits simultaneously for the temperature
and column density in each map pixel.  The mass calculation also
incorporates the \citet{vandishoek88} PDR photodissociation model of
CO, intended to account for the effect of the far--UV field, $G_0$,
incident on the cores.  The final CO column that we calculate
therefore should include the effect of UV dissociation; if the
UV--field is not strong, the application of this model can cause an
over-estimate of the column densities.

\begin{figure}
  \begin{center}
    \scalebox{0.44}{\includegraphics{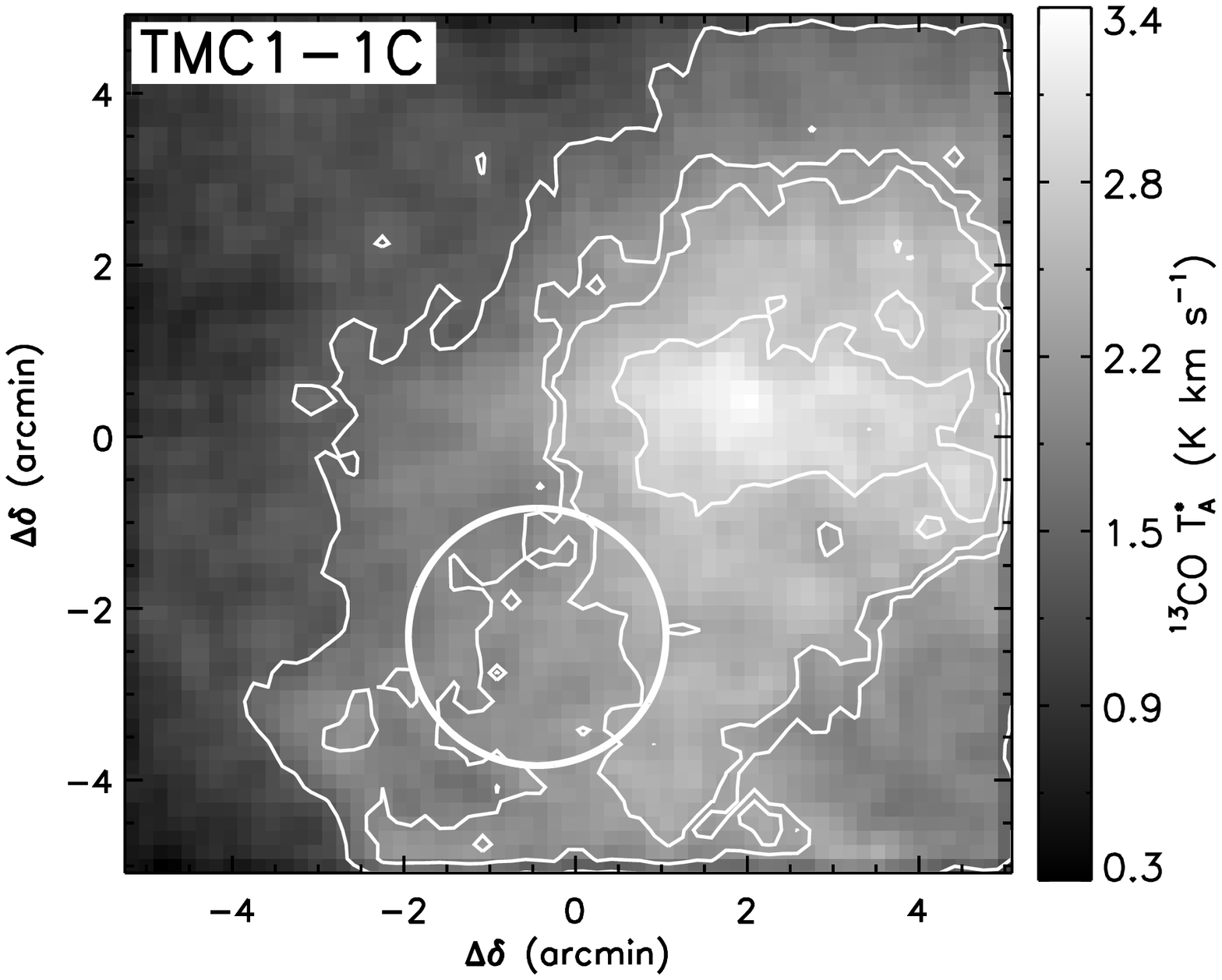}{\includegraphics{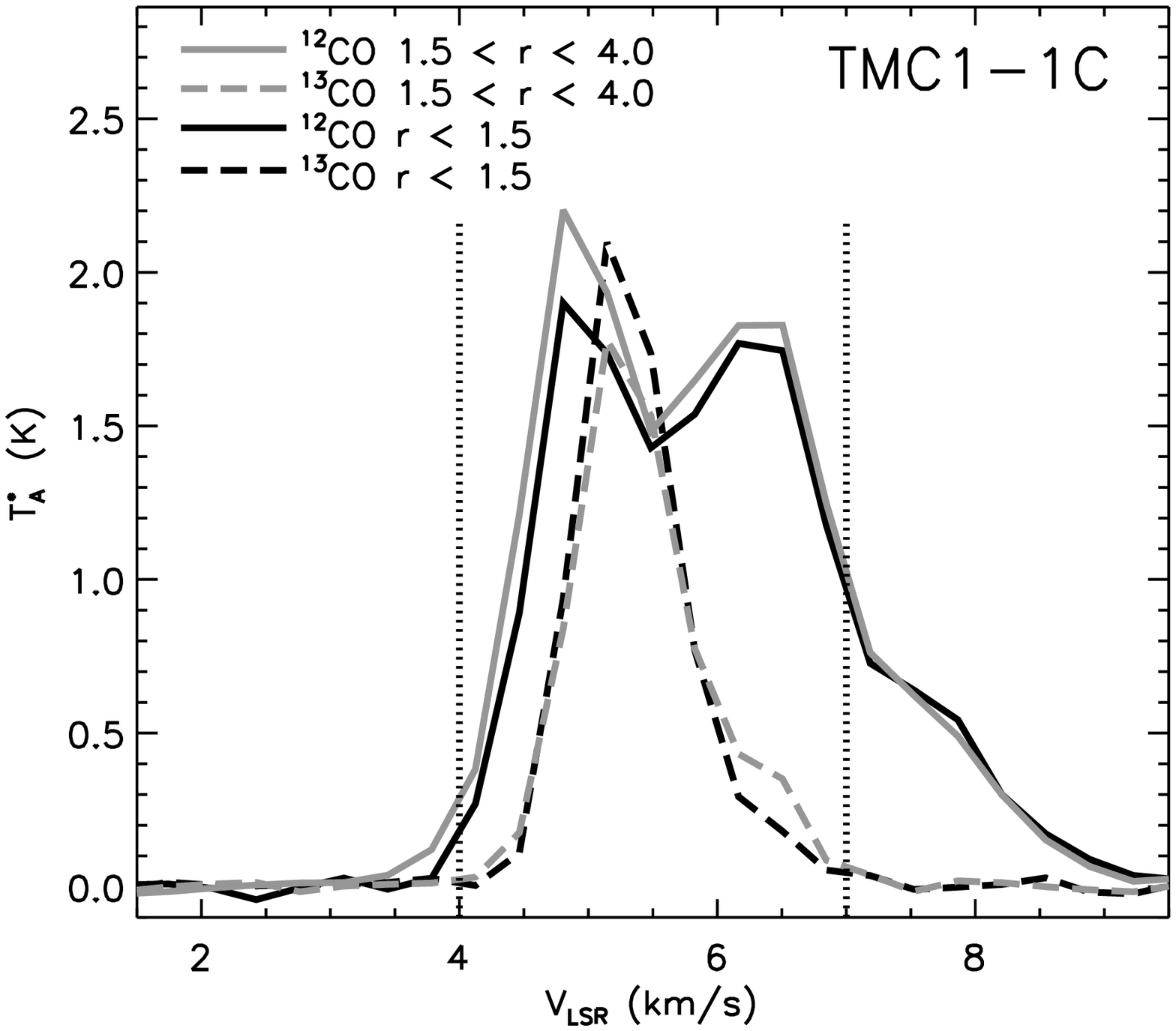}}}
    \scalebox{0.44}{\includegraphics{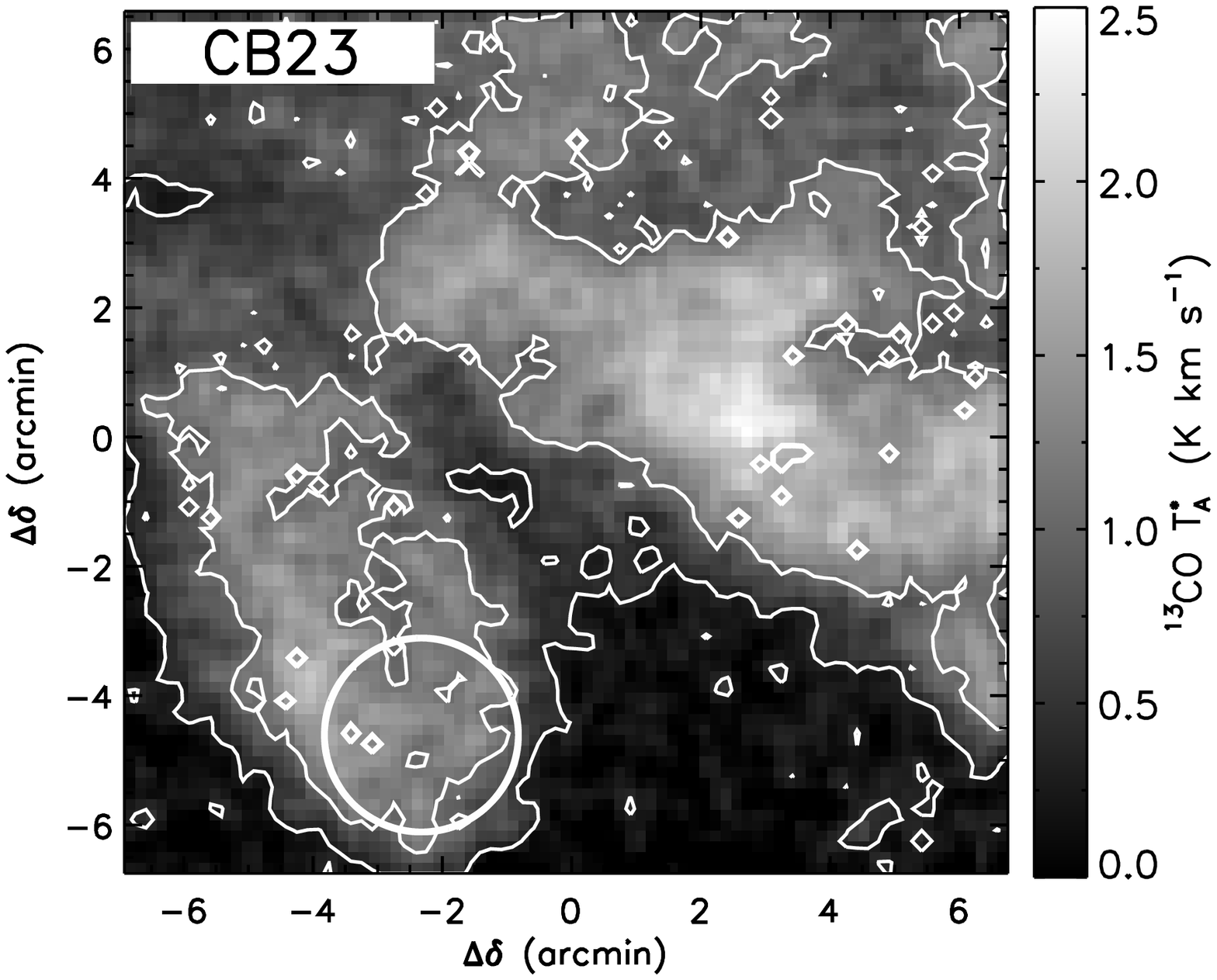}{\includegraphics{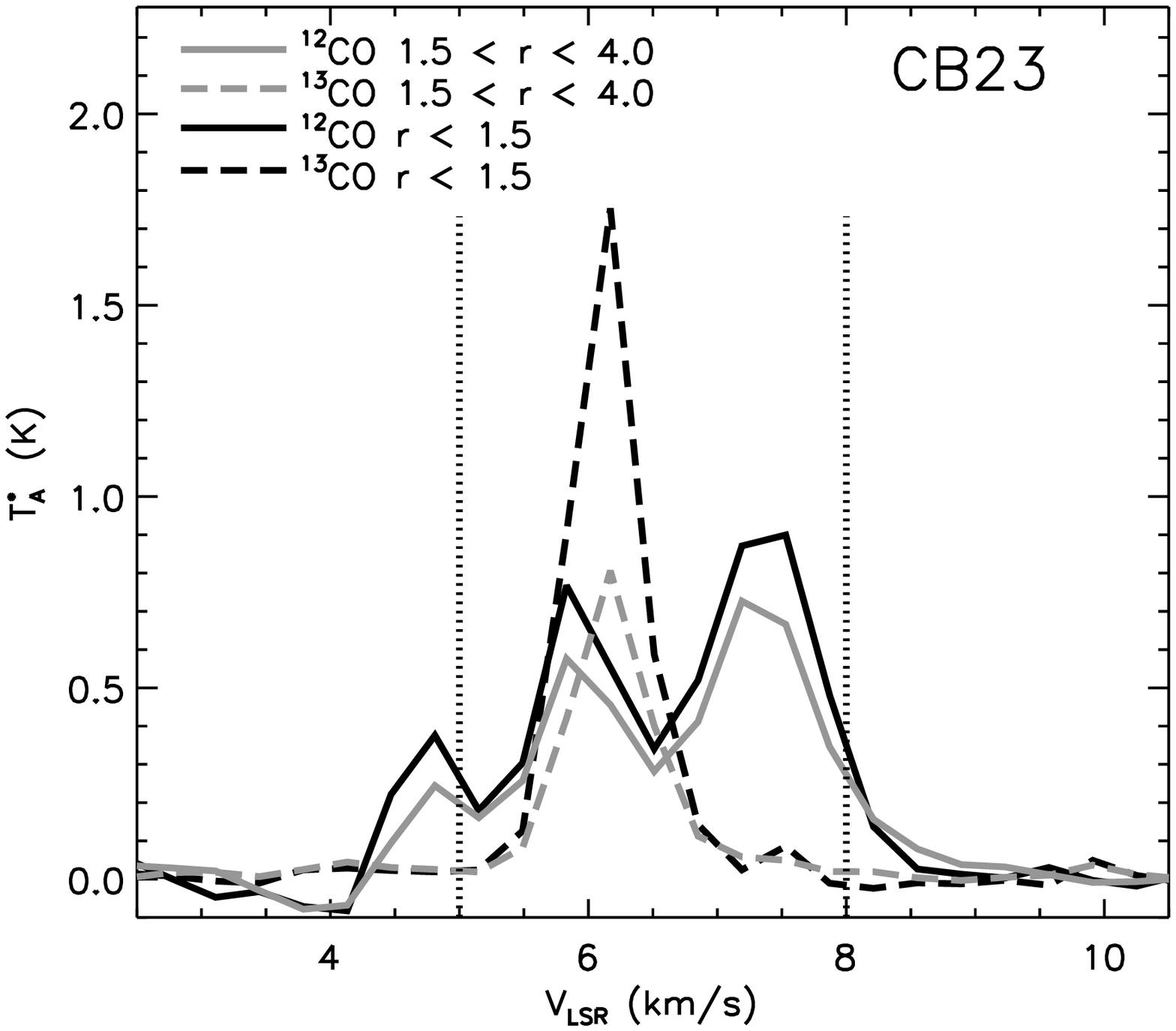}}}
    \scalebox{0.44}{\includegraphics{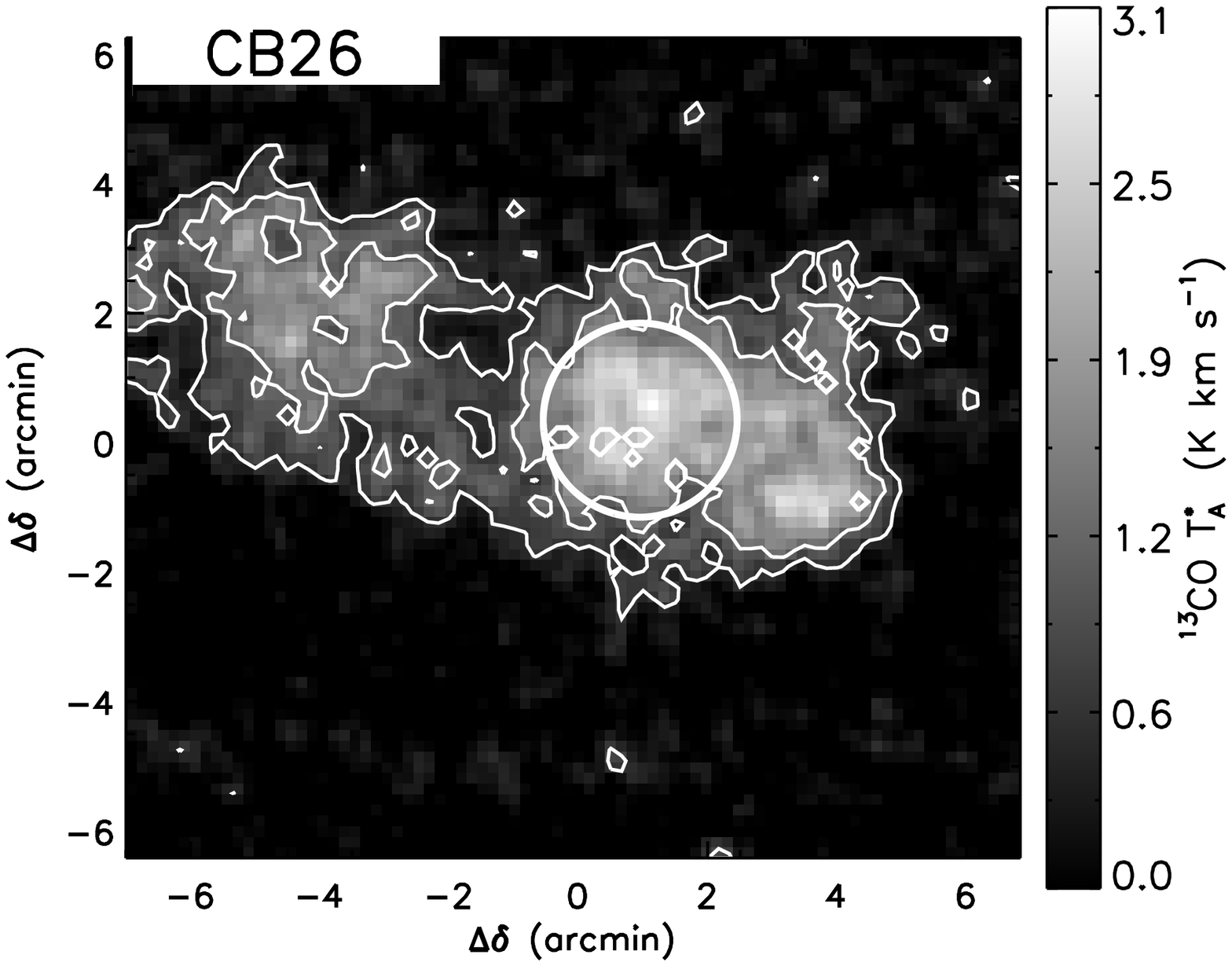}{\includegraphics{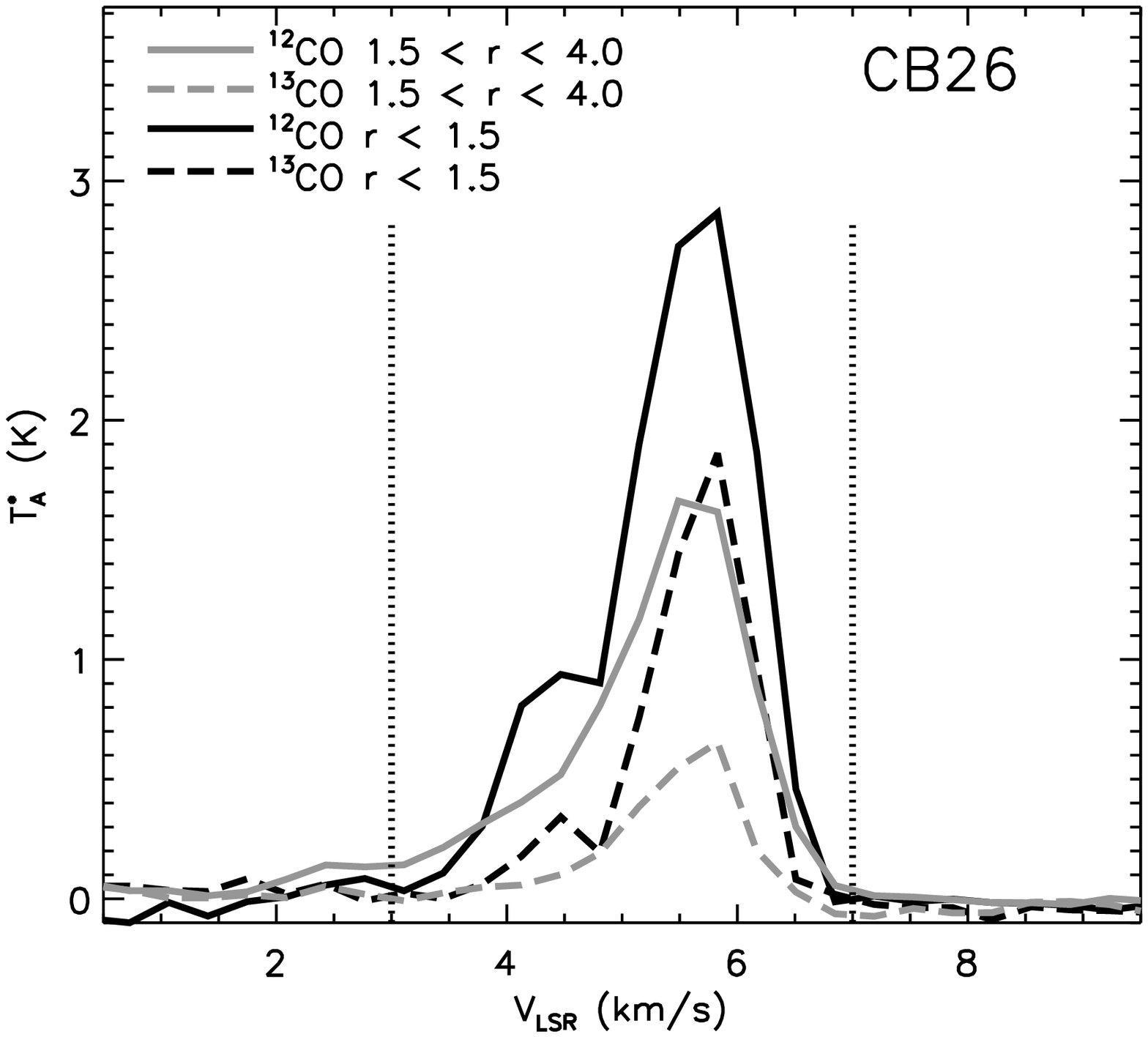}}}
    \caption{{\it Left Column --- } The grey scale image is $^{13}$CO
      (2-1) integrated intensity map.  Contours are the CO--model
      derived N(H) column density maps, which show good spatial
      agreement between the molecular map; contour levels are scaled
      to the maximum value of the column density in the region of the
      shadow.  Fractional contour levels for TMC-1C are
      $\{0.8,0.9,0.95,0.995\} \times 1.19 \times 10^{22}$~cm$^{-2}$,
      for CB23 are $\{0.6,0.8\} \times 1.09 \times 10^{22}$~cm$^{-2}$,
      and for CB26 are $\{0.5,0.6\} \times 1.49 \times
      10^{22}$~cm$^{-2}$.  The shadow is indicated by the small white
      circle, with a radius of $1\farcm5$; the large circle (r $=
      4\farcm0$) indicates the region in which we calculate mass
      M$_{\rm CO,4}$ (see Table~4 and text).  {\it Right Column --- }
      Mean $^{12}$CO and $^{13}$CO spectra for the regions indicated
      by the circles in the left panel.  {\it For full resolution
        figures contact stutz@mpia.de.}}
    \label{fig:cospec1}
  \end{center}
\end{figure}

\begin{figure}
  \begin{center}
    \scalebox{0.44}{\includegraphics{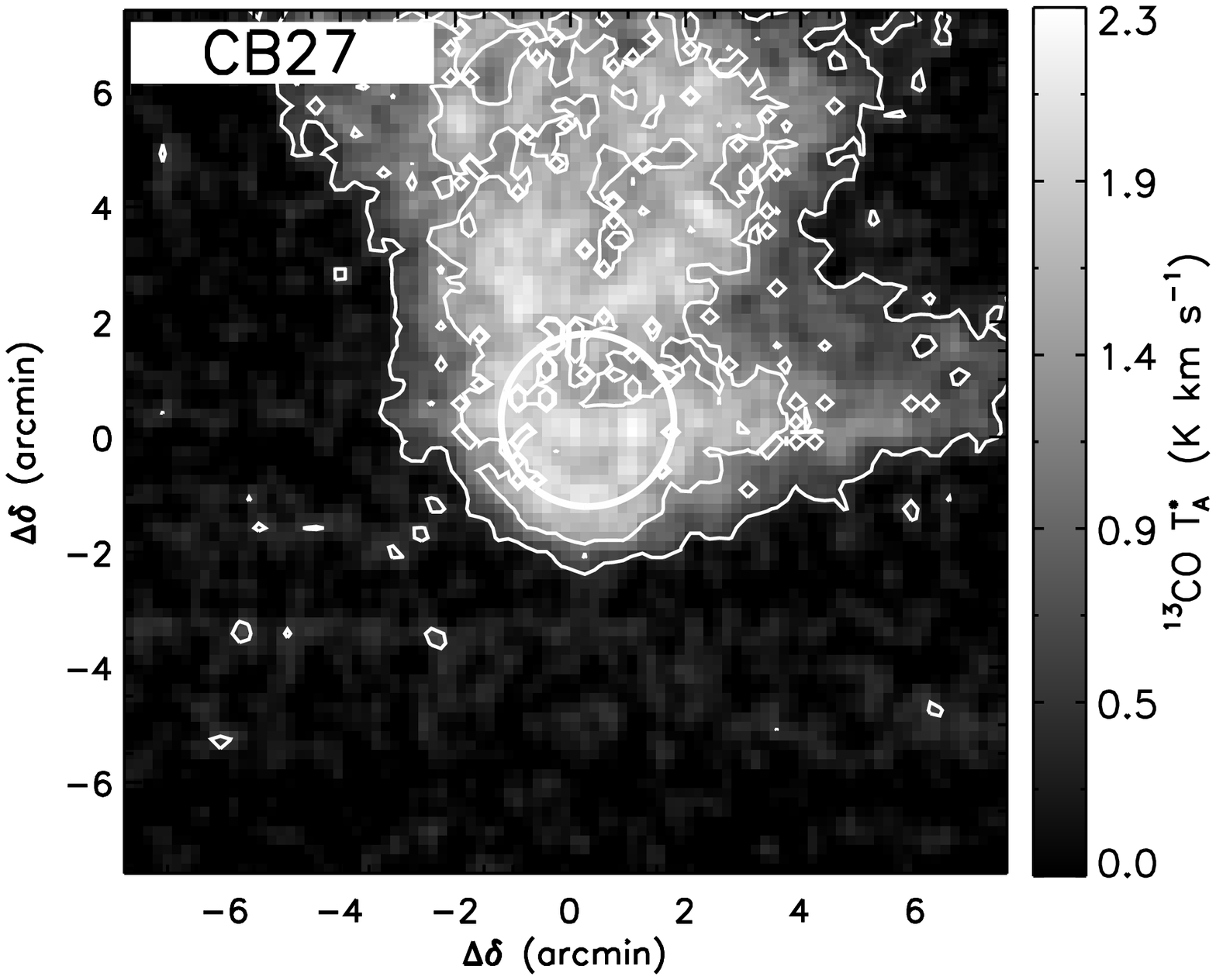}{\includegraphics{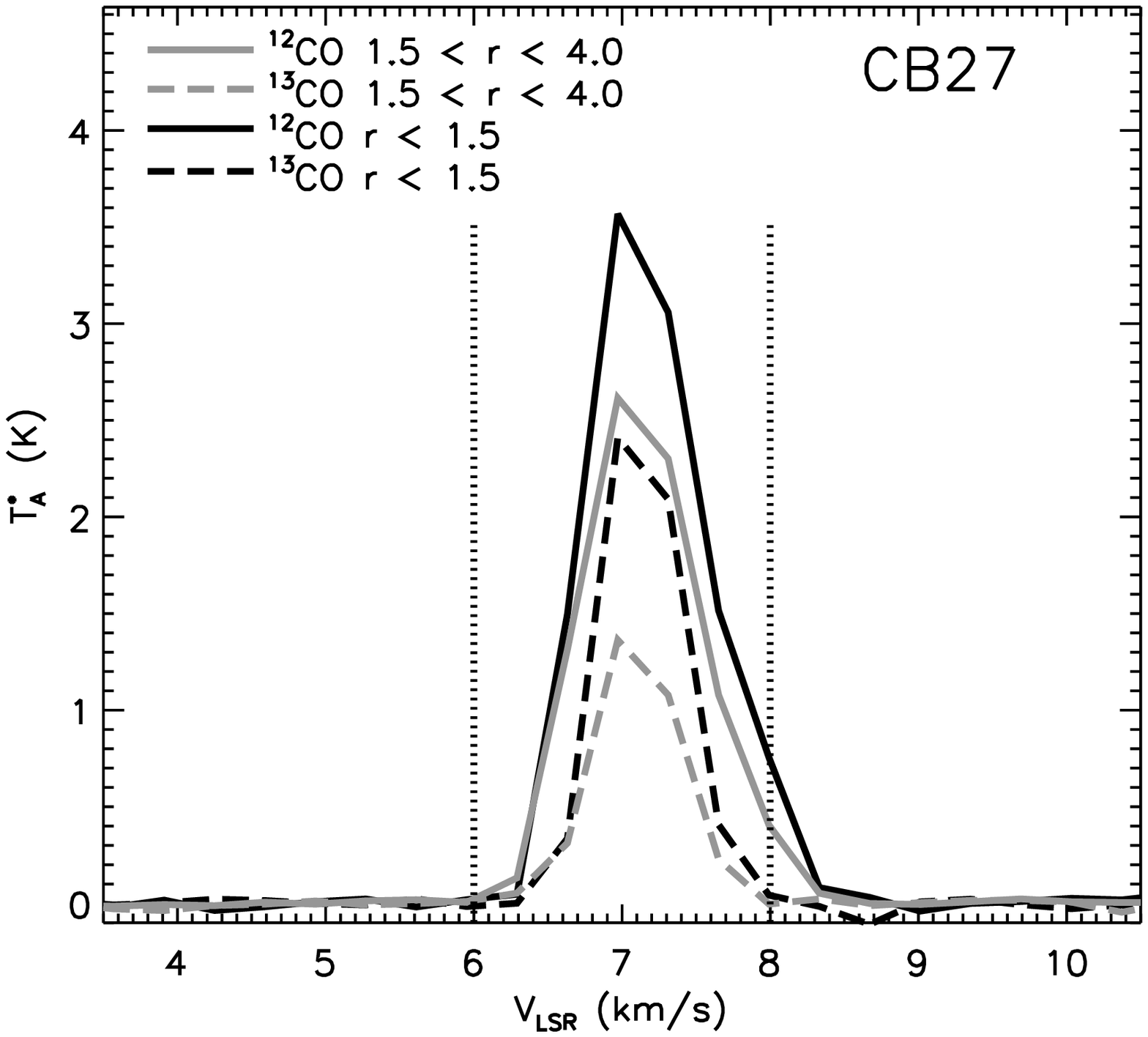}}}
    \scalebox{0.44}{\includegraphics{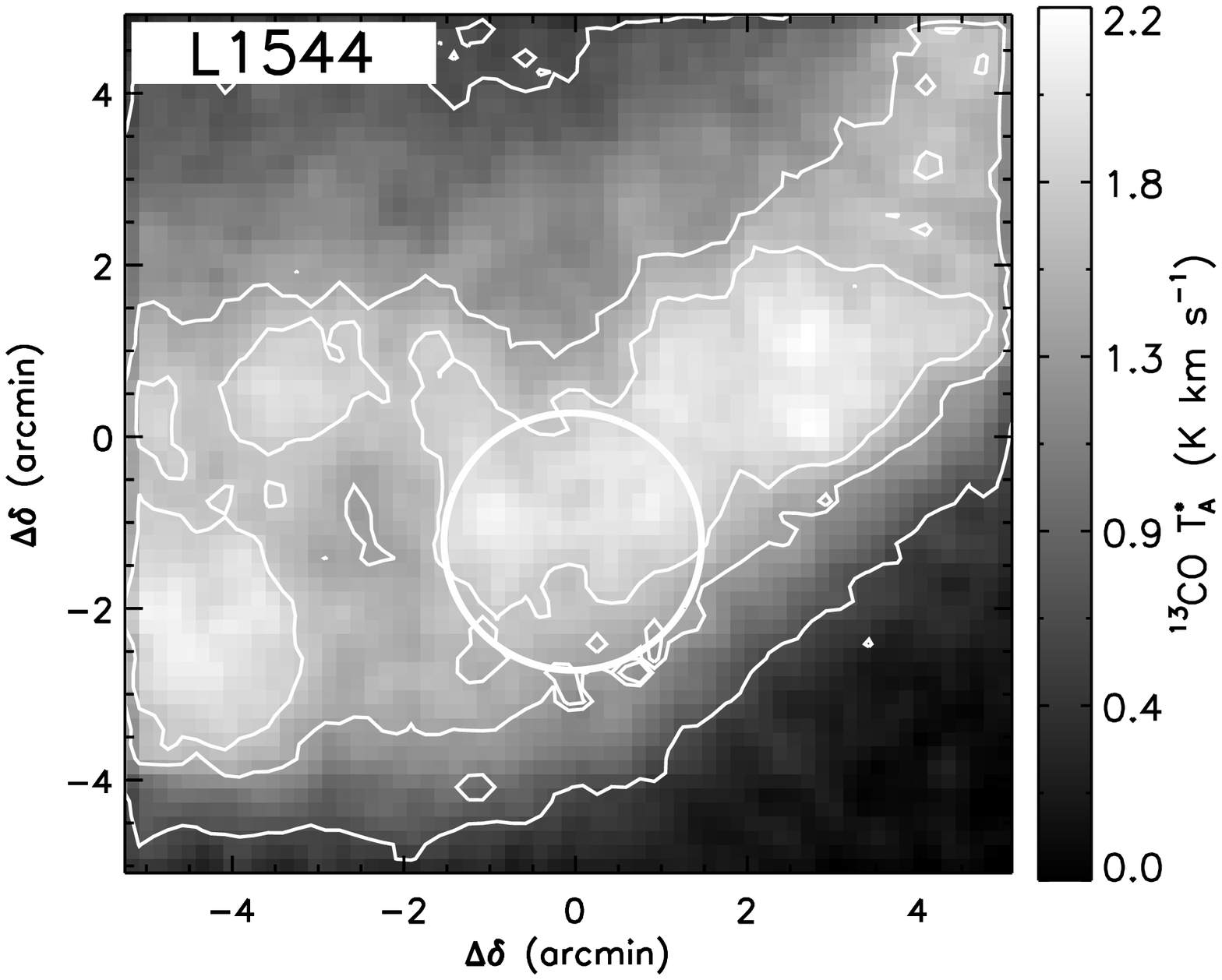}{\includegraphics{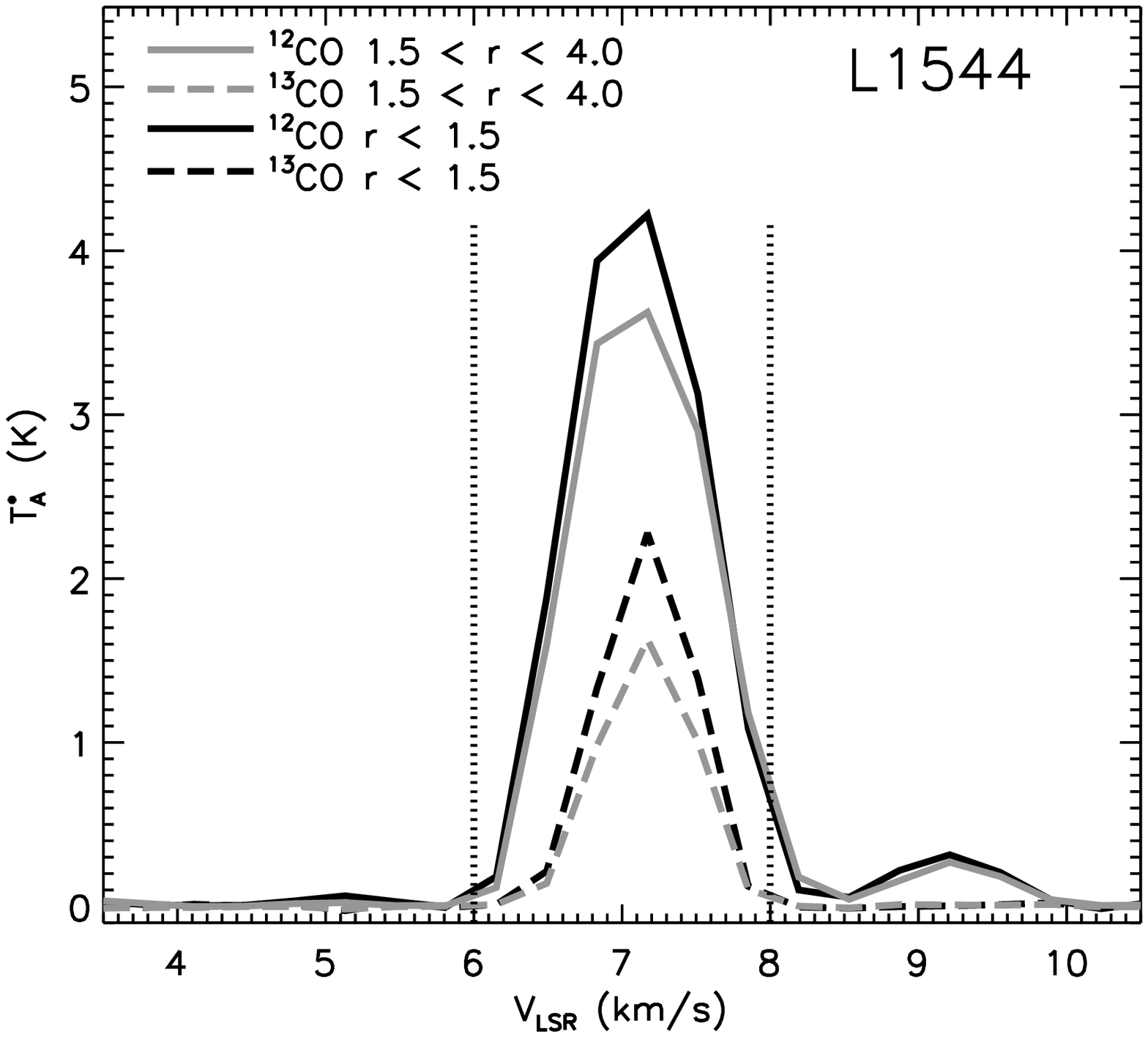}}}
    \scalebox{0.44}{\includegraphics{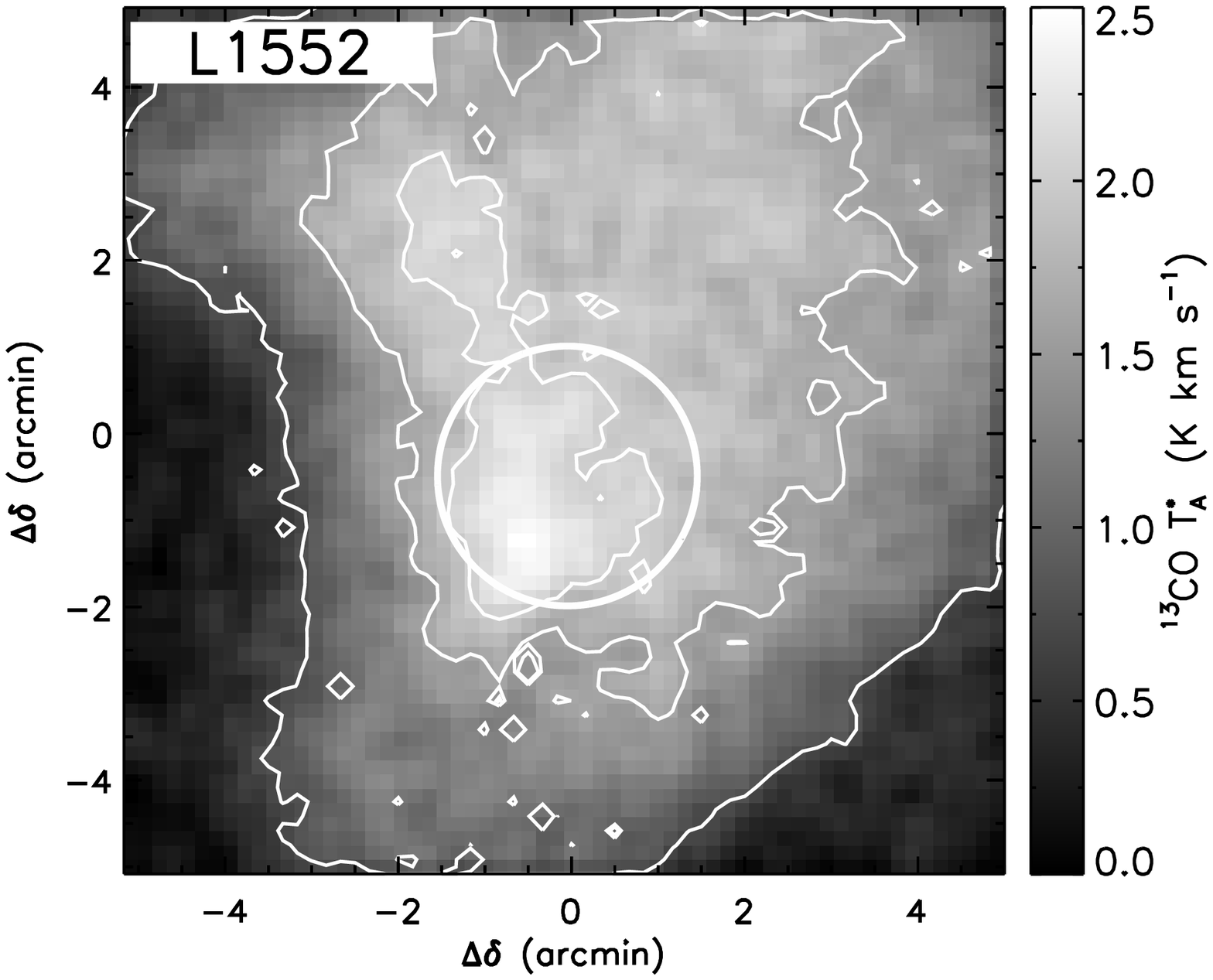}{\includegraphics{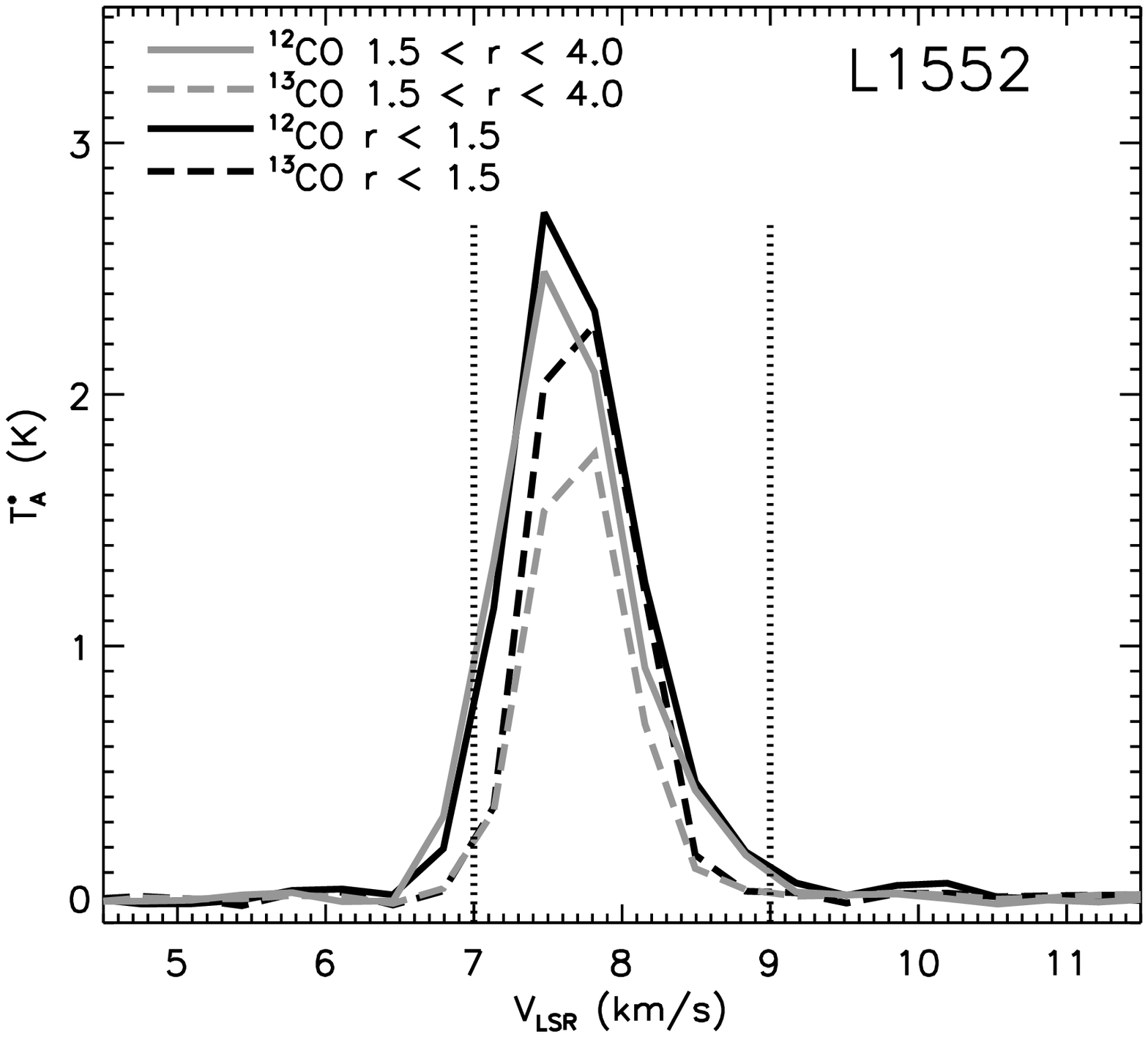}}}
    \caption{Same as Figure~\ref{fig:cospec1}. Fractional contour
	levels for CB27 are $\{0.5,0.7\} \times 1.40 \times 10^{22}$~cm$^{-2}$,
	for L1544 are $\{0.7,0.9,0.95\} \times 1.11 \times 10^{22}$~cm$^{-2}$,
	and L1552 are $\{0.7,0.9,0.95\} \times 1.15 \times 10^{22}$~cm$^{-2}$. }
    \label{fig:cospec2}
  \end{center}
\end{figure}

\begin{figure}
  \begin{center} 
    \scalebox{0.44}{\includegraphics{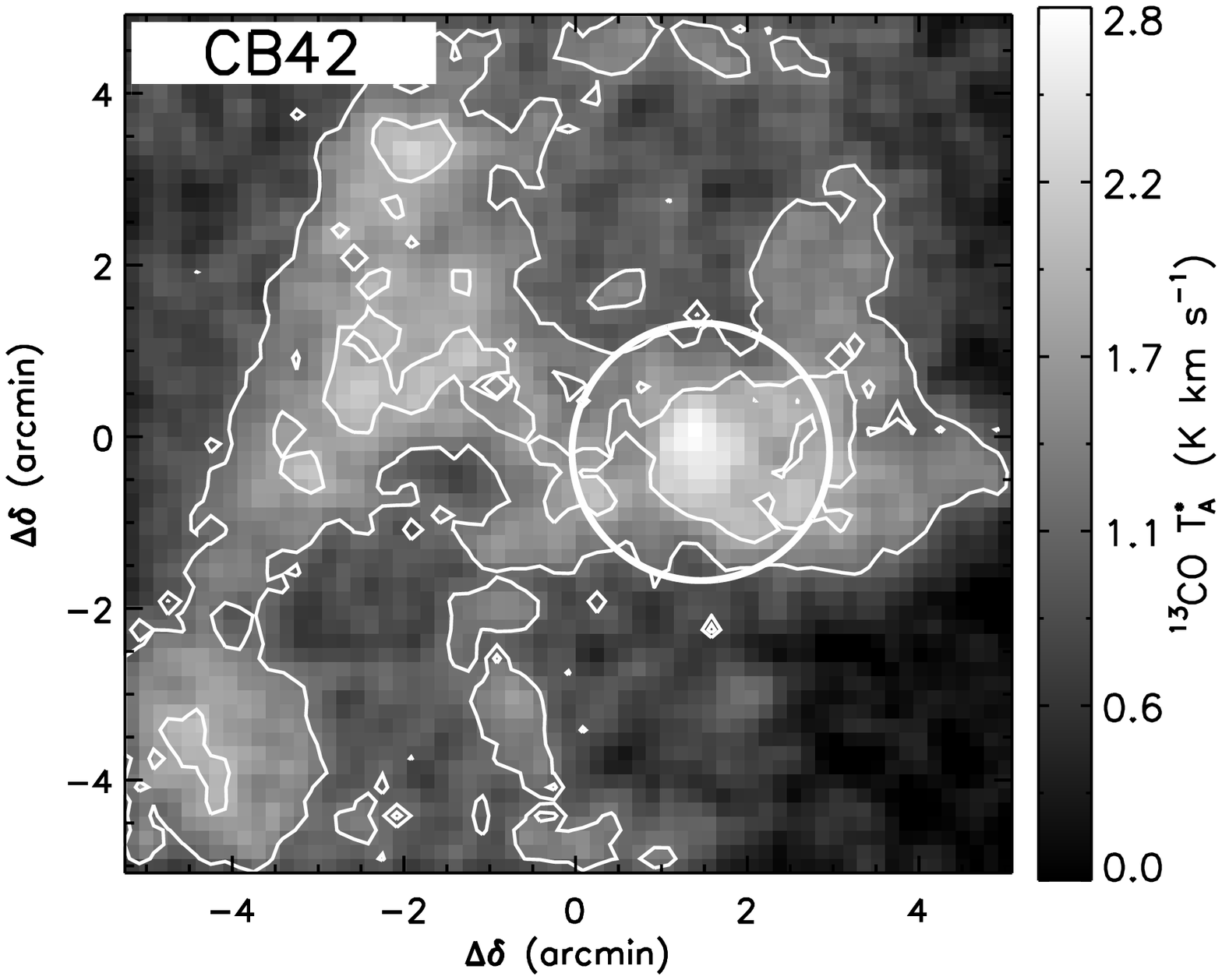}{\includegraphics{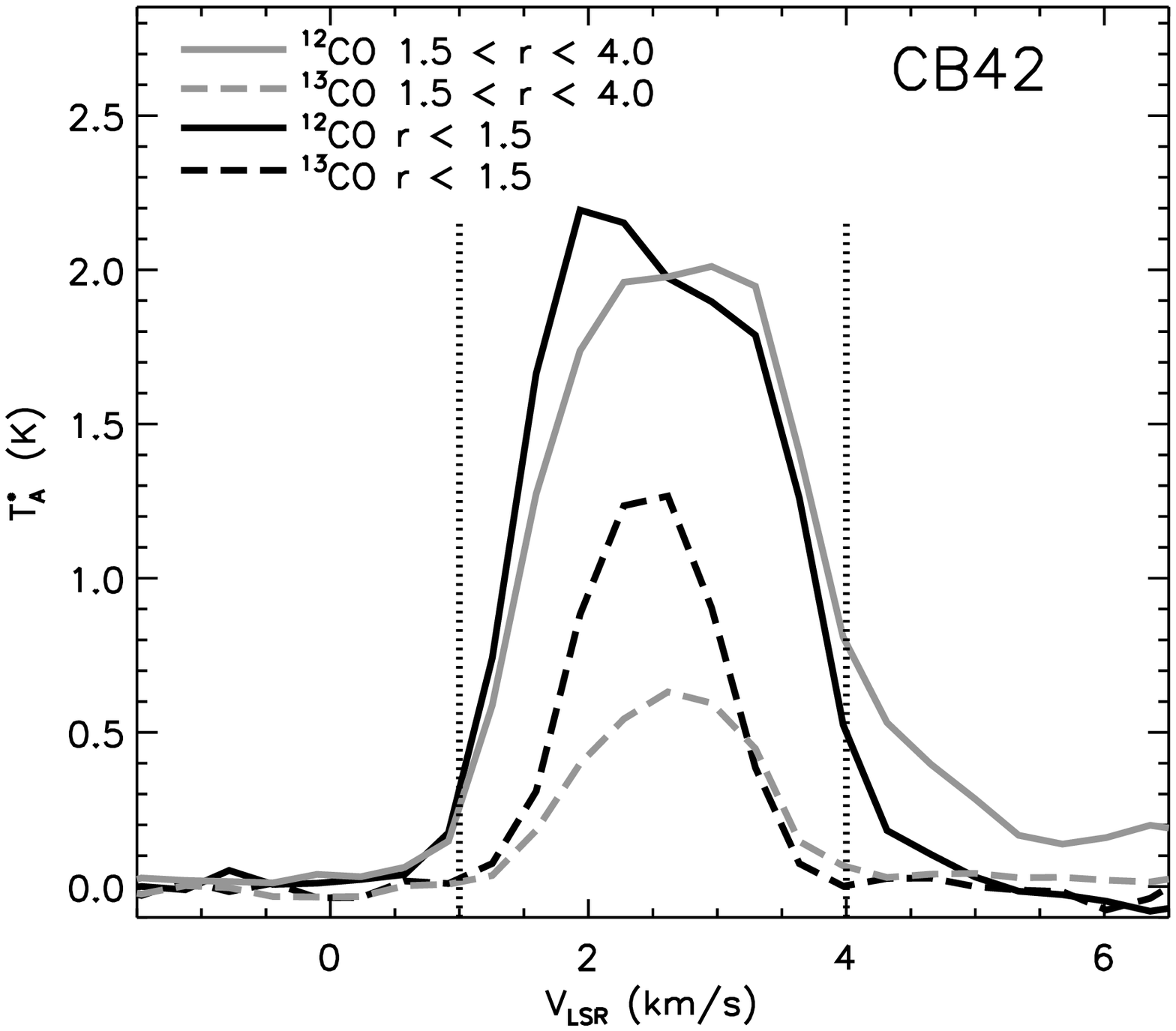}}}
    \scalebox{0.44}{\includegraphics{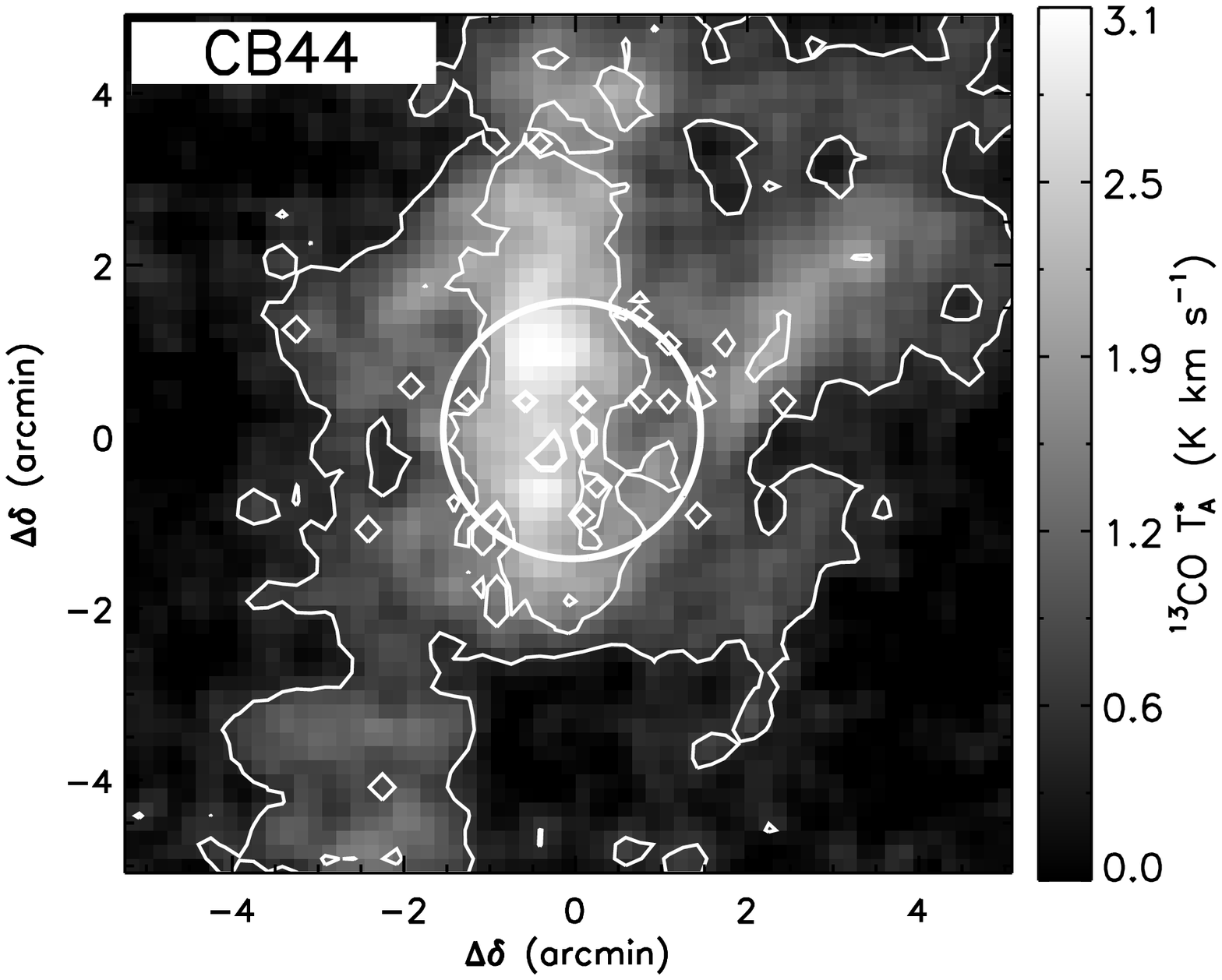}{\includegraphics{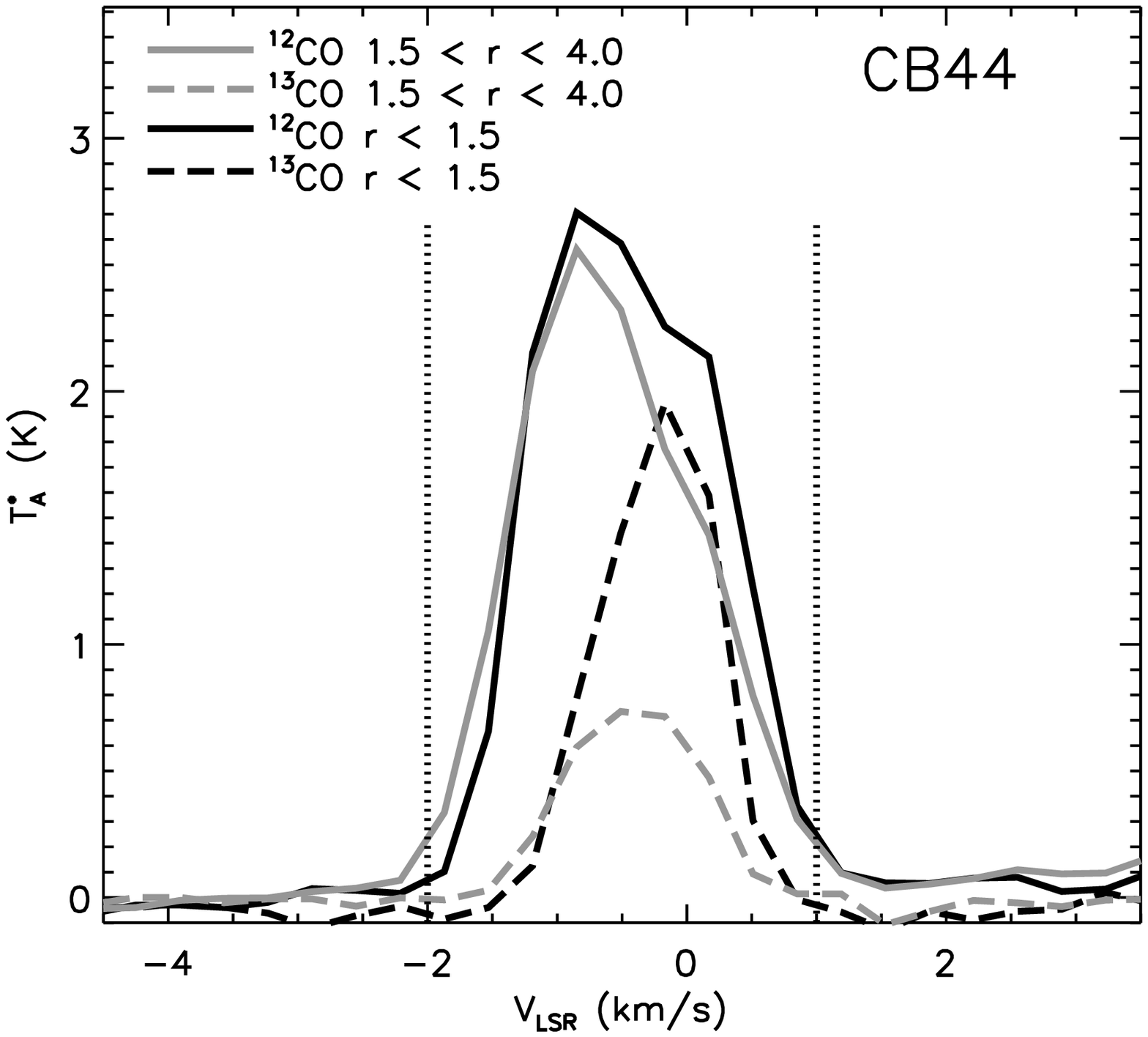}}}
    \scalebox{0.44}{\includegraphics{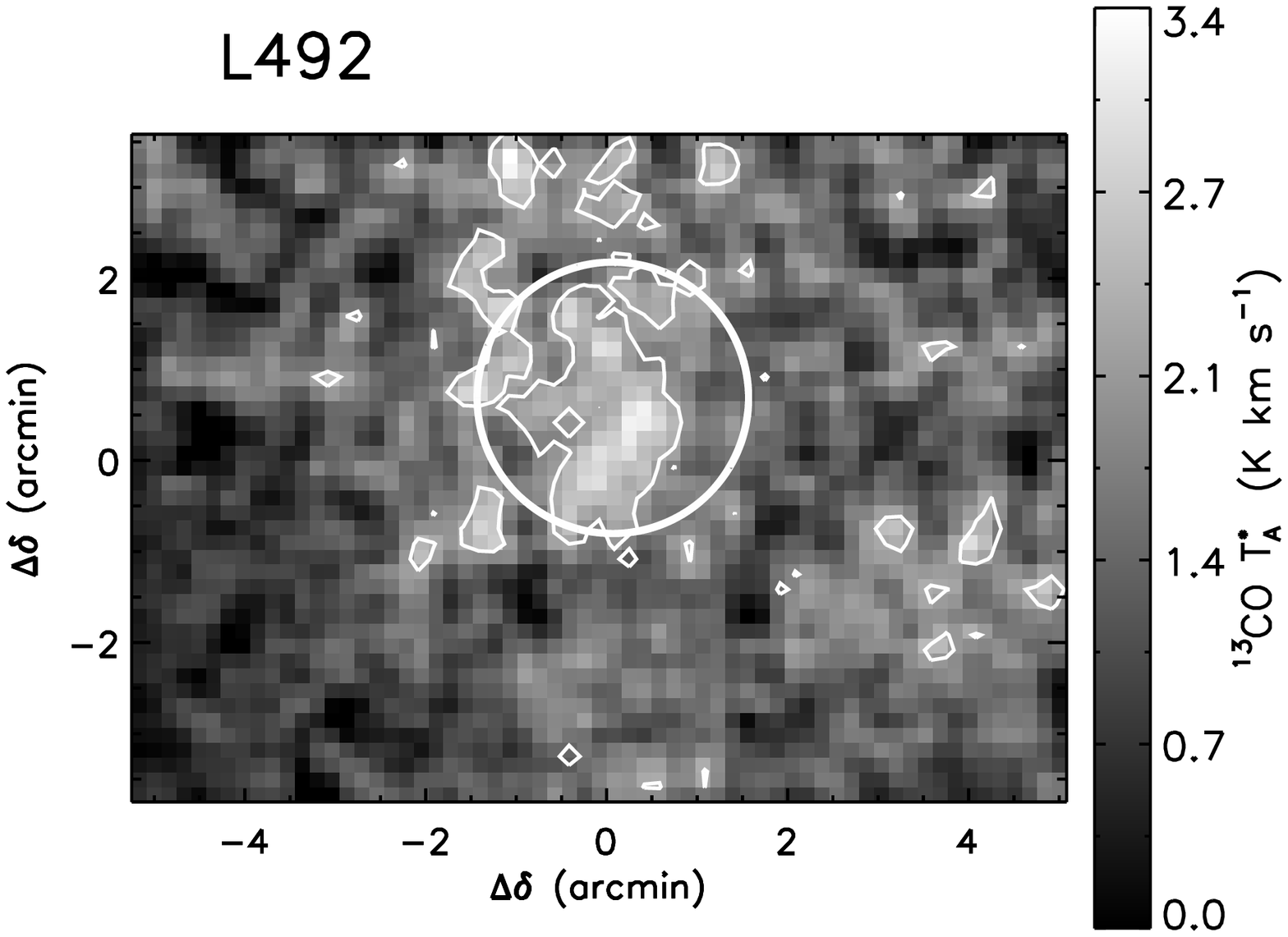}{\includegraphics{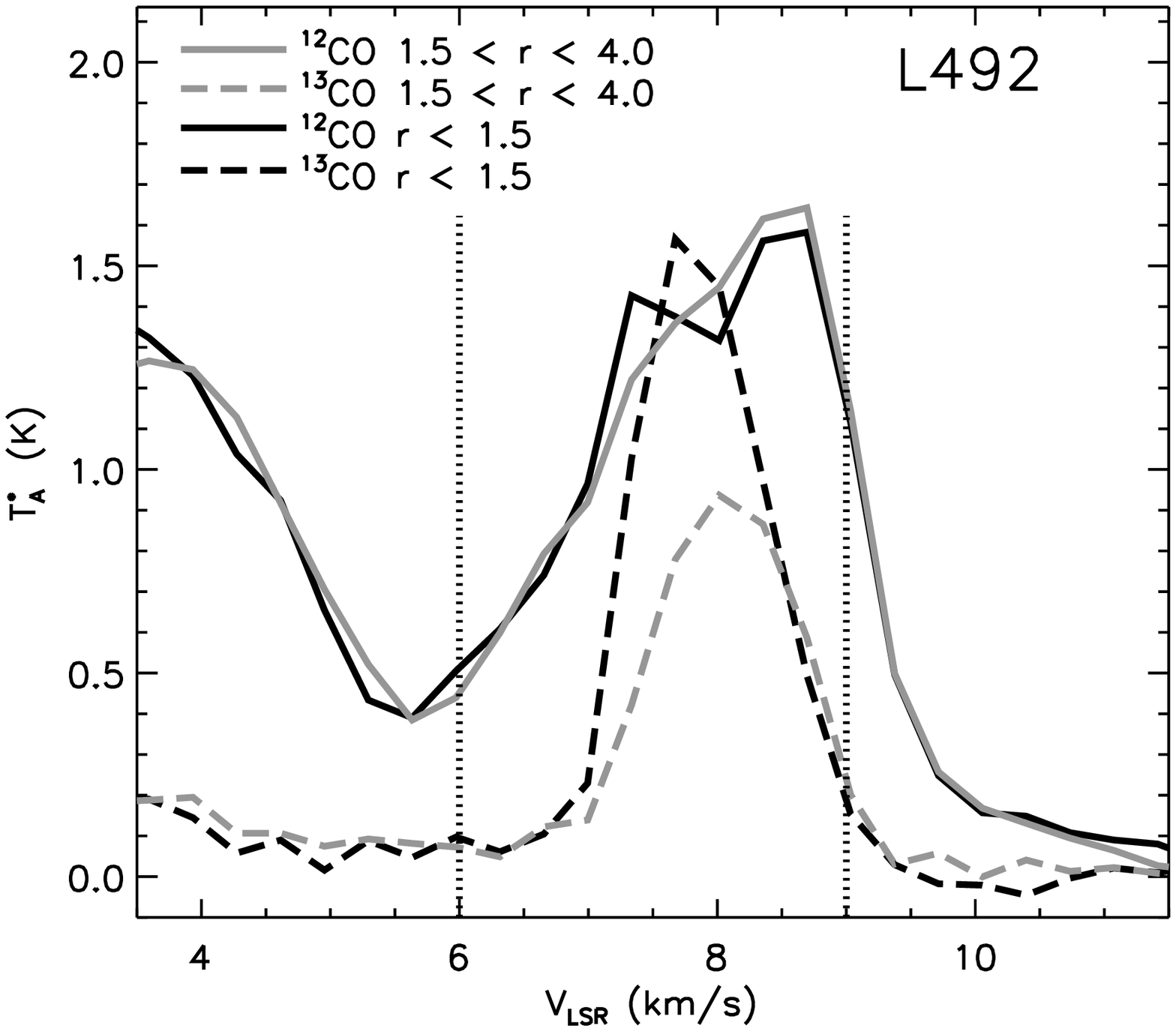}}}
    \caption{Same as Figure~\ref{fig:cospec1}. Fractional contour
	levels for CB42 are $\{0.75,0.85\} \times 1.19 \times
	10^{22}$~cm$^{-2}$, for CB44 are $\{0.4,0.55\} \times 1.86 \times
	10^{22}$~cm$^{-2}$, and L492 are $\{0.85\} \times 1.24 \times
	10^{22}$~cm$^{-2}$. }
    \label{fig:cospec3}
  \end{center}
\end{figure}

\begin{figure}
  \begin{center} 
    \scalebox{0.44}{\includegraphics{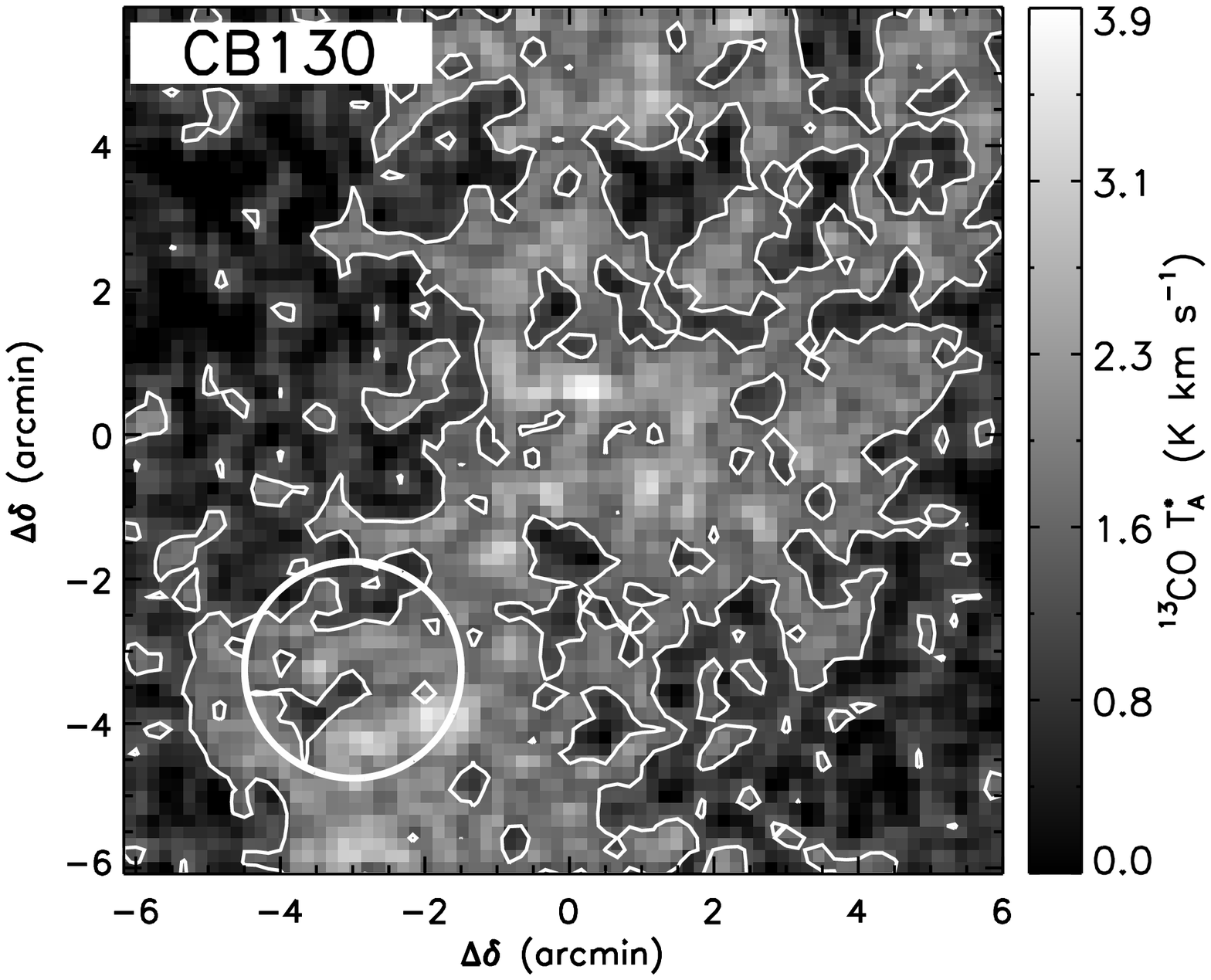}{\includegraphics{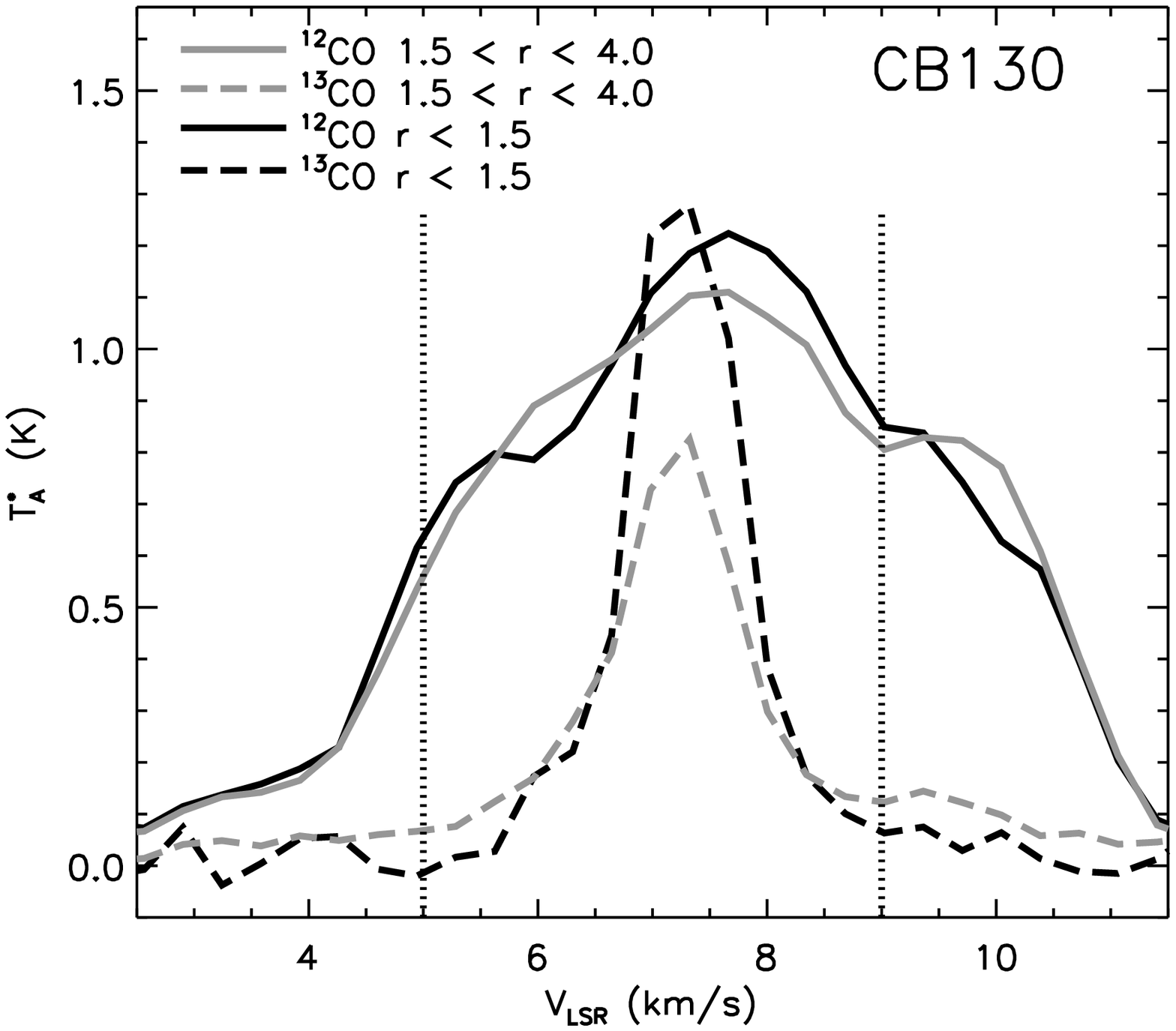}}}
    \scalebox{0.44}{\includegraphics{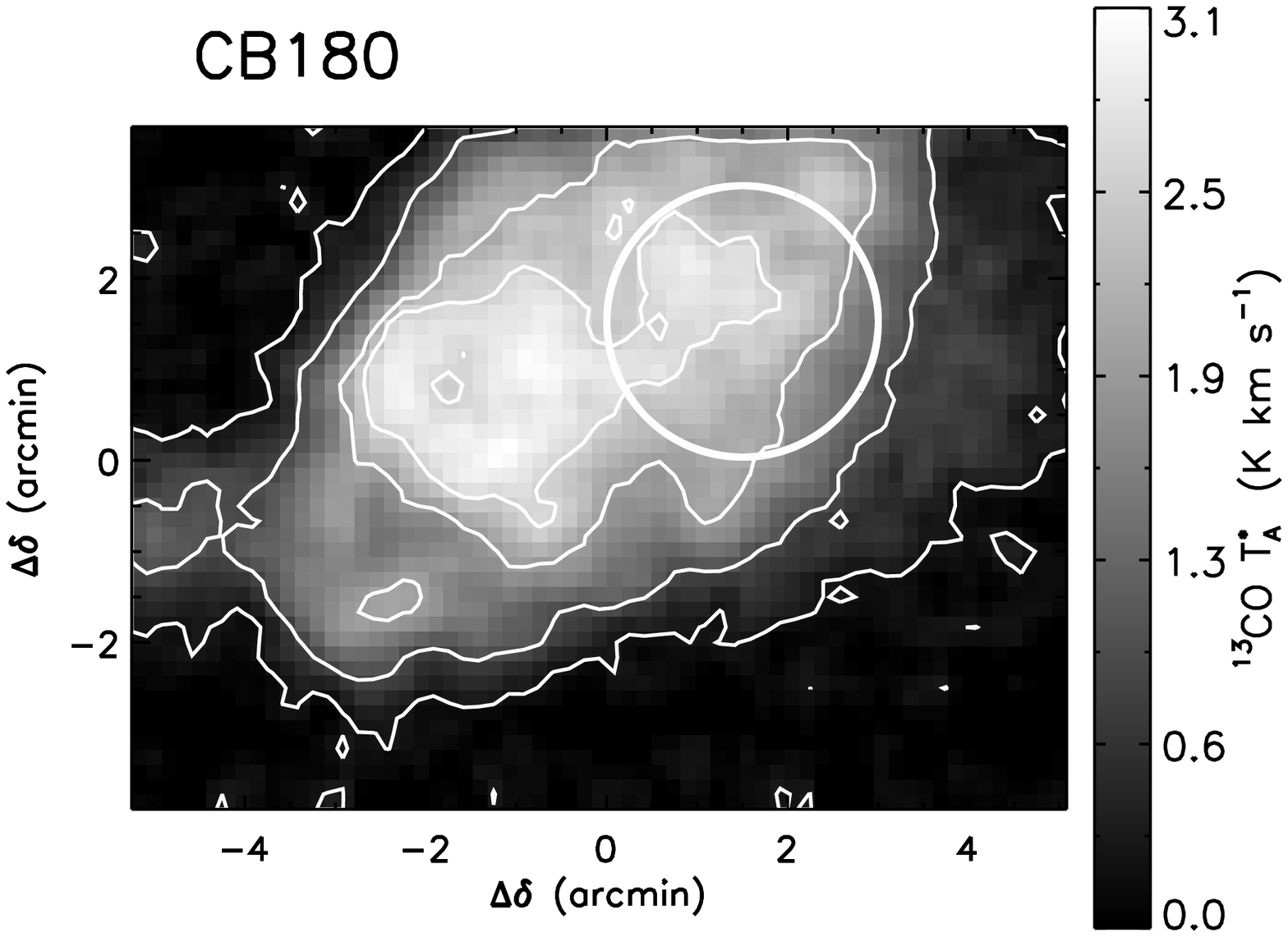}{\includegraphics{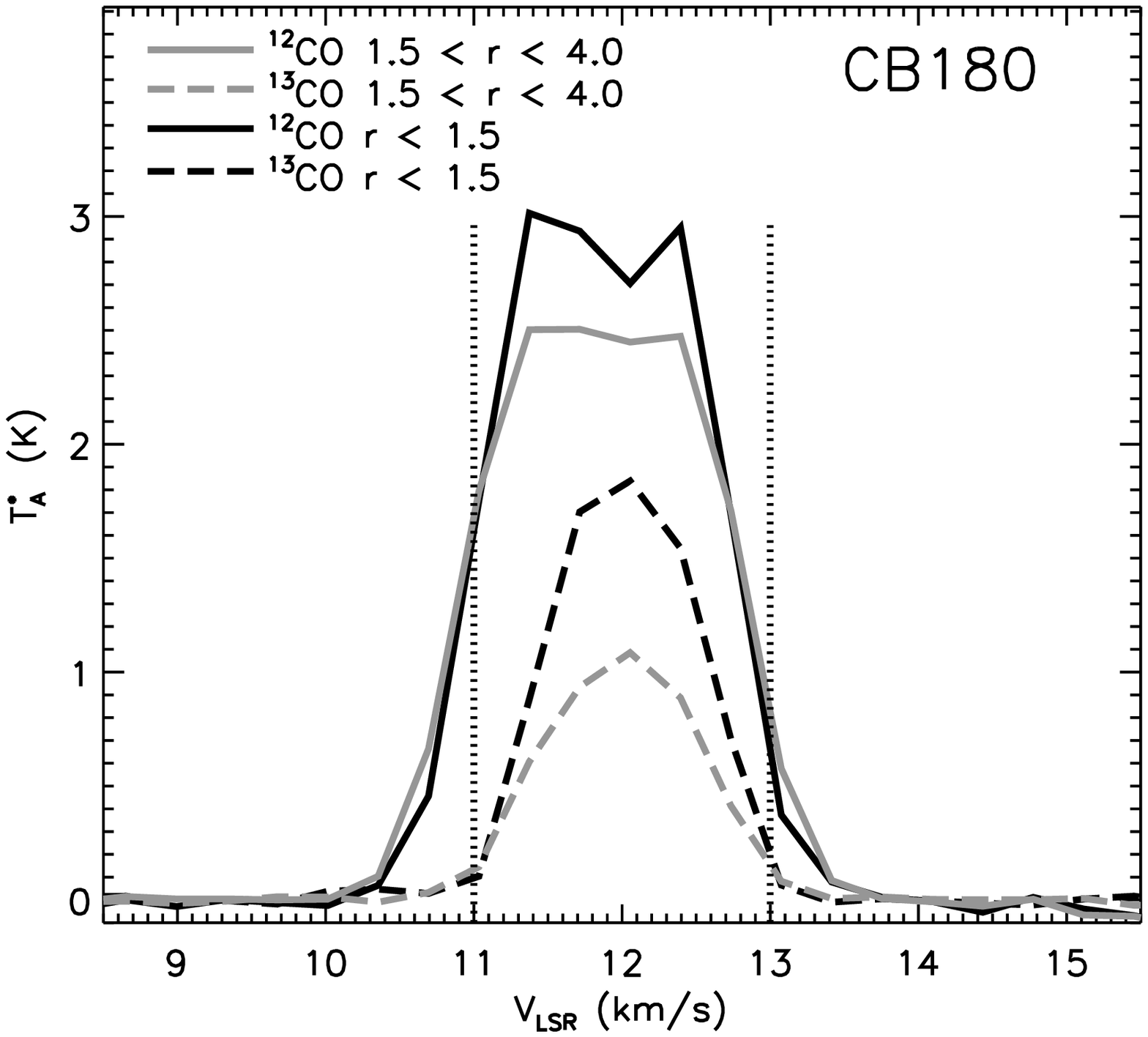}}}
    \scalebox{0.44}{\includegraphics{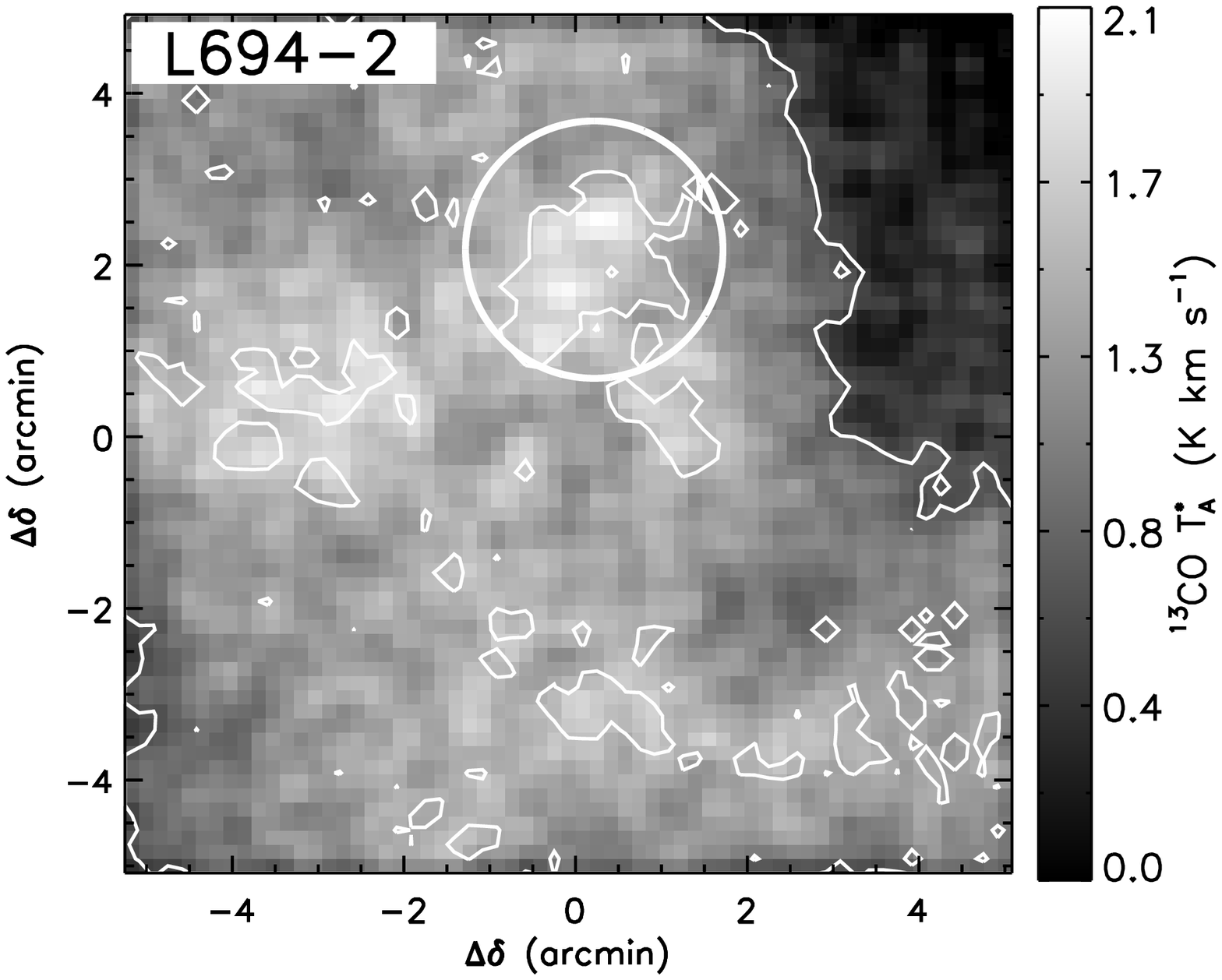}{\includegraphics{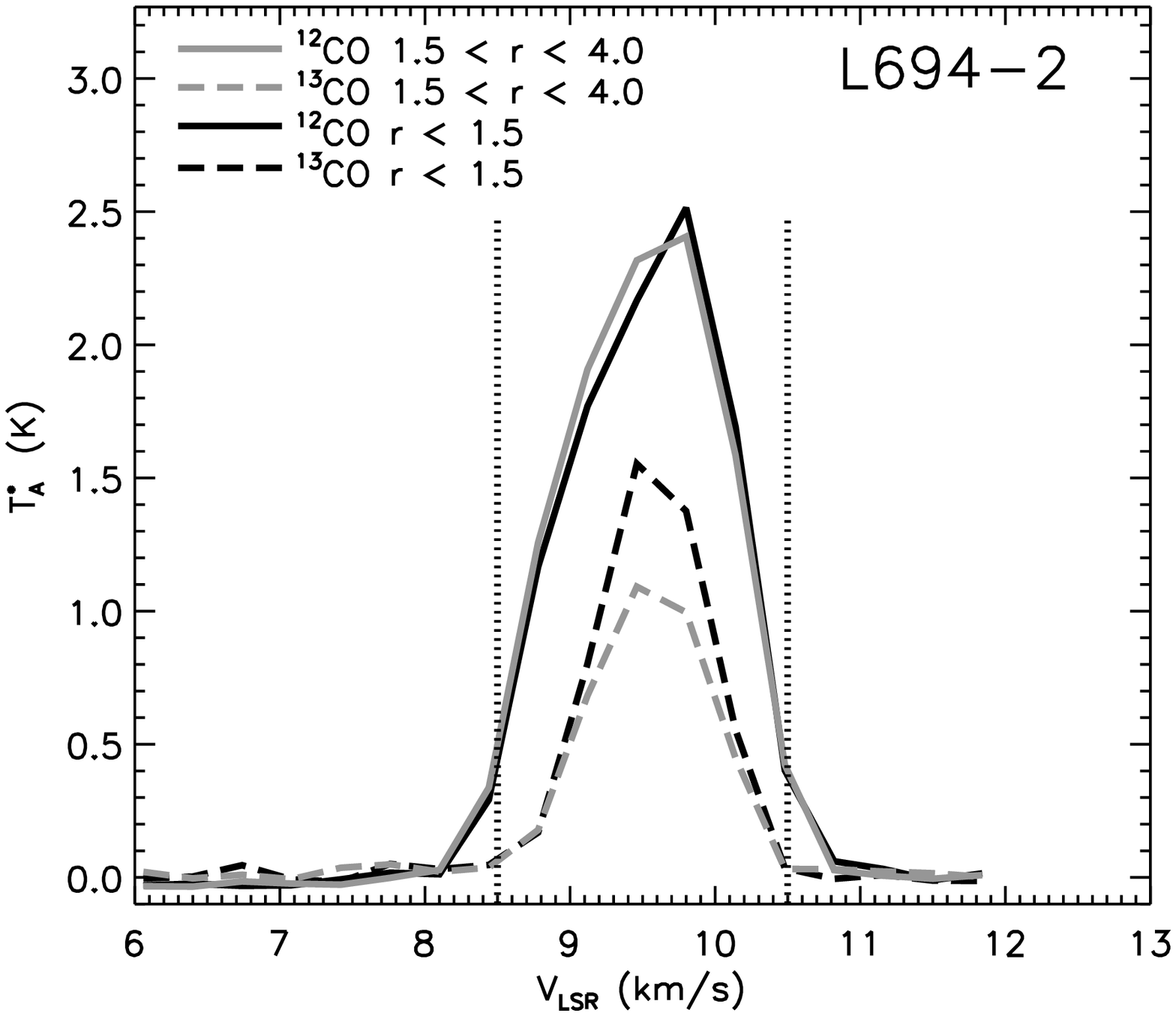}}}
    \caption{Same as Figure~\ref{fig:cospec1}.  Fractional contour
	levels for CB130 are $\{0.75\} \times 1.21 \times 10^{22}$~cm$^{-2}$,
	for CB180 are $\{0.5,0.7,0.9,0.97\} \times 1.21 \times
	10^{22}$~cm$^{-2}$, and L694-2 are $\{0.7,0.9\} \times 1.11 \times
	10^{22}$~cm$^{-2}$. }
    \label{fig:cospec4}
  \end{center}
\end{figure}

After calculating the total (atomic plus molecular) column N(H) in
each pixel, we derive masses using the distances collected from the
literature (listed in Table~3 and discussed in \S~3.2). We calculate
total cloud masses by integrating the $N(H)$~pixel distribution over a
$4\arcmin$ radius aperture centered on the shadow, indicated as
M$_{\rm CO,4}$ in Table~4.  The left column of
Figures~\ref{fig:cospec1} through \ref{fig:cospec4} show the $^{13}$CO
integrated intensity maps (grey scale) with the calculated N(H) column
density contours overlaid; we find good spatial agreement between the
two.  The right column of Figures~\ref{fig:cospec1} through
\ref{fig:cospec4} shows the $^{12}$CO and $^{13}$CO spectrum averaged
over the $r < 1\farcm5$ shadow region and in the $4\arcmin > r >
1\farcm5$ annulus where the dotted lines indicate the velocity range
used to calculate the column density N(H).  We also calculate
molecular core masses over a circular aperture of size equal to the
effective area of the core derived from the 24~\micron\ $\tau$ maps
(M$_{\rm CO}^{\rm core}$); the effective radii (R$_{\rm eff}^{\rm
  core}$) range from $1\farcm2$ to $1\farcm4$ for this sample of cores
and are listed in Table~4.

\subsection{Results}

We use our optical depth maps at 8 and 24~\micron, $\tau_8$ and
$\tau_{24}$, to derive core masses; we use the
$^{12}$CO and $^{13}$CO maps to measure the mass of material in the
larger environment surrounding the cores.  We then compare the Jeans
mass criterion and other information to assess the collapse status of
each core.  Here we summarize our results:

\noindent{\bf TMC1-1C:} Our analysis shows that this core is stable,
assuming only thermal support.  It is therefore highly unlikely that
this core will collapse soon given its current configuration. If the
mean density in the core increases by $\sim\!20\%$ the core will
exceed the Jeans mass criterion; if that happens and if magnetic and
turbulent support are negligible then eventually collapse will take
place.  \citet{kauffmann08} use the MAMBO bolometer array to observe
the thermal dust continuum emission at 1.2~mm.  In their study this
shadow is designated as TMC1-1C 3C, with a measured mass of $10.71 \pm
0.07$~\msun, in good agreement with our CO--measured mass of
11.4~\msun.  \citet{crapsi05} measure the N$_2$H$^+$ (2--1) and
N$_2$D$^+$ (2--1) line widths to be $0.21$ and $0.13$~km~s$^{-1}$,
respectively; they do not measure a large N$_2$D$^+$/N$_2$H$^+$ ratio
and do not classify this source as very advanced chemically or in
evolution.  \citet{caselli08} measure the ortho-H$_{2}$D$^+$
(1$_{1,0}$--1$_{1,1}$) emission, which is likely correlated with an
advanced chemical age, and also find that TMC1-1C does not appear to
be chemically evolved.  These results are consistent with our
conclusion that the core is probably stable.

\noindent{\bf CB23:} Our analysis shows that the 24~\micron\ shadow
mass exceeds the Jeans mass by $\sim\!25\%$, indicating that this core
is only marginally unstable, if only thermal support is considered.
The measured N$_{2}$H$^+$ line widths (see below) are narrow and
indicate that the turbulent contribution to the support is only of
order $\sim\!30\%$ \citep{hotzel02}.  A magnetic field of
$\sim\!120$~$\mu$G would be sufficient to support the 24~\micron\ mass
and to halt collapse \citep{stahler05}.  A magnetic field of this
level is on the high side of the range of field strengths measured for
other cores \citep{bergin07}, but remains plausible.  \citet{park04}
observe $^{12}$CO and $^{13}$CO emission; the line profiles appear
Gaussian. \citet{lee04} observe CS (3--2) and DCO$^+$ (2--1);
\citet{lee99} observe CS (2--1) and N$_{2}$H$^{+}$ (1--0).  A blue
asymmetry is observed in CS (2--1); a small amount of a blue asymmetry
is seen in CS (3--2) and DCO$^+$ (2--1) although it falls below the
\citet{lee04} threshold to be classified as a collapse candidate.
\citet{crapsi05} measure the N$_2$H$^+$ (1-0) and N$_2$D$^+$ (2-1)
line widths to be $0.17$ and $0.23$~km~s$^{-1}$, respectively but do
not classify this source as chemically or kinematically evolved.
\citet{sohn07} observe an infall asymmetry in the HCN (1--0) line
profiles of at least 2 of the 3 hyper--fine components, and possibly
in all three. We note that the \citet{crapsi05} and the \citet{sohn07}
observations were taken at locations separated by $\sim\!60\arcsec$,
however, both locations are well within the shadow feature.  Taken
with our results, these observations indicate that this core is
approaching collapse.

\noindent{\bf CB26:} Our analysis indicates that this core has a mass
that exceeds the Jeans criterion by a factor of $\sim\!3$.  The field
required to support the mass of CB26 is of order $200$~$\mu$G, much
larger than the upper limit of $74$~$\mu$G set by \citet{henning01} in
a nearby region of CB26.  The HCN (1-0) line widths of
$\sim\!0.6$--$0.9$~km~s$^{-1}$ measured by \citet{turner97} indicate
that the turbulent support is of order $\sim\!2$ to $4$ times the
thermal support; however, \citet{turner97} do not show their HCN
measurement and our CO spectra show 2 velocity components, which could
broaden the line widths (see Figure~\ref{fig:cospec1}).  We note that
\citet{turner97} do not list the coordinates of these observations and
we assume they are the same as those listed in \citet{clemens88},
which are about 1$\arcmin$ from the center of the shadow.  Without
higher density tracer line profiles from which to infer the velocity
field of the material, we consider it likely that CB26 is near
collapse.  The young T-Tauri star associated with this source has
received far more attention than the nearby starless core.  This YSO
has been studied in detail 
\citep[e.g.,][]{launhardt98,henning01,stecklum04,launhardt09,sauter09}.
It lies approximately $3\farcm5$ to the south--west of the shadow
center, and can be seen in Figure~\ref{fig:img3} as the brightest
24~\micron\ source.  Although it is not clear what evolutionary,
chemical, or other, relationship the shadow and the YSO might share,
it is interesting that \citet{henning01} measure relatively
well--ordered submillimeter dust polarization in a region
$\sim\!40\arcsec$ on a side towards the young star, perhaps indicating
that the magnetic field in the larger region, and therefore in the
shadow, might also be well behaved.  The polarization vectors point
roughly north--south, indicating that the magnetic field is oriented
in the east--west direction, the same direction as the longer axis of
the entire CB26 region.

\noindent{\bf CB27:} Our analysis indicates that this source is
stable, if supported only by thermal pressure.  With an additional
plausible level of turbulent and magnetic support, the core remains
stable.  This conclusion is supported by the molecular line
observations, which show no evidence of asymmetry.  \citet{lee04}
measure CS (3--2) and DCO$^+$ (2--1) line profiles that are
indistinguishable from Gaussian.  \citet{park04} observe $^{12}$CO and
$^{13}$CO Gaussian profiles as well.  \citet{sohn07} observe this
source to have a red asymmetry in all three HCN (1-0) line profiles.
\citet{gupta09} detect C$_6$H$^-$ emission with a line width of
$\sim\!0.15$~km~s$^{-1}$; \citet{roberts07} detect N$_2$H$^+$ and
N$_2$D$^+$ emission whose line profiles appear symmetric.

\noindent{\bf L1544:} We find that our 24~\micron\ mass is lower than
the Jeans criterion for collapse; however, the 70~\micron\ image
indicates that this mass estimate is a {\it lower limit}.  As can be
seen in Figure~\ref{fig:img5}, in the 70~\micron\ image there is a
hint of a dark region at the location of the 24~\micron\ shadow.  The
quality of the 70~\micron\ image does not allow for a robust analysis;
however, \citet{stutz09} show that in the case of starless core L429,
the 24~\micron\ shadow mass is underestimated compared to the
70~\micron\ shadow by a factor $\sim\!5$.  Taking this result, a
plausible corrected 24~\micron\ mass is $7$~\msun, a factor of
$\sim\!2.5$ larger than the Jeans mass, making the core much more
likely to be approaching collapse.  Many authors have observed this
source \citep[e.g.,][]{lee99,bacmann00, shirley00, bacmann02,
tafalla02, bacmann03, lee03, lee04, williams06, park04,
caselli08, gupta09}; the consensus is that this starless core is very
evolved and likely to be collapsing.  \citet{lee04} observe a CS
(3--2) profile with a double peak with a brighter blue peak, and a
DCO$^+$ (2--1) profile with a blue asymmetry.  \citet{crapsi05}
measure the N$_2$H$^+$ (2-1) and N$_2$D$^+$ (2-1) line widths to be
$0.32$ and $0.29$~km~s$^{-1}$, respectively; the N$_2$H$^+$ (3-2) and
N$_2$D$^+$ (3-2) were $0.19$ and $0.34$~km~s$^{-1}$, respectively.
They also find that this core has a large N$_2$D$^+$/N$_2$H$^+$ ratio
equal to 0.23 and classify it as a highly evolved core, based on
various observational chemical evolutionary probes.
\citet{williams06} compare the N$_2$H$^+$(1-0) emission from L1544 and
L694-2 and conclude that both cores will form stars in $\sim10^4$~yr.
\citet{sohn07} measure all three HCN (1-0) components to have two very
strong peaks, all with the blue peak higher than the red peak.
\citet{bacmann02} and \citet{bacmann03} find high degrees of CO
depletion and a large deuterium abundance; \citet{caselli08} measure a
large amount of ortho-H$_{2}$D$^+$ (1$_{1,0}$--1$_{1,1}$) emission,
one of the highest in their sample, and conclude that emission from
this species is correlated with CO depletion, deuterium fractionation,
and an advanced chemical evolutionary stage.  These results are
consistent with our conclusion that this core is likely to be
approaching collapse.

\noindent{\bf L1552:} In this case, our analysis shows that the
24~\micron\ shadow mass exceeds the Jeans mass by a wide margin, a
factor of $\sim\!8$.  This source also shows a clear detection of a
70~\micron\ shadow; applying a correction of a factor of 5 to the
24~\micron\ mass \citep[see L1544 discussion and][]{stutz09} makes this
core very unstable.  \citet{lee04} measure a blue asymmetry in both CS
(3--2) and DCO$^+$ (2--1).  The \citet{sohn07} HCN (1-0) profiles show
a mild blue asymmetry in the line profiles and report an N$_2$H$^+$
narrow line width of $0.22$~km~s$^{-1}$, indicating that turbulent
support is negligible.  These line measurements are consistent with
our conclusion that this core is near collapse.

\noindent{\bf CB42:} Our analysis indicates that this source is very
stable against collapse, with a Jeans mass that exceeds the shadow
mass by a factor $\sim\!10$.  \citet{turner97} measure an HCN line
width of $\sim\!0.82$~km~s$^{-1}$.

\noindent{\bf CB44:} Our analysis indicates that this core has a mass
that exceeds the Jeans criterion by a factor of $\sim\!12$.  Molecular
line observations of this source are sparse; however,
\citet{launhardt98} measure a CS (2-1) line width of
$1.5$~km~s$^{-1}$, with a profile showing some indication that it may
have a red asymmetry.  Observations of other (high density) tracers
are needed.  The magnetic field required to support the
24~\micron\ shadow mass is of order 190~$\mu$G, large compared to
observations of other cores \citep{bergin07}.  Based on this evidence
and without line observations of higher density tracers we consider it
likely that CB44 is approaching collapse.  

\noindent{\bf L492:} Our mass measurement indicates that this core may
be unstable to collapse.  Although the 24~\micron\ mass only
marginally exceeds the Jeans mass, by $\sim\!15\%$, this source has a
70~\micron\ shadow, which, as discussed above, indicates that the
24~\micron\ mass is a lower limit \citep{stutz09}.
\citet{kauffmann08} report a 1.2~mm dust mass of $18.35 \pm
0.17$~\msun, in reasonable agreement with our CO--measured mass of
23.6~\msun, accounting for our larger CO aperture ($4\arcmin$).  At
both 8~\micron\ and 24~\micron\ we observe two sub--components of this
core, oriented in the east-west direction; this structure is not
evident in the \citet{kauffmann08} MAMBO maps.  \citet{crapsi05}
measure the N$_2$H$^+$(2-1) and N$_2$D$^+$(2-1) line widths to be
$0.26$ and $0.22$~km~s$^{-1}$, respectively, at a position centered on
the western sub-clump.  The narrow line widths indicate that turbulence
does not play a significant role in the support of this core.  They
classify this source as marginally evolved because it exhibits only 3
out of the 8 evolutionary probes that they consider.  \citet{lee04}
measure a CS (3-2) with a blue asymmetry, DCO$^+$ (2-1) profile with a
blue asymmetry and possible self absorption; they classify this source
as an infall candidate.  \citet{sohn07} measure all three HCN (1-0)
components to have two strong peaks, all with the blue peak higher
than the red peak, and consider this source to be an infall candidate.
These results are consistent with our conclusion that this core is
likely to be approaching collapse.

\noindent{\bf CB130:} In the case of CB130, the 24~\micron\ shadow
mass is only marginally unstable.  A plausible level of turbulent
and/or magnetic support would halt collapse; for example, a magnetic
field of $\sim\!70$~$\mu$G will suffice to halt collapse.  There are
no molecular line observations at the location of the shadow.  The
source CB130-3, which lies $1\farcm5$ east of the shadow center, is
the closest object and was studied by \citet{park04} and
\citet{lee04}.  Inspection of the DSS red plate reveals what appears
to be a cusp, or knot, of material at the CB130-3 position.  However,
our 24~\micron\ maps indicate that a denser and compact region of the
CB130 complex is revealed by the 24~\micron\ shadow which is
completely enshrouded at near-IR and shorter wavelengths (see
Figure~\ref{fig:img10}).  At the CB130-3 position \citet{lee04}
measure a CS (3-2) line profile with a red asymmetry.  Without further
molecular line observations, we conclude that this source is probably
stable.

\noindent{\bf CB180:} Our analysis indicates that this source is
unstable, with a 24~\micron\ shadow mass that exceeds the Jeans mass
by a factor of $\sim\!2.6$.  \citet{caselli02} measure an N$_2$H$^+$
line width of $0.9$~km~s$^{-1}$ but do not show the spectrum of this
object; this width indicates hat the contribution of turbulence is
about 4 times that of the thermal support \citep{hotzel02}.  Given the
24~\micron\ shadow mass, a magnetic field of order $\sim\!70$~$\mu$G
would suffice to halt collapse.  Based on these pieces of evidence,
and without kinematical information from detailed line profiles, we
consider it likely that CB180 is close to collapse.

\noindent{\bf L694-2:} Our 24~\micron\ mass indicates that this core
is stable; however, as with L1544, L1552, and L492, this source has a
70~\micron\ shadow.  Correcting our 24~\micron\ mass by a factor 5, as
discussed above, yields a core mass that exceeds the Jeans mass by a
factor of $\sim\!4$.  Like L1544, this source has been studied in some
detail \citep[e.g.,][]{lee99, harvey03a, harvey03b, lee04, crapsi05,
williams06, lee07, sohn07, caselli08}.  \citet{crapsi05} measure the
N$_2$H$^+$(2-1) and N$_2$D$^+$(2-1) line widths to be $0.27$ and
$0.24$~km~s$^{-1}$, respectively; the N$_2$H$^+$(3-2) and
N$_2$D$^+$(3-2) were $0.31$ and $0.24$~km~s$^{-s}$, respectively.
They also find that this core has a large N$_2$D$^+$/N$_2$H$^+$ ratio
equal to 0.26.  Similar to L1544, this core is classified by
\citet{crapsi05} as highly evolved, both chemically and kinematically.
\citet{sohn07} measure HCO (1-0) double--peaked line profiles with a
blue assymetry.  \citet{caselli08} measure a large amount of
ortho-H$_{2}$D$^+$ (1$_{1,0}$--1$_{1,1}$) emission, one of the highest
in their sample, along with L1544 (and L429).  The line widths
indicate that the turbulent support is negligible, contributing of
order $\sim\!0.4\times$ the thermal support.  These results are
consistent with our conclusion that this core is approaching collapse.

Here we summarize results from two previous studies of cores CB190
\citep{stutz07} and L429 \citep{stutz09} which use similar techniques
applied to 24~\micron\ and 70~\micron\ shadows, respectively:

\noindent{\bf CB190:} \citet{stutz07} cary out a detailed study of
this core.  The 24~\micron\ shadows analysis indicates that the mass
of the core is twice the Jeans mass; however, the magnetic field
required to support the core material is about $90$~$\mu$G, a
plausible field strength for starless cores.  If the magnetic field is
in fact present in CB190 it will leak out of the core in
$\sim\!3$~Myr, an ambipolar diffusion time-scale \citep{stahler05}.
Turbulent pressure does not provide sufficient support against
collapse.  Millimeter line emission from various species is too faint
to constrain the kinematics in the core.  Without more line profile
information, we conclude that CB190 is near collapse.

\noindent{\bf L429:} \citet{stutz09} study the 70~\micron\ shadow cast
by this core and conclude that a plausible level of magnetic support
will not halt collapse.  Additionally, the molecular line widths are
narrow, indicating that the turbulent support is of order the thermal
support in this core.  \citet{stutz09} conclude that L429 is in a near
collapse state.  \citet{lee04} observe a blue asymmetry in the CS
(3-2) line, while \citet{crapsi05} and \citet{lee99} measure
symmetrical double peaked N$_2$H$^+$ (1-0) and CS (2-1) lines,
respectively.  \citet{sohn07} measure HCN (1-0) double peaked lines
but note that this source is one of two out of 85 cores in their study
with anomalous hyperfine line ratios.  More high-resolution line
observations are needed to clarify these results.  \citet{caselli08}
measure a large amount of ortho-H$_{2}$D$^+$ (1$_{1,0}$--1$_{1,1}$)
emission for this source, and find that this emission is correlated
with a large degree of central concentration, and a large amount
of CO depletion and deuteration, indicating that this source is
evolved.  Taken together, these pieces of evidence indicate L429 is
close to collapse.  

In this paper we emphasize how the 24~\micron\ shadow observations can
reveal structure not evident in previous studies.  We highlight the
24~\micron\ observations of CB130 and L492.  In the case of L492, we
observe what appear to be 2 sub--clumps in the shadow feature
(Figure~\ref{fig:img9}), oriented in a roughly east-west direction
separated by about $40\arcsec$, reported here for the first time.
Molecular line observations are centered on the western component
\citep[e.g.,][]{crapsi05} and indicate that this core is likely to be
collapsing.  The two components may be an example of core
fragmentation; high spatial resolution observations of this core
should be able to confirm if both sub--clumps are collapsing, and
to measure the relative radial velocity between them.  In the case of
CB130, we have discovered a 24~\micron\ shadow which is
$\sim\!1\farcm5$ west of the region previously selected for millimeter
line studies \citep[e.g.,][]{lee04,park04}; further observations taken
at the location of the 24~\micron\ shadow will confirm if this
core is near collapse.

\begin{figure}
  \begin{center}
    \scalebox{0.75}{\includegraphics{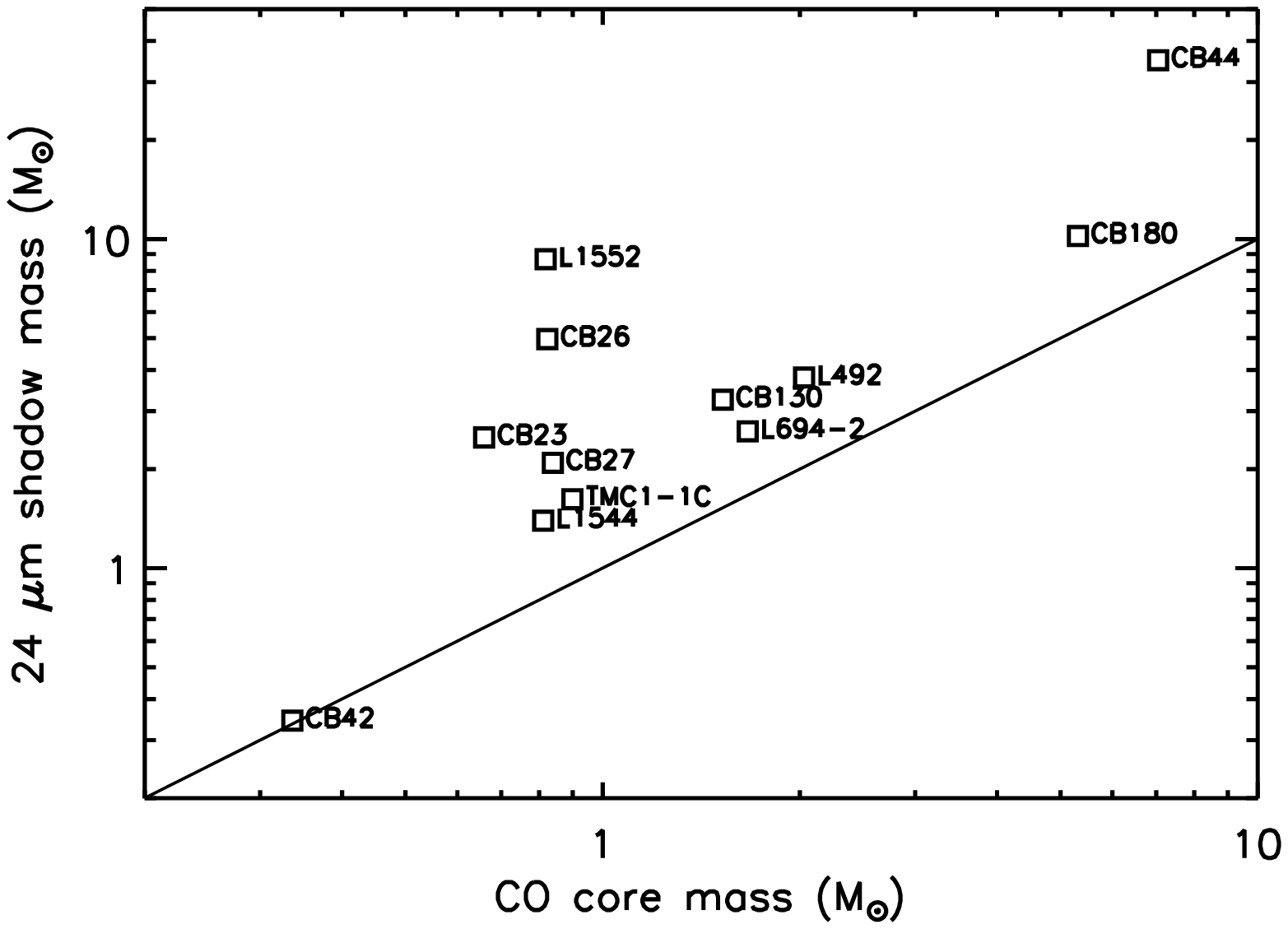}}
    \scalebox{0.75}{\includegraphics{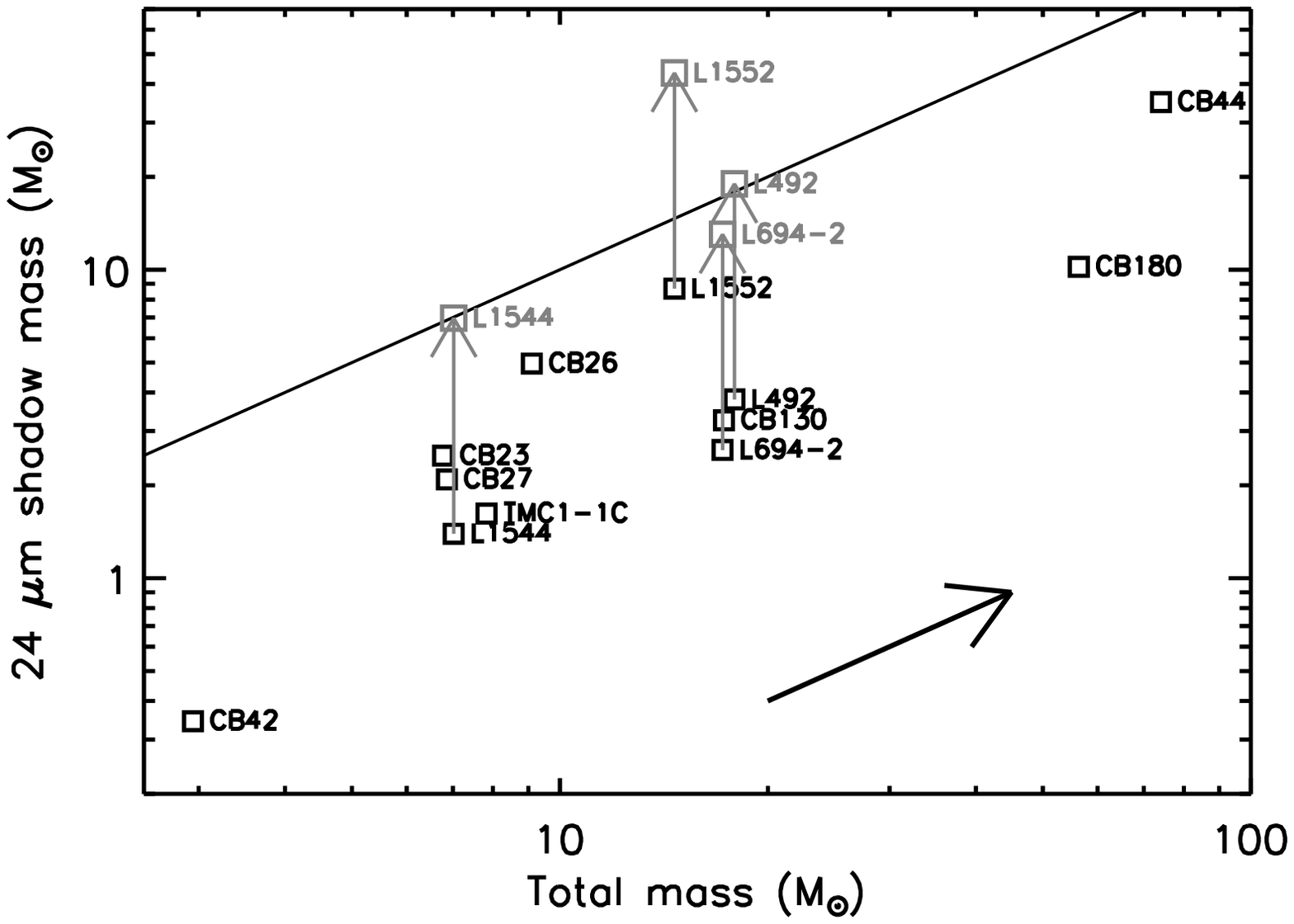}}
    \caption{Masses presented Table~4 are plotted here.  {\it Top
        panel --- } The CO core masses calculated over the area of the
      core are plotted versus the 24~\micron\ shadow masses.  {\it
        Bottom panel --- } The total masses (M$_{\rm CO,4}$ $+$
      M$_{24}^{\rm core}$ $-$ M$_{\rm CO}^{\rm core}$) are plotted
      versus the 24~\micron\ shadow masses.  The arrow indicates the
      effect of 50\% change in the assumed distance. In both panels
      the black curve indicates equality between the ordinate and the
      abscissa.}
    \label{fig:comass}
  \end{center}
\end{figure}

In the top panel of Figure~\ref{fig:comass} we compare the 24~\micron\
shadow masses (M$_{24}^{\rm core}$ in Table~4) to the CO--derived core
masses (M$_{\rm CO}^{\rm core}$ in Table~4).  In all cases, F${\rm
core} =$M$_{24}^{\rm core} / $M$_{\rm CO}^{\rm core} \geq 1.0$; this
discrepancy is likely due to the effect of freeze--out of CO in dense
regions \citep[e.g.,][]{bacmann02,tafalla02}; the effect will be
stronger that illustrated for cores with 70~\micron\ shadows (L1544,
L1552, L492, and L694-2, see above discussion) because the 24~\micron\
mass is likely to be underestimated.  In the lower panel of
Figure~\ref{fig:comass} we show the 24~\micron\ masses as a function
of the total molecular CO derived masses (integrated over a $4\arcmin$
radius).  Cores with 70~\micron\ shadows are marked with grey arrows
indicating a plausible mass correction of a factor of 5
\citep{stutz09}.  We tentatively detect a rough trend in core mass
with increasing envelope mass.  We caution that this trend may be
driven by distance uncertainties, small sample size, and incomplete
mass census.  Observations from upcoming facilities, e.g., SCUBA2,
which can image the sub--millimeter dust continuum emission, and {\it
Herschel}, which can image the 70~\micron\ shadows with a high degree
of spatial resolution ($\sim\!6\arcsec$) will provide further
constraints on the evolution of starless cores.

\section{Conclusions}

We study the 8~\micron, 24~\micron, and 70~\micron\ shadows cast by a
sample of 14 starless cores.  We derive 24~\micron\ core masses and
sizes; we apply a Jeans mass criterion, and attempt to account for
turbulent and magnetic support in the cores, in order to assess the
collapse state of each core.  We caution that distance uncertainties
can have a large effect on our Jeans mass analysis.  In addition, we
have obtained $^{12}$CO (2-1) and $^{13}$CO (2-1) OTF maps of the
cores; the molecular core masses we derive are always less than the
24~\micron\ masses.  This discrepancy is likely caused by freezeout
onto dust grains.  From this work we conclude that:

\noindent 1. 70\% (10/14) of cores selected to have prominent
24~\micron\ shadows seem to be approaching collapse, indicating that
this criterion selects dense, evolved cores.  

\noindent 2. 50\% (5/10) of cores that are classified as approaching
collapse have indications of 70~\micron\ shadows; the 70~\micron\ data
quality does not allow for a rigorous analysis of these shadows.  

\noindent 3. All cores with indications of 70~\micron\ shadows have
millimeter line profiles showing blue asymmetries, indicating that
these long--wavelength shadow features are produced by very evolved
cores.  These cores are all likely to be close to collapse.  

\noindent 4. Shadows at 24~\micron\ and especially at 70~\micron\
appear to be effective indicators of cores that are approaching
collapse.

\acknowledgments A.\ M.\ S.\ thanks Juna Kollmeier for her early input
and stimulating discussions throughout the course of this work.  The
authors thank Craig A.\ Kulesa for assistance with the molecular mass
models and Chris Walker for helpful comments.  We also thank the
anonymous referee for their helpful comments.  Support for this work
was provided through NASA contracts issued by Caltech/JPL to the
University of Arizona (1255094).  This work was also supported by the
National Science Foundation grant AST-0708131 to The University of
Arizona.  MK was supported by the KRF-2007-612-C00050 grant.

\clearpage 

\begin{deluxetable*}{llllrrcc}[b]
\tablewidth{0pt}
\tablecaption{{\it Spitzer} IRAC and MIPS data parameters}
\tablehead{
\colhead{Source} 
& \colhead{Other names}
& \colhead{R.A.} 
& \colhead{Decl.} 
& \colhead{$l$} 
& \colhead{$b$} 
& \colhead{MIPS AOR $^a$} 
& \colhead{IRAC AOR $^a$} \\
\colhead{} 
& \colhead{} 
& \colhead{$^{\rm h}$\phn$^{\rm m}$\phn$^{\rm s}$} 
& \colhead{$\degr$\phn$\arcmin$\phn$\arcsec$} 
& \colhead{$\degr$} 
& \colhead{$\degr$} 
& \colhead{} 
& \colhead{} \\
\colhead{(1)} 
& \colhead{(2)} 
& \colhead{(3)} 
& \colhead{(4)} 
& \colhead{(5)} 
& \colhead{(6)} 
& \colhead{(7)} 
& \colhead{(8)}
}
\startdata

TMC1-1C & B220 & 04 41 51.4 & $+$25 50 13 & 174.3 & $-$13.3 & 18155008 & 5085952  \\
CB23 & L1507 & 04 43 27.9 & $+$29 42 26 & 171.5 & $-$10.6 & 12026368 & 4914944  \\
CB26 & L1439 & 04 59 59.8 & $+$52 03 29 & 156.1 & $+$5.9 & 12020480 & 4916224  \\
CB27 & L1512 & 05 04 05.3 & $+$32 46 03 & 171.8 & $-$5.2 & 12021248 & 4916736  \\
L1544 & \nodata & 05 04 17.4 & $+$25 08 23 & 178.0 & $-$9.8 & 18155776 & 14610688  \\
L1552 & \nodata & 05 17 38.2 & $+$26 02 53 & 179.0 & $-$6.8 & 18156288 & 4917760  \\
CB42 & \nodata & 06 02 47.5 & $+$16 43 00 & 192.5 & $-$2.8 & 12024064 & \nodata  \\
CB44 & B227,L1570 & 06 07 31.0 & $+$19 31 00 & 190.6 & $-$0.4 & 12022272 & 4919296  \\
L492 & CB128 & 18 15 49.1 & $-$03 43 05 & 25.5 & $+$6.1 & 18159360 & 14603264  \\
CB130 & L507 & 18 16 14.3 & $-$02 30 16 & 26.7 & $+$6.6 & 18159616 & 5146368  \\
CB180 & B133,L531 & 19 06 15.1 & $-$06 55 25 & 28.4 & $-$6.4 & 12023808 & 4925696  \\
L694-2 & B143,CB200 & 19 41 07.4 & $+$10 58 59 & 48.4 & $-$5.8 & 18160128 & 14604288  \\

\enddata
\tablenotetext{a}{AOR = Astronomical Observation Request}
\tablecomments{Coordinates indicate the reference pixel coordinate for
  each listed observation.}
\end{deluxetable*}

\begin{deluxetable}{llllllccc}
\tablewidth{0pt}
\tablecaption{HHT OTF Mapping parameters}
\tablehead{
\colhead{Core} 
& \colhead{R.A.} 
& \colhead{Decl.} 
& \colhead{Obs. Date} 
& \colhead{Size} 
& \colhead{Line} 
& \colhead{T$_{sys}$ $^a$} 
& \colhead{RMS $^b$} 
& \colhead{Vel.$^c$} \\
\colhead{  } 
& \colhead{$^{\rm h}$\phn$^{\rm m}$\phn$^{\rm s}$} 
& \colhead{$\degr$\phn$\arcmin$\phn$\arcsec$} 
& \colhead{} 
& \colhead{$\arcmin$ $\times$ $\arcmin$} 
& \colhead{} 
& \colhead{K} 
& \colhead{K -- $T_A^*$} 
& \colhead{km s$^{-1}$} \\
\colhead{(1)} 
& \colhead{(2)} 
& \colhead{(3)} 
& \colhead{(4)} 
& \colhead{(5)} 
& \colhead{(6)} 
& \colhead{(7)} 
& \colhead{(8)} 
& \colhead{(9)} 
}
\startdata
TMC1     & 04 41 37.5 & $+$26 02 34 & 2009 Feb 01 & 10 $\times$ 10 & $^{12}$CO (2-1) & 203 & 0.15 & 2.0, 10.0  \\
  &   &   &   &   & $^{13}$CO (2-1) & 222 & 0.11 & 4.0, 7.0  \\
CB23     & 04 43 19.1 & $+$29 43 43 & 2008 Feb 07 & 13 $\times$ 13 & $^{12}$CO (2-1) & 241 & 0.22 & 4.0, 11.0  \\
  &   &   &   &   & $^{13}$CO (2-1) & 221 & 0.15 & 5.0, 8.0  \\
CB26     & 05 00 19.9 & $+$52 05 56 & 2009 Mar 02 & 14 $\times$ 12 & $^{12}$CO (2-1) & 263 & 0.17 & 1.7, 7.5  \\
  &   &   &   &   & $^{13}$CO (2-1) & 255 & 0.25 & 3.0, 7.0  \\
CB27B    & 05 04 10.5 & $+$32 43 00 & 2007 Jan 28 & 15 $\times$ 15 & $^{12}$CO (2-1) & 218 & 0.24 & 6.0, 8.3  \\
  &   &   &   &   & $^{13}$CO (2-1) & 231 & 0.25 & 6.0, 8.0  \\
L1544    & 05 04 16.9 & $+$25 11 50 & 2009 Feb 01 & 10 $\times$ 10 & $^{12}$CO (2-1) & 201 & 0.12 & 7.0, 10.0  \\
  &   &   &   &   & $^{13}$CO (2-1) & 220 & 0.10 & 6.0, 8.0  \\
L1552-SH & 05 17 38.6 & $+$26 05 29 & 2009 Feb 02 & 10 $\times$ 10 & $^{12}$CO (2-1) & 190 & 0.15 & 5.0, 10.0  \\
  &   &   &   &   & $^{13}$CO (2-1) & 207 & 0.13 & 7.0, 9.0  \\
CB42     & 06 02 34.2 & $+$16 57 04 & 2008 Apr 16 & 10 $\times$ 10 & $^{12}$CO (2-1) & 264 & 0.19 & -2.0, 9.0  \\
  &   &   &   &   & $^{13}$CO (2-1) & 249 & 0.18 & 1.0, 4.0  \\
CB44     & 06 07 29.2 & $+$19 27 33 & 2007 Jan 28 & 10 $\times$ 10 & $^{12}$CO (2-1) & 193 & 0.24 & -5.0, 3.0  \\
  &   &   &   &   & $^{13}$CO (2-1) & 207 & 0.26 & -2.0, 1.0  \\
L492     & 18 15 49.2 & $-$03 44 11 & 2009 Mar 01 & 10 $\times$ 7 & $^{12}$CO (2-1) & 299 & 0.24 & 0.0, 10.0  \\
  &   &   &   &   & $^{13}$CO (2-1) & 289 & 0.33 & 6.0, 9.0  \\
CB130-1  & 18 15 59.8 & $-$02 13 28 & 2009 Mar 01 & 12 $\times$ 12 & $^{12}$CO (2-1) & 261 & 0.19 & 1.0, 11.0  \\
  &   &   &   &   & $^{13}$CO (2-1) & 246 & 0.27 & 5.0, 9.0  \\
CB180    & 19 06 14.8 & $-$06 48 18 & 2008 Apr 18 & 10 $\times$ 7 & $^{12}$CO (2-1) & 209 & 0.15 & 9.0, 14.0  \\
  &   &   &   &   & $^{13}$CO (2-1) & 196 & 0.15 & 11.0, 13.0  \\
L694-2   & 19 41 05.3 & $+$10 54 54 & 2009 Mar 01 & 10 $\times$ 10 & $^{12}$CO (2-1) & 297 & 0.22 & 7.0, 11.0  \\
  &   &   &   &   & $^{13}$CO (2-1) & 295 & 0.18 & 8.5, 10.5  \\

\enddata
\tablenotetext{a}{Single side-band system temperature.}
\tablenotetext{b}{Final OTF map RMS per pixel per velocity channel.}
\tablenotetext{c}{Velocity range of CO emission.}
\tablecomments{The beam size (FWHM) for the HHT at the observed
  frequencies is $\sim\!32\arcsec$ FWHM; the velocity resolution for these
  observations is 0.34~km~s$^{-1}$.  Objects observed in 2008 and later
  were observed using the ALMA prototype mixer, see text for details.}
\end{deluxetable}

\begin{deluxetable}{lllccccc}
\tablewidth{0pt}
\tablecaption{Shadow parameters}
\tablehead{
\colhead{Source} 
& \colhead{R.A.} 
& \colhead{Decl.} 
& \colhead{Dist.$^a$} 
& \colhead{$f_{\rm DC,24}$$^b$} 
& \colhead{$f_{0,24}$} 
& \colhead{$f_{\rm DC,8}$$^c$} 
& \colhead{$f_{0,8}$} \\
\colhead{} 
& \colhead{$^{\rm h}$\phn$^{\rm m}$\phn$^{\rm s}$} 
& \colhead{$\degr$\phn$\arcmin$\phn$\arcsec$} 
& \colhead{pc} 
& \colhead{$\mu$Jy/arcsec$^2$} 
& \colhead{$\mu$Jy/arcsec$^2$} 
& \colhead{MJy/sr} 
& \colhead{MJy/sr} \\
\colhead{(1)} 
& \colhead{(2)} 
& \colhead{(3)} 
& \colhead{(4)} 
& \colhead{(5)} 
& \colhead{(6)} 
& \colhead{(7)} 
& \colhead{(8)}
}
\startdata

TMC1-1C & 04 41 39.1 & $+$26 00 19 & 140 & 1390.63 & 11.49 & 957.92 & 38.98  \\
CB23 & 04 43 29.5 & $+$29 39 11 & 140 & -4.51 & 3.15 & 1578.37 & 1.96  \\
CB26 & 05 00 12.6 & $+$52 06 02 & 140 & -5.60 & 2.56 & 1741.00 & 14.63  \\
CB27 & 05 04 08.6 & $+$32 43 23 & 140 & -7.67 & 5.82 & 903.63 & 3.88  \\
L1544 & 05 04 16.7 & $+$25 10 42 & 140 & 1555.05 & 12.36 & 4294.04 & 66.20  \\
L1552 & 05 17 38.8 & $+$26 05 05 & 140 & 1530.99 & 2.20 & 1456.65 & 17.89  \\
CB42 & 06 02 27.7 & $+$16 56 59 & 200 & -11.0 & 10.45 & \nodata & \nodata  \\
CB44 & 06 07 29.0 & $+$19 27 42 & 400 & -7.30 & 3.89 & 1089.62 & 28.00  \\
L492 & 18 15 48.6 & $-$03 45 44 & 225 & 1075.20 & 7.79 & 2402.96 & 26.16  \\
CB130 & 18 16 11.1 & $-$02 16 38 & 225 & 788.42 & 7.30 & 835.85 & 28.29  \\
CB180 & 19 06 08.4 & $-$06 52 56 & 400 & -5.78 & 4.60 & 865.37 & 30.85  \\
L694-2 & 19 41 04.1 & $+$10 57 10 & 230 & 627.72 & 6.22 & 1511.84 & 45.87  \\

\enddata

\tablenotetext{a}{A discussion of the adopted distances and associated
  references is included in \S~3.2.}

\tablenotetext{b}{$f_{\rm DC,24}$ indicated the 0.5 percentile pixel
  value in the 24~\micron\ image.}

\tablenotetext{c}{$f_{\rm DC,8}$ indicated the 0.1 percentile pixel
  value in the 8~\micron\ image.}

\tablecomments{The listed values of $f_{0,8}$ and $f_{0,24}$ are the
    $f_{\rm DC}$--subtracted quantities.  Coordinates indicate the
    approximate centers of the shadow features shown in
    Figures~\ref{fig:img1}--\ref{fig:img12}.  We also note that
    1~$\mu$Jy/arcsec$^2 = 23.5$~MJy/sr}
\end{deluxetable}

\begin{deluxetable}{lrrrccrcc}
\tablecaption{Core masses and sizes}
\tablewidth{0pt}
\tablehead{
\colhead{Source} 
& \colhead{M$_{24}^{\rm core}$} 
& \colhead{M$_{\rm CO}^{\rm core}$ $^{a}$} 
& \colhead{M$_{\rm CO,4}$ $^{b}$} 
& \colhead{R$_{\rm eff}^{\rm core}$ $^{c}$} 
& \colhead{R$_{\rm eff}^{\rm core}$ $^{c}$} 
& \colhead{F$^{ \rm core}$ $^{d}$}
& \colhead{M$_{\rm j}^{\rm core}$ $^{e}$}\\
\colhead{}
& \colhead{\msun}
& \colhead{\msun}
& \colhead{\msun}
& \colhead{arcmin}
& \colhead{pc}
& \colhead{}
& \colhead{\msun}\\
\colhead{(1)} 
& \colhead{(2)} 
& \colhead{(3)} 
& \colhead{(4)} 
& \colhead{(5)} 
& \colhead{(6)} 
& \colhead{(7)} 
& \colhead{(8)} 
}
\startdata

TMC1-1C & 1.6 & 0.9 & 7.1 & 1.4 & 0.06 & 1.8 & 2.7  \\
CB23 & 2.5 & 0.7 & 4.9 & 1.3 & 0.05 & 3.8 & 2.0  \\
CB26 & 5.0 & 0.8 & 5.0 & 1.4 & 0.06 & 6.0 & 1.6  \\
CB27 & 2.1 & 0.8 & 5.6 & 1.4 & 0.06 & 2.5 & 2.4  \\
L1544$^{g}$ & 1.4 & 0.8 & 6.4 & 1.4 & 0.06 & 1.7 & 2.8  \\
L1552$^{g}$ & 8.7 & 0.8 & 6.8 & 1.3 & 0.05 & 10.6 & 1.1  \\
CB42 & 0.3 & 0.3 & 2.9 & 1.2 & 0.04 & 1.0 & 3.0  \\
CB44 & 35.0 & 7.0 & 46.2 & 1.4 & 0.16 & 5.0 & 2.8  \\
L492$^{g}$ & 3.8 & 2.0 & 16.1 & 1.3 & 0.09 & 1.9 & 3.3  \\
CB130 & 3.2 & 1.5 & 15.6 & 1.2 & 0.08 & 2.1 & 3.1  \\
CB180 & 10.2 & 5.3 & 51.5 & 1.2 & 0.13 & 1.9 & 3.9  \\
L694-2$^{g}$ & 2.6 & 1.7 & 16.3 & 1.2 & 0.08 & 1.6 & 3.6  \\

\enddata

\tablenotetext{a}{CO mass integrated over R$_{\rm eff}^{\rm core}$,
  centered on the shadow coordinates.}
\tablenotetext{b}{CO mass integrated over an $4\arcmin$--radius
  aperture, centered on the shadow coordinates.}
\tablenotetext{c}{Effective radius of the 24~\micron\ shadow, see
  \S~4.2.1.}
\tablenotetext{d}{Ratio of M$_{24}^{\rm core}$ over M$_{\rm CO}^{\rm core}$.}
\tablenotetext{e}{Jeans mass assuming all cores are at 10~K, using
  the M$_{24}^{\rm core}$ and R$_{\rm eff}^{\rm core}$ to calculate the
  mean density in each core.}
\tablenotetext{f}{Rotational velocity required to support a core mass of
  M$_{24}^{\rm core}$ for a core size of R$_{\rm eff}^{\rm core}$}
\tablenotetext{g}{Although the quality of the 70~\micron\ data does
  not allow for robust measurements, these cores show strong
  indications of having 70~\micron\ shadows.  The presence of a
  70~\micron\ shadow implies that the cores are optically thick at
  shorter wavelengths; hence M$_{24}^{\rm core}$ should be regarded as
  a lower limit.}

\end{deluxetable}

\begin{deluxetable}{lccccccc}
\tablecaption{Stability analysis summary}
\tablewidth{0pt}
\tablehead{
\colhead{Source} 
& \colhead{Near Collapse} 
& \colhead{M$_{24}^{core}$/M$_j^{core}$} 
& \colhead{70~\micron~Shadow} 
& \colhead{B$^{a}$} 
& \colhead{E$_{turb}$/E$_{th}$$^{b}$} 
& \colhead{V$_{rot}$$^{c}$}
& \colhead{Chem. Evol. } \\
\colhead{}
& \colhead{}
& \colhead{}
& \colhead{}
& \colhead{$\mu$G}
& \colhead{}
& \colhead{km s$^{-1}$}
& \colhead{}\\
\colhead{(1)} 
& \colhead{(2)} 
& \colhead{(3)} 
& \colhead{(4)} 
& \colhead{(5)} 
& \colhead{(6)} 
& \colhead{(7)} 
& \colhead{(8)} 
}
\startdata

TMC1-1C & no & $\phn$0.6 & no  & $\phn$$\phn$$\phn$0  & $\phn$0$\phn$  & $\phn$0.4 & no  \\
CB23 & yes   & $\phn$1.3 & no  & $\phn$123            & $\phn$0.3      & $\phn$0.6 & no  \\
CB26 & yes   & $\phn$3.2 & no  & $\phn$214            & $\phn$2$\phn$  & $\phn$0.5 & \nodata \\
CB27 & no    & $\phn$0.9 & no  & $\phn$$\phn$$\phn$0  & $\phn$0$\phn$  & $\phn$0.6 & \nodata \\
L1544 & yes  & $>$0.5    & yes & $\phn$$>$66          & $\phn$0.4      & $>$0.4    & yes  \\
L1552 & yes  & $>$8.1    & yes & $>$424               & $\phn$0.2      & $>$0.2    & \nodata \\
CB42 & no    & $\phn$0.1 & no  & $\phn$$\phn$$\phn$0  & $\phn$0$\phn$  & $\phn$1.0 & \nodata \\
CB44 & yes   & $\phn$12. & no  & $\phn$188            & 11$\phn$       & $\phn$0.3 & \nodata \\
L492 & yes   & $>$1.1    & yes & $\phn$$>$71          & $\phn$0.3      & $>$0.8    & yes  \\
CB130 & no   & $\phn$1.0 & no  & $\phn$$\phn$$\phn$0  & \nodata        & $\phn$0.4 & \nodata \\
CB180 & yes  & $\phn$2.6 & no  & $\phn$$\phn$81       & $\phn$4        & $\phn$0.4 & \nodata \\
L694-2 & yes & $>$0.7    & yes & $\phn$$>$57          & $\phn$0.4      & $>$0.4    & yes \\
CB190 & yes  & $\phn$2.5 & no  & $\phn$$\phn$78       & $>$0.5         & $\phn$0.6 & \nodata \\
L429 & yes   & $>$1.0    & yes & $>$194               & $\phn$1$\phn$  & $>$0.4    & yes \\
\enddata

\tablenotetext{a}{Magnetic field strength required to halt collapse.}
\tablenotetext{b}{Calculated from line widths collected from the literature, see \S~4.3 for references.}
\tablenotetext{c}{Rotational velocity required for stability.}
\tablecomments{See \SS~4.3 for full discussion.}

\end{deluxetable}

\clearpage 

\clearpage

\clearpage

\clearpage

\clearpage

\clearpage

\clearpage

\clearpage

\clearpage

\clearpage

\clearpage

\clearpage

\clearpage

\clearpage

\clearpage

\clearpage

\clearpage

\clearpage

\clearpage

\clearpage

\clearpage

\clearpage

\clearpage

\clearpage

\clearpage

\clearpage

\clearpage

\clearpage

\begin{thebibliography}{}

\bibitem[Alves et al.(2001)]{alves01} Alves, J.~F., Lada, 
C.~J., \& Lada, E.~A.\ 2001, \nat, 409, 159 

\bibitem[Alves et al.(2008)]{alves08} Alves, F.~O.,
Franco, G.~A.~P., \& Girart, J.~M.\ 2008, \aap, 486, L13

\bibitem[Arquilla \& Goldsmith(1986)]{arquilla86} Arquilla,
R., \& Goldsmith, P.~F.\ 1986, \apj, 303, 356

\bibitem[Bacmann et al.(2000)]{bacmann00} Bacmann, A., Andr{\'e}, P.,
Puget, J.-L., Abergel, A., Bontemps, S., \& Ward-Thompson, D.\ 2000,
\aap, 361, 555

\bibitem[Bacmann et al.(2002)]{bacmann02} Bacmann, A., Lefloch, B.,
Ceccarelli, C., Castets, A., Steinacker, J., \& Loinard, L.\ 2002,
\aap, 389, L6

\bibitem[Bacmann et al.(2003)]{bacmann03} Bacmann, A., Lefloch, 
B., Ceccarelli, C., Steinacker, J., Castets, A., 
\& Loinard, L.\ 2003, \apjl, 585, L55 

\bibitem[Ballesteros-Paredes et al.(2003)]{ballesteros03}
  Ballesteros-Paredes, J., Klessen, R.~S., \& V{\'a}zquez-Semadeni,
  E.\ 2003, \apj, 592, 188

\bibitem[Bergin \& Tafalla(2007)]{bergin07} Bergin, E.~A.,
  \& Tafalla, M.\ 2007, \araa, 45, 339

\bibitem[Bok \& McCarthy(1974)]{bok74} Bok, B.~J., \&
McCarthy, C.~C.\ 1974, \aj, 79, 42

\bibitem[Burkert \& Bodenheimer(2000)]{burkert00} Burkert, A., \&
  Bodenheimer, P.\ 2000, \apj, 543, 822

\bibitem[Burkert \& Hartmann(2004)]{burkert04} Burkert, A.,
  \& Hartmann, L.\ 2004, \apj, 616, 288

\bibitem[Butler \& Tan(2009)]{butler09} Butler, M.~J., \&
Tan, J.~C.\ 2009, \apj, 696, 484

\bibitem[Caselli et al.(1999)]{caselli99} Caselli, P., Walmsley,
C.~M., Tafalla, M., Dore, L., \& Myers, P.~C.\ 1999, \apjl, 523, L165

\bibitem[Caselli et al.(2002)]{caselli02} Caselli, P.,
Benson, P.~J., Myers, P.~C., \& Tafalla, M.\ 2002, \apj, 572, 238

\bibitem[Caselli et al.(2008)]{caselli08} Caselli, P.,
Vastel, C., Ceccarelli, C., van der Tak, F.~F.~S., Crapsi, A., \&
Bacmann, A.\ 2008, \aap, 492, 703

\bibitem[Chapman et al.(2009)]{chapman09} Chapman, N.~L., Mundy, 
L.~G., Lai, S.-P., \& Evans, N.~J.\ 2009, \apj, 690, 496 

\bibitem[Chapman \& Mundy(2009)]{chapman09b} Chapman, N.~L.,
\& Mundy, L.~G.\ 2009, arXiv:0905.0655

\bibitem[Clemens \& Barvainis(1988)]{clemens88} Clemens,
D.~P., \& Barvainis, R.\ 1988, \apjs, 68, 257

\bibitem[Crapsi et al.(2005)]{crapsi05} Crapsi, A., Caselli, P.,
Walmsley, C.~M., Myers, P.~C., Tafalla, M., Lee, C.~W., \& Bourke,
T.~L.\ 2005, \apj, 619, 379

\bibitem[de Wit et al.(2005)]{dewit05} de Wit, W.~J.,
Testi, L., Palla, F., \& Zinnecker, H.\ 2005, \aap, 437, 247

\bibitem[Draine(2003)]{draine03} Draine, B.~T.\ 2003, \araa, 41, 241

\bibitem[Evans et al.(2001)]{evans01} Evans, N.~J., Rawlings,
J.~M.~C., Shirley, Y.~L., \& Mundy, L.~G.\ 2001, \apj, 557, 193

\bibitem[Fallscheer et al.(2009)]{fallscheer09} Fallscheer, C.,
Beuther, H., Zhang, Q., Keto, E., \& Sridharan, T.~K.\ 2009,
arXiv:0907.2232

\bibitem[Flaherty et al.(2007)]{flaherty07} Flaherty, K.~M., 
Pipher, J.~L., Megeath, S.~T., Winston, E.~M., Gutermuth, R.~A., Muzerolle, 
J., Allen, L.~E., \& Fazio, G.~G.\ 2007, \apj, 663, 1069 

\bibitem[Foster \& Goodman(2006)]{foster06} Foster, J.~B.,
  \& Goodman, A.~A.\ 2006, \apjl, 636, L105

\bibitem[Gregersen et al.(1997)]{gregersen97} Gregersen, E.~M., 
Evans, N.~J., II, Zhou, S., \& Choi, M.\ 1997, \apj, 484, 256 

\bibitem[Gregersen \& Evans(2000)]{gregersen00} Gregersen, E.~M., \&
Evans, N.~J., II 2000, \apj, 538, 260

\bibitem[Gordon et al.(2005)]{gordon05} Gordon, K.~D., et al.\ 
2005, \pasp, 117, 503 

\bibitem[Gupta et al.(2009)]{gupta09} Gupta, H., Gottlieb, 
C.~A., McCarthy, M.~C., \& Thaddeus, P.\ 2009, \apj, 691, 1494 

\bibitem[Gutermuth et al.(2008)]{guter08} Gutermuth, R.~A., et 
al.\ 2008, \apj, 674, 336 

\bibitem[Harvey et al.(2003a)]{harvey03a} Harvey, D.~W.~A.,
Wilner, D.~J., Lada, C.~J., Myers, P.~C., \& Alves, J.~F.\ 2003a, \apj,
598, 1112

\bibitem[Harvey et al.(2003b)]{harvey03b} Harvey, D.~W.~A., Wilner,
D.~J., Myers, P.~C., \& Tafalla, M.\ 2003b, \apj, 597, 424

\bibitem[Hayashi(1966)]{hayashi66} Hayashi, C.\ 1966, \araa,
4, 171

\bibitem[Heitsch et al.(2008)]{heitsch08} Heitsch, F., Hartmann,
  L.~W., Slyz, A.~D., Devriendt, J.~E.~G., \& Burkert, A.\ 2008, \apj,
  674, 316

\bibitem[Henning et al.(2001)]{henning01} Henning, T., Wolf,
S., Launhardt, R., \& Waters, R.\ 2001, \apj, 561, 871

\bibitem[Hilton \& Lahulla(1995)]{hilton95} Hilton, J., \& Lahulla,
  J.~F.\ 1995, \aaps, 113, 325

\bibitem[Hotzel et al.(2002)]{hotzel02} Hotzel, S., Harju,
J., \& Juvela, M.\ 2002, \aap, 395, L5

\bibitem[Juvela et al.(2008)]{juvela08} Juvela, M., Pelkonen, V.-M.,
  Padoan, P., \& Mattila, K.\ 2008, \aap, 480, 445

\bibitem[Kandori et al.(2005)]{kandori05} Kandori, R., et al.\ 2005,
\aj, 130, 2166

\bibitem[Kang et al.(2009)]{kang09} Kang, M., Bieging, J.~H., Kulesa,
  C.~A., \& Lee, Y.\ 2009, \apj, 701, 454

\bibitem[Kauffmann et al.(2008)]{kauffmann08} Kauffmann, J., Bertoldi,
F., Bourke, T.~L., Evans, N.~J., II, \& Lee, C.~W.\ 2008, \aap, 487,
993

\bibitem[Kawamura et al.(2001)]{kawamura01} Kawamura, A., Kun, M.,
Onishi, T., Vavrek, R., Domsa, I., Mizuno, A., \& Fukui, Y.\ 2001,
\pasj, 53, 1097

\bibitem[Kenyon et al.(1994)]{kenyon94} Kenyon, S.~J., Dobrzycka, D.,
\& Hartmann, L.\ 1994, \aj, 108, 1872

\bibitem[Kulesa et al.(2005)]{kulesa05} Kulesa, C.~A.,
Hungerford, A.~L., Walker, C.~K., Zhang, X., \& Lane, A.~P.\ 2005,
\apj, 625, 194

\bibitem[Kutner \& Ulich(1981)]{kutner81} Kutner, M.~L., \& Ulich,
B.~L.\ 1981, \apj, 250, 341

\bibitem[Lada \& Lada(2003)]{lada03} Lada, C.~J., \& Lada, E.~A.\
2003, \araa, 41, 57

\bibitem[Lada et al.(2008)]{lada08} Lada, C.~J., Muench, A.~A.,
Rathborne, J., Alves, J.~F., \& Lombardi, M.\ 2008, \apj, 672, 410

\bibitem[Lallement et al.(2003)]{lallement03} Lallement, R., Welsh,
B.~Y., Vergely, J.~L., Crifo, F., \& Sfeir, D.\ 2003, \aap, 411, 447

\bibitem[Launhardt et al.(1998)]{launhardt98} Launhardt, R., Evans,
N.~J., II, Wang, Y., Clemens, D.~P., Henning, T., \& Yun, J.~L.\ 1998,
\apjs, 119, 59

\bibitem[Launhardt et al.(2009)]{launhardt09} Launhardt, R.,
et al.\ 2009, \aap, 494, 147

\bibitem[Lee et al.(1999)]{lee99} Lee, C.~W., Myers, P.~C., 
\& Tafalla, M.\ 1999, \apj, 526, 788 

\bibitem[Lee et al.(2003)]{lee03} Lee, J.-E., Evans, N.~J., II,
Shirley, Y.~L., \& Tatematsu, K.\ 2003, \apj, 583, 789

\bibitem[Lee et al.(2004)]{lee04} Lee, C.~W., Myers, P.~C., 
\& Plume, R.\ 2004, \apjs, 153, 523 

\bibitem[Lee et al.(2007)]{lee07} Lee, S.~H., Park,
Y.-S., Sohn, J., Lee, C.~W., \& Lee, H.~M.\ 2007, \apj, 660, 1326

\bibitem[Lee et al.(2009)]{lee09} Lee, M.-Y., et al.\ 2009, 
arXiv:0908.2275 

\bibitem[Lehtinen \& Mattila(1996)]{lehtinen96} Lehtinen, K.,
  \& Mattila, K.\ 1996, \aap, 309, 570

\bibitem[Lemme et al.(1996)]{lemme96} Lemme, C., Wilson, T.~L.,
Tieftrunk, A.~R., \& Henkel, C.\ 1996, \aap, 312, 585

\bibitem[Mardones et al.(1997)]{mardones97} Mardones, D., Myers, 
P.~C., Tafalla, M., Wilner, D.~J., Bachiller, R., 
\& Garay, G.\ 1997, \apj, 489, 719 

\bibitem[Menten et al.(1987)]{menten87} Menten, K.~M., Serabyn, E.,
Guesten, R., \& Wilson, T.~L.\ 1987, \aap, 177, L57

\bibitem[Mundy et al.(1990)]{mundy90} Mundy, L.~G., Wootten, H.~A., \&
Wilking, B.~A.\ 1990, \apj, 352, 159

\bibitem[Myers(2005)]{myers05} Myers, P.~C.\ 2005, \apj, 623, 280

\bibitem[Nakajima et al.(2008)]{nakajima08} Nakajima, Y., Kandori, 
R., Tamura, M., Nagata, T., Sato, S., \& Sugitani, K.\ 2008, \pasj, 60, 731 

\bibitem[Narayanan et al.(1998)]{narayanan98} Narayanan, G.,
Walker, C.~K., \& Buckley, H.~D.\ 1998, \apj, 496, 292

\bibitem[Ossenkopf \& Henning(1994)]{ossenkopf94} Ossenkopf,
V., \& Henning, T.\ 1994, \aap, 291, 943

\bibitem[Padoan et al.(2006)]{padoan06} Padoan, P., Juvela, M., 
\& Pelkonen, V.-M.\ 2006, \apjl, 636, L101 

\bibitem[Park et al.(2004)]{park04} Park, Y.-S., Lee, C.~W., 
\& Myers, P.~C.\ 2004, \apjs, 152, 81 

\bibitem[Perault et al.(1996)]{perault96} Perault, M., et
al.\ 1996, \aap, 315, L165

\bibitem[Peretto \& Fuller(2009)]{peretto09} Peretto, N., \&
Fuller, G.~A.\ 2009, arXiv:0906.3493

\bibitem[Povich et al.(2009)]{povich09} Povich, M.~S., et al.\ 2009,
\apj, 696, 1278

\bibitem[Roberts \& Millar(2007)]{roberts07} Roberts, H., \&
Millar, T.~J.\ 2007, \aap, 471, 849

\bibitem[Sault et al.(1995)]{sault95} Sault R.J., Teuben P.J., Wright
M.C.H., 1995, in Astronomical Data Analysis Software and Systems IV,
ed. R. Shaw, H.E. Payne, J.J.E. Hayes, ASP Conf. Ser., 77, 433-436

\bibitem[Sauter et al.(2009)]{sauter09} Sauter, J., et al.\ 
2009, arXiv:0907.1074 

\bibitem[Shirley et al.(2000)]{shirley00} Shirley, Y.~L., Evans, 
N.~J., II, Rawlings, J.~M.~C., \& Gregersen, E.~M.\ 2000, \apjs, 131, 249 

\bibitem[Shirley et al.(2005)]{shirley05} Shirley, Y.~L., Nordhaus,
M.~K., Grcevich, J.~M., Evans, N.~J., Rawlings, J.~M.~C., \&
Tatematsu, K.\ 2005, \apj, 632

\bibitem[Shu et al.(1987)]{shu87} Shu, F.~H., Adams,
F.~C., \& Lizano, S.\ 1987, \araa, 25, 23

\bibitem[Sohn et al.(2007)]{sohn07} Sohn, J., Lee, C.~W., Park, Y.-S.,
Lee, H.~M., Myers, P.~C., \& Lee, Y.\ 2007, \apj, 664, 928

\bibitem[Stahler \& Palla(2005)]{stahler05} Stahler, S.~W.,
\& Palla, F.\ 2005, The Formation of Stars, by Steven W.~Stahler,
Francesco Palla, pp.~865.~ISBN 3-527-40559-3.~Wiley-VCH , January
2005.

\bibitem[Strai{\v z}ys et al.(2003)]{staizys03} Strai{\v
z}ys, V., {\v C}ernis, K., \& Barta{\v s}i{\= u}t{\.e}, S.\ 2003,
\aap, 405, 585

\bibitem[Stecklum et al.(2004)]{stecklum04} Stecklum, B., 
Launhardt, R., Fischer, O., Henden, A., Leinert, C., 
\& Meusinger, H.\ 2004, \apj, 617, 418 

\bibitem[Stutz et al.(2007)]{stutz07} Stutz, A.~M., et al.\ 
2007, \apj, 665, 466 

\bibitem[Stutz et al.(2008a)]{stutz08a} Stutz, A.~M., Papovich, 
C., \& Eisenstein, D.~J.\ 2008a, \apj, 677, 828 

\bibitem[Stutz et al.(2008b)]{stutz08b} Stutz, A.~M., et al.\ 
2008b, \apj, 687, 389 

\bibitem[Stutz et al.(2009)]{stutz09} Stutz, A.~M., Bourke, 
T.~L., Rieke, G.~H., Bieging, J.~H., Misselt, K.~A., Myers, P.~C., 
\& Shirley, Y.~L.\ 2009, \apjl, 690, L35 

\bibitem[Tafalla et al.(2002)]{tafalla02} Tafalla, M.,
Myers, P.~C., Caselli, P., Walmsley, C.~M., \& Comito, C.\ 2002, \apj,
569, 815

\bibitem[Tomita et al.(1979)]{tomita79} Tomita, Y., Saito, T., 
\& Ohtani, H.\ 1979, \pasj, 31, 407 

\bibitem[Torres et al.(2009)]{torres09} Torres, R.~M., Loinard, 
L., Mioduszewski, A.~J., \& Rodriguez, L.~F.\ 2009, arXiv:0903.5338 

\bibitem[Turner et al.(1997)]{turner97} Turner, B.~E., Pirogov, L., \&
Minh, Y.~C.\ 1997, \apj, 483, 235

\bibitem[van Dishoeck \& Black(1988)]{vandishoek88} van
Dishoeck, E.~F., \& Black, J.~H.\ 1988, \apj, 334, 771

\bibitem[Walker et al.(1986)]{walker86} Walker, C.~K., Lada, C.~J.,
Young, E.~T., Maloney, P.~R., \& Wilking, B.~A.\ 1986, \apjl, 309, L47

\bibitem[Walker et al.(1988)]{walker88} Walker, C.~K., Lada, C.~J.,
Young, E.~T., \& Margulis, M.\ 1988, \apj, 332, 335

\bibitem[Williams et al.(2006)]{williams06} Williams, J.~P.,
Lee, C.~W., \& Myers, P.~C.\ 2006, \apj, 636, 952

\end{thebibliography}
\end{document}